\documentclass[a4paper, oneside, 12pt, openany]{book}
\pdfoutput=1

\usepackage{packages}
\usepackage{macros}
\usepackage[scale=0.94]{inconsolata}

\numberwithin{equation}{section}
\numberwithin{table}{section}
\makeatletter
\let\c@equation\c@table
\makeatother

\newbool{shade-envs}
\setboolean{shade-envs}{true}

\newtheoremstyle{funny}
  {}
  {}
  {}
  {}
  {\bfseries}
  {.}
  { }
  {\thmname{#1} 
   \thmnumber{#2} \AgdaLogo 
   {\mdseries \thmnote{#3}}
  }

\ifthenelse{\boolean{shade-envs}}{%
  \colorlet{ShadeOfPurple}{blue!5!white}
  \colorlet{ShadeOfYellow}{yellow!5!white}
  \colorlet{ShadeOfGreen} {green!5!white}
  \colorlet{ShadeOfBrown} {brown!10!white}
  \colorlet{ShadeOfGray}  {gray!10!white}
}
{%
  \declaretheoremstyle[
      spaceabove=6pt,
      spacebelow=6pt,
      bodyfont=\normalfont,
      qed=\(\lozenge\)
  ]{definitionwithbox}
  \declaretheoremstyle[
      headfont=\itshape,
      bodyfont=\normalfont,
      qed=\(\lozenge\)
      ]{remarkwithbox}
}

\ifthenelse{\boolean{shade-envs}}{%
  \declaretheorem[sibling=equation]{theorem}
  \declaretheorem[sibling=theorem]{lemma,proposition,corollary}
  \declaretheorem[sibling=theorem,style=definition]{definition}
  \declaretheorem[sibling=theorem,style=definition]{example}
  \declaretheorem[sibling=theorem,style=remark]{remark}
  \theoremstyle{funny}
  
  \theoremstyle{funny}
  
  \theoremstyle{funny}
  
  \newlength{\normalparindent}
  \AtBeginDocument{\setlength{\normalparindent}{\parindent}}
  \tcbset{shadedenv/.style={
      colback={#1},
      frame hidden,
      enhanced,
      breakable,
      boxsep=0pt,
      left=2mm,
      right=2mm,
      add to width=1.1mm,
      enlarge left by=-0.6mm,
      before upper={\setlength{\parindent}{\normalparindent}}}
  }
  \newcommand{\setenvcolor}[2]{%
      \tcolorboxenvironment{#1}{shadedenv={#2}}
      \addtotheorempreheadhook[#1]{\tikzcdset{background color=#2}}
  }
  \setenvcolor{theorem}{ShadeOfPurple}
  \setenvcolor{lemma}{ShadeOfPurple}
  \setenvcolor{lemmanp}{ShadeOfPurple}
  \setenvcolor{proposition}{ShadeOfPurple}
  \setenvcolor{corollary}{ShadeOfPurple}
  \setenvcolor{definition}{ShadeOfYellow}
  \setenvcolor{definitionnp}{ShadeOfYellow}
  \setenvcolor{example}{ShadeOfGreen}
  \setenvcolor{examplenp}{ShadeOfGreen}
  \setenvcolor{remark}{ShadeOfBrown}
  \setenvcolor{proof}{ShadeOfGray}
}{
  \declaretheorem[sibling=equation]{theorem}
  \declaretheorem[sibling=theorem]{lemma,proposition,corollary}
  \declaretheorem[sibling=theorem,style=definitionwithbox]{definition}
  \declaretheorem[sibling=theorem,style=definitionwithbox]{example}
  \declaretheorem[sibling=theorem,style=remarkwithbox]{remark}
  }

\begin{document}

\frontmatter

\newgeometry{total={180mm,267mm}} 
\begin{titlepage}
\begin{center}
  \vspace*{\stretch{0.5}}

  \large 

  {\Huge\textsc{Constructive and Predicative\\ Locale Theory\\ in Univalent Foundations}\par}

  \vspace{\stretch{0.2}}

  by

  \vspace{\stretch{0.2}}

  {\huge\textsc{Ayberk Tosun}}

  \vspace{\stretch{0.5}}

  A thesis submitted to the University of Birmingham for the degree of\\
  \textsc{Doctor of Philosophy}

  \vfill

  \flushright{
    School of Computer Science\\
    College of Engineering and Physical Sciences\\
    University of Birmingham\\\phantom{force-newline}\\

    \begin{tabular}{rr}
      Submitted:    &    30 November 2024\\
      Defended:     &    27 February 2025\\
      Accepted:     &    3  June     2025\\
    \end{tabular}%
  }
\end{center}
\end{titlepage}

\restoregeometry%

\chapter{Abstract}

We develop locale theory constructively and predicatively in \VUF{}
(also known as homotopy type theory or \VHoTTUF{}),
with a particular focus on the theory of spectral and Stone locales.
The \emph{constructivity} of our foundational setting means that we do not use
any classical principles such as the axiom of choice (or any of its weaker
forms), the law of excluded middle, or any form of the limited principle of
omniscience.
In the context of \VUF{}, \emph{predicativity} refers specifically to the
development of mathematics without the use of Voevodsky's propositional resizing
axioms.
The traditional approach to the predicative development of point-free topology
is to work with presentations of locales known as formal topologies.
Here, we take a different approach:
we work directly with frames and locales,
keeping careful track of the universes involved and
adopting certain size assumptions to ensure that the theory is amenable to
predicative development.
Although it initially appears that many fundamental constructions of locale
theory rely on impredicativity,
we show that these can be circumvented under rather natural size assumptions.
Our development here is inspired by \VTDJMHE{}'s constructive and predicative
development of domain theory in \VUF{}.

We first lay the groundwork for the predicative development of locale theory.
We then orient this towards an investigation of the theory of spectral and Stone
locales, using \VTheUniAx{} and \VSRPrinciple{} to ensure a predicatively
well-behaved notion of spectral locale.
We also develop Stone duality in the context of spectral locales, showing that
there is a categorical equivalence between the type of large, locally small, and
small-complete spectral locales and the type of small distributive lattices.
Moreover, we exhibit the category of Stone locales as a coreflective subcategory
of the category of spectral locales and spectral maps, using the construction
known as the \emph{patch locale} of a spectral locale (the localic manifestation
of the so-called constructible topology).
Finally, we investigate the topology of algebraic \VDCPO{}s and Scott domains in
this constructive and predicative framework for locale theory.
We develop the Scott locale of a Scott domain, show that it forms a spectral
locale, and then proceed to investigate its patch.
Using this, we obtain a topological characterization of de Jong's notion of
sharp element: we establish a correspondence between the sharp elements of a
Scott domain and the points of the patch of its Scott locale.

Our development is completely formalized and has been machine-checked using the
\VAgda{} proof assistant.


\chapter{Acknowledgements}\label{chap:acknowledgements}
\markboth{Acknowledgements}{Acknowledgements}

First and foremost, I would like to express my heartfelt thanks to my
supervisor \VMHEFullName{} for his support and guidance.
\VMHEFirstName{} not only proposed the main ideas underlying my PhD project, but
he has also patiently helped me navigate my way around the rocky (yet beautiful)
landscape of constructive mathematics.
My PhD started at the height of the \textsc{COVID}-19 pandemic in July 2020, and
has sometimes been a rather turbulent journey as a result.
I am particularly grateful to \VMHEFirstName{} for supporting me during these
challenging times.

I would also like to thank Eric Finster and Bas Spitters for serving as my
thesis examiners and for their careful reading of this thesis. Their feedback
has been invaluable during the preparation of the corrected version. I am also
grateful to Paul B.\ Levy for serving as my viva chair.

During my PhD, I have been fortunate to be part of the Theory Group at
Birmingham,
whose engaging research environment has significantly shaped my development as a
researcher.
Over the years, I have had the privilege of working alongside many amazing
friends among the group's PhD students:
Todd Ambridge,
George Kaye,
Avgerinos Delkos,
Paaras Padhiar,
Bruno da Rocha Paiva.
Equally important has been the support and mentorship of other group members,
especially Anupam Das, Paul B.\ Levy, Sonia Marin, Eric Finster, Abhishek De,
Paul Taylor, and Steve Vickers.
I would like to thank Paul and Steve in particular for serving in my thesis
group.

Among the members of the Theory Group, three people deserve special mention: my
colleagues Igor \IALastName{} and Tom \VTDJLastName{}, and my second supervisor
Vincent Rahli.
Their support, friendship, and scientific guidance have been instrumental to the
completion of this thesis.

I owe a huge debt of gratitude to Vincent for uncountably many things: his
mentorship, unwavering support, and for many supervision meetings where he
patiently listened to my excited monologues about whatever subject happened to
have my interest at the time. I consider myself genuinely lucky to have been his
student.
Tom's doctoral investigation on the predicative development of domain theory in
\VUF{} inspired the work presented in this thesis.
Furthermore, I have significantly benefited from Tom's pellucid expositions of
his research,
most notably his meticulously crafted \VAgda{} formalizations in \VTypeTopology{}.
At a more personal level, I would like to thank Tom for the support he has given
me over the years, and for serving the r\^{o}le of a big brother to me.
Finally, I would like to record my indebtedness to my friend Igor, not only for
always enthusiastically sharing his expertise on locale theory with me, but also
for his contributions
to the predicative proof of the universal property of the patch locale (the
subject of \cref{sec:ump-of-patch}).

Eric and Tom have taken the initiative to organize the \textsc{ASSUME} seminar
series, which has brought together the Theory Group at Birmingham and the
Functional Programming Lab at Nottingham.
I have greatly benefited from these meetings and had many insightful
conversations with various \textsc{ASSUME} participants, most notably
Josh Chen and Jacob Neumann from the University of Nottingham.
I thank Eric and Tom for their efforts in organizing \textsc{ASSUME}.

As a PhD student at the School of Computer Science, I have been lucky to have
had the support of Felipe Orihuela\nobreakdash-Espina, who has done an excellent
job as the Research Students Tutor.
Furthermore, Jamie Hough from Research Support deserves special recognition for
his diligence and commitment his r\^{o}le.

Throughout my undergraduate years, my teachers Mattox Beckman, Jos\'{e}
Meseguer, and Grigore Ro\c{s}u nurtured my curiosity and supported my studies. I
am particularly grateful to José for always patiently answering my random
questions and giving me useful advice on how to study mathematics. During my
master’s studies at Chalmers University of Technology, the encouragement of my
teachers Thierry Coquand, Peter Dybjer, Nils A. Danielsson, and Jesper Cockx
motivated me to delve deeper into type theory and constructive mathematics. I am
thankful to Jesper in particular for supervising my independent study of domain
theory, which is the origin of my interest in point-free topology. The influence
of Thierry’s mentorship and support on me has been incalculable: he introduced
me to formal and point-free topology, and patiently supervised my studies when I
was an MSc student taking baby steps in research. My master’s thesis work
on formal topology with him can be considered the beginning of the research
programme behind this thesis.

Although I wanted to limit these acknowledgements to academic colleagues and
collaborators, I feel it would be unfair not to make some exceptions for some
friends and family members who have stood by my side over the years.
Firstly, I am eternally grateful to my parents Aynur and Fuat for everything
they have done for me.
Furthermore, I would like to thank my family in Chicago, namely Banu, Belma,
O\u{g}uzhan, Rana, Hayriye, and \"{O}mer Temizel for being my family away from
home during my undergraduate years in Illinois.
Equally important has been the support of some of our family friends:
T\"{u}lay and Fikret Bulut, Yurdanur and Tacettin Varol, Elifnur Ceylan,
and G\"{u}ler Apayd\i{}n.
Moreover, the friendship, support, and encouragement of some of my close friends
has meant a great deal to me:
Tolga Akta\c{s}, \"{O}mer Ery\i{}lmaz, Kardelen Kara\c{c}or, Cumhur (Joomy)
Korkut, Tunahan Moral\i{}, Furkan~E.\ Pehlivanl\i{}, and Onur\ T\"{u}rk\"{o}z.
I am indebted to Joomy specifically for encouraging my study of type theory and
\textsc{Agda}.
Finally, I would like to thank my girlfriend Riya for her love and support,
especially during the hectic couple of months during which I undertook the
writing of this thesis.

\vspace{2\baselineskip}

\begin{flushright}
  Ayberk Tosun\\
  30 November 2024
\end{flushright}


\setcounter{tocdepth}{2}
\tableofcontents

\mainmatter%

\chapter{Introduction}

The aim of this thesis is to develop point-free topology constructively and
predicatively in \VUF{},
with a particular focus on the theory of
\indexedCE{spectral}{spectral locale} and \indexedCE{Stone locales}{Stone locale}.

Point-free topology~\cite{ptj-ss,ptj-the-point,picado-and-pultr}
is an approach to topology where the fundamental object of study is taken to be
a \indexed{locale} rather than a topological space.
The latter is a set of \indexedCE{points}{point} equipped with a topology, which
is a class of subsets of the set of points required to be closed under finite
intersection and arbitrary union.
A locale, in contrast, is a notion of space that is characterized solely by an
abstract lattice of opens, without any stipulation that this lattice consist of
the subsets of some set of points.
The lattice-theoretic structure abstracting the behaviour of the class of open
sets is called a \indexed{frame}.

The point-free approach to topology arises from the remarkable insight that
nontrivial topological questions can be entertained solely through the algebraic
study of frames.
The origin of this idea can be traced back to the work of
Stone~\cite{stone-1934, stone-1936} on what is now called
\indexed{Stone duality}.

The primary motivation for adopting a point-free approach to topology is
that it makes it possible to develop topology in a constructive foundational
setting.
It is well known that some crucial theorems of point-set topology are
fundamentally classical.
The prime example is the \indexedp{Tychonoff theorem},
which not only uses the \indexedp{\VAC{}} in its proof,
but is in fact known to be \emph{equivalent} to it~\cite{kelley-tychonoff-ac}.
Although this shows that there can be no hope of proving the Tychonoff theorem
constructively in the point-set setting,
it turns out to be
constructively provable~\cite{coq-compact-spaces-2003,sjv-roads}
if one adopts the point of view that locales are the subject matter of topology.
Point-free topology thus holds a particular significance in the foundational
setting of constructive mathematics, in addition to being also useful in
classical mathematics.

This thesis is concerned with the development of locale theory in the
foundational setting of \VUTT{},
with a particular focus on the theory of spectral and Stone locales.
\VCapitalizedUTT{} (also called homotopy type theory or \VHoTTUF{})
is a recent refinement of
\VMLTT{}~\cite{pml-tt-1975,pml-tt-1982,pml-bibliopolis},
emanating from
the view of types as
\(\infty\)-groupoids~\cite{hofmann-streicher,warren-fmcs,uppsala-awodey,voevodsky-very-short}.
In addition to having applications in synthetic homotopy theory and higher topos
theory, \VUTT{} also serves as a \emph{practical} foundation for mathematics,
and turns out to be particularly useful for the development of constructive
mathematics.

\paragraph{Constructive mathematics}
The discipline of \indexed{constructive mathematics} has its roots in the ideas
of L.E.J.\ Brouwer~\cite{brouwer-unreliability}, who rejected the use of
\indexedCE{classical principles}{classical principle} in accordance with
his epistemological view of
mathematics~\cite[\textsection 3.1]{sep-constructive-mathematics}.
Brouwer's vision of constructive mathematics was an \indexed{anti-classical}
one: he worked with axioms contradicting the \indexedp{law of excluded middle}.

Later on, Bishop~\cite{bishop-constructive-analysis} developed a neutral form of
constructive mathematics.
In the Bishop school of constructive mathematics, classical principles are
\indexedCE{taboos}{taboo}:
they are neither provable nor disprovable.
This view of mathematics is strictly more general than classical mathematics, as
the possibility of adopting classical axioms is always left open.
Bishop argued that constructive mathematics can also be \emph{practical},
buttressing this view through his constructive development of a significant part
of 20\textsuperscript{th}-century analysis
(as explained in \cite[\textsection 3.3]{sep-constructive-mathematics}).

In this thesis, we adopt a similar neutral approach based on \VUF{},
where constructivity means the development of mathematics without
the law of excluded middle,
the \indexedp{limited principle of omniscience}~\cite{bishop-constructive-analysis}
(including its weaker forms),
or the \indexedp{\VAC{}} (including countable choice).

\paragraph{Type theory}
\indexedpCE{Type theory}{type theory}, in the sense of
\indexedp{Martin-L\"of Type Theory}~\cite{pml-tt-1975,pml-tt-1982,pml-bibliopolis},
is a family of foundational systems for mathematics.
It was originally developed~\cite{pml-tt-1975}
to clarify the language of Bishop-style constructive mathematics,
through a \emph{deepening} of the so-called
\indexedp{Curry-Howard correspondence}~\cite{wah-formulae}.
This enables the view of constructive proofs in mathematics as the terms of a
suitable programming language with a rich type system.

One practical advantage of the type-theoretical view of mathematics
is the importation of techniques from the discipline of programming language
theory to mathematical foundations.
Turning mathematical foundations into a problem in programming languages gives
rise to the development of computer implementations of type theory, called
\indexedCE{proof assistants}{proof assistant}.
Some noteworthy examples of type-theoretical proof assistants are:
\textsc{NuPRL}~\cite{nuprl-book},
\VCoq{}~\cite{coq-huet-coc},
\VAgda{}~\cite{norell-agda}, and
\textsc{Lean}~\cite{lean}.
One of the major contributions of this thesis is the development of a locale
theory library in the \VAgda{} proof assistant, as part of \MHELastName{}'s
open-source \VTypeTopology{} library~\cite{type-topology}.

\paragraph{Homotopical view of types} Homotopy type theory~\cite{hott-book} is a
recent refinement of \VMLTT{}, whose origins can be traced back to Hofmann and
Streicher's groupoid interpretation of type theory~\cite{hofmann-streicher}.
Exploiting this interpretation in order to develop a new form of type theory
turned into a research programme of its own around the late 2000s and early
2010s, starting with the works of
Voevodsky~\cite{voevodsky-very-short} as well as
Awodey and Warren~\cite{awodey-warren-2009}.
Two main novelties that spring from the homotopical interpretation of type
theory are
\indexed{\VTheUniAx{}} and \indexedCE{higher inductive types}{higher inductive type}.

Historically, the treatment of extensionality principles as postulates in
\VMLTT{} has been a deficiency of it as a foundational system.
Voevodsky's formulation of the \VUnivalenceAxiom{} identified a single
\indexedp{extensionality axiom} that implies \emph{all} other extensionality
axioms, the prime examples being functional and propositional extensionality.
This paved the way for the invention of a type theory in which such axioms can
be given computational rules through the assignment of a computational
interpretation to the \VTheUniAx{}.

The problem of finding such a type theory remained open until the invention of
\VCTT{}~\cite{cubical-type-theory} in 2018.
In \VCTT{}, the \VUnivalenceAxiom{} is a theorem rather than a postulate.
This is achieved through a fundamentally new design: \VCTT{} has a built-in
interval object along with primitive operations for manipulating higher
dimensional paths.
\VCTTCapitalized{} has attracted widespread interest in the \VHoTTUF{}
community, and
multiple computer implementations of it have been developed.
Among them, \textsc{Cubical Agda}~\cite{cubical-agda} has become the leading
implementation, with its standard library~\cite{agda-cubical-library} being one
of the major libraries of formalized univalent mathematics at the time of
writing.

\paragraph{Predicative mathematics}

The term \indexed{predicative mathematics} is used by various researchers to
refer to closely related, though not identical, concepts.
In the context of \VUF{}, predicativity refers to the development of mathematics
without the use of
\indexedCE{propositional resizing axioms}{propositional resizing axiom}~\cite[Axiom~3.5.5]{hott-book},
which were originally proposed as rules by
Voevodsky~\cite{voevodsky-resizing}.
The avoidance of these axioms amounts to the restriction that propositions
involving large quantifications over a universe are not admitted as propositions
in that universe.
It has been noted~\cite[\textsection 3~Notes]{hott-book} that the propositional
resizing axioms bear a close resemblance to Russell's notorious
\indexedp{axiom of reducibility}~\cite[241-244]{br-logic-based-on-tt},
originating in the context of the
\indexedpCE{theory of types}{theory of types (\`{a} la Russell)}~\cite[Appendix~B]{russell-1903}.
Our use of the term predicativity is thus in the same sense as in the tradition
of Russell.

In the context of our use of \VUTT{} as a constructive foundation,
we have a concrete motivation for taking a predicative approach to mathematics:
finding computational rules for the propositional resizing axioms is an open
problem at the time of
writing~\cite[27]{cubical-type-theory}.
Our commitment to predicative mathematics is thus a further form of our
commitment to constructive mathematics, as we would like to refrain from the use
of postulates without a well-understood computational behaviour.
Furthermore, it has also been shown~\cite{uemura-cubical-assemblies} that there
are models of \VUTT{} in which propositional resizing fails.

\section{Overview and summary of contributions}

In \cref{sec:outline-and-contributions}, we provide a chapter-wise outline of
the thesis, and explain the contributions presented in each chapter.
In \cref{sec:publications}, we list the publications containing parts of the
work presented here.

Almost all of the results presented in this thesis have been fully formalized
using the \VAgda{} proof assistant~\cite{agda,norell-agda}, as part of
\MHELastName{}'s \VTypeTopology{}~\cite{type-topology} library.
All main results have been formalized, and the unformalized exceptions mostly
relate to tangential developments or examples.
For salient definitions, lemmas, theorems, and examples,
we provide links to the corresponding constructions in the HTML rendering of our
\VAgda{} formalization.
For this, we use the hyperlinked \VAgda{}
icon~\AgdaLink{Locales.index.html} (pronounced \emph{chicken}).

\subsection{Thesis outline and list of contributions}
\label{sec:outline-and-contributions}

\paragraph{\cref{chap:foundations}: Univalent foundations}

We provide a standard presentation of \VUF{}, mostly following
the HoTT Book~\cite{hott-book}
and
\MHELastName{}'s tutorial introduction \cite{mhe-intro-uf}.
Our exposition has also benefited from the presentations of
\VTDJLastName{}~\cite[\textsection 2]{tdj-thesis} and
\VCMKLastName{}~\cite[\textsection I.2]{cmk-thesis}.
This chapter contains no new results and is merely a summary of the
type-theoretical foundation in which we work.

\paragraph{\cref{chap:basics}: Basics of point-free topology}

The main contribution of this chapter is to develop the groundwork for locale
theory in the context of constructive and predicative \VUF{}, working explicitly
with universes.
The univalent study of frames (\cref{defn:frame}) and nuclei on them
(\cref{sec:sublocales})
started in the author's
MSc thesis~\cite{tosun-msc} as well as his work with
Coquand~\cite{coq-tosun-book-chapter} on formal topology.
The main novelty of the predicative approach here is the use of small locale
bases (\cref{sec:bases}) and the use of the posetal \VAFT{}
(\cref{sec:aft}).
As we explain in \cref{sec:related-work}, our predicative approach to locales is
closely related to formal topology, though it differs in certain key aspects.

Parts of the work presented here have been published in~\cite{patch-mfps}
and~\cite{patch-mscs},
although the account given in this chapter contains several improvements and
simplifications.

\paragraph{\cref{chap:spec-and-stone}: Spectral and Stone locales}
We present a predicative treatment of the theory of
\indexedpCE{spectral}{spectral locale} and \indexedpCE{Stone locales}{Stone locale}
in this chapter.
It is here that we start to make use of the foundational novelties of \VUTT{}.
For instance, we give a definition of spectrality
in \cref{defn:spectral-locale}
that makes use of \VTheUniAx{}.
Furthermore, in \cref{thm:spec-characterization},
we prove the logical equivalence of various characterizations of the notion of
spectral locale using the \indexedp{\VSRPrinciple{}}.
This relies on \cref{lem:split-support}, which entails a strengthening of
\VTDJMHE{}'s characterization of
\indexedpCE{algebraic \VDCPO{}s}{algebraic \VDCPO{}}~\cite[Lemma~4.8.3]{tdj-thesis}.

In \cref{sec:zero-dimensional}, we undertake a brief detour into the predicative
treatment of \VZeroDimensional{} and regular locales in our setting,
and pin down the predicatively well-behaved notions of \VZeroDimensional{} and
regular locales.
This finds application in \cref{sec:stone}, where we develop the theory of Stone
locales (i.e.\ locales that are compact and \VZeroDimensional{}).

After this, we proceed to establish
\indexedp{Stone duality}
for spectral locales in \cref{sec:spec-duality}.
We show that a frame is spectral if and only if it is isomorphic to the frame of
\emph{small} ideals over a \emph{small} distributive lattice,
which justifies that our definition of spectrality from
\cref{defn:spectral-locale} captures the correct category of spectral locales.
We then extend these two maps to functors and show that they form a categorical
equivalence.

A preliminary version of our work in this chapter has been published in
\cite{patch-mfps}.
\cref{thm:spec-characterization,thm:answer-stone} characterizing spectral and
Stone locales have appeared in a more recent version of our
work~\cite{patch-mscs}.
These provide a substantive simplification of our predicative account of
spectral locales.
In particular, it is remarked at the end of \cite{patch-mfps} that the
formalization of \cite[Lemma~7.11]{patch-mfps} was work in progress at the time
of writing.
The development
in \cite[Theorem~in~\textsection 4.1]{patch-mscs} allows this proof to be completed.

\paragraph{\cref{chap:patch}: The patch locale}

We contribute a constructive and predicative construction of the patch locale of
a spectral locale along with a proof of its universal property.

The construction of the patch locale is given in
\cref{sec:frame-of-scott-continuous-nuclei}.
Although this is based on earlier work by
\MHELastName{}~\cite{mhe-properly-injective,mhe-patch-short,mhe-patch-full},
it contains some new ingredients, since the development in \emph{op.\ cit.}\
uses various impredicative constructions that are not available in our
predicative setting.
These new ingredients are the following:
\begin{enumerate}
  \item The patch locale is defined by the \indexedp{frame of \VScottContinuous{} nuclei}
    in \cite{mhe-patch-short}. In order to prove that this is indeed a frame,
    one starts with the \indexedpCE{frame of all nuclei}{frame of nuclei}, and then exhibits the
    \indexedp{\VScottContinuous{} nuclei}
    as a subframe.
    This procedure, however, does not seem to be possible in the context of our
    work, as it is not clear whether all nuclei can be shown to form a frame
    predicatively.
    We thus construct the frame of \VScottContinuous{} nuclei \emph{directly},
    which requires predicative reformulations of all standard facts about it
    inherited from the frame of all nuclei.
  \item \MHELastName{}'s description of the patch locale makes use of the notion
    of \indexed{open nucleus}, which does not exist a priori in a predicative
    setting.
    This is solved by our use of the posetal \VAFT{} from \cref{sec:aft}.
  \item \MHELastName{}'s proofs from \cite{mhe-patch-short,mhe-patch-full}
    make use of the notion of \indexed{perfect map}, which again does not
    exist a priori in a predicative setting.
    This is because the right adjoint of a frame homomorphism is not a priori
    available in a predicative context.
\end{enumerate}

The proof of the universal property is given in \cref{sec:ump-of-patch}.
The predicative proof that this construction satisfies its universal property is
new as \MHELastName{}'s proof relies on the existence of the frame of all
nuclei, which does not seem possible to construct in an impredicative setting
(see \cref{sec:frame-of-nuclei-impredicatively}).

A preliminary version of the results presented here have first been published
in~\cite{patch-mfps}.
Our more recent work \cite[\textsection 8]{patch-mscs} contains a new and
simplified proof of the universal property of the patch locale,
which has been developed in collaboration with Igor~\IALastName{}.

\paragraph{\cref{chap:scott}: Topology of Scott domains}

In this chapter, we focus our investigation of point-free topology on the
topology of
algebraic \VDCPO{}s and \indexedpCE{Scott domains}{Scott domain}.
In \cref{sec:scott-locale}, we contribute a predicative construction of the
\indexedp{Scott locale} of an \indexedp{algebraic \VDCPO{}}
and show that it is predicatively well-behaved i.e.\ that it is large, locally
small, and small-complete.
For the special case of Scott domains, we also construct a small strong base for
the \indexedp{Scott topology} in \cref{sec:spec-scott-locale-over-scott-domain}.

We then study the topology of Scott domains using the framework we built in
\cref{chap:spec-and-stone}.
In \cref{sec:spec-scott-locale-over-scott-domain}, we give a constructive and
predicative proof of the fact that the Scott topology of a Scott domain is
always a \indexed{spectral locale}, which yields a rich class of examples of
spectral locales that can be predicatively constructed.

In addition to these, we also study the patch locale of the spectral Scott
locale of a Scott domain, using the results developed in \cref{chap:patch}.
We show that the \indexedCE{sharp elements}{sharp element}~\cite{tdj-sharp-elements}
of a Scott domain coincide with the \indexedCE{spectral points}{spectral point}
of its Scott locale,
which in turn coincide with the points of the patch of the Scott locale in
consideration.
This is an illustrative application of the universal property of the patch
locale from \cref{sec:ump-of-patch}.

Results presented in this chapter have not been published, but are planned to be
submitted for publication in the not-too-distant future.
Furthermore, we also believe that the equivalence between sharp elements and the
points of patch presented here will allow us to clarify,
in a constructive setting,
the relationship between the patch locale of the Scott locale of a Scott domain
and the \indexed{Lawson topology} on the domain in consideration. The extension
of our work in this direction is work in progress at the time of writing.

\subsection{Publications}
\label{sec:publications}

Parts of the work presented in this thesis have been published
in \cite{patch-mfps} and \cite{patch-mscs}.

\printbibliography[keyword=ayberkt,heading=none]


\chapter{Foundations}
\label{chap:foundations}

Our development takes place in the foundational setting of
\VUTT{}\index{univalent type theory}, which is
\VMLTT{}\index{{Martin-L\"of Type Theory}}~\VCiteTT{}
extended with features that enable the development of
\indexed{univalent mathematics}.
In this chapter, we present the fundamental ideas of \VUF{}~\cite{hott-book} and
discuss the details of our foundational setting.

Type theory differs fundamentally from set theory as a mathematical foundation.
In set theory, one works with the untyped notion of \emph{set}, out of which all
mathematical constructions are built. The rules governing the construction of
sets are given by the \emph{axioms} of set theory. In type theory, on the other
hand, one works with \emph{types} which come equipped with rules governing the
construction of their inhabitants. These rules are used to carry out the
mathematical constructions of interest. This means that mathematical
constructions in type theory do not exist in isolation, but \emph{only as the
inhabitants} of some type. In other words, all mathematical constructions
carried out in type theory are fundamentally typed.

This typed approach to mathematical foundations yields a useful foundation for
constructive mathematics.
For example, the view of equality as a type in itself, which is a hallmark
feature of \VMLTT{}, paves the way to the formulation of the \indexed{\VUniAx{}}
and \defineCE{higher inductive types}{higher inductive type} --- concepts that
cannot quite be formulated in set theory (as explained by
\MHELastName{}~\cite{mhe-univalence}).

The organization of this chapter is as follows.
\begin{description}[leftmargin=!,labelwidth=\widthof{Section 7.77:}]
  \item[\cref{sec:mltt}:]
    We present the basics of \VMLTT{}, which is the starting point of the \VUTT{}
    that we use.
  \item[\cref{sec:universes}:]
    We discuss the universe structure of our type theory.
  \item[\cref{sec:w-types}:] We present \(\WTySym\)-types, which give a formal
    account of \emph{inductively generated} types.
  \item[\cref{sec:curry-howard}:] We explain the Curry-Howard view of
    propositions as types in \VMLTT{}.
  \item[\cref{sec:id-type}:] We present the technical details of the identity
    type and explain its homotopical interpretation,
    which underlies the key features of \VUTT{}.
  \item[\cref{sec:decidable-types-and-sets}:]
    We discuss the notion of decidability for a type and present
    Hedberg's theorem.
  \item[\cref{sec:funext}:] We discuss function extensionality
    as well as invertible and left-cancel\-lable maps,
    and explain the problem that their na\"ive formulations are not properties.
  \item[\cref{sec:embeddings-and-equivalences}:]
    We present the notions of embedding and equivalence, which refine the notions
    of left-cancellable and invertible maps to the finer-grained setting of
    higher groupoids.
  \item[\cref{sec:hlevels}:] We explain the stratification of types with
    respect to their homotopy levels, which lies at the heart of \VUF{}.
  \item[\cref{sec:univalence}:] We present the \VUniAx{}.
  \item[\cref{sec:hit}:] We present the notion of higher inductive type.
  \item[\cref{sec:uf-logic}:] We revisit the na\"{i}ve form of the
    propositions-as-types principle from \cref{sec:curry-howard} and discuss
    its refinement in \VUTT{}.
  \item[\cref{sec:predicativity}:]
    We present the theory of small and locally small types, which underpins the
    notion of \emph{predicativity}, and its lack thereof, in our foundational
    setting. We then explain the propositional resizing axioms, embodying
    what we mean by \emph{impredicativity}, and discuss our motivations for avoiding
    them.
  \item[\cref{sec:set-replacement}]
    Finally, we discuss the \VSRPrinciple{} in \VUTT{} and its presentation
    as a higher inductive type.
    Besides propositional truncations, \VSRPrinciple{} is the only other
    higher inductive type that we use in this thesis.
\end{description}

\section{Basics of \VMLTT{}}\label{sec:mltt}

We give a high-level summary of the basics of \VMLTT{} and refrain from formally
presenting it in full detail. For a complete formal presentation, we refer the
reader to~\cite[Appendix~A]{hott-book}.

Type theory is a formal system based on \indexedCE{judgements}{judgement}.
In the type-theoretical development of mathematics, we are concerned with the
two basic judgements:
\begin{enumerate}
  \item
    The \define{typing judgement},
    \nomenclature{\(a \oftype A\)}{%
      Typing judgement expressing that term $a$ is of type $A$%
    }%
    written $a \oftype A$,
    expressing that term $a$ has type $A$.
  \item%
    \nomenclature{\(\DefnEqTyped{A}{a}{b}\)}{%
      equality judgement expressing that term $a$ and $b$ are definitionally
      equal inhabitants of type $A$\index{definitionally equal}%
    }%
    The \define{equality judgement},
    written $\DefnEqTyped{A}{a}{b}$,
    expressing that terms $a$ and $b$ are equal inhabitants of type $A$.
\end{enumerate}

Types in type theory are defined by specifying what constitutes evidence for
their typing, equality, and formation judgements
(as explained by \VPMLLastName{}~\cite{pml-meanings}).
More specifically, this consists in giving their
\begin{itemize}
  \item \emph{formation rules}\index{formation rule},
    explaining how to form the type;
  \item \emph{introduction rules}\index{introduction rule},
    explaining how to build inhabitants of the type;
  \item \emph{elimination rules}\index{elimination rule},
    explaining how to use the inhabitants of the type;
  \item \emph{computation rules}\index{computation rule},
    specifying the computational behaviour of the eliminator on the inhabitants
    of the type.
\end{itemize}

\subsection{Basic types}
\label{sec:basic-types}

The starting point of our foundational setup is \VMLTT{} with the following
types:
\begin{enumerate}
  \item\label{item:empty}
    The \define{empty type} $\EmptyTy$.%
    \nomenclature{\(\EmptyTy\)}{empty type with universe left implicit}
  \item
    \nomenclature{\(\UnitTy\)}{The unit type}%
    The \define{unit type} $\UnitTy$.
  \item\label{item:nat}
    \nomenclature{\(\NatTy\)}{The type of natural numbers}%
    The type $\NatTy$ of \define{natural numbers}.
  \item\label{item:list}
    \nomenclature{\(\ListTy{A}\)}{The type of lists (or words) over type $A$}%
    The type former $\ListTy{\blank}$, giving the type $\ListTy{A}$ of
    \emph{lists}, or \emph{words},\index{list}\index{list type} over every type~$A$.
  \item\label{item:prod}%
    \nomenclature{\(\ProdTy{A}{B}\)}{product type of types $A$ and $B$}%
    The type former $\ProdTy{(\blank)}{(\blank)}$,
    giving the \define{product type}\/ $\ProdTy{A}{B}$, for every pair of
    types $A$ and $B$.
  \item\label{item:sum}%
    \nomenclature{\(\SumTy{A}{B}\)}{The sum type of types $A$ and $B$}%
    The type former $\SumTy{(\blank)}{(\blank)}$,
    giving the \define{sum type} $\SumTy{A}{B}$, for every pair of
    types $A$ and $B$.
  \item
    \nomenclature{\(\SigmaType{a}{A}{B(a)}\)}{%
      Dependent sum type of an $A$-indexed family of types $B$%
    }%
    The type former $\SigmaTypeSym(\blank, \blank)$, giving the
    \define{dependent sum type}\/ $\SigmaTypeSym(A, B)$,
    for every type $A$ and every $A$-indexed family of types $\FamEnum{a}{A}{B_a}$.
  \item
    \nomenclature{\(\PiTy{a}{A}{B(a)}\)}{%
      Dependent product type of an $A$-indexed family of types $B$%
    }%
    The type former $\PiTySym(\blank, \blank)$,
    giving the \define{dependent product type}\/ $\PiTySym(A, B)$,
    for every type~$A$ and every $A$-indexed family of types
    $\FamEnum{a}{A}{B_a}$.
  \item\label{item:id-ty}%
    \nomenclature{\(\IdTyAlt{A}{a}{b}\)}{%
      type of identifications (or equalities) of inhabitants $a$ and $b$
      of type $A$
    }%
    The type former $\IdTyAlt{(\blank)}{\blank}{\blank}$, giving the type
    $\IdTyAlt{A}{a}{b}$ of \emph{identifications}\index{identification} or
    \emph{equalities}\index{equality} between every pair of inhabitants
    $a, b : A$, for every type $A$.
\end{enumerate}

\begin{remark}\label{rmk:w-types}
  Types (\ref{item:nat}) and (\ref{item:list}) above
  are examples of \emph{inductive types}, the formal
  account of which is given using the general mechanism of $\WTySym$-types.
  These types are the only instances of $\WTySym$-types that we are interested
  in so we prefer to present them directly.
  Nevertheless, we discuss their representation using $\WTySym$-types in
  \cref{sec:w-types}.
\end{remark}

\subsubsection{Notation}

Given a type $A$ and an $A$-indexed family of types $\FamEnum{a}{A}{B_a}$ , we
denote the dependent product type and the dependent sum type by, respectively,
\begin{center}
  $\PiTy{x}{X}{Y_x}$ \quad and \quad $\SigmaType{x}{X}{Y_x}$.
\end{center}
\nomenclature{\(\PiTy{x}{X}{Y_x}\)}{dependent product type}
The dependent product type is sometimes called the \define{dependent function type}
or the~\VPiType{}. If the type family $\FamEnum{a}{A}{B_a}$ in consideration is
constant, we abbreviate $\PiTy{a}{A}{B_a}$ to $\ArrTy{A}{B}$, and call this the
\define{function type}, or the \emph{non-dependent function type} if additional
clarity is required.

Given some $a : A$ and $b : B(a)$, the inhabitant of
$\SigmaType{a}{A}{B_a}$ is denoted by the tuple notation $\Pair{a}{b}$.
Inhabitants of the $\PiTySym$-type are constructed using
\define{$\lambda$-abstraction}, which are denoted $\PiLambda{x}{t}$
or $x \mapsto t$.

The unique constructor of the unit type is denoted $\TTUnit$. The two
constructors of $\NatTy$ are denoted $\NatZero$ and $\NatSuccSym$.
The constructors of the sum type $\SumTy{A}{B}$ are denoted
$\SumInlSym(a)$ and $\SumInrSym(b)$.
For the special case $\TwoTySym \is \SumTy{\UnitTySym}{\UnitTySym}$,
we define the more suggestive notation of
$\BoolZero \is \SumInlSym(\TTUnit)$ and $\BoolOne \is \SumInrSym(\TTUnit)$.
We denote the empty list by $\emptyl$, and for the prepending of an element $x$
onto a list $s$, we write $x \consl s$. We use the same notation for the
concatenation of lists $s$ and $t$ and write $s \append t$.

For identity types, we often use the more suggestive notation
$\IdTyWithType{A}{a}{b}$ for $\IdTyAlt{A}{a}{b}$. If the type is clear from the
context, we further abbreviate this to $\IdTy{a}{b}$.
Given some inhabitant $a \oftype A$, the constructor of its identity with itself
is denoted
$\Refl{a} : \IdTyWithType{A}{a}{a}$.
\nomenclature{\(\Refl{a}\)}{%
  construction of the identity type $\IdTyWithType{A}{a}{a}$%
}
Moreover, we carefully distinguish between judgemental equality and the identity
type, refraining from using the symbols $\DefnEq{(\blank)}{(\blank)}$ and
$\IdTy{(\blank)}{(\blank)}$ in place of each other.
The former always refers to judgemental equality and the latter always to the
identity type. Finally, we note that we use $(\blank) \is (\blank)$ for the
introduction of new definitional equalities, following the convention of
\cite{hott-book}.
When defining new functions, we often abbreviate $f \is \PiLambda{x}{t}$ to
$f(x) \is t$, following the standard convention.

\subsection{Axioms in type theory}\label{sec:axioms}

An \define{axiom} in the context of type theory is an inhabitant of a particular
type that is declared to exist, independently of the type's introduction and
elimination rules.
Working with axioms in type theory is problematic for at least two reasons:
\begin{enumerate}
  \item Extending one's foundational system with axioms is potentially
    dangerous, as it poses the risk of introducing inconsistency into the
    system.
  \item The \emph{computational content} of axioms in type theory are not
   specified, as they are not built using introduction and elimination principles
   which all have specified computational behaviour. This means that the
   computational interpretation of a proof using an axiom is unclear, which
   violates the main desideratum of the type-theoretic view of mathematics: to
   elucidate the computational content of constructive proofs.
\end{enumerate}

In our development, we will use several principles that we will call axioms,
though our use of this term will be slightly different.
In this thesis, we axiomatize our foundational setting instead of committing
to a concrete implementation of type theory.
Therefore, an ``axiom'' for us will simply be an assumption about our
foundational setting, and our use of this term does not imply that we are
committing to the view of their inhabitants as uninterpreted constants.

We will circumvent the above problems by
only using axioms that follow either from \VUniAx{} or the existence of certain
well-behaved higher inductive types.
Both of these are known to be consistent and to admit computational
interpretations.
The axioms we use are all theorems in, for instance,
\VCTT{}~\cite{cubical-type-theory}.

\section{Universes}
\label{sec:universes}

We explained in the introduction to this chapter that all terms of \VMLTT{} are
inherently typed. This invites the question of how to assign types to types
themselves. Such types that govern the structure of types are called
\defineCE{universes}{universe}. For the purposes of our predicative development
of locale theory, universes are of paramount importance, as they provide a
notion of \emph{size}\index{size (of a type)} for types.
This underpins the \indexed{predicativity} of our foundational setting.

\subsection{Universe operations}

The universe structure in our type-theoretical setting closely follows the type
system of the \VAgda{} proof assistant~\cite{agda,norell-agda} and the
presentation of \MHELastName{}~\cite[\textsection 2.4]{mhe-intro-uf}.
Universes form a hierarchy of increasingly larger universes. We do not assume
that this hierarchy is, or is not, cumulative.\index{cumulativity}
This hierarchy is formally given by the following operations:
\begin{itemize}
  \item \textbf{Ground universe}.\nomenclature{\(\UZero\)}{ground universe}
    There is a basic universe \(\mkern+2mu\UZero\), called the
    \define{ground universe}, which forms the base level of the \indexedp{universe hierarchy}.
  \item \textbf{Successor operation.}
    There is an operation $\USucc{(\blank)}$ that gives the
    \define{successor universe}, denoted \VUni{\USucc{\UU}},
    of every universe \VUni{\UU}.
    \nomenclature{\(\USucc{\UU}\)}{Successor universe of universe $\UU$}%
  \item \textbf{Maximum operation.} %
    \nomenclature{\(\UMax{\UU}{\VV}\)}{Maximum of universes $\UU$ and $\VV$}%
    There is an operation $\UMax{(\blank)}{(\blank)}$ giving the least upper
    bound $\UMax{\UU}{\VV}$ of every pair of universes $\UU$ and $\VV$.
\end{itemize}

We reserve the calligraphic letters $\UU, \VV, \WW, \ldots$ for variables
ranging over universes. Furthermore, we note that we sometimes use the
term \emph{\VUni{\UU}-type} for a type inhabiting universe\/ \VUni{\UU},
when additional clarity is required.

\begin{remark}
  In \cref{sec:mltt}, we used the notation $\FamEnum{a}{A}{B_a}$ for an
  $A$-indexed family of types, without explaining its precise representation in
  type theory. After having introduced universes, we can make this precise: a
  family $\FamEnum{a}{A}{B_a}$ of types living in universe \VUni{\UU}
  is simply a function $B \oftype A \to \UU$.
\end{remark}

\subsection{Universes rules for basic types}

Under this universe structure, we assume that the types given in \cref{sec:mltt}
are assigned universes as follows:
\begin{enumerate}
  \item The empty type $\EmptyTy$ and the unit type $\UnitTy$ have a copy in
    every universe. When additional clarity is needed, we denote their universes by
    using subscripts and write $\EmptyTy_{\UU}$ and $\UnitTy_{\UU}$.
    \nomenclature{\(\EmptyTy_{\UU}\)}{empty type in universe $\UU$}
    \nomenclature{\(\UnitTy_{\UU}\)}{Unit type in universe $\UU$}
  \item The type $\NatTy$ has a copy in every universe \VUni{\UU}.
  \item For every type $X \oftype \UU$, the type $\ListTy{X}$ inhabits
    the universe \VUni{\UU}.
  \item For every type $X \oftype \UU$ and every pair of inhabitants
    $x, y \oftype X$, the identity type $\IdTyWithType{X}{x}{y}$ inhabits the
    universe \VUni{\UU}.
  \item For every pair of types $X \oftype \UU$ and $Y \oftype \VV$,
    the sum type $\SumTy{X}{Y}$ and the product type $\ProdTy{X}{Y}$ both
    inhabit the uni\-verse~$\UMax{\UU}{\VV}$.
  \item For every pair of types $X \oftype \UU$ and type family
    $Y \oftype X \to \VV$, the dependent product
    $\PiTy{x}{X}{Y(x)}$ and the dependent sum $\SigmaType{x}{X}{Y(x)}$ inhabit
    the universe $\UMax{\UU}{\VV}$.
\end{enumerate}

If the inductively defined types above were defined using $\WTySym$-types, as
discussed in \cref{rmk:w-types}, it would be possible to obtain these universe
rules as a special case of the universe rule for $\WTySym$-types
(\cref{defn:w-type-universe}).
However, we prefer to present these directly and refrain from involving the
generality of $\WTySym$-types.

\section{\texorpdfstring{\(\WTySym\)-types}{W-types}}
\label{sec:w-types}

We mentioned in \cref{rmk:w-types} that some of the basic types we gave in
\cref{sec:basic-types} are examples of
\indexedCE{inductively generated types}{inductively generated type}.
The formal account of the general notion of an inductively generated type is
given by the mechanism of
\VWType{}s~\cite[\textsection I.15]{pimltt},
which we briefly summarize in this section.

\subsection{Definition of \texorpdfstring{\(\WTySym\)-types}{W-types}}

\begin{definition}
  The \VWType{} former $\WTySym_{(\blank),(\blank)}$
  gives a type $\WTySym_{A,B}$,
  for every type~$A \oftype \UU$
  and
  every $A$-indexed family of types $B \oftype A \to \VV$.
  This type has a single constructor
  \(%
    \mathsf{sup}
    \oftype
    \PiTy{a}{A}{\paren{\ArrTy{(B(a) \to \WTySym_{A,B})}{\WTySym_{A,B}}}}
    \text{,}%
  \)
  and has the following induction principle:
  for every type family $Y \oftype \WTySym_{A,B} \to \TT\!\text{,}$
  if it is the case that,
  for every $a \oftype A$ and $f \oftype \ArrTy{B(a)}{\WTySym_{A,B}}$, the
  type
  \begin{equation*}
    \paren{\PiTy{b}{B(a)}{Y(f(b))}} \ImplSym Y(\mathsf{sup}(a, f))\text{,}
  \end{equation*}
  is inhabited, then we have \(\PiTy{w}{\WTySym_{A,B}}{Y(w)}\).
\end{definition}

The witness of the above induction principle is denoted $\mathsf{wrec}$.%
\nomenclature{\(\mathsf{wrec}\)}{eliminator of the \(\WTySym\)-type}

\begin{definition}[Universe rule for \VWType{}s]\label{defn:w-type-universe}
  \VWType{}s are assigned universes by the following rule:
  given some type $A \oftype \UU$ and a type family \(B \oftype \ArrTy{A}{\VV}\),
  the type \(\WTySym_{A,B}\) inhabits the universe \VUni{\UMax{\UU}{\VV}}.
\end{definition}

The key idea in the above definition of \VWType{} is that every inductively
generated type has two components:
\begin{itemize}
  \item a type of \indexedCE{constructor names}{constructor name},
  \item a type family of \emph{subtree selectors} for each constructor.
\end{itemize}
This is exactly what the \VWType{} former encodes:
the type $A$ is the type of constructor names,
whereas the type family $B \oftype \ArrTy{A}{\VV}$
gives the type of subtree labels for each constructor $a \oftype A$.
The introduction principle $\mathsf{sup}$ then takes
(1) a constructor~$a \oftype
A$, and (2) a function encoding how to construct a smaller inhabitant
of~$\WTySym_{A,B}$, for each~$b \oftype B(a)$,
and returns an inhabitant of $\WTySym_{A,B}$.

\subsection{Examples of \texorpdfstring{\VWType{}s}{W-types}}
\label{sec:wtype-examples}

Let us look at an example to make things a bit more concrete.

\begin{example}\label{example:nat-as-an-indexed-wtype}
  The type of natural numbers is an illustrative example of a \VWType{}.
  We show how the copy of the type $\NatTy$ can be constructed
  in a given universe \VUni{\UU}.
  Since the type $\NatTy$ has two constructors, we pick $A \is \TwoTy{\UU}$.
  The two constructors have types~$\NatZero\oftype\NatTy$ and
  $\NatSuccSym \oftype \Endomap{\NatTy}$.
  The former has zero arguments whereas the latter takes one subtree.
  To encode these, we define
  the type family~$B\oftype\ArrTy{A}{\UU}$ as follows:
  \begin{align*}
    B(\BoolZero) \quad&\is\quad \EmptyTy_{\UU}\\
    B(\BoolOne)  \quad&\is\quad \UnitTySym_{\UU}
  \end{align*}
  Using the $\mathsf{sup}$ constructor of $\WTySym_{A,B}$,
  we then encode the two constructors as
  \begin{align*}
    \begin{array}{rl}
      \NatZero' &\oftype \quad \WTySym_{A,B}\\
      \NatZero' &\is \;\mathsf{sup}({\BoolZero}, \EmptyTy\text{-elim})
    \end{array}
    & \qquad
    \begin{array}{rl}
     \NatSuccSym'    &\oftype \quad \ArrTy{\WTySym_{A,B}}{\WTySym_{A,B}}\\
     \NatSuccSym'(w) &\is \;\mathsf{sup}({\BoolOne}, \TTUnit \mapsto w)
    \end{array}
  \end{align*}
\end{example}

\begin{example}
  For every type $X \oftype \UU$,
  the type $\ListTy{X}$ can be encoded as a \VWType{}.
  We define $A_X \is \SumTy{\UnitTy}{X}$, as we have one constructor for the empty
  list and $X$-many cons constructors.
  We define the type family $B \oftype \ArrTy{A_X}{\UU}$ as
  \begin{align*}
    B(\BoolZero) \quad&\is\quad \EmptyTy_{\UU}\\
    B(\BoolOne)  \quad&\is\quad \UnitTySym_{\UU}
  \end{align*}
  The $\mathsf{empty}$ and $\mathsf{cons}$ constructors are then encoded as
  \begin{align*}
    \begin{array}{rl}
      \mathsf{empty} \quad&\oftype\quad \WTySym_{{A_X},B}\\
      \mathsf{empty} \quad&\is\; \mathsf{sup}({\BoolZero}, \EmptyTy\text{-elim})
    \end{array}
    & \qquad
    \begin{array}{rl}
      \mathsf{cons}       &\quad\oftype\quad \ArrTy{X}{\ArrTy{\WTySym_{{A_X},B}}{\WTySym_{{A_X},B}}}\\
      \mathsf{cons}(x, w) &\quad\is\; \mathsf{sup}(\SumInrSym(x), \TTUnit \mapsto w)
    \end{array}
  \end{align*}
  The type family $X \mapsto \WTySym_{{A_X},B}$ is then equivalent to the type
  family $\ListTySym \oftype \Endomap{\UU}$.
\end{example}

We also note here that there is a generalization of \VWType{}s, called
\indexedCE{indexed\/ \VWType{}s}{indexed \VWType{}},
allowing the en\-coding of \emph{inductively defined families of types} in
a similar manner.
We do not present indexed \VWType{}s here as we will not need them in our
development.

\section{The Curry-Howard interpretation of logic}
\label{sec:curry-howard}

Set theory is a foundational system with two layers: one starts with the logical
layer of first-order logic, and then uses it to formulate the axioms of a theory
of sets, such as Zermelo-Fraenkel set theory. In contrast, logic has no special
status in type theory; it is obtained by assigning a logical reading to a
certain fragment of the types.

This idea springs from the so-called \emph{Curry-Howard interpretation of
logic}\index{{Curry-Howard interpretation}}~\cite{wah-formulae} by which a type
$A$ is viewed as a proposition, and the typing judgement $a : A$ is read as
expressing that term $a$ constitutes a proof for proposition $A$.
Under this interpretation, the following logical readings are assigned to types:
\begin{itemize}
  \item The empty type $\EmptyTy$ is interpreted as falsum $\bot$.
  \item The unit type $\UnitTySym$ is interpreted as truth $\top$.
  \item The product type $\ProdTy{A}{B}$ is interpreted as logical conjunction
    $A \wedge B$.
  \item The sum type $\SumTy{A}{B}$ is interpreted as logical disjunction
    $\Disj{A}{B}$.
  \item The function type $\ArrTy{A}{B}$ is interpreted as logical implication
    $\Impl{A}{B}$.
  \item A type family $B \oftype X \to \UU$ is interpreted as a predicate on
    type $X$.
  \item Negation of a proposition $A$ is accordingly given by the type
    $\ArrTy{A}{\EmptyTy}$.
  \item The dependent product type $\PiTy{a}{A}{B(a)}$ is interpreted as the
    universal quantifier~$\Forall{a}{A}{B(a)}$.
  \item The dependent sum type $\SigmaType{a}{A}{B(a)}$ is interpreted as the
    existential quan\-ti\-fier~${\Exists{a}{A}{B(a)}}$.
  \item If there exist functions $f \oftype X \to Y$ and $g \oftype Y \to X$
    between types $X$ and $Y$,
    they are viewed as \emph{logically equivalent} propositions.
\end{itemize}
Under this logical reading of types, to assert that proposition $A$ holds is to
assert that type $A$ is inhabited. We will hereafter use this interpretation
tacitly, and will not explicitly assert types to be inhabited if it is clear
that we are taking their view as propositions. For example, we will simply say
``$B$ whenever $A$'' instead of ``$B$ is inhabited whenever $A$ is inhabited''.

\begin{example}\label{example:every-nat-is-either-zero-or-succ}
The logical statement
\(\Forall{n}{\NatTy}{\IdTy{n}{\NatZero} \vee \Exists{m}{\NatTy}{\IdTy{n}{\NatSuccSym(m)}}},\)
expressing that every natural number is either zero or the successor of some
natural number, is interpreted under the Curry-Howard principle as
\begin{equation*}
  \PiTy{%
    n
  }{%
    \NatTy
  }{%
    \SumTy{%
      \paren{\IdTy{n}{\NatZero}}
    }{%
      \paren{\SigmaType{m}{\NatTy}{\IdTy{n}{\NatSuccSym(m)}}}
    }
  }.
\end{equation*}
\end{example}

In some type theories, there is a designated universe for those types viewed as
propositions. The prime example is the
\indexedp{Calculus of Constructions}~\cite{coq-huet-coc} (\VCoC{} for short).
In \VCoC{}, there is a universe $\mathsf{Prop}$ of types that are interpreted as
propositions.
However, this notion of \indexedp{being a proposition} is a judgement rather
than an internal statement in \VCoC{}.

One of the salient novelties of \VUTT{} is that it makes it possible to work
with an internal notion of propositional type,
which allows one to \emph{define the type of propositional types} inside the
type theory itself.
This gives rise to a refined interpretation of the Curry-Howard principle.
In \cref{sec:uf-logic}, we will discuss this refined view of logic in \VUTT{},
and will contrast it with the na\"{i}ve form that we presented in this section.

\section{The identity type}
\label{sec:id-type}

We briefly mentioned the identity type in \cref{sec:mltt}.
In this section, we revisit it to focus on its technical details and to discuss
its \emph{homotopical interpretation}.

\subsection{Formal definition}

\begin{definition}[The identity type]\label{defn:id-type}
  The \define{identity type} of a type $X$ is the inductive type family
  $\IdTyWithType{X}{(\blank)}{(\blank)}$ generated by the single constructor:
  \begin{equation*}
    \Refl{(\blank)} \oftype \PiTy{x}{X}{\IdTyWithType{X}{x}{x}}.
  \end{equation*}
  Its elimination principle, often called the \define{$\JSym$-rule}, states
  that, for every family
  \[C \oftype \PiTy{x, y}{X}{\paren{\IdTyWithType{X}{x}{y} \to \UU}}\text{,}\]
  and every function
  \(c \oftype \PiTy{x}{X}{C(x, x, \Refl{x})}\),
  we have an inhabitant
  \begin{equation*}
    f \oftype \PiTy{x, y}{X}{\PiTy{p}{\IdTyWithType{X}{x}{y}}{C(x, y, p)}}
  \end{equation*}
  satisfying the \indexedp{computation rule} $\DefnEq{f(x, x, \Refl{x})}{c(x)}$.
\end{definition}

Given a type $X$ and type family $C$ as in the above definition, the witness of
this elimination principle is denoted
\begin{equation*}
  \JId{C}{X}
   \oftype \ArrTy{%
       \paren{\PiTy{x}{X}{C(x, x, \Refl{x})}}%
     }{
       \PiTy{x, y}{X}{\PiTy{p}{\IdTyWithType{X}{x}{y}}{C(x, y, p)}}
     }\text{.}
\end{equation*}%
\nomenclature{\(\JSym\)}{eliminator of the identity type}
In practice, it turns out to be rather cumbersome to work directly with this
elimination principle.
The standard practice in type theory is to explicitly prove that the identity
type satisfies some basic properties using the \(\JSym\)-rule, and then work
only with these properties.
We address this next.

\subsection{Transport and action on paths}

We now prove two immediate corollaries of the elimination principle $\JSym$:
the so-called \indexed{transport} and \indexed{action on paths} lemmas.
In practice, working with these corollaries turns out to be much more convenient
than working directly with the $\JSym$-rule.

\begin{lemma}[Transport]\label{lem:transport}
  For every type $X \oftype \UU$ and type family $Y \oftype X \to \VV$,
  we have an inhabitant
  \begin{equation*}
    \TransportSym^{Y} \oftype \PiTy{x, x'}{X}{\IdTy{x}{x'} \to Y(x) \to Y(x')},
  \end{equation*}
  satisfying the computation
  rule\/ \(\DefnEq{\TransportSym^Y(x, x, \Refl{x})}{\FnIdSym}\),
  for every\/ $x \oftype X$.
\end{lemma}
\begin{proof}
  Consider a type $X \oftype \UU$ and type family $Y \oftype \ArrTy{X}{\VV}$.
  Define
  \nomenclature{\(\TransportSym_{Y}\)}{inhabitant of the transport lemma (from \cref{lem:transport}}
  \begin{align*}
    C(x_1, x_2, p) &\is Y(x_1) \to Y(x_2),\\
    c(x)           &\is \mathsf{id}.
  \end{align*}
  The function $\TransportSym^{Y}$ is then defined as
  $\TransportSym^{Y}(x, x', p, y) \is \JId{X}{C}(c, x, x', p, y)$.
\end{proof}

The transport lemma is sometimes called the type-theoretic principle of the
\indexed{indiscernibility of identicals}.
It says that, if two terms of a type are equal, then
no predicate (i.e.\ type family)
can distinguish them.

As a further corollary of the transport lemma, we obtain the following.
\begin{lemma}\label{lem:ap}
  For every function $f \oftype \ArrTy{X}{Y}$, we have an inhabitant
  \nomenclature{\(\ActOnPathSym_{f}\)}{action of function $f$ on paths (from \cref{lem:ap})}
  \begin{equation*}
    \ActOnPathSym_{f} : \IdTyWithType{X}{x}{x'} \to \IdTyWithType{Y}{f(x)}{f(x')}.
  \end{equation*}
\end{lemma}
\begin{proof}
  Let $f : X \to Y$.
  Define the type family $Z(x_1, x_2) \is \IdTyWithType{Y}{f(x_1)}{f(x_2)}$.
  We then use \cref{lem:transport} to define
  \(\ActOnPathSym_{f}(x, x', p) \is \TransportSym^{Z(x, \blank)}(x, x', p, \Refl{f(x)})\).
\end{proof}

The term $\ActOnPathSym$ is an abbreviation of the term
\emph{\textbf{a}ction on \textbf{p}aths},
springing from the homotopical interpretation of the identity type.
We will discuss in the next section.

\begin{lemma}\label{lem:id-is-symmetric-and-transitive}
  For every type\/ $X$,
  the identity type\/ $\IdTyWithType{X}{(\blank)}{(\blank)}$
  is symmetric and transitive.
\end{lemma}
\begin{proof}
  Let $X \oftype \UU$ be a type.

  For symmetry, consider two inhabitants $x, y \oftype X$ and let
  $p \oftype \IdTyWithType{X}{x}{y}$. We need to construct
  an inhabitant of type $\IdTyWithType{X}{y}{x}$.
  We define the type family $P(z) \is \IdTyWithType{X}{z}{x}$ and appeal
  to $\TransportSym$.
  It is easy to see that we have an inhabitant
  \begin{equation*}
    \TransportSym^{P}(x, y, p, \Refl{x}) \oftype \IdTyWithType{X}{y}{x}\text{.}
  \end{equation*}

  For transitivity, let $x, y, z \oftype X$, and consider proofs
  $p \oftype \IdTyWithType{X}{x}{y}$ and $q \oftype \IdTyWithType{X}{y}{z}$.
  We need to construct an inhabitant of type $\IdTyWithType{X}{x}{z}$.
  We define $P(w) \is \IdTyWithType{X}{x}{w}$.
  The transport lemma then gives us
  \begin{equation*}
    \TransportSym^{P}(y, z)
    \oftype
    \IdTyWithType{X}{y}{z} \to \IdTyWithType{X}{x}{y} \to \IdTyWithType{X}{x}{z}\text{,}
  \end{equation*}
  using which we can easily construct an inhabitant of
  the type $\IdTyWithType{X}{x}{z}$
  as
  \begin{equation*}
    \TransportSym^{P}(y, z, q, p) \oftype \IdTyWithType{X}{x}{z}\text{.}%
  \end{equation*}%
\end{proof}

\subsection{Identity type of \VSigmaType{}s}
\label{sec:id-for-product-types}

Consider two pairs $(x_1, y_1)$ and $(x_2, y_2)$ inhabiting
the dependent sum type $\SigmaType{x}{X}{Y(x)}$
for some $X \oftype \UU$ and type family $Y \oftype \ArrTy{X}{\VV}$.
It is natural to expect that the identity type \[\IdTy{(x_1, y_1)}{(x_2, y_2)}\]
be inhabited if and only if the identity types between the two components are
inhabited.
This indeed turns out to be the case in type theory, although some care has to
be taken due to the dependency in $Y(x)$.

\begin{lemma}\label{lem:id-sigma-logical-equivalence}
  Let $X \oftype \UU$
  and let $Y \oftype \ArrTy{X}{\VV}$.
  Consider two pairs $\Pair{x_1}{y_1}, \Pair{x_2}{y_2}$,
  inhabiting the type $\SigmaType{x}{X}{Y(x)}$.
  We have that the identity type $\IdTy{(x_1, y_1)}{(x_2, y_2)}$ is logically
  equivalent to the type
  \begin{equation*}
    \SigmaType{p}{\IdTyWithType{X}{x_1}{x_2}}{\IdTy{\TransportSym^Y(x_1, x_2, p, y_1)}{y_2}}\text{.}
  \end{equation*}
\end{lemma}
\begin{proof}
  Let $X \oftype \UU$ and $Y \oftype \ArrTy{X}{\VV}$.
  Consider two inhabitants $\Pair{x_1}{y_1}, \Pair{x_2}{y_2}$ of type
  $\SigmaType{x}{X}{Y(x)}$.

  ($\Rightarrow$) By path induction,
    it suffices to show that the implication holds for $\Refl{\Pair{x_1}{y_1}}$.
    The pair \(\Pair{\Refl{x_1}}{\Refl{y_1}}\) then inhabits the type
    \[\SigmaType{p}{\IdTyWithType{X}{x_1}{x_1}}{\IdTy{\TransportSym^Y(x_1, x_1, p, y_1)}{y_1}}\text{.}\]

  ($\Leftarrow$)
    By the induction principle of the \(\SigmaTypeSym\)-type and path induction,
    we need to show the implication for
    \(\Refl{x_1} \oftype \IdTy{x_1}{x_1}\) and
    $\Refl{y_1} \oftype \IdTy{\TransportSym^Y(x_1, x_1, \Refl{x_1}, y_1)}{y_1}$.
    The constructor \(\Refl{\Pair{x_1}{y_1}}\) inhabits
    the type $\IdTy{\Pair{x_1}{y_1}}{\Pair{x_1}{y_1}}$.
\end{proof}

We will later show that the above logical equivalence can be extended to a full
\indexed{equivalence of types}.
As pinning down the right notion of type equivalence will require some
discussion, however, we postpone this fact for now.

\subsection{Homotopical interpretation of the identity type}
\label{sec:id-homotopy}

The key insight that gives rise to \VUM{} arises from the following semantic
interpretation of types:
\begin{enumerate}
  \item a type $X$ is viewed as a space (up to homotopy equivalence), and
    under this view,
  \item the identity type $\IdTyWithType{X}{x}{y}$ can be interpreted as the
    type of \indexedCE{paths}{path} from point $x$ to point $y$ in space $X$.
\end{enumerate}
In this context,
\begin{itemize}
  \item transitivity of equality corresponds to \indexed{path composition},
  \item symmetry of equality corresponds to \indexed{path inversion},
  \item the $\JSym$-rule is interpreted as a form of \indexed{path induction}.
\end{itemize}
To highlight this view, we denote the inhabitant of $\IdTyWithType{X}{y}{x}$,
obtained from some $p \oftype \IdTyWithType{X}{x}{y}$ by
symmetry (\cref{lem:id-is-symmetric-and-transitive}),
by $\PathInv{p}$.
Similarly, given two proofs of identity
\(p \oftype \IdTyWithType{X}{x}{y}\)
and
\(q \oftype \IdTyWithType{X}{y}{z}\),
we denote the inhabitant of $\IdTyWithType{X}{x}{z}$, obtained by
transitivity of identity (\cref{lem:id-is-symmetric-and-transitive}),
by
\begin{equation*}
  \PathComp{p}{q} \oftype \IdTyWithType{X}{x}{z}\text{.}
\end{equation*}%
\nomenclature{\(\PathComp{p}{q}\)}{composition of paths \(p\) and \(q\)}

In algebraic topology, spaces are classified with respect to the complexity of
their homotopy groups. The above observation, that the identity type can be read
as the type of paths, paves the way to the importation of this technique into
type theory,
giving rise to the idea of classifying types with respect to the complexity of
their identity types.
However, some caution needs to be exercised to ensure that one's
type-theoretical setting is suitable for such an interpretation.

\subsubsection{Uniqueness of identity types}
\label{sec:uip}

Up until this point, the only means of constructing inhabitants of the identity
type that we have seen is the $\ReflSym$ constructor from \cref{defn:id-type}.
Therefore, there appears to be just one way to construct a path: using
$\Refl{x}$ to construct a proof $\IdTyWithType{X}{x}{x}$.
In other words, the proof structure of identity types appears, prima facie, to
be trivial.

The statement that there is at most one way to prove equality is known as
the \VAxiomK{}~\cite{ts-habilitation}.
\begin{definition}[The \VAxiomK{}]\label{defn:axiom-K}
  Given a type $X$, define
  \begin{equation*}
    \SatisfiesAxiomK{X}
    \quad\is\quad
    \PiTy{x, y}{X}{\PiTy{p,\, q}{\IdTyWithType{X}{x}{y}}{\IdTy{p}{q}}}\text{.}
  \end{equation*}
  \defineCE{The \VAxiomK{}}{\VAxiomK{}} then says that every type\/ $X$
  satisfies \(\SatisfiesAxiomK{X}\).
\end{definition}
The \VAxiomK{} is known to be \emph{independent} of, but consistent with, the
basic type theory that we presented in \cref{sec:mltt}.
We therefore have the option to adopt it or not.
Adopting this axiom results in a type theory in which a rich homotopical
interpretation of the identity type is not possible, since this axiom asserts
that all types have \emph{trivial path types}.

This is the bifurcation point at which the vision of univalent type theory
emerges:
by refraining from adopting the \VAxiomK{}, we keep open the option of
\emph{adopting certain other axioms that contradict it}
by providing new ways of constructing inhabitants of the identity type.
These are:
\begin{enumerate}
  \item The \indexed{univalence axiom}, which allows us to construct nontrivial
    inhabitants of the identity type~$\IdTyWithType{\UU}{A}{B}$, between types
    $A, B \oftype \UU$.
  \item \indexedCE{Higher inductive types}{higher inductive type},
    which are types that come equipped with identity constructors other than
    $\ReflSym$.
\end{enumerate}
In \cref{example:universes-are-not-sets}, we will look at a concrete
counterexample to the \VAxiomK{}, constructed using \VUniAx{}.

\subsubsection{First look at homotopy levels}

In a type-theoretical setting where nontrivial identity types are possible,
types can be sensibly classified with respect to the complexity of their
identity types.
This leads to the hierarchy of
\emph{homotopy levels} of types~\cite[\textsection 7.1]{hott-book}.

Types with the simplest possible homotopy structure are the contractible types.
\begin{definition}[Contractible type~{\cite[Definition~3.11.1]{hott-book}}]\label{defn:contractible-type}
  A type $X$ is called \define{contractible}
  (also called \define{singleton})
  if it has exactly one inhabitant, i.e.\ the type
  \begin{equation*}
    \IsContr{X}
      \quad\is\quad
        \SigmaType{c}{X}{\PiTy{x}{X}{\IdTyWithType{X}{x}{c}}}
  \end{equation*}\nomenclature{\(\IsContr{X}\)}{type $X$ is contractible}%
  is inhabited.
\end{definition}
This unique inhabitant is sometimes called
the \define{centre of contraction}
of type $X$.

After contractible types, we have the propositional types, which are types that
have \emph{at most} one element:
\begin{definition}\label{defn:prop}
  A type $X$ is called a \define{proposition}
  (also called \define{subsingleton})
  if it has at most one inhabitant, i.e.\ it satisfies
  \begin{equation*}
    \IsProp{X} \quad\is\quad \PiTy{x, y}{X}{\IdTyWithType{X}{x}{y}}\text{.}
  \end{equation*}
\end{definition}

We define the type of propositional types as
\begin{equation*}
  \hprop{\UU} \quad\is\quad \SigmaType{X}{\UU}{\IsProp{X}}\text{.}
\end{equation*}%
\nomenclature{\(\hprop{\UU}\)}{type of propositions in universe \VUni{\UU}}
Throughout the \VHoTTUF{} literature, this is sometimes denoted
$\mathsf{hProp}_{\UU}$.

\begin{example}
  The unit type $\UnitTy$ is contractible,
  with centre $\TTUnit \oftype \UnitTy$.
  It follows easily from the elimination principle of $\UnitTy$ that every
  inhabitant of it is equal to $\TTUnit$.
  The empty type $\EmptyTy$ is a proposition but is not contractible since it
  has no inhabitants.
\end{example}

\begin{example}\label{example:contractibility-initial-and-terminal}
  For every type $X$,
  the types $\ArrTy{X}{\UnitTySym}$ and $\ArrTy{\EmptyTy}{X}$ are
  both contractible (assuming function extensionality).
  The former has the function
  $!_{\UnitTySym}(x) \is \TTUnit$ as its centre of contraction:
  it is easy to see that every $f \oftype \ArrTy{X}{\UnitTySym}$
  is equal to $!_{\UnitTySym}$, since
  $f(x) \IdTySym \TTUnit \IdTySym\ !_{\UnitTySym}(x)$.
  As to the latter, there is a function~$\ArrTy{\EmptyTy}{X}$ given by the
  induction principle of the type $\EmptyTy$, which is vacuously equal
  to every other such map.
\end{example}

\begin{example}\label{lem:contractible-types-are-propositions}
  Every contractible type is a proposition, since every pair of inhabitants
  are equal to the centre of contraction and thus to each other.
\end{example}

\begin{lemma}\label{lem:inhabited-proposition-iff-contractible}
  A type\/ $X$ is contractible
  if and only if
  it is an inhabited proposition.
\end{lemma}

The next homotopy level is that of \emph{sets}: types with trivial identity
types.

\begin{definition}\label{defn:hset}
  A type $X$ is called a \defineCE{set}{set (homotopy level)}
  (or, more explicitly, \define{homotopy set})
  if it satisfies $\SatisfiesAxiomK{X}$.
  In other words, a set is a type $X$ with the property that,
  for every pair of inhabitants $x, y \oftype X$,
  the identity type~\(\IdTyWithType{X}{x}{y}\) is a proposition.
\end{definition}

Later on, we will discuss the general definition of this hierarchy, defining
the generalized notion of an
\indexed{\(n\)-type}~\cite[Definition~7.1.1]{hott-book}.
For now, however, this preliminary explanation of the core ideas of \VUTT{} will
be sufficient to start developing the basics of our foundational setting.

\subsection{Structure vs.\ property}
\label{sec:str-vs-prop}

If a type $X$ is a proposition in the sense of \cref{defn:prop},
we think of it as expressing a \indexed{property}.
The idea is that propositional types are naturally \indexed{proof irrelevant},
as the propositionality of a type amounts to the assertion that there cannot be
two distinct proofs of it.
If a type $X$ is not a proposition, this means that it bears nontrivial proof
\indexed{structure}.
A careful distinction between structure and property has to be maintained in
\VUTT{}.

\begin{example}
  We translated the logical statement from \cref{example:every-nat-is-either-zero-or-succ}
  into type theory as
  \begin{equation}\label{eqn:statement-translation}
    \PiTy{%
      n
    }{%
      \NatTy
    }{%
      \SumTy{%
        \paren{\IdTy{n}{\NatZero}}
      }{%
        \paren{\SigmaType{m}{\NatTy}{\IdTy{n}{\NatSuccSym(m)}}}
      }
    }\text{,}
    \tag{\dag}
  \end{equation}
  following the Curry-Howard view of propositions as types.
  For every $n \oftype \NatTy$, the type
  \begin{equation*}
    \SumTy{\paren{\IdTy{n}{\NatZero}}}{\paren{\SigmaType{m}{\NatTy}{\IdTy{n}{\NatSuccSym(m)}}}}\text{,}
  \end{equation*}
  is a \emph{property} of the type of natural numbers.
  This implies that the type (\ref{eqn:statement-translation}) is also a
  property.
\end{example}

Let us now look at an example of a type, obtained through the Curry-Howard
principle, that is \emph{not} a proposition.
\begin{example}\label{ex:root}
  Let
  \(\alpha \oftype \Endomap{\NatTy}\)
  and consider the statement
  \(\ExistsType{i}{\NatTy}{\IdTy{\alpha(i)}{\NatZero}}\text{,}\)
  expressing that $\alpha$ has a root.
  Translating this into type theory through
  Curry-Howard principle gives us
  \begin{equation*}
    \SigmaType{i}{\NatTy}{\paren{\IdTy{\alpha(i)}{\NatZero}}}\text{.}
  \end{equation*}
  This is, however, structure and \emph{not} property, since a sequence
  $\alpha \oftype \Endomap{\NatTy}$ can have distinct roots.
  It should therefore be thought of as the \emph{type of roots} of $\alpha$
  rather than a property expressing that $\alpha$ has a root.
\end{example}

This example justifies our use of the adjective ``na\"{i}ve'' for the
propositions-as-types principle that we presented in \cref{sec:curry-howard}.
In a context where we are carefully stratifying types with respect to their
homotopy levels, it is misleading that translations of
logical statements yield types that are not propositions.
In \cref{sec:uf-logic}, we will revisit this problem and explain how \VUTT{}
provides us with the means to circumvent it.

In our development, we will use the propositional extensionality axiom, stating
that logically equivalent properties are \emph{equal}.

\begin{definition}[%
  \AgdaLink{UF.Subsingletons.html\#7815}
  \textsc{Axiom}: Propositional extensionality%
]\label{defn:propext}
  A universe\/ \VUni{\UU} is said to satisfy \define{propositional extensionality}
  if, for every pair of propositions $P, Q \oftype \hprop{\UU}$ with
  $\Impl{P}{Q}$ and $\Impl{Q}{P}$,
  we have an inhabitant of the identity type $\IdTy{P}{Q}$.
  The \emph{propositional extensionality axiom} states that all universes
  satisfy propositional extensionality.
\end{definition}

\section{Decidable types and sets}
\label{sec:decidable-types-and-sets}

In this section, we discuss the notion of a decidable type,
and then present Hedberg's Theorem~\cite{hedberg-coherence,keca}, which states
that all decidable types are sets in the sense of \cref{defn:hset}.

\subsection{Decidability of a type}
\label{sec:decidability}

In logic, the \indexed{law of excluded middle} asserts that every proposition
is either true or false.
If we apply the na\"{i}ve propositions-as-types translation to this, we obtain
\begin{equation*}
  \PiTy{A}{\UU}{\paren{\SumTy{A}{\neg A}}}\text{.}
\end{equation*}
There is no reason for the above type to be inhabited in general as we are
working in a constructive foundational setting.
Those types that satisfy this are called \indexedCE{decidable types}{decidable type}.

\begin{definition}\label{defn:decidable-type}
  A type $X$ is said to be \define{decidable} if we have an inhabitant
  of the sum type \(\SumTy{X}{\neg X}\)\text{.}
\end{definition}

\begin{definition}
  A type $X$ is said to have \define{decidable equality}
  (or to be \defineCE{discrete}{discrete (type)})
  if the identity type $\IdTyWithType{X}{x}{y}$ is decidable,
  for every pair of inhabitants $x, y \oftype X$.
\end{definition}

\begin{example}\label{example:props-have-decidable-equality}
  Every proposition $P$ (in the sense of \cref{defn:prop})
  has decidable equality, as the identity type $\IdTyWithType{P}{p}{q}$
  is a decidable type by virtue of being always inhabited.
\end{example}

\begin{lemma}\label{lem:decidable-props-are-the-booleans}
  A proposition \(P \oftype \hprop{\UU}\) is decidable if and only if
  \(P \IdTySym \TrueProp{\UU}\) or \(P \IdTySym \FalseProp{\UU}\),
  assuming propositional and functional extensionality.
\end{lemma}
\begin{proof}
  The fact that \(\TrueProp{\UU}\) and \(\FalseProp{\UU}\) are
  \indexedpCE{decidable propositions}{decidable proposition} is immediate.
  Let \(P \oftype \hprop{\UU}\) be a decidable proposition.
  If \(P\) is inhabited, it is obvious that \(\TruePropSym \ImplSym P\).
  One can thus easily construct a logical equivalence \(P \BiImplSym \TruePropSym\),
  and use propositional extensionality to conclude that the underlying types
  are equal.
  If \(P\) is not inhabited, which is to say we have an inhabitant of \(\neg P\),
  we have \(P \ImplSym \FalsePropSym\) by construction.
  Again, by propositional extensionality,
  we have that the underlying types of the propositions in consideration are
  equal.
  Moreover, since being proposition is a proposition
  (by \cref{lem:being-prop-is-prop}, assuming function extensionality),
  we may appeal to \cref{lem:id-sigma-logical-equivalence} to conclude
  that \(P \IdTySym \TruePropSym\), in the former case, and
  that \(P \IdTySym \FalsePropSym\), in the latter case.
\end{proof}

\begin{definition}[Na\"{i}ve LEM]\label{defn:naive-lem}
  The \indexedp{na\"{i}ve law of excluded middle} states that every type is decidable.
\end{definition}

This axiom is independent of, but consistent with, the base type theory from
\cref{sec:basic-types}.
It is, however, inconsistent with \VTheUniAx{} as shown in
\cite[Corollary~3.2.7]{hott-book}.

\begin{example}\label{lem:nat-decidable-equality}
  The type\/ $\NatTy$ has decidable equality, witnessed by the type family:
  \begin{align*}
    \text{\textsf{nateq}}\Pair{\NatZero}{\NatZero}       \quad&\is\quad \UnitTy\\
    \text{\textsf{nateq}}\Pair{\NatSuccSym(m)}{\NatZero} \quad&\is\quad \EmptyTy\\
    \text{\textsf{nateq}}\Pair{\NatZero}{\NatSuccSym(n)} \quad&\is\quad \EmptyTy\\
    \text{\textsf{nateq}}\Pair{\NatSuccSym(m)}{\NatSuccSym(n)} \quad&\is\quad
                                                                      \text{\textsf{nateq}}\Pair{m}{n}
  \end{align*}
  It is easy to see that $\text{\textsf{nateq}}(m, n)$ is
  either $\UnitTy$ or $\EmptyTy$.
  Furthermore, it can be shown by straightforward induction that
  $(\IdTyWithType{\NatTy}{m}{n}) \IdTySym \text{\textsf{nateq}}\Pair{m}{n}$.
\end{example}

\subsection{Sethood of decidable types}
\label{sec:hedberg}

There is a well-known theorem in type theory,
due to Michael Hedberg~\cite{hedberg-coherence},
stating that every type with decidable equality is a set.
This is perhaps the most useful proof technique for establishing that certain
types are sets.
Here, we will present Hedberg's Theorem through a recent reformulation of it
due to \citeauthor{keca}~\cite{keca}.
To be more specific, our presentation here follows the \VAgda{} development of
\MHELastName{}~\cite{type-topology-hedberg}.

\begin{lemma}\label{lem:path-decomposition}
  For every type\/ $X$, inhabitant\/ $x \oftype X$,
  and\/ $X$-indexed family
  \begin{equation*}
    \varphi \oftype \PiTy{y}{X}{\Endomap{\IdTyWithType{X}{x}{y}}}
  \end{equation*}
  of endofunctions on the identity type, we have the equality
  \[p \IdTySym \PathComp{\PathInv{\varphi_x(\Refl{x})}}{\varphi_y(p)}\text{,}\]
  for every\/ $y \oftype X$ and every
  proof of identity\/ $p \oftype \IdTyWithType{X}{x}{y}$.
\end{lemma}
\begin{proof}
  Straightforward path induction.
\end{proof}

\begin{definition}[Weakly constant function]
  A function $f \oftype \ArrTy{X}{Y}$ is called \define{weakly constant}
  if $\IdTyWithType{Y}{f(x)}{f(x')}$,
  for every pair of inhabitants $x, x' \oftype X$.
\end{definition}

\begin{lemma}[Hedberg's Lemma]\label{lem:hedberg}
  For every type\/ $X$ and every\/ $x \oftype X$,
  if there is a family of
  weakly constant endofunctions
  \[\varphi \oftype \PiTy{y}{X}{\paren{\Endomap{\IdTyWithType{X}{x}{y}}}}\text{,}\]
  then the identity type $\IdTyWithType{X}{x}{y}$ is a proposition,
  for every $y \oftype X$.
\end{lemma}
\begin{proof}
  Let $X$ be a type and consider an inhabitant $x \oftype X$.
  Suppose that, for every $y \oftype X$, there is a weakly constant
  endofunction
  \begin{equation*}
    \varphi_y \oftype \Endomap{\IdTyWithType{X}{x}{y}}\text{.}
  \end{equation*}
  To see that the type $\IdTyWithType{X}{x}{y}$ is a proposition
  for every $y \oftype X$,
  consider some $y \oftype X$ and
  let $p, q \oftype \IdTyWithType{X}{x}{y}$.
  \begin{align*}
    p \quad&\IdTySym\quad \PathComp{\PathInv{\varphi(\Refl{x})}}{\varphi(p)} &[\text{\cref{lem:path-decomposition}}]\\
      \quad&\IdTySym\quad \PathComp{\PathInv{\varphi(\Refl{x})}}{\varphi(q)} &[\varphi\ \text{is weakly constant}]\\
    \quad&\IdTySym\quad q                                        &[\text{\cref{lem:path-decomposition}}]
  \end{align*}
\end{proof}

\begin{theorem}[{Hedberg~\cite{hedberg-coherence}}]
  Every type with decidable equality is a set.
\end{theorem}
\begin{proof}
  Let $X$ be a type with decidable equality.
  We need to show that the identity type~$\IdTyWithType{X}{x}{y}$ is a
  proposition, for every~$x, y \oftype X$.
  It suffices, by \cref{lem:hedberg}, to construct a family
  \begin{equation*}
    \PiTy{x, y}{X}{\Endomap{\IdTyWithType{X}{x}{y}}}
  \end{equation*}
  of weakly constant endofunctions.
  Let $x, y \oftype X$.
  We show that there is a weakly constant endofunction on
  $\IdTyWithType{X}{x}{y}$ by case analysis on its proof of decidability.
  We have two cases:
  either we have a proof of $\neg (\IdTyWithType{X}{x}{y})$
  or we have some inhabitant~$p \oftype \IdTyWithType{X}{x}{y}$.
  In the former case, it is easy to see that the obvious map
  $\ArrTy{\IdTyWithType{X}{x}{y}}{\ArrTy{\EmptyTy}{\IdTyWithType{X}{x}{y}}}$
  is weakly constant.
  In the latter case, the mapping \(x \mapsto p\),
  which sends everything to the identity proof \(p\),
  constitutes a weakly constant endomap.
\end{proof}

\begin{example}
  The type $\NatTy$ of natural numbers is a set since
  it has decidable equality
  as shown in \cref{lem:nat-decidable-equality}.
\end{example}

\begin{example}\label{example:props-are-sets}
  Every proposition is a set, since every proposition has decidable equality
  by \cref{example:props-have-decidable-equality}.
\end{example}

We also note the following facts as useful corollaries of the above examples.

\begin{lemma}\label{lem:being-prop-is-prop}
  For every type\/ $X$, the type\/ $\IsProp{X}$ is a proposition.
\end{lemma}
\begin{proof}
  Follows from
  \cref{lem:homotopy-closure-under-pi} and \cref{example:props-are-sets}.
\end{proof}

\begin{lemma}\label{lem:being-contractible-is-a-proposition}
  For every type\/ $X$,
  the type\/ $\IsContr{X}$,
  expressing that\/ $X$ is contractible,
  is a proposition
\end{lemma}
\begin{proof}
  Let $X$ be a type and consider two proofs
  $\Pair{c_1}{\varphi_1}$, $\Pair{c_2}{\varphi_2}$
  that it is contractible.
  By \cref{lem:id-sigma-logical-equivalence}, it suffices to construct
  a proof $p \oftype \IdTy{c_1}{c_2}$
  and a proof \[q \oftype \IdTy{\TransportSym(c_1, c_2, p, \varphi_1)}{\varphi_2}\text{.}\]
  The function $\varphi_2$ gives a proof $\varphi_1(c_2) \oftype \IdTy{c_1}{c_2}$.
  Since propositions are sets by \cref{example:props-are-sets},
  we must have that
  $\TransportSym(c_1, c_2, \varphi_1(c_2), \varphi_1) \IdTySym \varphi_2$
  by function extensionality.
\end{proof}

\section{Function extensionality and invertible functions}
\label{sec:funext}

In this section, we discuss the function extensionality axiom as well as the
notions of invertible and left-cancellable maps.

\subsection{The function extensionality axiom}

It is reasonable to expect that, in type theory, functions satisfy the principle
of \indexed{function extensionality} with respect to the identity type. That is
to say, given a pair of functions~$f, g \oftype \ArrTy{X}{Y}$,
the identity type $\IdTyWithType{(\ArrTy{X}{Y})}{f}{g}$ should be
logically equivalent to the type
\begin{equation*}
  \PiTy{x, x'}{X}{\IdTyWithType{Y}{f(x)}{f(x')}}\text{,}
\end{equation*}
expressing that functions $f$ and $g$ are extensionally equal.
Unfortunately, however, this is not provable~\cite[Theorem~3.17]{ts-habilitation}
in the base type theory that we presented in \cref{sec:mltt}.
The usual solution for this is to postulate function extensionality in the form
of an \indexed{axiom} in type theory (as we discussed in \cref{sec:axioms}).

To discuss this axiom, we start by defining the pointwise equality of two
functions.
\begin{definition}[Pointwise equality]\label{defn:ext-eq}
  Two dependent functions $f, g \oftype \PiTy{x}{X}{Y(x)}$ are said to be
  \define{extensionally equal} or \define{pointwise equal} if the type
  \begin{equation*}
    \ExtEq{f}{g} \quad\is\quad \PiTy{x}{X}{\IdTyWithType{Y(x)}{f(x)}{f(x')}}
  \end{equation*}
  \nomenclature{\(\ExtEq{f}{g}\)}{extensional equality of functions $f$ and $g$}%
  is inhabited.
\end{definition}

Under the homotopical interpretation of the identity type, this is precisely a
homotopy between $f$ and $g$.

\begin{definition}[\textsc{Axiom}: Na\"{i}ve function extensionality]\label{defn:funext-naive}
  The \define{na\"{i}ve function extensionality} axiom states that
  $\ExtEq{f}{g}$ implies $\IdTy{f}{g}$,
  for every pair of functions $f, g \oftype \ArrTy{X}{Y}$.
\end{definition}

Similar to the examples discussed in \cref{sec:str-vs-prop},
the above formulation is the na\"{i}ve form of the function extensionality
axiom.
As it is possible for the codomains of the functions in consideration to have
nontrivial identity types,
they can have different homotopies witnessing their extensional equality,
meaning this is \emph{structure} and not \emph{property} in general.
We will give examples of such higher groupoids in
\cref{sec:universe-is-not-a-set} and in \cref{sec:hit}.

\subsection{Quasi-inverses}
\label{sec:quasi-inverses}

The interpretation of pointwise equality as homotopy
suggests the following notion of homotopy equivalence between types.

\begin{definition}[Quasi-inverse~{\cite[\textsection 4.1]{hott-book}}]\label{defn:qinv}
  A function $g \oftype \ArrTy{Y}{X}$ is said to be
  the \define{quasi-inverse}
  of a function $f \oftype \ArrTy{X}{Y}$
  if there are homotopies
  \begin{equation*}
    \ExtEq{\Comp{g}{f}}{\FnId{X}}
    \quad\text{and}\quad
    \ExtEq{\Comp{f}{g}}{\FnId{Y}}\text{.}
  \end{equation*}
  Given a function $f \oftype \ArrTy{X}{Y}$,
  its type of quasi-inverses is defined to be%
  \nomenclature{\(\QInv{f}\)}{the type of quasi-inverses of function $f$}%
  \begin{equation*}
    \QInv{f} \quad\is\quad
    \SigmaType{%
      g
    }{%
      \ArrTy{Y}{X}
    }{%
      \ProdTy{
        \paren{\ExtEq{\Comp{g}{f}}{\FnId{X}}}
      }{
        \paren{\ExtEq{\Comp{f}{g}}{\FnId{Y}}}
      }
    }\text{.}
  \end{equation*}
\end{definition}

\begin{definition}\label{defn:invertibility}
  A function $f \oftype \ArrTy{X}{Y}$ is said to be \define{invertible}
  if we have an inhabitant of $\QInv{f}$.
\end{definition}

Once again, this is the na\"{i}ve form of the concept in consideration, and will
not be our official definition. The reason is that, since extensional equality
is structure and not property, there could be distinct witnesses of the fact
that a function $g \oftype \ArrTy{Y}{X}$ constitutes inverse to some function
$f \oftype \ArrTy{X}{Y}$.
Nevertheless, it will be convenient to have pinned down with these na\"{i}ve
forms for two reasons:
\begin{enumerate}
  \item They are \emph{logically equivalent} to their refined forms.
  \item For the homotopy level of sets (in the sense of \cref{defn:hset}), they
    are \emph{equivalent} to their refined versions.
\end{enumerate}
In \cref{sec:embeddings-and-equivalences},
we will refine the notion of invertible map to that of \indexed{equivalence},
which is a property for arbitrary types and not just sets.

\subsection{Left-cancellable maps}

Yet another useful albeit na\"{i}ve notion is that of
a \emph{left-cancellable map}.

\begin{definition}[Left-cancellability~{\cite[139]{hott-book}}]
\label{defn:left-cancellability}
  A function $f \oftype \ArrTy{X}{Y}$ is called \define{left-cancellable} if
  we have a map
  \begin{equation*}
    \ArrTy{\IdTyWithType{Y}{f(x)}{f(x')}}{\IdTyWithType{X}{x}{x'}}\text{,}
  \end{equation*}
  for every pair of inhabitants $x, x' \oftype X$.
\end{definition}

Similar the notion of invertible map, left-cancellability is not a proposition
in general, but is always a proposition when the domain is a set.
In \cref{sec:embeddings}, we will refine this to
the notion of \indexed{embedding},
analogously to our refinement of invertible maps to equivalences.

\section{Equivalences and embeddings}
\label{sec:embeddings-and-equivalences}

We presented the notions of
\indexedp{invertible} and \indexedp{left-cancellable} maps
in \cref{sec:quasi-inverses}, and discussed the fact that these are not
properties for arbitrary \(n\)-types.
The problem here is that the notion of invertible function is the appropriate
notion of equivalence only for \emph{sets}, and it does not generalize to
arbitrary \(n\)-types.
In this section, we will look at the refined versions of these notions that are
well-behaved for all types.

\subsection{Equivalences}

The notion of equivalence given in \cref{defn:equivalence} was first formulated
by Voevodsky.

\begin{definition}[{\cite[Definition~4.2.4]{hott-book}}]
  The \define{fibre} of a function $f \oftype \ArrTy{X}{Y}$
  over a point $y \oftype Y$
  is the type
  \begin{equation*}
    \Fibre{f}{y} \quad\is\quad \SigmaType{x}{X}{\IdTyWithType{Y}{f(x)}{y}}\text{.}
  \end{equation*}
\end{definition}

\begin{definition}[{\cite[Definition~4.4.1]{hott-book}}]\label{defn:equivalence}
  A function $f \oftype \ArrTy{X}{Y}$ is called an \define{equivalence} if%
  \nomenclature{\(\IsEquiv{f}\)}{function $f$ is an equivalence}%
  \[\IsEquiv{f} \quad\is\quad \PiTy{y}{Y}{\IsContr{\Fibre{f}{y}}}\text{.}\]%
\end{definition}

We denote by $\Equiv{X}{Y}$ the type of equivalences between
types $X$ and $Y$.%
\nomenclature{\(\Equiv{X}{Y}\)}{type of equivalences between $X$ and $Y$}
Furthermore, given an equivalence \(e : \Equiv{X}{Y}\),
we denote by \(e^{-1}\) its inverse, and by \(\ulcorner e \urcorner\), its
underlying map \(\ArrTy{X}{Y}\).

\begin{example}\label{example:idequiv}
  For every type\/ $X$,
  the identity function \(\FnId{X} \oftype \ArrTy{X}{X}\) on it
  obviously forms an equivalence.
  We denote this by\/ $\mathsf{idequiv}_X \oftype \Equiv{X}{X}$.
\end{example}

\begin{lemma}\label{lem:being-equiv-is-prop}
  Being an equivalence is a property.
  That is to say,
  the type\/ $\IsEquiv{f}$ is a proposition,
  for every function\/~$f \oftype \ArrTy{X}{Y}$.
\end{lemma}
\begin{proof}
  Follows directly from \cref{lem:homotopy-closure-under-pi} and
  \cref{lem:being-contractible-is-a-proposition}.
\end{proof}

In \cref{sec:quasi-inverses}, we explained that being invertible is not a
proposition for arbitrary types.
The two notions are thus not equivalent.
They are, however, \emph{logically} equivalent.

\begin{lemma}[\AgdaLink{UF.Equiv.html\#17408}]
  A function\/ $f \oftype \ArrTy{X}{Y}$ is \indexedp{invertible}
  if and only if it is an equivalence.
\end{lemma}

Having a well-behaved notion of equivalence for \(n\)-types is crucial
in \VUTT{},
since it enables the formulation of various extensionality axioms.
In \cref{sec:function-extensionality-revisited}, we will present the official
formulation of the function extensionality axiom using the above notion of
equivalence, thereby circumventing the na\"{i}vety involved in the formulation
we gave in \cref{defn:funext-naive}.
Furthermore, this notion of equivalence is also crucial to the statement of
\VTheUniAx{}, since the formulation using invertible maps is provably false
as shown in \cite[Exercise~4.6.(iii)]{hott-book}.

\subsection{Embeddings}
\label{sec:embeddings}

As we refined the notion of invertible map to that of equivalence above, we now
refine the notion of
left-cancellable map (\cref{defn:left-cancellability})
to that of an embedding.

\begin{definition}\label{defn:embedding}
  A function \(f \oftype \ArrTy{X}{Y}\)
  is called an \define{embedding}
  if
  \begin{equation*}
    \IsEmbedding{f} \quad\is\quad \PiTy{y}{Y}{\IsProp{\Fibre{f}{y}}}\text{.}
  \end{equation*}
\end{definition}

\begin{lemma}
  For every function\/ \(f \oftype \ArrTy{X}{Y}\!\), the type\/
  $\IsEmbedding{f}$ is a proposition.
\end{lemma}
\begin{proof}
  Follows immediately from \cref{lem:homotopy-closure-under-pi}
  and the fact that
  being proposition is a proposition (\cref{lem:being-prop-is-prop}).
\end{proof}

\begin{lemma}[\AgdaLink{UF.Embeddings.html\#3221}]
  Every equivalence is an embedding.
\end{lemma}
\begin{proof}
  See \cite[\textsection 3.26]{mhe-intro-uf}.
\end{proof}

\begin{lemma}[\AgdaLink{UF.Embeddings.html\#5107}]
  Every embedding is a left-cancellable map.
\end{lemma}
\begin{proof}
  See \cite[\textsection 3.26]{mhe-intro-uf}.
\end{proof}

\subsection{Function extensionality revisited}
\label{sec:function-extensionality-revisited}

In \cref{sec:funext}, we formulated what we called
the \emph{na\"{i}ve} function extensionality axiom, and promised to give the
official definition later.
Having pinned down the well-behaved notion of equivalence in
\cref{defn:equivalence},
we are now ready to revisit function extensionality as promised.

\begin{definition}[\(\mathsf{happly}\)]
  For every pair of dependent functions $f, g \oftype \PiTy{x}{X}{Y(x)}$,
  we define the map\index{happly}
  \[\mathsf{happly}_{f,g} \oftype \ArrTy{\IdTy{f}{g}}{\ExtEq{f}{g}}\]
  as follows. Let $f, g \oftype \PiTy{x}{X}{Y(x)}$,
  and define the type family $E(h) \is \ExtEq{f}{h}$; we define
  \begin{equation*}
    \mathsf{happly}_{f,g}(p)
      \quad\is\quad \TransportSym^E(f, g, p, \lambda x.\,\Refl{x})\text{.}
  \end{equation*}
\end{definition}

\begin{definition}[%
  \textsc{Axiom}: Function extensionality {\cite[Axiom~2.9.3]{hott-book}}%
]\label{defn:funext}
  The \define{function extensionality axiom} states that,
  for every pair of functions $f, g \oftype \PiTy{x}{X}{Y(x)}$,
  the map \(\mathsf{happly}_{f,g}\) is an equivalence.
\end{definition}

This is clearly a proposition since being an equivalence is a proposition as
shown in \cref{lem:being-equiv-is-prop}.
The above \emph{official} formulation of function extensionality is logically
equivalent to the na\"{i}ve formulation from \cref{defn:funext-naive},
even though the above is formulated for dependent functions and might thus seem
stronger at first glance.
This fact was first observed by Voevodsky as explained in
\cite[\textsection 3.18]{mhe-intro-uf}.

\begin{lemma}
  Function extensionality is logically equivalent to both of the following
  statements:
  \begin{itemize}
    \item \indexedp{na\"{i}ve function extensionality}
      from \cref{defn:funext-naive},
    \item the dependent version of na\"{i}ve function extensionality i.e.\ the
      statement that, for every pair of dependent
      functions\/ $f, g \oftype \PiTy{x}{X}{Y(x)}$, we have that
      \begin{equation*}
        \ExtEq{f}{g}\ \text{implies}\ \IdTy{f}{g}\text{.}
      \end{equation*}
  \end{itemize}
\end{lemma}
\begin{proof}
  See \cite[\textsection 3.18]{mhe-intro-uf}.
\end{proof}

\section{Homotopy levels of types}
\label{sec:hlevels}

In \cref{sec:id-homotopy}, we explained that the homotopical interpretation of
the identity type gives rise to a classification of types with respect to the
complexity of their identity types.
We also discussed the examples of contractible and propositional types as well
as sets.
In this section, we present the notion of an \(n\)-type, which generalizes this
classification to an arbitrary homotopy level $n \oftype \NatTy$.
This stratification of types lies at the heart of \VHoTTUF{}.

\subsection{Definition of the homotopy hierarchy}
\label{sec:defn-homotopy-levels}

We define a predicate
$\IsOfHLevelSym \oftype \ArrTy{\UU}{\ArrTy{\NatTy}{\UU}}$,
expressing what it means for a type~$X\oftype\UU$ to be of some homotopy level
$n \oftype \NatTy$.

\begin{definition}[{\emph{cf.}\ \cite[Definition~7.1.1]{hott-book}}]
  The \define{homotopy level} of a type $X \oftype \UU$ is defined by induction
  as follows:
  \begin{align*}
    \IsOfHLevel{X}{\NatZero}       &\quad\is\quad \text{\textsf{is-contractible}}(X)\\
    \IsOfHLevel{X}{\NatSuccSym(n)} &\quad\is\quad \PiTy{x, y}{X}{\IsOfHLevel{(\IdTyWithType{X}{x}{y})}{n}}%
    \text{.}
  \end{align*}
\end{definition}

A type $X$ is called an \define{$n$-type} if we have that $X$ is of h-level $n$.

\begin{remark}
  Note that
  the \VHoTTBook{}~\cite[Definition~7.1.1]{hott-book}
  uses the indexing convention
  of starting from $-2$,
  which differs from our indexing convention above.
\end{remark}

We can now reconstruct the homotopy levels that we have looked at as special
cases of this.

\begin{lemma}[{\cite[Lemma~3.11.10]{hott-book}}]\label{lem:hlevel-one-iff-prop}
  For every type\/ $X$, the type\/
  $\IsOfHLevel{X}{1}$ is inhabited
  if and only if type\/~$X$ is a proposition.
\end{lemma}
\begin{proof}
  Let $X$ be a type.

  ($\Rightarrow$) This direction is easy to see.
    Suppose $X$ satisfies $\IsOfHLevel{X}{1}$, which is to say that,
    for every pair of inhabitants $x, y \oftype X$,
    the type $\IdTyWithType{X}{x}{y}$ is contractible.
    It immediately follows that $\IdTyWithType{X}{x}{y}$ is inhabited,
    since contractible types are obviously inhabited.

    ($\Leftarrow$) Suppose that $X$ is a proposition,
    meaning the type $\IdTyWithType{X}{x}{y}$ is inhabited,
    for every pair of inhabitants $x, y \oftype X$.
    Since propositions are sets by \cref{example:props-are-sets}, we also know
    that the type $\IdTyWithType{X}{x}{y}$ is a proposition.
    It follows from \cref{lem:inhabited-proposition-iff-contractible}
    that the identity type $\IdTyWithType{X}{x}{y}$ is contractible.
\end{proof}

\begin{corollary}
  For every type\/ $X$,
  we have that\/ $X$ is a set if and only\/ $\IsOfHLevel{X}{2}$.
\end{corollary}

\subsection{Basic properties of homotopy levels}
\label{sec:hlevels-properties}

We now collect some key properties of the hierarchy of homotopy types.
Our goal here is to summarize the theory of $n$-types to the extent that we
need it in our development.
We thus omit many interesting and crucial facts about the hierarchy of homotopy
levels.
The reader interested in a detailed treatment is referred to
\cite[\textsection 7.1]{hott-book}.

As contractible types are propositions by
\cref{lem:contractible-types-are-propositions},
it follows from \cref{lem:hlevel-one-iff-prop} that, for every type $X$, we have
\begin{equation*}
  \IsOfHLevel{X}{0} \ImplSym \IsOfHLevel{X}{1}\text{.}
\end{equation*}
Furthermore, \cref{example:props-are-sets} means that, for every type $X$, we have
\begin{equation*}
  \IsOfHLevel{X}{1} \ImplSym \IsOfHLevel{X}{2}\text{.}
\end{equation*}
In fact, this inclusion into the next homotopy level can be generalized to an
arbitrary homotopy level $n$:
\begin{lemma}[{\cite[Theorem~7.17]{hott-book}}]
  Homotopy levels are cumulative.
  That is to say,
  for every type\/ $X$ and every\/ $n \oftype \NatTy$, we have that
  \begin{equation*}
    \IsOfHLevel{X}{n} \ImplSym \IsOfHLevel{X}{\NatSuccSym(n)}\text{.}
  \end{equation*}
\end{lemma}
\begin{proof}
  Let $X$ be a type.
  We proceed by induction.

  \emph{Base case}: $\NatZero$.
    We have already shown that contractible types are propositions in
    \cref{lem:contractible-types-are-propositions}, and that being proposition
    is equivalent to having a
    homotopy level of~$1$ in \cref{lem:hlevel-one-iff-prop}.

  \emph{Inductive step}: $\NatSuccSym(n')$.
    Assume inductively that every $n'$-type is also an $(n'+1)$-type,
    and suppose $X$ is an $(n'+1)$-type.
    This is to say that the identity type $\IdTyWithType{X}{x}{y}$
    is an $n'$-type, for every pair of inhabitants $x, y \oftype X$.
    By the inductive hypothesis, this implies that the identity type
    $\IdTyWithType{X}{x}{y}$
    is also an $(n'+1)$-type, which means the type~$X$ is an $(n'+2)$-type
    as needed.
\end{proof}

\begin{lemma}[{\cite[Corollary~7.1.5]{hott-book}}]
  For every $n$-type\/ $X$ and every type\/ $Y$,
  if\/ $\Equiv{X}{Y}$ then\/ $Y$ is also
  an\/ $n$-type.
\end{lemma}

\subsubsection{Closure properties}

Homotopy levels are closed under $\SigmaTypeSym$-types.

\begin{lemma}[{\cite[Theorem~7.1.8]{hott-book}}]%
\label{lem:homotopy-closure-under-sigma}
  For every type\/ $X \oftype \UU$, family\/ $Y \oftype \ArrTy{X}{\UU}$,
  and\/ $n \oftype \NatTy$,
  if\/ $X$ is an\/ $n$-type and each\/ $Y_x$ is an\/ $n$-type,
  then the dependent sum\/ $\SigmaType{x}{X}{Y_x}$ is also an\/ $n$-type.
\end{lemma}

Similarly, the dependent function type $\PiTy{x}{X}{Y_x}$ is an $n$-type,
whenever the family~$Y$ consists of $n$-types,
irrespective of the homotopy level of the domain\/ $X$.

\begin{lemma}[{\cite[Theorem~7.1.9]{hott-book}}]
\label{lem:homotopy-closure-under-pi}
  For every type\/ $X \oftype \UU$, family\/ $Y \oftype \ArrTy{X}{\UU}$,
  and every\/ $n \oftype \NatTy$,
  if each\/ $Y_x$ is an\/ $n$-type,
  then the dependent product type\/ $\PiTy{x}{X}{Y_x}$ is also an\/ $n$-type.
\end{lemma}

\section{The univalence axiom}
\label{sec:univalence}

We have discussed two extensionality axioms so far:
\indexedp{propositional extensionality} and \indexedp{function extensionality}.
Since logical equivalence is the special case of \indexed{type equivalence} at
the homotopy level of $n \IdTySym 1$,
propositional extensionality can be interpreted as stating that the identity
type of propositions coincide with the type of equivalences between them.
Similarly, \indexedp{function extensionality} gives a characterization of the
identity type of the function type,
as it implies that there is an equivalence $\Equiv{(\ExtEq{f}{g})}{\IdTy{f}{g}}$.

The \emph{\VUnivalenceAxiom{}} can be construed as a generalized
\indexedp{extensionality axiom},
relating the identity types of arbitrary \(n\)-types to the types of
equivalences between them.
Somewhat surprisingly, \VTheUniAx{} gives \emph{all} standard extensionality
axioms as corollaries: propositional extensionality and function extensionality
both follow from it.

\subsection{Formal definition of \VTheUniAx{}}

To state \VTheUniAx{}, we first define the map $\IdToEqvSym$ from the identity
type of two types into their type of equivalences.

\begin{definition}[\AgdaLink{UF.Equiv.html\#3374}]\label{lem:idtoeqv}
  For every pair of types $X, Y \oftype \UU$,
  we define a map
  \begin{equation*}
    \IdToEqvSym \oftype \ArrTy{\IdTyWithType{\UU}{X}{Y}}{\Equiv{X}{Y}}
    \text{.}
  \end{equation*}%
  \nomenclature{\(\IdToEqvSym\)}{%
    the canonical map from \(\IdTyWithType{\UU}{X}{Y}\) into \(\Equiv{X}{Y}\)%
  }%
  Let $X$ and $Y$ be a pair of types.
  We define the type family
  $E(Z) \is \Equiv{X}{Z}$,
  and use \cref{lem:transport} to define
  \begin{equation*}
    \IdToEqvSym(p) \quad\is\quad \TransportSym^{E}(X, Y, p, \mathsf{idequiv}_X)\text{,}
  \end{equation*}
  where $\mathsf{idequiv}$ refers to the identity equivalence from
  \cref{example:idequiv}.
  The judgemental equality for $\TransportSym$ gives
  $\DefnEq{\IdToEqvSym(\Refl{x})}{\FnIdSym}$.
\end{definition}

\begin{definition}[\AgdaLink{UF.Univalence.html\#412} Univalent universe]
  A universe\/ $\UU$ is called \defineCE{univalent}{univalent universe} if,
  for every pair of types\/ $X, Y \oftype \UU$,
  the map $\IdToEqvSym \oftype (\IdTyWithType{\UU}{X}{Y}) \to (\Equiv{X}{Y})$
  defined above is an \indexedp{equivalence}.
\end{definition}

\begin{definition}[\AgdaLink{UF.Univalence.html\#494} The \VUnivalenceAxiom{}]{}
  The \defineCE{\VUnivalenceAxiom{}}{univalence axiom} states that all universes are univalent.
\end{definition}

We will usually speak of \emph{the} \VUnivalenceAxiom{}, although in practice we
will be assuming a specific universe to be univalent.
We will explicitly specify the universe for which we are assuming univalence
only when extra clarity is required.

The \VUnivalenceAxiom{} is known to be consistent with the base \VMLTT{} that we
presented in \cref{sec:basic-types}.
This was first shown by
Voevodsky's \indexed{simplical sets model}~\cite{kapulkin-lumsdaine-simplicial}
which validates \VAxUA{}.

\begin{theorem}
  Every univalent universe \VUni{\UU} satisfies propositional extensionality
  and function extensionality.
\end{theorem}
\begin{proof}
  It is easy to see that \VUniAx{} implies propositional extensionality,
  as it is clear that equivalence specializes to logical equivalence
  at the homotopy level of $n \IdTySym 1$.
  The proof that it implies function extensionality is due to Voevodsky.
  A proof can be found in \cite[\textsection 4.9]{hott-book}.
\end{proof}

\subsection{Universes are not sets, assuming univalence}
\label{sec:universe-is-not-a-set}

When we discussed the \VAxiomK{} in \cref{sec:id-homotopy}, we mentioned that
it is inconsistent with \VAxUA{} since \VTheUniAx{} gives us the means to
construct nontrivial inhabitants of identity types.
We now look at our first example of a type that is not a set under the
assumption of \VAxUA{}.

\begin{example}\label{example:universes-are-not-sets}
  Universes are \emph{not} sets.
  To see this, consider an arbitrary universe~\VUni{\UU}.
  We take the copy $\TwoTy{\UU}$ of the type of Booleans in universe $\UU$.
  We have a proof
  \[\Refl{\TwoTySym} \oftype \IdTyWithType{\UU}{\TwoTySym}{\TwoTySym}\text{,}\]
  that this type is equal to itself.
  We then have an equivalence
  \[\IdToEqvSym(\Refl{\TwoTySym}) \oftype \Equiv{\TwoTySym}{\TwoTySym}\text{.}\]
  Both sides of this equivalence can be seen to be the identity map by
  the computation rule for $\TransportSym$ (\cref{lem:transport}).
  Using \VTheUniAx{}, we can construct a distinct proof that it is equal
  to itself. Consider the function
  \begin{align*}
    \mathsf{swap}(\BoolZero) \quad&\is\quad \BoolOne\\
    \mathsf{swap}(\BoolOne)  \quad&\is\quad \BoolZero
  \end{align*}
  It is easy to see that this type also gives an
  equivalence of $\TwoTy{\UU}$ with itself, and thus a path
  $p \oftype \IdTyWithType{\UU}{\TwoTySym}{\TwoTySym}$ by univalence.
  If we have that $\IdTy{p}{\Refl{x}}$, however,
  it follows that $\mathsf{swap} \IdTySym \FnIdSym$,
  which is a contradiction since it implies $\BoolZero \IdTySym \BoolOne$.
\end{example}

\subsection{The \VSIP{}}
\label{sec:sip}

It is common practice in mathematics to regard two algebraic structures to be
\emph{the same} when they are isomorphic.
Thanks to the \VUnivalenceAxiom{},
we can obtain a precise mathematical account of this informal practice
in \VUF{}, actually proving that isomorphic structures are equal.

This is achieved through the so-called \indexed{\VSIP{}},
which can be thought of as a generalized extensionality axiom for mathematical
structures.
The key idea is to define a generalized notion of isomorphism of structures,
subject to some conditions ensuring that it is well behaved.
The \VSIP{} then describes the identity type of mathematical structures in terms
of this generalized notion of isomorphism.

There are various forms of the \VSIP{} in the literature.
Three well-known formulations are due to
\citeauthor{coq-nad}~\cite{coq-nad},
Aczel~\cite[\textsection 9.8]{hott-book},
and \MHELastName{}~\cite[\textsection 3.33]{mhe-intro-uf}.
In this section, we present the \VSIP{} of
\MHELastName{} from \emph{loc.\ cit}.
We provide a high-level overview and do not delve into the full technical
details.

When defining a mathematical structure in \VUF{}, we usually start by expressing
the structure in consideration as a type family.
Take the notion of \indexedp{monoid} as an example.
To define the type of monoids, we define a type family
\begin{equation*}
  \mathsf{Monoid}\text{-}\mathsf{Structure} \oftype \UU \to \UU\text{,}
\end{equation*}
that takes a type $X \oftype \UU$ and expresses the type of
monoid structures on it.
This may be defined as a \VSigmaType{}, equipping the carrier type $X$ in
consideration with a unit element~$\epsilon \oftype X$ and a binary operation
$\mathsf{op} \oftype \ArrTy{X}{\ArrTy{X}{X}}$, subject to the monoid axioms:
\begin{align*}
  &\mathsf{Monoid}\text{-}\mathsf{Structure}(X) \is\\
    &\qquad
    \SigmaType{\epsilon}{X}{\SigmaType{\mathsf{op}}{X \to X \to X}{%
        \text{\,\textsf{satisfies-monoid-axioms}}(\epsilon, \mathsf{op})%
    }}\text{.}
\end{align*}
The type of monoids would then be defined as
\begin{equation*}
  \mathsf{Monoid}_{\UU}
    \quad\is\quad
      \SigmaType{X}{\UU}{\mathsf{Monoid}\text{-}\mathsf{Structure}(X)}\text{.}
\end{equation*}

In \MHELastName{}'s formulation of the \VSIP{}
from \cite[\textsection 3.33]{mhe-intro-uf},
the isomorphism of two mathematical structures is expressed by imposing certain
conditions on an \indexed{equivalence} between the carrier types of the
structures in consideration.
Consider two monoids $M$ and $N$.
An equivalence \(e \oftype \Equiv{|M|}{|N|}\)
between their carrier types
is said to be homomorphic if the underlying map
$\ulcorner e \urcorner \oftype \ArrTy{M}{N}$ is a homomorphism of monoids.

We generalize this state of affairs by considering a mathematical structure
as given by a type family \(S \oftype \ArrTy{\UU}{\UU}\).
A notion of isomorphism for $S$-structures is given by a family
\begin{equation*}
  \iota_{s_A,s_B} \oftype \Equiv{A}{B} \to \UU\text{,}
\end{equation*}
for each pair of $S$-structures
$(A, s_A)$ and $(B, s_B) \oftype \SigmaType{X}{\UU}{S(X)}$.
The idea is that $\iota$ expresses what it means for an equivalence to preserve
the $S$-structure.
\MHELastName{}'s \VSIP{} then stipulates the following conditions to ensure that
such a notion of isomorphism is well-behaved:
\begin{enumerate}
  \item For every type $A$ and structure $s \oftype S(A)$,
    the identity equivalence $\mathsf{idequiv}_A$
    (from \cref{example:idequiv})
    satisfies~\(\iota_{s_A,s_A}(\mathsf{idequiv}_A)\).
  \item For every type $A$, and every pair of $S$-structures
    $s_A$ and $t_A$ on it,
    the canonical map
    \(\ArrTy{\IdTy{s_A}{t_A}}{\iota_{s_A,_A}(\mathsf{idequiv}_X)}\)
    is an equivalence.
\end{enumerate}
A structure $S \oftype \ArrTy{\UU}{\UU}$ is called a
\indexed{standard notion of structure}
when it is equipped with a type family $\iota$ satisfying the above conditions.
The \VSIP{} can then be precisely stated as follows:

\begin{theorem}[\AgdaLink{UF.SIP.html\#2226} {\cite[\textsection 3.33]{mhe-intro-uf}}]
  Let\/ $S \oftype \UU \to \VV$ and let\/ $A, B \oftype \UU$.
  Assuming that universe\/ \VUni{\UU} is univalent,
  if\/ $S$ is a standard notion of structure as given by some\/ $\iota$,
  we then have
  \[\Equiv{(\IdTy{(A, s_A)}{(B, s_B)})}{((A, s_A) \cong_\iota (B, s_B))}\text{,}\]
  where $(A, s_A) \cong_\iota (B, s_B)$
  denotes the type of $\iota$-isomorphisms between $(A, s_A)$ and $(B, s_B)$.
\end{theorem}

We will make use of the \VSIP{} in \cref{sec:spec-duality}, where we will
construct an equivalence between the type of large, locally small, and
small-complete spectral locales and the type of small distributive lattices.

\section{Higher inductive types}
\label{sec:hit}

We previously mentioned the notion of an \indexed{inductively generated type},
and explained in \cref{sec:w-types} that the formal account of inductive types
is given by the mechanism of \VWType{}s.
The avoidance of the \VAxiomK{},
which amounts to a generalization from sets to \(n\)-types,
gives us the freedom to generalize the idea of an inductively generated type
to the setting of \(n\)-types as well.
This generalization gives rise to the notion of \indexed{higher inductive type}.

An inductive type is generated by a set of constructors, which inductively
generate the type in consideration.
A higher inductive type is generated by a set of $n$-constructors for each
homotopy level $n \oftype \NatTy$.
For the definition of a type $X$ as a higher inductive type,
\begin{itemize}
  \item the $0$-constructors are the usual constructors generating the inhabitants
    of $X$,
  \item the $1$-constructors augment the path type $\IdTyWithType{X}{x}{y}$ of
    $X$ through the free addition of nontrivial path constructors,
  \item the $2$-constructors augment the path type
    $\IdTyWithType{(\IdTyWithType{X}{x}{y})}{p}{q}$
    of the path type itself by adding path constructors between the path
    constructors.
\end{itemize}
By generalizing this to an arbitrary $n$, we obtain the notion of
$n$-constructor.

In this thesis, we will use only two higher inductive types.
We thus refrain from presenting the complete mathematical account of the general
notion of a higher inductive type, which is a rather technical and intricate
question.
We will instead look at some easy examples and then present the specific higher
inductive types that we will be using in our work.

\subsection{The circle type}

\begin{definition}[{\cite[\textsection 6.4]{hott-book}}]
  The \define{circle} (denoted \(\CircleTy\)) is the higher inductive type
  generated by the following $n$-constructors:
  \begin{itemize}
    \item a single $0$-constructor $\CircleBase \oftype \CircleTy$,
    \item a single $1$-constructor
      $\CircleLoop \oftype \IdTyWithType{\CircleTy}{\CircleBase}{\CircleBase}$.
  \end{itemize}
  This type has the following \indexedp{induction principle}:
  for every type family $P \oftype \ArrTy{\CircleTy}{\UU}$, if there is
  \begin{itemize}
    \item a point $b \oftype P(\CircleBase)$, and
    \item a path $l \oftype \IdTy{\CircleBase}{\CircleBase}$,
  \end{itemize}
  there is a function $f \oftype \PiTy{x}{\CircleTy}{P(x)}$ satisfying the
  equalities
  \begin{itemize}
    \item $f(\CircleBase) \IdTySym b$, and
    \item $\ActOnPath{f}{\CircleLoop} \IdTySym l$.
  \end{itemize}
\end{definition}

\begin{lemma}
  The paths\/
  \(\Refl{\CircleBase}, \CircleLoop \oftype \IdTyWithType{\CircleTy}{\CircleBase}{\CircleBase}\)
  are not equal to each other.
\end{lemma}
\begin{proof}[Proof sketch]
  If these two paths are equal, then it can be shown by the induction principle
  of the circle that
  all types are sets,
  which is not the case as we showed in \cref{example:universes-are-not-sets}.
  See \cite[Lemma~6.4.1]{hott-book} for the full proof.
\end{proof}

\begin{corollary}
  The circle type is not a set, which means it is neither
  contractible nor a proposition.
\end{corollary}

From a homotopical perspective, the fact that $\CircleLoop$ is distinct from
$\Refl{\CircleBase}$ \emph{detects} that the circle type contains a hole.

\subsection{Propositional truncation}
\label{sec:propositional-truncation}

We now look at another example of a higher inductive type that is essential for
the use of \VUTT{} as a mathematical foundation:
the \emph{propositional truncation} of a type.

\begin{definition}\label{defn:propositional-truncation}
  The \define{propositional truncation} is given
  by the type former \[\TruncTy{\blank} \oftype \ArrTy{\UU}{\UU}\text{.}\]
  For every type $X \oftype \UU$, the type $\TruncTy{X}$ is defined by the
  following \(n\)-constructors:
  \begin{itemize}
    \item A single \(0\)-constructor \(\TruncTm{\blank} \oftype X \to \TruncTy{X}\).
    \item A single \(1\)-constructor
      \(\mathsf{trunc} \oftype \PiTy{x, y}{\TruncTy{X}}{\IdTyWithType{X}{x}{y}}\).
  \end{itemize}
  This gives the induction principle
  \begin{equation*}
    \PiTy{P}{\VV}{\IsProp{P} \to (X \to P) \to (\TruncTy{X} \to P)}%
    \text{.}
  \end{equation*}
\end{definition}%
\nomenclature{\(\TruncTy{X}\)}{%
  propositional truncation of type \(X\)%
}%
\nomenclature{\(\TruncTm{(\blank)}\)}{%
  constructor of the propositional truncation type%
}
Observe that the $\mathsf{trunc}$ constructor here says exactly that the type
$\TruncTy{X}$ is a proposition.

\begin{lemma}
  For every pair of types\/ $X \oftype \UU$, and\/ $Y \oftype \VV$,
  we have a function\/ $\ArrTy{\TruncTy{X}}{\TruncTy{Y}}$,
  whenever we have a function\/ $\ArrTy{X}{Y}$.
\end{lemma}
\begin{proof}
  Immediate by the recursion principle of propositional truncation since
  $\TruncTy{Y}$ is a proposition.
\end{proof}

It is easy to see that this defines a functor.
This functor is in fact the \indexed{reflector} of types in a universe\/ $\UU$
into its full subcategory\/ $\hprop{\UU}$ of propositions.

\begin{lemma}
  The propositional truncation\/ $\TruncTy{\blank} \oftype \ArrTy{\UU}{\hprop{\UU}}$
  is the left adjoint to the
  inclusion\/ \(\hprop{\UU} \xhookrightarrow{~\projISym~}\UU\).
\end{lemma}

In \cref{sec:str-vs-prop}, we discussed the distinction between structure and
property.
We will frequently encounter situations where a type expressing some statement
of interest will be structure and \emph{not} a property. We already gave an
example of this in \cref{ex:root}.
The propositional truncation is particularly useful in such situations where we
can obtain the property of interest simply by truncating the structure in
consideration.

\section{Propositions-as-types revisited}\label{sec:uf-logic}

Now that we have introduced higher inductive types and presented the mechanism
of propositional truncation, we are ready to present the \emph{refined} form of
the propositions-as-types interpretation as we promised in
\cref{sec:curry-howard}.

In \cref{ex:root}, we gave an example illustrating that it may be misleading to
think of $\SigmaTypeSym$-types as asserting the \indexed{existence} of objects
in type theory.
This is because a proof of a $\SigmaTypeSym$-type \emph{specifies} the object in
consideration, and is thus structure and not property (in general).
There is a similar problem with interpreting the sum type $\SumTy{P}{Q}$ as the
logical disjunction of propositions $P$ and $Q$. A proof of a sum type contains
not only the information that one of the disjuncts holds, but also the
information of \emph{which disjunct} holds.
It is thanks to propositional truncation that we can form the propositional
versions of these, and thereby view all logical statements as the denotations of
propositional types.

\subsection{The refined propositions-as-types translation}

We showed in \cref{lem:homotopy-closure-under-sigma} that propositions are
closed under product types, meaning we have a function
\begin{equation*}
  (\blank) \meet (\blank) \oftype \hprop{\UU} \to \hprop{\VV} \to \hprop{\UMax{\UU}{\VV}}\text{.}
\end{equation*}
Furthermore, we established in \cref{lem:homotopy-closure-under-pi} that the
dependent product type gives a function
\begin{equation*}
  \PiTySym(\blank, \blank)
  \oftype
  \PiTy{X}{\UU}{\paren{(X \to \hprop{\VV}) \to \hprop{\UMax{\UU}{\VV}}}}
  \text{,}
\end{equation*}
taking a domain of quantification $X \oftype \UU$ and a \VUniHyphen{\VV}valued
predicate on it,
and returning a proposition in universe $\UMax{\UU}{\VV}$.

\begin{table}[h!]
  \footnotesize
  \centering
  \begin{tabular}{c c}
    \toprule
    Logical statement                & Proposition\\\midrule
    Falsum \(\bot\)                  & Empty type \(\EmptyTy\)\\
    Truth \(\top\)                   & Unit type \(\UnitTySym\)\\
    Conjunction \(P \meet Q\)        & Product \(\ProdTy{P}{Q}\)\\
    Disjunction \(P \vee Q\)         & Truncated sum $\TruncTy{\SumTy{P}{Q}}$\\
    Implication \(P \ImplSym Q\)     & Function type \(\ArrTy{P}{Q}\)\\
    Predicate on type $X$            & Family $\ArrTy{X}{\HPropSym}$ of propositions\\
    Universal quantification
    \(\forall_{(x \oftype X)}\, P(x)\) & Dependent product $\PiTy{x}{X}{P(x)}$\\
    Existential quantification
    \(\ExistsType{x}{X}{P(x)}\)      & Truncated dependent sum $\TruncTy{\SigmaType{x}{X}{P(x)}}$\\\bottomrule
  \end{tabular}
  \caption{summary of the refined propositions-as-types principle.}
  \label{fig:refined-curry-howard}
\end{table}

Thanks to the existence of propositional truncations, we can now consider the
truncated sum type,
which gives a function
\(%
  \TruncTy{\SumTy{(\blank)}{(\blank)}}
  \oftype
  \hprop{\UU} \to \hprop{\VV} \to \hprop{\UMax{\UU}{\VV}}
  \text{,}%
\)
and the truncated $\SigmaTypeSym$-type former, which is of type
\[%
  \TruncTy{\SigmaTypeSym(\blank, \blank)}
  \oftype
  \PiTy{X}{\UU}{\paren{(X \to \hprop{\VV}) \to \hprop{\UMax{\UU}{\VV}}}}%
  \text{.}
\]
We think of truncated \VSigmaType{}s as asserting
the \indexed{unspecified existence} of an object.
Using these, we can give a complete translation of logical statements into the
type~$\HPropSym$ of propositions.
We give a summary of this refined form of the propositions-as-types principle in
\cref{fig:refined-curry-howard}.

As an example, we refine
the \indexedp{na\"{i}ve law of excluded middle} (\cref{defn:naive-lem})
through the refined propositions-as-types translation.

\begin{definition}[Law of excluded middle (\textsf{LEM})]
  The \define{law of excluded middle} for universe \VUni{\UU}
  asserts that, for every proposition \(P \oftype \hprop{\UU}\),
  we have an inhabitant of the disjunction $\TruncTy{\SumTy{P}{\neg P}}$.
\end{definition}

We will hereafter use the shorthand notation $\ExistsType{x}{X}{P(x)}$ and $P
\vee Q$ for truncated \VSigmaType{}s and sum types respectively.

\subsection{Specified vs.\ unspecified existence}

As explained before, the distinction between structure and property is a
hallmark feature of \VUTT{}.
Now that we have presented both of the notions of \emph{specified} and
\emph{unspecified} existence that we will be working with, we adopt some
vernacular conventions to distinguish between structure and property.
\begin{itemize}
  \item We refrain from saying ``$x$ is $Y$'' if the type family $Y$ in
    consideration does not consist of propositions.
  \item For the anonymous inhabitation $\TruncTy{A}$ of a type $A$, we say that
    \emph{$A$ is inhabited}.
  \item We say \emph{specified inhabitant} of type $A$ to contrast it with the
    anonymous inhabitation $\TruncTy{A}$. Similarly, we say there is a
    \emph{specified} or \emph{chosen} element to emphasize that we are using
    $\SigmaTypeSym$ instead of $\exists$.
  \item For truncated $\SigmaTypeSym$-types, we use the terminologies
    \emph{there is some} and \emph{there exists some}.
    We also use the phrase
    \emph{there is a specified/unspecified},
    when extra clarity is required.
  \item Unlike the \VHoTTBook{}~\cite{hott-book}, we do not use the adjective
    \emph{mere} for unspecified existence.
\end{itemize}

\subsection{Subsets}
\label{sec:subsets}

As $\HPropSym$ is the (large) type of types that we view as properties,
we think of $\HPropSym$-valued functions as
\indexedCE{predicates}{predicate} or \indexedCE{subsets}{subset}.

\begin{definition}[\AgdaLink{UF.Powerset.html\#350} {\VUniHyphen{\VV}subset}]
  A \define{$\VV$-subset}\index{subset}
  (or \define{\VUniHyphen{\VV}valued subset})
  of a type $X \oftype \UU$ is
  a predicate $S \oftype \ArrTy{X}{\hprop{\VV}}$.
  The \VUniHyphen{\VV}powerset of $X$,
  denoted $\Pow{\VV}{X}$,
  is the type of all \VUniHyphen{\VV}subsets of $X$.%
  \nomenclature{\(\Pow{\VV}{X}\)}{type of \VUniHyphen{\VV}valued subsets of \(X\)}
\end{definition}

\begin{remark}\label{rmk:powerset-is-large}
  It is useful to observe here that,
  for every type $X \oftype \UU$,
  the type
  $\Pow{\VV}{X} \DefnEqSym \ArrTy{X}{\hprop{\VV}}$
  lives in universe \VUni{\UMax{\UU}{\USucc{\VV}}}.
  In particular, the type $\Pow{\UU}{X}$ of all
  \VUniHyphen{\UU}subsets lives in universe \VUni{\USucc{\UU}}.
\end{remark}

The situation here is very similar to topos logic.
In topos theory, one works with a notion of \indexedp{subobject classifier} $\Omega$,
which is axiomatized through a universal property stating that the morphisms
$\mathsf{Hom}(X, \Omega)$
are in bijection with the subobjects of~$X$.

In \VUTT{}, the type $\hprop{\UU}$ plays the r\^{o}le of the subobject
classifier as it classifies \indexedCE{subtypes}{subtype}, which we will discuss
in detail later.
Unlike topos logic, however,
its construction is completely internal to the type theory;
the fact that it classifies subobjects is a theorem and not an axiom.

Another difference with topos logic here is that $\hprop{\UU}$ is not admitted
as a type living in universe \VUni{\UU}, since it quantifies over the types in
universe \VUni{\UU}. This brings us to the subject of predicativity, which we
will discuss soon in \cref{sec:predicativity}.

\subsubsection{Kuratowski finite subsets}

In constructive mathematics, there are several distinct notions of \indexedp{finiteness}.
The two salient ones among them are
\indexed{Bishop finiteness} and \indexedCE{Kuratowski finiteness}{Kuratowski finite}.
We now briefly explain the latter,
as it will be of interest to us in our development of locale theory.

We denote by $\mathsf{Fin}(n)$ the type with exactly $n$ inhabitants,
\nomenclature{\(\mathsf{Fin}(n)\)}{the finite type with exactly \(n\) inhabitants}
which can be constructed in our setting as the sum type consisting of
$n$-many copies of the unit type $\UnitTySym$.
Furthermore, given a subset $P \oftype \ArrTy{X}{\hprop{\UU}}$,
we denote by $\mathbb{T}(P)$ the type $\SigmaType{x}{X}{P(x)}$
i.e.\ the type of all inhabitants of $x$ falling in $P$.

\begin{definition}[Kuratowski finiteness]
\label{defn:kuratowski-finiteness}~

  \begin{itemize}
    \item (\AgdaLink{Fin.Kuratowski.html\#1009})
      A type $X$ is called \define{Kuratowski finite} if
      there exists a surjection $\mathsf{Fin}(n) \twoheadrightarrow X$,
      for some $n \oftype \NatTy$.
      This is called an \define{enumeration} of $X$.
    \item (\AgdaLink{UF.Powerset-Fin.html\#838})
      Given a type $X \oftype \UU$, a subset
      $P \oftype \ArrTy{X}{\hprop{\UU}}$
      is called \emph{Kuratowski finite} if
      the type $\mathbb{T}(P)$ is Kuratowski finite.
  \end{itemize}
\end{definition}

We denote by $\mathsf{KFin}(X)$ the type of Kuratowski finite subsets of a type $X$.%
\nomenclature{\(\mathsf{KFin}(X)\)}{the type of Kuratowski finite subsets of \(X\)}

\begin{lemma}[\AgdaLink{UF.Powerset-Fin.html\#9015} Kuratowski finite induction {\cite[Lemma~4.2.10]{tdj-thesis}}]\label{lem:kuratowski-induction}
  For every type\/ $X \oftype \UU$, and every property\/
  $P \oftype \mathsf{KFin}(X) \to \hprop{\VV}$
  of the collection of Kuratowski finite subsets of\/ $X$,
  to show that\/ $P$ holds for all Kuratowski finite subsets of\/ $X$,
  it suffices to show the following:
  \begin{itemize}
    \item The empty set satisfies\/ $P$.
    \item All singleton sets satisfy\/ $P$.
    \item The property\/ $P$ is closed under binary unions.
  \end{itemize}
\end{lemma}

\section{Predicativity and propositional resizing}
\label{sec:predicativity}

We now start studying the notion of \emph{size} of a type,\index{size (of a type)}
which captures what we mean by the predicativity of our foundational setting.
We start by defining the notions of \VUniSmall{\UU}
and locally \VUniSmall{\UU} types.

\begin{definition}[{\cite{rijke-the-join}}]\label{defn:smallness}
  A type $X \oftype \UU$ is called \define{\VUniSmall{\VV}}\index{small type}
  if
  \begin{equation*}
    \IsVSmall{\VV}{X}
      \quad\is\quad
        \SigmaType{Y}{\VV}{\Equiv{X}{Y}}\text{.}
  \end{equation*}
\end{definition}
\nomenclature{\(\IsVSmall{\VV}{X}\)}{%
  the type expressing that type \(X\) is \VUniHyphen{\VV}-small%
}

\begin{definition}[{\cite{rijke-the-join}}]\label{defn:local-smallness}
  A type $X \oftype \UU$ is called \define{locally \VUniSmall{\VV}} if
  the identity type $\IdTyWithType{X}{x}{y}$ is \VUniSmall{\VV},
  for every pair of inhabitants $x, y \oftype X$.
  In formal notation, we denote this by
  $\mathsf{is}\text{-}\mathsf{locally}\text{-}\VV\text{-}\mathsf{small}(X)$.
\end{definition}

\begin{example}
  Assuming \indexedp{propositional extensionality},
  the type $\hprop{\UU}$ is locally \VUniHyphen{\UU}small.
  This type lives in $\USucc{\UU}$ implying that the identity type
  $\IdTy{P}{Q}$ lives in $\USucc{\UU}$.
  However, an equivalent \VUniHyphen{\UU}small copy of the identity
  type $\IdTy{P}{Q}$ is always given by the type
  $\ProdTy{(\Impl{P}{Q})}{(\Impl{Q}{P})}$,
  which is small since the implication between two \VVSmall{\UU}
  propositions is always \VVSmall{\UU}.
\end{example}

It is natural to ask whether the type
$\IsVSmall{\VV}{X}$
is a proposition.
This turns out to be the case, but somewhat surprisingly, its propositionality
\emph{crucially relies} on \VTheUniAx{}\index{univalence axiom} as captured by
the lemma below.

\begin{lemma}[\AgdaLink{MGS.Equivalence-Induction.html\#1856}]%
\label{lem:being-small-is-prop}
  The following are equivalent:
  \begin{enumerate}
    \item For every pair of universes \VUni{\UU} and \VUni{\VV}, the
      type\/ $\IsVSmall{\VV}{X}$ is a proposition, for every
      type\/ $X \oftype \UU$.
    \item The univalence axiom holds.
  \end{enumerate}
\end{lemma}
\begin{proof}
  A proof can be found in \cite{type-topology-mgs-equivalence-induction}.
\end{proof}

The fact that being small is a property under the assumption of \VAxUA{} will
find an important application in our development of the predicative theory of
spectral locales. To be more specific, this will ensure in
\cref{prop:spectral-is-prop} that one of our definitions of being a spectral
locale (\cref{defn:spectral-locale}) is a property and not structure.

\subsection{Subsets and families}
\label{sec:subsets-and-families}

In \cref{sec:subsets}, we defined a subset of a type as a predicate on the type
in consideration.
We now look at an alternative description of subsets through
\indexedCE{embeddings}{embedding}.

\begin{definition}[\AgdaLink{Slice.Family.html\#286} {\VUniHyphen{\VV}family}]%
\label{defn:family}
  A \define{\VUniHyphen{\VV}family}
  (or \define{\VUniSmall{\VV} family})
  on a type $X$
  is a function $\ArrTy{I}{X}$
  whose domain $I$ lives in \VUni{\VV}.
\end{definition}

We denote by $\mathsf{Fam}_{\VV}(X)$ the type of \VUniHyphen{\VV}families on
some type \(X\).%
\nomenclature{\(\mathsf{Fam}_{\VV}(X)\)}{%
  the type of \VUniHyphen{\VV}families on type \(X\)%
}
Also, we often use the notation~\(\FamEnum{i}{I}{x_i}\) for a family
\(x \oftype I \to X\).
Moreover, we sometimes use
the comprehension notation~\(\{ x_i \mid i \oftype I, P(i) \}\)
for families of the form \(x \oftype (\SigmaType{i}{I}{P(i)}) \to X\), where
the index type is a \VSigmaType{}.

\begin{definition}[Subfamily]
  Given a \VUniHyphen{\VV}family \(\FamEnum{i}{I}{x_i}\) and
  a map \(\alpha \oftype \ArrTy{J}{I}\), where \(J \oftype \VV\),
  the composition \(x \circ \alpha \oftype \ArrTy{J}{X}\)
  is called the \define{subfamily} obtained by reindexing along~\(\alpha\).
\end{definition}

It is convenient to use the family notation for reindexing maps themselves,
thus writing \(\FamEnum{j}{J}{x_{i_j}}\) for reindexed families.
Accordingly, we will simply speak of a subfamily~\(\FamEnum{j}{J}{x_{i_j}}\)
of a family \(\FamEnum{i}{I}{x_i}\), and will refrain from working explicitly
with a reindexing map \(\alpha : \ArrTy{I}{J}\).

\begin{definition}[\AgdaLink{UF.Embeddings.html\#4622} {\VUniHyphen{\VV}subtype}]\label{defn:subtype}
  A \define{\VUniHyphen{\VV}subtype} of a type $X$ is a \VUniHyphen{\VV}family
  on \(X\)
  that is an \indexedp{embedding}.
\end{definition}

We denote the type of \VUniHyphen{\VV}subtypes by $\EmbFam{\VV}{X}$.
\nomenclature{\(\EmbFam{\VV}{X}\)}{%
  the type of \VUniHyphen{\VV}subtypes of type \(X\)%
}

\begin{lemma}\label{lem:subtype-classification}
  Assuming \VTheUniAx{},
  the type\/ $\hprop{\UU}$ classifies\/ the\/ \VUniHyphen{\UU}subtypes
  of\/ \VUniHyphen{\UU}types.
  In other words, for every type\/ $X \oftype \UU$,
  we have an equivalence \[\Equiv{\EmbFam{\UU}{X}}{\Pow{\UU}{X}}\text{.}\]
\end{lemma}
\begin{proof}
  A proof can be found in \cite[\textsection 3.31]{mhe-intro-uf}.
\end{proof}

A family that is not necessarily an embedding can be thought of as
containing \emph{repetitions}. In other words, if a family is not an embedding,
this means that the index set does not provide a unique representation for the
elements of the subset in consideration.
Whilst the embedding families correspond exactly to subsets,
it turns out to be convenient in practice to drop the embedding requirement and
represent subtypes by families that may have repetitions.
In \cref{sec:set-replacement}, we will discuss \VSRPrinciple{}, which will allow
us to remove such repetitions when the need arises.
This procedure works even in a constructive and predicative setting, provided
that \VSRPrinciple{} is available.
Moreover, in \cref{sec:bases}, we will revisit this question and discuss the
applications of this to our development of locale theory.

\subsection{Propositional resizing axioms}
\label{sec:resizing}

We will frequently encounter situations where we will be working with
a subset \[\ArrTy{X}{\hprop{\USucc{\UU}}}\]
given by large propositions (with respect to some base universe \VUni{\UU}),
and we would like it to be small\index{small (type)} i.e.\ a \VUniHyphen{\UU}valued one.
We previously explained that a hallmark feature of \VUTT{} is
the \emph{internal} definition of the notion of property,
using which we define the type $\hprop{\UU}$
of \VUniHyphen{\UU}propositions.
This is in contrast with the Calculus of Constructions~\cite{coq-huet-coc}%
\index{Calculus of Constructions}
where the universe $\mathsf{Prop}$ of propositions can be made impredicative by
construction thanks to the fact that being proposition is a judgement in
\VCoC{}.

At this point of our foundational setting, there arises the question of how one
can do impredicative mathematics in a setting where we work with $\hprop{\UU}$
as our subtype classifier.
Voevodsky's proposal to address this problem was to consider the so-called
\emph{propositional resizing} rules~\cite[10]{voevodsky-resizing}.
There are slightly different forms of these rules, but in summary,
they state that every proposition $P \oftype \hprop{\UU}$
can be \emph{resized down} to $\hprop{\UZero}$.

Although the resizing rules are convenient,
it is not known whether type theory extended with them remains strongly
normalizing.
This is a long-standing open problem in \VUTT{}.
Because of this, the resizing rules are studied in the form of an axiom.

\begin{definition}[%
  {\emph{cf.}\ \cite[Definition~2]{tdj-mhe-aspects} and \cite[Axiom~3.5.5]{hott-book}}%
]
  ~
  \begin{itemize}
    \item The \define{propositional $(\UU, \VV)$-resizing axiom} says that
      every proposition $P \oftype \hprop{\UU}$ is \VVSmall{\VV}.
    \item The \define{global propositional resizing} axiom says that, for every
      pair of universes \VUni{\UU} and \VUni{\VV},
      the propositional $(\UU, \VV)$-resizing axiom holds.
    \item The \define{$\HPropSym\text{-}(\UU,\VV)$-resizing axiom} states that,
      for every pair of universes \VUni{\UU} and \VUni{\VV},
      the type $\hprop{\UU}$ is \VUniSmall{\VV}.
    \item
      The type $\HPropSym^{\neg\neg}_{\UU}$ denotes the type of propositions
      $P \oftype \hprop{\UU}$ satisfying $\Impl{\neg\neg P}{P}$.
      The \define{$\HPropSym_{\neg\neg}\text{-}(\UU,\VV)$-resizing axiom}
      states that the type $\HPropSym^{\neg\neg}_{\UU}$ is \VVSmall{\VV}.
  \end{itemize}
\end{definition}

\begin{remark}
  Notice that the law of excluded middle immediately gives $\HPropSym$-resizing.
  The law of excluded middle, considered for a universe \VUni{\UU}, amounts to the
  assertion that
  $\hprop{\UU}$ is equivalent to the type $\TwoTy{\UU}$.
  Since the type $\TwoTySym$ has a copy in every universe,
  this automatically allows us to resize $\hprop{\UU}$ as we wish.
  This is to say that, one cannot do predicative mathematics in a classical
  setting.
  To put it another way,
  \emph{predicative mathematics is a stricter branch of constructive mathematics}.
\end{remark}

\begin{lemma}
  For every pair of universes \VUni{\UU} and \VUni{\VV},
  we have that\/ {$\HPropSym\text{-}(\UU,\VV)$-resizing} implies
  the propositional \((\UU, \VV)\)-resizing axiom.
  Furthermore, if it is the case that
  propositional $(\VV, \UU)$-resizing holds,
  we have that propositional $(\UU, \VV)$-resizing
  implies $\HPropSym\text{-}(\UU,\USucc{\VV})$-resizing
\end{lemma}
\begin{proof}
  See
  \cite[Proposition~3]{tdj-mhe-aspects}
  and
  \cite[\textsection 3.36.4]{mhe-intro-uf}.
\end{proof}

In the form of axioms, all of the above are independent of, but consistent with
\VUTT{}.
When we speak of ``the propositional resizing axiom'' without further
specification, we will be referring to the global propositional resizing axiom
from the above definition.

By \indexed{impredicative mathematics} in the context of \VUF{}, we mean
mathematics developed using any of the above forms of resizing.
At this point, we recall our discussion of axioms in type theory from
\cref{sec:axioms}:
an axiom in type theory is an inhabitant of a type that is declared to exist
independently of the type's introduction rules, which are ensured to have
well-defined computational behaviour.

It is exactly at this point that the motivation for the \indexed{predicativity}
of our foundational setting emerges.
At the time of writing, finding computational rules for the propositional
resizing axioms above is an open problem.
We previously explained that our foundational setting is an axiomatization of a
class of univalent type theories, as we are assuming that certain axioms can be
fulfilled instead of taking them as postulates in a fixed, concrete
implementation of a type theory.
To obtain computational applications, we could always take a concrete
implementation of a type theory fulfilling these axioms,
such as \VCTT{}~\cite{cubical-type-theory} in which all of the axioms we use can
be given computational rules.
This is owing to the fact that \VCTT{}~\cite{cubical-type-theory}
provides a computational interpretation for \VTheUniAx{} as well as higher
inductive types (at least for the ones we use here).

There is no known type theory (to the author's knowledge), at the time of
writing, that gives a computational interpretation to the propositional resizing
axioms.
In accordance with our motivation to use only axioms whose computational
behaviours are well-understood, we \emph{refrain from using all forms} of the
propositional resizing axioms above. We thus commit to predicative mathematics,
as a further form of our commitment in this thesis to constructive mathematics.

\subsubsection{Propositional resizing and propositional truncations}

Another motivation for considering the propositional resizing axioms is that
they give an alternative way to construct propositional truncations.
The idea here is that we can take the double negation of a type
$\ArrTy{(\ArrTy{X}{\FalsePropSym})}{\FalsePropSym}$,
which will always be a proposition.
This proposition can be thought of as an approximation to the propositional
truncation of $X$. Unfortunately, however, it does not satisfy the universal
property of propositional truncation unless the \indexedp{law of excluded middle}
holds~\cite[\textsection 7]{keca}.

By quantifying over all propositions, however, one can obtain a large
type
\begin{equation}\label{eqn:prop-trunc-impredicative}
  \PiTy{P}{\hprop{\UU}}{\ArrTy{(\ArrTy{X}{P})}{P}}\text{,}
\end{equation}
which does give the propositional truncation of $X$, although it cannot be
admitted as a proposition in the same universe as it involves a large
quantification.
\MHELastName{}~\cite[\textsection 36.5]{mhe-intro-uf} shows that, if
propositional $(\UU, \UZero)$-resizing holds for every universe \VUni{\UU}, then
the above construction satisfies the universal property of truncation in
every universe.

For the purposes of our work, we will not need the details of the construction
here.
We mention this use of propositional resizing in passing as it is an
illustrative example of how higher inductive types can circumvent the need for
impredicativity by playing the r\^{o}le of resizing principles.
We refer the reader \cite[\textsection 7]{keca}
and \cite[\textsection 2.11.1]{tdj-thesis}
for a more detailed discussion.

\section{\VTTheSRPrinciple{}}
\label{sec:set-replacement}

We now consider a principle that will find important applications in our
development of the theory of spectral and Stone locales in
\cref{chap:spec-and-stone}.
Similar to how the higher inductive type of propositional truncation circumvents
the need to work with the impredicative encoding
from (\ref{eqn:prop-trunc-impredicative}),
\VTheSRPrinciple{} serves as a resizing principle that will allow us to avoid
impredicativity at certain points of our development of locale theory.

\begin{definition}[\textsc{Axiom}: Set replacement~{\cite[Axiom~18.1.8]{rijke-intro-to-hott}}]%
\label{defn:set-replacement-principle}
  The \define{set replacement principle}
  states that, for every function $f \oftype \ArrTy{X}{Y}$, if we have that
  \begin{itemize}
    \item $Y$ is a set,
    \item $X$ is \VUniSmall{\UU}, and
    \item $Y$ is locally \VUniSmall{\VV},
  \end{itemize}
  then the type $\mathsf{image}(f)$ is \VUniSmall{(\UMax{\UU}{\VV})}.
\end{definition}

It is useful to explicitly record the following special case.

\begin{corollary}\label{cor:set-replacement-principle}
  Let\/ $X \oftype \UU$ be a type and\/ $Y \oftype \USucc{\UU}$, a set.
  For every function\/ $f \oftype \ArrTy{X}{Y}$,
  to show that\/ $\mathsf{image}(f)$ is \VUniSmall{\UU},
  it suffices to show that\/ $Y$ is locally\/ \VUniSmall{\UU}.
\end{corollary}

An inhabitant of the set replacement principle can be constructed as a higher
inductive type
(as shown by \citeauthor{rijke-the-join}~\cite{rijke-the-join}),
and it is thus a theorem in a foundational setting where all higher inductive
types are available (such as
\VCTT{}~\cite{cubical-type-theory}).
As explained in \cref{sec:axioms}, we will simply assume that our foundational
setting satisfies this principle as we do not commit to a concrete
implementation of \VUTT{}.

We also note a relevant result due to
\VTDJLastName{}~\cite[Theorem~2.11.24]{tdj-thesis},
showing that
\VSRPrinciple{} is logically equivalent to the existence of small set
quotients,
which is another construction requiring the use of higher inductive types.
For us, \VTDJLastName{}'s result means that the above axiom is satisfied in any
foundational setting where small set quotients are available.


\chapter{Basics of point-free topology}
\label{chap:basics}

As we explained in the introduction, point-free topology is an approach to
topology where the fundamental object of study is taken to be a \emph{locale}
rather than a topological space. The latter is a set of \emph{points} equipped
with a \emph{topology}, which is a class of subsets of the points that is
required to be closed under finite intersection and arbitrary union. A
\indexedp{locale}, in contrast, is a notion of space that is characterized
solely by an abstract lattice of opens. The lattice-theoretic structure
abstracting the behaviour of the class of open sets is called a \emph{frame},
and the remarkable insight that nontrivial topological questions can be
entertained by studying frames as spaces can be traced back to the work of
Stone~\cite{stone-1934, stone-1936}.

The motivation for taking a point-free approach to topology is twofold:
\begin{enumerate}
  \item\label{itm:locale-constructive}
    Locales are a constructively better-behaved notion of space than topological
    spaces.
  \item The category of locales satisfies some desirable properties that
    the category of topological spaces does not, even in a classical setting.
\end{enumerate}
For our purposes, the primary motivation is (\ref{itm:locale-constructive}) as
we would like to develop topology in a constructive foundational setting. The
prime example of locales being constructively better behaved is the localic form
of the Tychonoff Theorem: its point-set form is equivalent to the
\VAC{}~\cite{kelley-tychonoff-ac} whereas its localic form is constructively
provable~\cite{coq-compact-spaces-2003, sjv-roads}. Furthermore, there are
numerous theorems in classical mathematics characterizing when a locale can be
represented as the frame of opens of a topological space (i.e.\ is
\indexed{spatial}). These show that the class of spatial locales encompasses a
variety of interesting spaces. However, a common feature of such theorems is
that they turn out to be equivalent to various classical principles, meaning the
point-free approach to such spaces is essential to their constructive treatment.

In this chapter, we develop the basics of locale theory in the setting of
constructive and predicative \VUF{}, whose technical details we presented in
Chapter~\ref{chap:foundations}. The organization of this chapter is as follows:
\begin{description}[leftmargin=!,labelwidth=\widthof{Section 7.77:}]
  \item[\cref{sec:frames}:] We present the notions of frame and locale, and
    address the technical details of their predicative development.
  \item[\cref{sec:sublocales}:] We discuss sublocales.
  \item[\cref{sec:compactness}:] We discuss compactness and
    \VTheWayBelowRelation{}.
  \item[\cref{sec:clopens-and-wi}:] We discuss the notion of clopen in point-free
    topology as well as \VTheWellInsideRelation{}.
  \item[\cref{sec:bases}:] We present various notions of bases for locales,
    which play a pivotal role in our development of the theory of spectral
    locales.
  \item[\cref{sec:classes}:] We present predicatively well-behaved definitions of
    some interesting classes of locales, including spectral, Stone,
    zero-dimensional, and regular locales.
  \item[\cref{sec:aft}:] We present a predicative form of the \VAFT{}
    for frames.
\end{description}

\section{Definition of the notion of frame}
\label{sec:frames}

Our type-theoretic definition of a frame, in its most general form, is
parameterized by three universes:
\begin{enumerate}
  \item the universe where the carrier set lives,
  \item the universe where the truth values of the order live,
  \item the universe for join-completeness i.e.\ for the index types of families
    on which the join operation is defined.
\end{enumerate}
We adopt the convention of using the universe variables
\VUni{\UU}, \VUni{\VV}, and \VUni{\WW} for these respectively.

\begin{definition}[{\AgdaLink{Locales.Frame.html\#10215}} Frame]\label{defn:frame}
  A \defineCE{$(\UU, \VV, \WW)$-frame}{frame} $L$ consists of:
  \begin{itemize}
  \item a type $| L | \oftype \UU$,
  \item a partial order $(\blank) \le (\blank) \oftype | L | \to | L | \to \hprop{\VV}$,
    \item a top element $\FrmTopSym \oftype | L |$,
    \item an operation $(\blank) \meet (\blank) \oftype | L | \to | L | \to | L |$ giving the
      greatest lower bound $x \meet y$ of every pair of elements $x, y \oftype | L |$,
    \item an operation $\bigvee(\blank) \oftype (\ArrTy{I}{| L |}) \to | L |$ giving the
      least upper bound $\bigvee \alpha$ of
      every \VUniHyphen{\WW}family $\alpha \oftype I \to | L |$,
  \end{itemize}
  such that binary meets distribute over arbitrary joins, i.e.
  $x \wedge \bigvee_{i \oftype I} y_{i} \IdTySym \bigvee_{i \oftype I} x \wedge y_{i}$
  for every $x \oftype | L |$
  and \VUniHyphen{\WW}family $\FamEnum{i}{I}{y_i}$ in $| L |$.
\end{definition}

As we have done in the definition above, we adopt the shorthand notation
$\FrmJoin_{i \oftype I} x_i$
\nomenclature{\(\FrmJoin_{i \oftype I} x_i\)}{join of a family $\FamEnum{i}{I}{x_i}$}
for the join of a family $\FamEnum{i}{I}{x_i}$. We also adopt the usual abuse of
notation and write $L$ instead of $| L |$.

We also note here that the universe generality involved in \cref{defn:frame}
will not be needed, a point we will discuss in detail in \cref{sec:primer}.
Nevertheless, it will be useful to have formulated the notion of frame at this
level of generality as this will allow us to discuss \emph{why} it will not be
needed.

The reader might have observed that we have not imposed a sethood condition on
the carrier type in this definition. The reason for this is that sethood follows
automatically from the antisymmetry of the partial order on the type.
\begin{lemma}\label{lem:antisymmetry-gives-sethood}
  Every type $X$ equipped with a partial order is a set.
\end{lemma}
\begin{proof}
  Let $X$ be a type equipped with a partial order $(\le)$. By \cref{lem:hedberg},
  it suffices to construct a family of constant endomaps on the identity types of
  $X$. Let $x, y : X$. We define a map $\IdTyWithType{X}{x}{y} \to
  \IdTyWithType{X}{x}{y}$ as the composition of the maps:
  \begin{center}
  \begin{tikzpicture}
    \node (A) at (0,0) {$\IdTyWithType{X}{x}{y}$};
    \node (B) at (4,0) {$\ProdTy{(x \leq y)}{(y \leq x)}$,};

    \draw[->, line width=0.6pt]
      ([yshift=0.1cm]A.east) to node[midway, above] {$f$} ([yshift=0.1cm]B.west);
    \draw[->, line width=0.6pt]
      ([yshift=-0.1cm]B.west) to node[midway, below] {$g$}([yshift=-0.1cm]A.east);
  \end{tikzpicture}
  \end{center}
  where $f$ is the obvious map obtained by reflexivity and path induction on the
  equality proof, whereas the map $g$ is exactly the antisymmetry condition.
  Since the order is proposition-valued, the type $\ProdTy{(x \leq y)}{(y \leq x)}$
  is a proposition and hence the composite $\Comp{g}{f}$ is a constant
  map.
\end{proof}

Our presentation of this proof follows the \VAgda{} proof
\AgdaLink{UF.Hedberg.html\#3631}
of \MHELastName{}~\cite{type-topology-hedberg}.

\subsection{Local smallness of frames}

In \cref{defn:local-smallness}, we presented the notion of a locally small type.
In the presence of a semilattice structure,
the local smallness of a type can be characterized in terms of the order as
follows:

\begin{lemma}[\AgdaLink{Locales.Frame.html\#69901}]\label{prop:local-smallness-equiv}
  For every\/ $(\UU, \VV, \WW)$-frame\/ $L$, every pair of
  elements\/ $x, y \oftype L$, and
  every universe\/ \VUni{\TT},
  we have that\/ $x \le y$ is \VVSmall{\TT} if and only if the carrier
  of\/ $L$ is a locally\/ \VUniSmall{\TT} type.
\end{lemma}
\begin{proof}
  Let $x, y \oftype L$ and let \VUni{\TT} be a universe.
  It is a standard fact of lattice theory that
  $x \le y \leftrightarrow \IdTy{x \meet y}{x}$.

  ($\Rightarrow$) If the frame is locally \VVSmall{\TT} in the sense that $x \le y$ is
  \VVSmall{\TT} for every two $x, y : L$, then the identity type $\IdTy{x}{y}$ must
  also be \VVSmall{\TT} since it is equivalent to the conjunction
  $\ProdTy{(x \le y)}{(y \le x)}$.

  ($\Leftarrow$) Conversely, if the carrier of $L$ is a locally \VVSmall{\TT}
  type, then $x \le y$ must be \VVSmall{\TT} since it is equivalent to the identity
  type $\IdTy{x \wedge y}{x}$.
\end{proof}

\begin{remark}
  As we will discuss in detail in \cref{sec:primer}, we will always be
  interested in \emph{large and small-complete frames}, that is, frames where
  the join-completeness
  universe is below the carrier universe. The lemma above is best understood in
  this special case, where it says that the carrier is locally small if and only
  if the order has small truth values.
\end{remark}

\subsection{Primer on predicative lattice theory}
\label{sec:primer}

The preliminary definition of the notion of frame from
Definition~\ref{defn:frame} is highly general; all three universes involved in
the definition are permitted to be distinct. It is worth
considering whether this level of generality is sensible. In order to address
this question in full detail, we now provide a brief summary of \VTDJMHE{}'s
investigation~\VCitePredAsp{} of predicative aspects of lattice theory in
\VUF{}.

Frames may be viewed as certain cocomplete
\indexedCE{$(-1)$-categories}{$(-1)$-category} (also called \indexed{posetal} or
\indexedCE{thin}{thin (category)} categories). Under this view, a $(\UU, \VV, \WW)$-frame is a
category whose objects live in universe $\UU$, whose (trivial) Hom-sets live in
universe $\VV$, and whose colimits are over $\WW$-diagrams. In category theory,
there is a well-known (classical) result due to Freyd~\cite[Chapter~3,
Exercise~D]{freyd-abelian} stating that every complete small category is a
posetal. This is a so-called no-go theorem showing that the study of non-posetal
complete categories \emph{must be} limited to those that are large and
\VSmallComplete{}.

\begin{remark}
  We note here that we will be using the term \emph{complete lattice} instead
  of \emph{cocomplete lattice}.
  Although the latter is technically more accurate, it is nonstandard and
  somewhat confusing in the context of lattice of theory.
\end{remark}

Although complete, small lattices can be constructed impredicatively, there are
analogous no-go theorems for them in a predicative foundational setting. It can
be shown that such lattices cannot be constructed \emph{without} using some form
of impredicativity. The first result of this form is due to
Curi~\cite{curi-peculiar, curi-compactification}, who showed that the existence
of a complete, small lattice is not provable in Aczel and Rathjen's
Constructive%
\footnote{%
  Note that in the context of set theory, the term \emph{constructive}
  corresponds to what we call \emph{constructive and predicative}. The term
  \emph{intuitionistic} is used for what we call constructive i.e.\ mathematics
  without the use of classical principles.
}
ZF set theory~\cite{aczel-rathjen-book}, which is a predicative system.

\VTDJMHE{}~\VCitePredAsp{} proved an analogue of Curi's result in the setting of
\VUF{}. Their result is more direct and is in the style of reverse constructive
mathematics: they give a \indexed{resizing taboo}, showing directly that the
existence of a nontrivial complete and small lattice implies a form of
propositional resizing. For the sake of self-containment, we now provide a
high-level overview of this result, and for the technical details we refer
the reader to \cite[\textsection 6]{tdj-thesis} and
\cite[\textsection 4]{tdj-mhe-small-types}.

The aforementioned resizing taboo applies to a class of preorders much more
general than join-complete lattices.
\begin{definition}[\AgdaLink{Locales.Frame.html\#20304} Directed family]
  Let \(X\) be a type equipped with a preorder.
  A family \(\FamEnum{i}{I}{x_i}\) is called
  \define{directed}\index{directed family} if
  \begin{itemize}
    \item the index type \(I\) is inhabited, and
    \item for every pair of indices $i, j \oftype I$,
      there is some \(k : I\) such that \(x_k\) is an upper bound
      of \(x_i\) and \(x_j\).
  \end{itemize}
\end{definition}

\begin{definition}[{\cite[Examples~4.3]{tdj-mhe-small-types}}]
\label{defn:completeness-conditions}
  A $(\UU, \VV)$-preorder $X$ is called
  \begin{itemize}
    \item \define{$\WW$-sup-complete} if it has joins of $\WW$-families,
    \item \define{$\WW$-bounded-complete} if it has joins of bounded
      $\WW$-families,
    \item \define{$\WW$-directed-complete} if it has joins of directed
      $\WW$-families (i.e.\ all ``directed joins'').
      \index{directed join}
  \end{itemize}
\end{definition}

When formulated in terms of $\WW$-sup-complete lattices above, a
{$(\UU, \VV, \WW)$-frame} is a $(\UU, \VV)$-preorder that is (1) $\WW$-sup-complete, (2) has
finite meets, and (3) in which the finite meets distribute over the $\WW$-joins.

\begin{remark}
  Note that \TDJLastName{}~\cite{tdj-thesis} uses the notation
  $\WW\text{-DCPO}_{\UU, \VV}$ (see \cite[Definition~3.2.11]{tdj-thesis}) for
  what we called a $\WW$-directed-complete $(\UU, \VV)$-preorder above.
  Furthermore, the order and the carrier universes are often left implicit so
  that one simply says $\WW\text{-DCPO}$. We will also leave these two universes
  implicit, and will hence speak of $\WW$-sup-complete preorders.
\end{remark}

\begin{remark}
  The term \emph{complete small lattice} is not to be confused with
  \emph{small-complete lattice}. To avoid possible sources of confusion, we always
  use the hyphenated form for the latter.
\end{remark}

Each of the completeness conditions from \cref{defn:completeness-conditions}
implies a useful technical notion of completeness called
$\DeltaFam{\WW}$-completeness:

\begin{definition}[%
  \AgdaLink{OrderedTypes.DeltaCompletePoset.html\#1380}
  $\DeltaFam{\WW}$-complete poset~{\cite[Definition~6.2.1]{tdj-thesis}}%
]
  A poset $X$ is called \define{$\DeltaFam{\WW}$-complete},
  for some universe \VUni{\WW},
  if the family $\DeltaFam{x,y,P} \oftype \SumTy{\UnitTySym}{P} \to X$,
  defined as
  \begin{align*}
    \delta_{x,y,P}(\SumInlSym(\star)) &\is x\\
    \delta_{x,y,P}(\SumInrSym(p))     &\is y\text{,}
  \end{align*}
  admits a join in $X$, for every $x \le y$ and
  every \VUniHyphen{\WW}proposition $P$.
\end{definition}

The \VAgda{} formalization of the above notion is due to \VTypeTopology{}
contributor Ian Ray.

\VTDJMHE{} also use a positive reformulation of a poset being nontrivial, which
they define through a positive formulation of the strictly less than relation.
They call this
\emph{strictly below}~(see \cite[Definition~4.8]{tdj-mhe-small-types} for the
definition).
\begin{itemize}
  \item A poset $X$ is \define{nontrivial}
    \cite[Definition~4.4]{tdj-mhe-small-types} if there are specified
    $x, y : X$ such that $x \neq y$ and $x \le y$.
  \item A $\DeltaFam{\WW}$-complete poset $X$ is
    \define{positive}~\cite[Definition~4.13]{tdj-mhe-small-types} if there are
    specified $x, y : X$ with $x$ \emph{strictly below} $y$.
\end{itemize}

Note that the above notion of positivity is formulated for
$\DeltaFam{\WW}$-complete posets.
The resizing taboo of interest is then formulated as below.

\begin{theorem}[{\cite[Theorem~4.23]{tdj-mhe-small-types}}]\label{thm:tdj-taboo}~
  \begin{enumerate}
    \item There is a nontrivial small $\delta_{\WW}$-complete poset if and only
      if\/ $\Omega_{\neg\neg}$-resizing holds in universe $\WW$.
    \item There is a positive small\/ $\delta_{\WW}$-complete poset if and only
      if\/ $\Omega$-resizing holds in universe \VUni{\WW}.
  \end{enumerate}
\end{theorem}

Because $\DeltaFam{\WW}$-completeness is implied by the completeness notions
introduced above, we have the following corollary:

\begin{corollary}~
  \begin{enumerate}
    \item There is a nontrivial, small\/ \VUniHyphen{\UU}sup-lattice,
      \VUniHyphen{\UU}dcpo, or
      $\UU$-bounded-complete lattice if and only if\/ $\Omega_{\neg\neg}$-resizing
      holds in universe \VUni{\UU}.
    \item There is a positive, small\/ \VUniHyphen{\UU}sup-lattice,\/
      \VUniHyphen{\UU}dcpo,
      or\/ \VUniHyphen{\UU}bounded-complete
      lattice if and only if\/ $\Omega$-resizing holds in universe\/ \VUni{\UU}
  \end{enumerate}
\end{corollary}

This implies that \emph{there are no nontrivial small frames} unless a form of
resizing holds:

\begin{corollary}\label{cor:no-small-frames}
  There is a positive (resp.\ nontrivial) small frame if and only if
  \VOmegaResizing{} (resp.\ \VOmegaNotNotResizing) holds.
\end{corollary}

\subsection{Categories of frames and locales}
\label{sec:frm-cat}

\cref{cor:no-small-frames} implies that attention \emph{must be} restricted to
large and small-complete frames in a predicative foundational setting.
Furthermore, we know that a universe larger than \VUni{\USucc{\UU}} is never
needed for the order, since \cref{prop:local-smallness-equiv} establishes that
the order can always be resized to \VUni{\USucc{\UU}}.
For the purposes of our work, it is in fact sufficient to consider only those
frames that are locally small (i.e.\ which have \VUniHyphen{\UU}small values) as
this class encompasses most frames that are encountered in practice. In light of
these, we hereby restrict attention to $(\USucc{\UU}, \UU, \UU)$-frames, which
we refer to as \emph{large, locally small, and small-complete} frames.

\paragraph{\textsc{Convention.}}
We fix a \emph{base} universe \VUni{\UU}, and refer to types that have
equivalent copies in \VUni{\UU} as \emph{small types}. In contrast, we refer to
types in \VUni{\USucc{\UU}} as \emph{large types}. Accordingly,
\begin{center}
  \emph{we hereafter take frame/locale to mean $(\USucc{\UU}, \UU, \UU)$-frame/locale},
\end{center}
unless otherwise specified.\index{frame}

\begin{remark}
  Even though large, locally small, and small-complete frames are sufficient for
  the work presented in this thesis, there do exist large and small-complete
  frames that cannot be locally small unless a form of resizing holds. We give a
  new resizing taboo showing this in \cref{sec:examples}.
\end{remark}

Having established the above convention, we are now ready to define the category
of frames, and thereby the category of locales.

\begin{definition}[%
  \AgdaLink{Locales.ContinuousMap.FrameHomomorphism-Definition.html\#2251}
  Frame homomorphism%
]
  Let $K$ and $L$ be two frames.
  A function $h : |K| \to |L|$ is called a \define{frame homomorphism} if it
  preserves the top element, binary meets, and joins of small families. We denote
  by $\Frm_{\UU}$%
  \nomenclature{\(\Frm_{\UU}\)}{%
    the category of large, locally small, and small-complete frames over base
    universe~\VUni{\UU}%
  }
  the category of frames and their homomorphisms (over our base
  universe $\UU$). We often drop the subscript~\VUni{\UU} when the base universe
  is clear from context.
\end{definition}

A \define{locale} \AgdaLink{Locales.Frame.html\#59154} is a frame considered in
the opposite category, denoted $\Loc_{\UU} \is \opposite{\Frm_{\UU}}$.
\nomenclature{\(\Loc_{\UU}\)}{%
  the category of large, locally small, and small-complete locales over base
  universe~\VUni{\UU}%
}
We adopt
the notational conventions of Mac~Lane and Moerdijk~\cite{sheaves} and
Johnstone~\cite[Definition~1.2.1]{elephant-vol-1}; to highlight that we view the
category of locales as a category of spaces, we
\begin{itemize}
  \item use the letters $X, Y, Z, \ldots$ (or sometimes $A, B, C, \ldots$) for locales,
  \item denote the frame corresponding to a locale $X$ by $\opens{X}$.
\end{itemize}
For variables that range over the frame of opens of a locale $X$, we use the
letters $U, V, W, \ldots$ We use the letters $f$ and $g$ for continuous maps $X \to Y$
of locales. A \define{continuous map} $f : X \to Y$ is a frame homomorphism $f^* :
\opens{Y} \to \opens{X}$. A \define{homeomorphism} of locales is a frame isomorphism
between their defining frames.

\subsection{Impredicative vs.\ predicative locale theory}
\label{sec:implications}

The traditional development of locale theory takes place in an impredicative
foundational setting, where it is possible to construct nontrivial complete,
small lattices. In fact, the standard practice in an impredicative setting is to
leverage this capability as to develop the theory by working \emph{only} with
complete small lattices.
Using the propositional resizing axiom, one achieves this as follows:
\begin{enumerate}
  \item Consider a complete small frame $L$ in universe $\UU$.
  \item For every subset $P \oftype L \to \hprop{\UU}$,
    the equivalence from \cref{lem:subtype-classification}
    is used,
    leveraging the smallness of $| L |$,
    to take the corresponding \VUniHyphen{\UU}family
    whose join can then be computed in $L$.
  \item For more general subsets $P : L \to \hprop{\VV}$, a copy $P' : L \to
    \hprop{\UU}$ can always be obtained by further appeal to propositional
    resizing.
\end{enumerate}
Due to the availability of the above method in the presence of propositional
resizing, the traditional development of locale theory focuses on complete small
frames.


As a result of this focus on complete small frames, many fundamental
constructions in the traditional development of the subject are applicable only
within the context of complete small frames. An example of this is the
construction of
\indexedCE{Heyting implications}{{Heyting implication}}
in frames, which is crucial to many vital concepts of locale theory, including
fixsets~\cite[Proposition~1.1.13.(iv)]{elephant-vol-1},
open sublocales~\cite[Examples~1.1.16.(a)]{elephant-vol-1},
the double-negation nucleus~\cite[Examples~1.1.16.(c)]{elephant-vol-1},
among others.

Heyting implications are constructed for complete, small lattices as
\begin{equation*}
  \Implication{x}{y} \quad\is\quad \FrmJoin \setof{ z \in L \mid z \meet x \le y },
\end{equation*}
which involves a quantification over the entirety of the carrier set. When the
frame in consideration has a large carrier set, this family is large, and its
join is hence not available in our predicative setting.

Similarly, arbitrary meets can also be constructed from arbitrary joins when
working with complete, small lattices, by defining the meet of a family to the
be join of the set of lower bounds of the family in consideration.
Unfortunately, however, the collection of lower bounds of a family is a large
one, meaning this join does not exist in general in a predicative setting.
Because of this, the meaning of the term \indexed{complete lattice} in a
predicative setting is not quite the same as its meaning in impredicative
foundations: meet-complete and join-complete lattice mean different things.
Many constructions of locale theory rely on the existence of arbitrary meets,
none of which will be a priori available in our predicative foundational
setting.

Our solution for such problems will be to restrict attention to locales with
\emph{small} bases\index{base}\index{small base}, drawing inspiration from the
literature on locally presentable categories~\cite{adamek-rosicky-book}.
We will also show that certain rich classes of locales, such as the category of
spectral locales, admit small bases by default. We return to this issue
in \cref{sec:bases,sec:aft}.

\subsection{Basic examples of locales}

We now look at three basic examples of locales:
\begin{enumerate}
  \item the terminal locale $\LocTerm{\UU}$ (of the category $\Loc_{\UU}$),
  \item the discrete locale over a (small) set $X$,
  \item the \VSierpinski{} locale, which is the free frame on one generator.
\end{enumerate}

\subsubsection{The terminal locale}

The \define{terminal locale} \AgdaLink{Locales.InitialFrame.html\#8553} (denoted
$\LocTerm{\UU}$) is the locale defined by the initial frame $\hprop{\UU}$:

\begin{example}[{\AgdaLink{Locales.InitialFrame.html\#2041}} Initial frame]\label{defn:terminal-locale}
  For every universe $\UU$, the (large) type $\hprop{\UU}$ forms a poset under
  the order defined by implication. This poset is locally small since each
  proposition $\Impl{P}{Q}$ is small.
  Note that the antisymmetry of this order is exactly \indexedp{propositional extensionality}.
  Furthermore, this poset forms a frame:
  the top element is the true proposition $\TrueProp{\UU}$,
  the meet operation is given by the conjunction of two propositions, and
  the join of a family of propositions $\FamEnum{i}{I}{P_i}$ is defined as
  \begin{equation*}
    \FrmJoin_{i \,\oftype\, I} P_i \is \ExistsType{i}{I}{P_i},
  \end{equation*}
  which is a small proposition for every small family of propositions.

  The fact that these operations define meet and join operations is easy to see.
  For the distributivity law, it is easily checked that
  $(P \meet \ExistsType{i}{I}{Q_i}) \leftrightarrow \ExistsType{i}{I} (P \meet Q_i)$, for
  every $P \oftype \hprop{\UU}$ and every family $\FamEnum{i}{I}{Q_i}$,
  using the recursion principle of propositional truncation.
\end{example}

Before proving the universal property of the initial frame, we first establish the following lemma:
\begin{lemma}[\AgdaLink{Locales.InitialFrame.html\#9382}]%
\label{lem:initial-frame-proto-base}
  Let $\beta : \TwoTy{\UU} \to \hprop{\UU}$ be the function defined as
  $\beta(\BoolZero) \is \FalsePropSym$,
  $\beta(\BoolOne)  \is \TruePropSym$.
  Every $P \oftype \hprop{\UU}$ is equal to
  \begin{enumerate}
    \item the join $\FrmJoin_{p \oftype P} \TrueProp{\UU}$, and
    \item the join of the directed family
      \(\setof{ \beta(b) \mid b : \TwoTySym, \beta(b) \le P }\),
      which is formally given by the map
      \((\SigmaType{b}{\TwoTy{\UU}}{\beta(b) \le P}) \xrightarrow{~\beta \circ \projISym~} \hprop{\UU}\).
  \end{enumerate}
\end{lemma}
\begin{proof}
  (1) is easy to see since $P \IdTySym \TruePropSym$ if and only if $P$ holds.
  For (2), we show that
  \[\IdTy{P}{\FrmJoin \setof{ \beta(b) \mid b \oftype \TwoTySym, \beta(b) \le P }}\]
  by antisymmetry. If $P$ holds then $P \IdTySym \TruePropSym$
  and hence $\beta(\BoolOne) \DefnEqSym \top \le P$.
  For the other direction, observe that $P$ is an upper bound of the family
  since $\bot \DefnEqSym \beta(\BoolZero)$ is trivially below $P$ and
  $\top \le P$ implies that $P$ holds.
  The family  $\setof{ \beta(b) \mid b : \TwoTySym, \beta(b) \le P }$ is
  directed since it is always inhabited by $\beta(\BoolZero)$, and for every pair
  of Booleans
  $b_1, b_2 \oftype \TwoTySym$ with $\beta(b_1) \le P$ and $\beta(b_2) \le P$,
  the disjunction $\Disj{b_1}{b_2}$ is either $b_1$ or $b_2$.
\end{proof}

\begin{lemma}[\AgdaLink{Locales.InitialFrame.html\#8232}]
  The frame\/ $\hprop{\UU}$ is the initial object in the category\/ $\Frm_{\UU}$.
\end{lemma}
\begin{proof}
  Let $L$ be a frame.
  By \cref{lem:initial-frame-proto-base}, $P : \hprop{\UU}$ can be expressed as
  the small join $\FrmJoin_{p \oftype P} \TruePropSym$.
  This implies that every frame homomorphism $h \oftype \hprop{\UU} \to L$
  is necessarily unique as it must satisfy the equality:
  \begin{equation*}
    h(P) = h\paren{\FrmJoin_{p \oftype P} \TruePropSym}
         = \FrmJoin_{p \oftype P} h(\TruePropSym)
         = \FrmJoin_{p \oftype P} \FrmTop{L}.
  \end{equation*}
  We accordingly define the unique map $\hprop{\UU} \to L$ as above.
  It is easy to see that this defines a frame homomorphism.
\end{proof}

\begin{remark}
  The universes in the above proof can be generalized, but as explained
  in \cref{sec:primer}, we focus only on the category $\Frm_{\UU}$ in this
  thesis.
  The reader interested in the generalization of universes is referred to the
  \VAgda{} formalization.
\end{remark}

\subsubsection{The discrete locale}

The
\define{discrete locale}
\AgdaLink{Locales.DiscreteLocale.Definition.html}
over a small type $X \oftype \UU$ is the locale defined by the powerset frame
$\Pow{\UU}{X}$, which we define below.

\begin{example}[Powerset frame]%
\label{ex:powerset-frame}
  Let $X \oftype \UU$ be a small type.
  The \define{powerset} $\ArrTy{X}{\hprop{\UU}}$ is a poset under the order
  defined by subset inclusion, assuming propositional and function extensionality.
  This poset is locally small thanks to the smallness of $X$, since it implies that
  the quantification over $X$ is small. Moreover, this poset forms a frame:
  \begin{itemize}
    \item the top open is given by the full subset $x \mapsto \top_{\UU}$,
    \item the meet of two subsets $S, T : X \to \hprop{\UU}$ is given by set
      intersection, defined as
      \[S \cap T \is x \mapsto \Conj{x \in S}{x \in T},\]
    \item the join of a family of subsets $\FamEnum{i}{I}{S_i}$ is given by set
      union, defined as
      \begin{equation*}
        \paren{\FrmJoin_{i : I} S_i} \is x \mapsto \ExistsType{i}{I}{x \in S_i},
      \end{equation*}
      which is a small subset for every small family of subsets.
  \end{itemize}
\end{example}

We will often write $X$ for the full subset $x \mapsto \TruePropSym$.

As we mentioned in the above example, the smallness of type $X$ is crucial for
the local smallness of the frame $\Pow{\UU}{X}$, since the inclusion order
$S \subseteq T \is \PiTy{x}{X}{\Impl{x \in S}{x \in T}}$
is, in general, not small without the use of a form of resizing.
The resizing taboo below proves this.

\begin{lemma}
  If the powerset lattice over a large proposition\/ $P$ is locally small,
  then the negation\/ $\neg P$ is a small proposition.
\end{lemma}
\begin{proof}
  Let $P \oftype \hprop{\USucc{\UU}}$.
  Recall that in a powerset frame $\PowSym(X)$,
  we denote the top subset $x \mapsto \top$ by $X$, which implies that we denote
  the top subset of $\PowSym(P)$ by $P$ itself.
  We have that \(P \subseteq \emptyset\)
  if and only if $\neg P$ holds. Assuming the local smallness of the lattice
  $\PowSym(P)$, the type
  \[P \subseteq \emptyset\]
  is small, and hence so is $\neg P$.
\end{proof}

If $\neg P$ is small, then obviously so is $\neg\neg P$ meaning we have:

\begin{corollary}\label{cor:local-smallness-resizing-taboo}
  If the powerset frame over every large proposition is locally small, then
  we have \VOmegaNotNotResizing{} from\/ \VUni{\USucc{\UU}} to\/ \VUni{\UU},
  for every universe\/ \VUni{\UU}.
\end{corollary}

Given a large type $X \oftype \USucc{\UU}$, note that the local smallness of the
powerset lattice $\Pow{\UU}{X}$ is a consequence of propositional resizing,
since the inclusion $S \subseteq T$ is always a proposition. Whether the local
smallness of this lattice is \emph{equivalent} to a well-known form of resizing
is unclear to the author at the time of writing.

We will look at the \VSierpinski{} locale next, which is an example of a frame
that is locally small and small-complete \emph{despite} being defined on a large
type.

\subsubsection{The \VSierpinski{} locale}
\label{sec:sierpinski}

\begin{example}[{\AgdaLink{Locales.Sierpinski.Definition.html\#8322}}
  \VSierpinski{} frame]\label{defn:sierpinski}
  A subset $S \oftype \hprop{\UU} \to \hprop{\UU}$ of
  \VUniHyphen{\UU}propositions is called
  \define{\VScottOpen{}}
  if it preserves joins of \indexedp{directed}\index{directed family} families
  of propositions.
  The set of \VScottOpen{} subsets forms a poset under the inclusion order,
  which can be shown to be locally small.
  This poset forms a small-complete frame and the locale it defines is called the
  \define{\VSierpinski{} locale}.
\end{example}
\begin{proof}
  It is easy to see that the full subset is \VScottOpen{}.
  Given \VScottContinuous{} subsets $S, T : \hprop{\UU} \to \hprop{\UU}$, to
  see that the intersection $S \cap T$ is \VScottContinuous{}, consider a directed
  family of propositions $\FamEnum{i}{I}{P_i}$. We have
  \begin{align*}
    (S \cap T)\paren{\FrmJoin_{i : I}P_i}
    \quad&=\quad \Conj{\paren{\FrmJoin_{i : I}S(P_i)}}{\paren{\FrmJoin_{i : I}T(P_i)}} \tag{Scott-continuity}\\
    \quad&=\quad \FrmJoin_{i, j : I} S(P_{i}) \meet T(P_{j})\\
    \quad&=\quad \FrmJoin_{i : I} S(P_{i}) \meet T(P_{i})\tag{\dag}
  \end{align*}
  For the step $(\dag)$, the $\ge$ direction is obvious. The $\le$ direction follows
  from the directedness of the families $\FamEnum{i}{I}{S(P_i)}$ and
  $\FamEnum{i}{I}{T(P_i)}$.

  To see that (small) unions of \VScottOpen{} subsets are \VScottOpen{}, let
  $\FamEnum{i}{I}{S_i}$ be a family of \VScottOpen{} subsets and let
  $\FamEnum{j}{J}{P_j}$ be a directed family of propositions.
  \begin{align*}
    \paren{\FrmJoin_{i : I} S_i}\paren{\FrmJoin_{j : J}{P_j}}
    \quad&=\quad \ExistsType{i}{I}{S_i\paren{\FrmJoin_{j : J}{P_j}}}\\
    \quad&=\quad \ExistsType{i}{I}{\paren{\FrmJoin_{j : J}{S_i(P_j)}}}      \tag{Scott-continuity of $S_i$}\\
    \quad&=\quad \ExistsType{i}{I}{\ExistsType{j}{J}{S_i(P_j)}}\\
    \quad&=\quad \ExistsType{j}{J}{\ExistsType{i}{I}{S_i(P_j)}}\\
    \quad&=\quad \FrmJoin_{j : J}{\paren{\paren{\FrmJoin_{i : I} S_i}(P_j)}}
  \end{align*}

  We also need to show that the pointwise order
  $\PiTy{P}{\hprop{\UU}}{\Impl{S(P)}{T(P)}}$ is \VUniHyphen{\UU}small, for every
  pair of \VScottOpen{} subsets $S$ and $T$.
  We define a small version of the order above as
  $S \subseteq_s T \is \PiTy{P}{\hprop{\UU}}{\Impl{S(P)}{T(P)}}$,
  and show that $S \subseteq_s T$ is logically equivalent to $S \subseteq T$. The nontrivial
  direction is $\Impl{S \subseteq_s T}{S \subseteq T}$. Assume $S \subseteq_s T$ and let $P : \hprop{\UU
  }$. By \cref{lem:initial-frame-proto-base}, we know that
  $P$ can be expressed as the directed join
  $\FrmJoin \setof{ \beta(b) \mid b : \TwoTy{\UU}, \beta(b) \le P }$.
  The Scott openness of $S$ and $T$ implies that
  \begin{align*}
    S\paren{\FrmJoin \setof{ \beta(b) \mid b : \TwoTy{\UU}, \beta(b) \le P }}
    &\quad=\quad
    \FrmJoin \setof{ S(\beta(b)) \mid b : \TwoTy{\UU}, \beta(b) \le P }\\
    &\quad\le\quad
    \FrmJoin \setof{ T(\beta(b)) \mid b : \TwoTy{\UU}, \beta(b) \le P }\\
    &\quad=\quad
    T\paren{\FrmJoin \setof{ \beta(b) \mid b : \TwoTy{\UU}, \beta(b) \le P }}.
  \end{align*}
\end{proof}

We denote the \VSierpinski{} locale at universe \VUni{\UU}
by $\LocSierp_{\UU}$.
If the universe is clear from context, drop the subscript and write $\LocSierp$.
\nomenclature{\(\LocSierp\)}{\VSierpinski{} locale}

Given a small set $A$, a frame $K$ is said to be \define{freely generated} by
$A$ if there is a function~$i : A \to |K|$ such that, for every frame $L$ and
every function $f : A \to |L|$, there is a unique frame homomorphism $h : K \to L$
satisfying $\IdTy{f}{\Comp{|h|}{i}}$, as captured by the diagram:
\[
\begin{tikzcd}
  {|K|} \arrow[r, "|h|"]                     & {|L|} & & K \arrow[r, dashed, "h"] & L \\
  A     \arrow[u, "i"] \arrow[ur, "f", swap] & ~
\end{tikzcd}
\]

The universal property of the \VSierpinski{} locale $\LocSierp$ is that its
frame $\opens{\LocSierp}$ is freely generated by the unit type $\OneSym$.
Since functions $\ArrTy{\OneSym}{|L|}$ are exactly the opens of frame~$L$,
this universal property can be simplified as given below.
\begin{definition}[Universal property of Sierpi\'{n}ski]%
\label{defn:sierp-universal-property}
  A locale \(X\) has the \define{universal property of Sierpi\'nski}
  if it has a specified open $U$
  satisfying the property that,
  for every locale \(Y\) and every open \(V \oftype \opens{Y}\),
  there is a map \(f \oftype \ArrTy{Y}{X}\)
  satisfying \(\IdTy{V}{f^*(U)}\).
\end{definition}

\begin{proposition}[\AgdaLink{Locales.Sierpinski.UniversalProperty.html\#3139}]
\label{prop:sierpinski-universal-property}
  The \VSierpinski{} frame,
  which we constructed in \cref{defn:sierpinski},
  satisfies the above universal property.
  The specified open is given by the identity map.
\end{proposition}

Conceptually, the universal property above states that the \VSierpinski{} locale
\emph{classifies the opens} of every locale.
We postpone the proof that our construction satisfies this property until
\cref{sec:universal-property-of-sierpinski},
as the proof requires some technical preparation.

\subsubsection{The frame of $\neg\neg$-stable propositions}

\begin{example}\label{example:not-not-stable}
  We denote by $\hprop{\UU}^{\neg\neg}$
  the type of propositions $P \oftype \hprop{\UU}$
  satisfying $\neg\neg P \ImplSym P$, which are called the
  \indexed{\(\neg\neg\)-stable} propositions.
  These form a frame when ordered under implication,
  with meets given by conjunction as
  in \cref{defn:terminal-locale}.
  The join, however, is different from the one in \cref{defn:terminal-locale}.
  Given a family $\FamEnum{i}{I}{P_i}$ of \(\neg\neg\)-stable propositions,
  its join in $\hprop{\UU}^{\neg\neg}$ is defined to be
  \begin{equation*}
    \neg\neg \paren{\FrmJoin_{i \oftype I} P_i}
    \DefnEqSym
    \neg\neg \ExistsType{i}{I}{P_i}\text{.}
  \end{equation*}
\end{example}

We denote by $\LocTerm{\UU}^{\neg\neg}$ the locale defined by the above frame.
\nomenclature{\(\LocTerm{\UU}^{\neg\neg}\)}{%
  the locale defined by the frame of \(\neg\neg\)-stable propositions.%
}

The above frame can be described more naturally as a \indexed{sublocale} of the
terminal locale, namely the one given by
the double-negation nucleus~\cite[\textsection II.2.4]{ptj-ss}:
\[\neg\neg \oftype \hprop{\UU} \to \hprop{\UU}\text{.}\]
Unfortunately, we have to postpone this description at this point,
as we have not presented the notions of sublocale and nuclei yet.
We will address this in the next section.

\section{Sublocales}\label{sec:sublocales}

There are several equivalent ways to represent the notion of \emph{sublocale} in
the traditional development of locale theory
\cite[Proposition~1.1.13]{elephant-vol-1}:
\begin{enumerate}
  \item Regular frame quotients of $\opens{X}$ i.e.\ isomorphism classes
    of surjective frame homomorphisms into $\opens{X}$.
  \item Frame congruences on $\opens{X}$.
  \item Nuclei on $\opens{X}$.
  \item Fixsets of $\opens{X}$, which are subsets that are (1) closed under
    arbitrary meets and (2) form exponential ideals with respect to
    Heyting implication.
\end{enumerate}

Some of the notions of sublocale above are not predicatively available. For
example, it is not clear how fixsets\index{fixset} would be defined in our
setting since they presuppose the existence of arbitrary meets and Heyting
implications in the frame in consideration, neither of which is predicatively
available as we explained in \cref{sec:implications}.

For the purposes of this thesis, we will take nuclei as the primary notion of
sublocale.

\begin{definition}[Nucleus]\label{defn:nucleus}
  A \define{nucleus on a locale} $X$ is an endofunction $j : \opens{X} \to
  \opens{X}$ that is inflationary, preserves binary meets, and is idempotent.
\end{definition}

We define
\(\Nucleus{X} \is \SigmaType{j}{\ArrTy{\opens{X}}{\opens{X}}}{\IsNucleus{j}}\).
\nomenclature{\(\Nucleus{X}\)}{type of nuclei over locale $X$}
Furthermore, we adopt the standard convention of using the letters
$j, k, l \ldots$
for variables ranging over nuclei.

\begin{example}[\AgdaLink{Locales.Nucleus.html\#5468} Closed nucleus]%
\label{example:closed-nucleus}
  For every locale $X$ and every open $U : \opens{X}$, the map
  \begin{equation*}
    \ClosedNucleus{U} \is V \mapsto U \vee V
  \end{equation*}
  is a nucleus.
  It is
  inflationary, since $V \le \ClosedNucleus{U}(V) \DefnEqSym U \vee V$,
  for every pair of opens $U, V : \opens{X}$.
  It is also idempotent since
  \begin{equation*}
    \ClosedNucleus{U}(\ClosedNucleus{U}(V)) \DefnEqSym U \vee (U \vee V)
                                      \DefnEqSym U \vee V.
  \end{equation*}
  The preservation of binary meets follows from dual distributivity.
\end{example}

\begin{lemma}[{\cite[Proposition~3.27]{tosun-msc}}]\label{lem:sublocale}
  For every locale $X$ and nucleus $j : \opens{X} \to \opens{X}$, the type
  \[\SigmaType{U}{\opens{X}}{\paren{ \IdTy{U}{j(U)} }}\]
  of fixed points of $j$ forms a frame, the locale defined by which is denoted
  $X_j$.
  This satisfies the property that the inclusion
  $\opens{X_j} \hookrightarrow \opens{X}$
  is the right adjoint of the map
  \begin{align*}
    \opens{X} &\to \opens{X_j}\\
    U         &\mapsto j(U)
  \end{align*}
\end{lemma}

\begin{lemma}
  For every nucleus $j \oftype \opens{X} \to \opens{X}$,
  the type \(\SigmaType{U}{\opens{X}}{\paren{ \IdTy{U}{j(U)} }}\)
  is equivalent to the image of $j$.
\end{lemma}
\begin{proof}
  Standard fact of locale theory. See, for example, \cite[49]{ptj-ss}.
\end{proof}

\begin{definition}
  A locale $Y$ is called a \define{sublocale} of locale $X$ if there is some
  nucleus $j \oftype \opens{X} \to \opens{X}$ such that $Y$ is homeomorphic to $X_j$.
\end{definition}

Conceptually, the notion of nucleus axiomatizes the \indexedp{monad} arising
from the adjunction of a frame quotient
$q \oftype \opens{X} \twoheadrightarrow \opens{Y}$
with its right adjoint $q_* : \opens{Y} \to \opens{X}$.
Defining $j_q \is \Comp{q_*}{q}$, the $j_q$-fixed opens are the representatives
of the equivalence class given by the quotient $q : \opens{X} \twoheadrightarrow \opens{Y}$.

\begin{remark}
  We will also work with inflationary and binary-meet-preserving functions that
  are not necessarily idempotent. Such functions are called
  \defineCE{prenuclei}{prenucleus}.
  We
  also note that, to show a prenucleus $j$ to be idempotent, it suffices to show
  $j(j(U)) \le j(U)$ as the other direction follows from inflationarity. In fact,
  the notion of a nucleus could be defined as a prenucleus satisfying the
  inequality $j(j(U)) \le j(U)$, but we define it as in
  Definition~\ref{defn:nucleus} for the sake of simplicity and make implicit use
  of this fact in our proofs of idempotency.
\end{remark}

\begin{lemma}\label{lem:prenuclei-are-monotone}
  Every prenucleus is a monotone map.
\end{lemma}
\begin{proof}
  It is a well-known fact of lattice theory that meet preservation implies
  monotonicity~\cite[Lemma~2.8]{davey-priestley-book}.
\end{proof}

\section{Compactness and the way-below relation}
\label{sec:compactness}

\subsection{Compact opens}

\indexedpCE{Compactness}{compactness} is a fundamental topological property. In point-set
topology, a subspace $Y \subseteq X$ of a space is \define{compact} if every
open cover of $Y$ has a finite subcover. As this is a property involving only
the open subsets of the space $X$, it can be readily translated into the
point-free setting to define the notion of \indexedp{compact open}
in a locale. Our type-theoretic definition below closely follows
Johnstone~\cite[Paragraph~II.3.1]{ptj-ss}, except we refine it by working with
Kuratowski finite families.

We first do some preparation.

\begin{lemma}[%
  \AgdaLink{Locales.Compactness.Properties.html\#11037}%
]%
\label{lem:comp-char-lem}
  For every directed family of opens $\FamEnum{i}{I}{U_i}$, every
  \indexedp{Kuratowski finite} set of elements of the family has an upper bound
  in the family.
\end{lemma}
\begin{proof}
  Let $\FamEnum{i}{I}{U_i}$ be a directed family of opens.
  We show by Kuratowski finite subset induction
  (\cref{lem:kuratowski-induction})
  that every Kuratowski finite set of elements of the family
  $\FamEnum{i}{I}{U_i}$ has an upper bound in the family.

  \emph{Base case (1)}: the empty subset. The family is inhabited by the
  assumption of directedness and any member of the family is vacuously an upper
  bound of the empty set.

  \emph{Base case (2)}: the singleton subset.
  We need to show that the singleton subset $\{ U_i \}$ has an upper bound in
  the family $\FamEnum{i}{I}{U_i}$, for every $i : I$.
  This is immediate since every open $U_i$ is in the family by definition and is
  an upper bound of the singleton containing itself.

  \emph{Induction step}: binary unions. Let $F_1$ and $F_2$ be two subsets
  of elements of the family and assume by induction that they have upper
  bounds $U_1$, $U_2$ in the family.
  We need to show that the union $F_1 \cup F_2$ has an upper bound in the family
  $\FamEnum{i}{I}{U_i}$.
  By the directedness of the family, there must be some $U_3$ in the family that
  is above both $U_1$ and $U_2$. This clearly implies that $U_3$ is an upper bound
  of the subset $F_1 \cup F_2$.
\end{proof}

\begin{definition}\label{defn:closure-under-finite-joins}
  Let $X$ be a locale and consider a family $\FamEnum{i}{I}{U_i}$
  of opens.
  The family $U^{\uparrow} \oftype \ListTy{I} \to \opens{X}$, defined as
  \begin{equation*}
    U^{\uparrow}(i_0, \ldots, i_{n-1}) \is U_{i_0} \vee \cdots \vee U_{i_{n-1}}\text{,}
  \end{equation*}
  is called the \emph{closure of $\FamEnum{i}{I}{U_i}$ under finite joins}.%
  \index{closure under finite joins}
\end{definition}

\begin{lemma}[%
  \AgdaLink{Locales.Compactness.Properties.html\#16259}
  {\emph{cf.}\ \cite[\textsection II.3.1]{ptj-ss}}%
]%
\label{prop:compactness}
  For every open $U$ of a locale $X$, the following are equivalent:
  \begin{enumerate}
    \item\label{itm:is-compactp} For every family of opens $\FamEnum{i}{I}{V_i}$
      with $U \le \FrmJoin_{i : I} V_i$, there is a Kuratowski finite subfamily
      $\FamEnum{j}{J}{V_{i_j}}$ such that $U \le \FrmJoin_{j : J}{V_{i_j}}$.
    \item\label{itm:is-compactpp}
      For every family of opens $\FamEnum{i}{I}{V_i}$ with
      $\IdTy{U}{\FrmJoin_{i : I} V_i}$, there is a Kuratowski finite subfamily
      $\FamEnum{j}{J}{V_{i_j}}$ such that $\IdTy{U}{\FrmJoin_{j : J}{V_{i_j}}}$.
    \item\label{itm:compact-dirfam}
      For every directed family $\FamEnum{i}{I}{V_i}$ with
      $U \le \FrmJoin_{i : I}{V_i}$,
      there is an index $i : I$ such that $U \le V_i$.
  \end{enumerate}
\end{lemma}
\begin{proof}
  We start with the equivalence
  (\ref{itm:is-compactp}) $\BiImplSym$ (\ref{itm:is-compactpp}).
  For the forward implication, suppose some open $U$ satisfies
  Condition (\ref{itm:is-compactp}) and let $\FamEnum{i}{I}{V_i}$ be a family
  of opens with $\IdTy{U}{\FrmJoin_{i : I} V_i}$.
  It must be the case that there is a Kuratowski finite subfamily
  $\FamEnum{j}{J}{V_{i_j}}$ such that $U \le \FrmJoin_{j : J}{V_{i_j}}$. It
  remains to show that $\FrmJoin_{j : J}{V_{i_j}} \le U$, but we already have
  $\FrmJoin_{j : J} V_{i_j} \le \FrmJoin_{i : I} V_i \IdTySym U$.
  For the backward implication, let $U$ be an open satisfying Condition
  (\ref{itm:is-compactpp}) and consider a family of opens
  $\FamEnum{i}{I}{V_{i}}$
  with $U \le \FrmJoin_{i : I} V_i$. This means
  \begin{equation*}
    U \IdTySym U \meet \FrmJoin_{i : I} V_i,
  \end{equation*}
  so by frame distributivity we have
  \begin{equation*}
    \IdTy{U}{\FrmJoin_{i : I} U \meet V_i}.
  \end{equation*}
  By the assumption of Condition (\ref{itm:is-compactpp}), this means that there
  is a finite subcover \[\IdTy{U}{\FrmJoin_{j : J}{U \meet V_{i_j}}}.\]
  This implies,
  again by distributivity, that $\IdTy{U}{U \meet {\FrmJoin_{j : J}{V_{i_j}}}}$,
  which holds if and only if $U \le \FrmJoin_{j : J} V_{i_j}$ so we are done.

  For the equivalence
  (\ref{itm:compact-dirfam}) $\Leftrightarrow$ (\ref{itm:is-compactp}),
  we first address
  the implication (\ref{itm:compact-dirfam}) $\ImplSym$ (\ref{itm:is-compactp}). Let
  $U$ be an open satisfying (\ref{itm:compact-dirfam}) and let
  $\FamEnum{i}{I}{V_i}$ be a family with $U \le \FrmJoin_{i : I} V_i$.
  We consider the closure of $\FamEnum{i}{I}{V_i}$ under finite
  joins, as defined above.
  Because directification preserves the join,
  we have that $U \le \FrmJoin_{s \oftype \ListTy{I}} V^{\uparrow}_s$, and we therefore know,
  by the assumption of Condition (\ref{itm:compact-dirfam}), that there exists
  a list of indices $i_0, \ldots, i_{n-1}$ satisfying
  $U \le V_{i_0} \vee \cdots \vee V_{i_{n-1}}$.
  Membership in the list $V_0, \ldots, V_{i_{n-1}}$ defines a Kuratowski finite
  subfamily of $\FamEnum{i}{I}{V_i}$, an upper bound of which is in the family
  and is above $U$.
  For the backward implication, let $U$ be an open satisfying
Condition~(\ref{itm:is-compactp}) and consider a directed family
  $\FamEnum{i}{I}{V_i}$ such that $U \le \FrmJoin_{i : I} V_i$. There must be a
  Kuratowski finite subfamily $\FamEnum{j}{J}{V_{i_j}}$ with
  $U \le \FrmJoin_{j : J} V_{i_j}$. By \cref{lem:comp-char-lem}, this subfamily
  must have an upper bound $V_k$ in $\FamEnum{i}{I}{V_{i}}$, meaning we have
  \[U \le \FrmJoin_{j : J} V_{i_j} \le V_k.\]
\end{proof}

\begin{definition}[\AgdaLink{Locales.Compactness.Definition.html\#1610} Compactness]%
\label{defn:compactness}~
  \begin{itemize}
  \item An open $U : \opens{X}$ is called \define{compact} if it satisfies the
    equivalent conditions from \cref{prop:compactness}.
  \item A locale $X$ is called \defineCE{compact}{compact locale} if its
    top open $\One{X}$ is compact.
  \end{itemize}
\end{definition}

In the formalization, we take Condition~(\ref{itm:compact-dirfam}) of
\cref{prop:compactness} to be our primary definition of compactness.

\begin{example}\label{ex:term-loc-compact}
  The terminal locale $\LocTerm{\UU}$ is compact.
  It follows directly from the definition of the join that,
  for every family of $\UU$\mbox{-}propositions $\FamEnum{i}{I}{P_i}$,
  we have that $\TruePropSym \le \FrmJoin_{i : I} P_i$
  if and only if some $P_i$ holds i.e. $\TruePropSym \le P_i$ for some $i : I$.
\end{example}

\begin{example}\label{ex:disc-loc-not-compact}
  The discrete locale on $\mathbb{N}$ is \textbf{not} compact, since this would mean
  by \cref{prop:compactness} that $\mathbb{N}$ is a Kuratowski finite type.
\end{example}

\begin{example}\label{example:sierpinski-is-compact}
  The \VSierpinski{} locale is compact.
  Let $\FamEnum{i}{I}{S_i}$ be a directed family of \VScottOpen{} subsets
  of $\hprop{\UU}$
  with $\TruePropSym \le \bigcup_{i \oftype I} S_i$, which is to say
  that the union $\bigcup_{i \oftype I} S_i$ contains every proposition
  in $\hprop{\UU}$.
  There must then be a \VScottOpen{} subset $S_i$
  with $\FalsePropSym \in S_i$.
  By the monotonicity of \VScottOpen{} subsets, $S_i$ must then contain
  every proposition since $\FalsePropSym$ implies every
  \VUniHyphen{\UU}proposition. Hence we have $\TruePropSym \subseteq S_i$.
\end{example}

\begin{lemma}[\AgdaLink{Locales.Compactness.Definition.html\#3298}]%
\label{lem:compact-fin-join-closure}
  Compact opens are closed under finite joins, that is,
  the bottom open $\FrmBotSym$ is always compact
  and the join of two compact opens is compact.
\end{lemma}
\begin{proof}
  Well-known fact of locale theory \cite[Lemma~II.3.2]{ptj-ss}.
\end{proof}

We denote by $\CompactOpens{X}$ the type of compact opens of a locale $X$. In
other words, we define
\begin{equation*}
  \CompactOpens{X} \is \SigmaType{U}{\opens{X}}{U\ \text{is compact}}.
\end{equation*}%
\nomenclature{\(\CompactOpens{X}\)}{the type of compact opens of locale $X$}

\subsection{The way-below relation}

In point-free topology, there is a generalization of compactness called the
\emph{way-below relation}\index{way-below relation}.
Whilst an open $U : \opens{X}$ is compact if every open cover of $U$ contains a
finite subcover for $U$, it is said to be way below some other open $V$
if every open cover of~$V$ has a finite subcover for $U$. In other words, the
statement that $U$ is way below $V$ (written $U \WayBelow V$) expresses that
$U$ is \emph{compact relative to} the open $V$.

\begin{definition}[\AgdaLink{Locales.WayBelowRelation.Definition.html\#680} Way-below relation]\label{defn:way-below}
  We say that an open $U$ of a locale $X$ is \define{way below} an open $V$,
  written $U \WayBelow V$,
  if for every directed family $\FamEnum{i}{I}{W_i}$ with
  $V \le \FrmJoin_{i : I} W_{i}$, there is some $i : I$ with $U \le W_i$.%
  \nomenclature{\(U \WayBelow V\)}{\(U\) is way below \(V\)}
\end{definition}

\begin{lemma}
  The statement $U \WayBelow V$ is a proposition, for every pair of opens
  $U$~and~$V$.
\end{lemma}

Note that for every open $U : \opens{X}$, the proposition $U \WayBelow U$ is
judgementally equal to
Condition (\ref{itm:compact-dirfam}) from \cref{prop:compactness}.

\begin{lemma}~
  \begin{itemize}
  \item \AgdaLink{Locales.WayBelowRelation.Properties.html\#1192}
    For every pair of opens\/ $U, V : \opens{X}$, if\/ $U \WayBelow V$
    then\/ $U \le V$.
    \item \AgdaLink{Locales.WayBelowRelation.Properties.html\#1967}
      If\/ $U \WayBelow V$ and\/ $V \le W$, then\/ $U \WayBelow W$, for
      every\/ $U, V, W : \opens{X}$.
    \item If\/ $U \le V$ and\/ $V \WayBelow W$, then $U \WayBelow W$,
      for every\/ $U, V, W : \opens{X}$.
  \end{itemize}
\end{lemma}
\begin{proof}
  See \cite[Lemma~VII.2.2]{ptj-ss}. Proofs are unchanged in our foundational
  setting.
\end{proof}

Finally, we also note that \cref{lem:compact-fin-join-closure} generalizes to
the way-below relation.

\begin{lemma}
  For every open $U : \opens{X}$,
  \begin{enumerate}
    \item\label{itm:bot-wb} \AgdaLink{Locales.WayBelowRelation.Properties.html\#837}
      the bottom open $\FrmBot{X}$ is way below $U$, and
    \item\label{itm:vee-wb} \AgdaLink{Locales.WayBelowRelation.Properties.html\#2379}
      for every pair of opens $V$ and $W$, if $V \WayBelow U$ and
      $W \WayBelow U$ then $V \vee W \WayBelow U$.
  \end{enumerate}
\end{lemma}
\begin{proof}~

  (\ref{itm:bot-wb}) Let $\FamEnum{i}{I}{U_i}$ be a directed family such that
  $\FrmBot{X} \le \FrmJoin_{i : I} U_i$.
  The family must be inhabited by some $U_k$ since it is directed, and so we
  have $\FrmBot{X} \le U_k$.

  (\ref{itm:vee-wb}) Let $U_1, U_2$ be a pair of opens of $X$ with $U_1 \WayBelow U$ and
  $U_2 \WayBelow U$. Let $\FamEnum{i}{I}{V_i}$ be a family of opens such that
  $U \le \FrmJoin_{i : I} V_i$. There must be $i, j : I$ such that $U_1 \le V_i$ and
  $U_2 \le V_j$. Since the family $\FamEnum{i}{I}{V_i}$ is directed, there must
  be some $k : I$ with $U_1 \le V_i \le V_k$ and $U_2 \le V_j \le V_k$. It follows that
  $U_1 \vee U_2 \le V_k$.
\end{proof}

\section{Clopens and the well inside relation}\label{sec:clopens-and-wi}

\subsection{Clopens}\label{sec:clopens}

We now look at the point-free analogue of the notion of clopen subset from
point-set topology.

\begin{definition}[%
  \AgdaLink{Locales.Clopen.html\#896}
  Clopen%
]
  An open $U \oftype \opens{X}$ of a locale $X$ is called a \define{clopen}
  if it is a
  complemented element of the lattice $\opens{X}$.
  That is to say, there is a specified $U'$ satisfying
  \begin{equation*}
    \IdTy{U \meet U'}{\FrmBot{X}} \quad\text{and}\quad \IdTy{U \vee U'}{\FrmTop{X}}.
  \end{equation*}
\end{definition}

Note that it is not necessary to use propositional truncation in the above
definition:

\begin{lemma}\label{lem:clopenness-prop}
  For every open $U : \opens{X}$, the type expressing that $U$ is clopen is a
  proposition.
\end{lemma}
\begin{proof}
  Follows directly from the fact that Boolean complements are unique.
\end{proof}

\begin{example}
  The clopens of the terminal locale $\LocTerm{\UU}$ are exactly the decidable
  propositions
  (i.e.\ $\TrueProp{\UU}$ and $\FalseProp{\UU}$,
  as shown by \cref{lem:decidable-props-are-the-booleans}).
\end{example}

We record some elementary facts about clopens for the sake of self-containment.

\begin{lemma}\label{lem:clopen-closure}
  In every locale\/ \(X\),
  \begin{itemize}
    \item the top open\/ $\FrmTop{X}$ and the bottom
      open\/ $\FrmBot{X}$ are clopen,
    \item clopens are closed under binary joins and binary meets.
  \end{itemize}
\end{lemma}
\begin{proof}
  It is easy to see that $\FrmTop{X}$ and $\FrmBot{X}$ are clopen, since
  they clearly complement each other.
  Given two clopens $F_1, F_2 : \opens{X}$, with complements $F_1'$ and $F_2'$,
  the meet $F_1 \meet F_2$ (resp.\ the join $F_1 \vee F_2$) is complemented by
  the join $F_1' \vee F_2'$ (resp.\ the meet $F_1' \meet F_2'$).
\end{proof}

\subsection{The well-inside relation}
\label{sec:well-inside}

Similar to how the way-below relation generalizes the notion of a compact open,
the \indexed{well-inside relation} generalizes the notion of a clopen. The
statement that $U$ is well-inside $V$ (written $U \WellInside V$) expresses that
the open $U$ is clopen relative to the open $V$.

\begin{definition}[\AgdaLink{Locales.WellInside.html\#1122} The well-inside relation]%
\label{defn:well-inside}
  We say that an open $U$ of a locale $X$ is
  \define{well inside}\index{{the well-inside relation}}
  an open $V$, written $U \WellInside V$,
  \nomenclature{\(U \WellInside V\)}{$U$ is well inside $V$}
  if there is some open $W$ with $U \wedge W = \Zero{X}$ and $V \vee W = \One{X}$.
  Such a $W$ is called a meet-complement of $U$ and a join-complement of $V$.
\end{definition}

We note here that the use of unspecified existence is essential in the above
definition, since the untruncated version is structure and not property. The
example below illustrates this:

\begin{example}[\AgdaLink{Locales.CompactRegular.html\#4559}]
  Consider the terminal locale $\LocTerm{\UU}$, which has the unit
  proposition~$\TrueProp{\UU}$ as its top open, and the empty proposition $\FalseProp{\UU}$
  as its bottom open. We have that $\FalsePropSym \WellInside \TruePropSym$,
  but this can be witnessed by two distinct propositions. The
  proposition~$\FalsePropSym$ is a meet-complement of itself since
  $\FalsePropSym \meet \FalsePropSym \IdTySym \FalsePropSym$ and also
  a join-complement of $\TruePropSym$ since
  $\TruePropSym \vee \FalsePropSym = \TruePropSym$. It is also the case that
  $\TruePropSym$ is a meet-complement of $\FalsePropSym$ and a join-complement
  of itself.
\end{example}

\begin{lemma}[\AgdaLink{Locales.CompactRegular.html\#5963}]%
\label{lem:wi-prop-1}
  Given opens $U, V, W$ of a locale,
  \begin{enumerate}
  \item if $U \WellInside V$ and $V \le W$, then $U \WellInside W$;
  \item if $U \le V$ and $V \WellInside W$, then $U \WellInside W$.
  \end{enumerate}
\end{lemma}
\begin{proof}
  See \cite[Lemma~III.3.1.1.(iii)]{ptj-ss}. The proof is unchanged in our
  foundational setting.
\end{proof}

In compact locales, the well-inside relation implies the way-below relation.

\begin{lemma}\label{lem:well-inside-implies-way-below}
  In every compact locale\/ $X$, if\/ $U \WellInside V$ then\/ $U \WayBelow V$,
  for every pair of opens\/ $U, V \oftype \opens{X}$.
\end{lemma}
\begin{proof}
  The proof is well-known. See \cite[303]{ptj-ss}.
\end{proof}

Since every open $U : \opens{X}$ is compact if and only if $U \WayBelow U$ and
is clopen if and only if $U \WellInside U$, we immediately obtain the following
corollary:

\begin{corollary}\label{cor:clopen-implies-compact-in-compact-locale}
  In every compact locale, clopens are compact.
\end{corollary}

What about the converse of this implication? In \cref{sec:zd-sub-regular}, we
will prove that the way-below relation implies the well-inside relation in
\emph{regular locales}\index{regular locale}, as we define them in
\cref{defn:regular-locale}.

\section{Bases}\label{sec:bases}

In the context of point-set topology,
a \define{base} (or \define{basis}) for a topology $X$~\cite[12]{engelking-topology}
is a collection $\mathcal{B}$ of open subsets of $X$ such that every open $U \in
\opens{X}$ can be expressed as the union of a subcollection of $\mathcal{B}$.
These open sets given by $\mathcal{B}$ are called
the \defineCE{basic opens}{basic open}.

In the context of point-free topology, this notion of base is abstracted by the
order-theoretic notion of a \indexed{generating lattice}. In
\cite[548]{elephant-vol-1}, for example, a base for a locale\index{base for a
locale} $X$ is defined as a subset $\mathcal{B} \subseteq \opens{X}$ such that every element of
the frame $\opens{X}$ is expressible as a join of members of $\mathcal{B}$. This
definition can be readily translated into the context of type theory:

\begin{definition}\label{defn:int-base}
  Let $X$ be a locale. A family of opens $\FamEnum{i}{I}{B_i}$ is said to
  form a base for $X$ if, for every $U : \opens{X}$, there is a specified
  directed subfamily $\FamEnum{j}{J}{B_{i_j}}$ such that
  $\IdTy{U}{\FrmJoin_{j : J} B_{i_j}}$. This is called the
  \define{basic covering family} for the open $U$.
\end{definition}

Note that we did not require the collection of basic opens to be given by an
\indexedp{embedding} family in the above definition. Technically, the precise translation
of the notion of generating lattice would include this requirement since subsets
correspond to embedding families. However, it turns out to be quite useful for
our purposes to consider a variant of this notion where we permit repetitions in
the family of basic opens. We call this an
\defineCE{intensional~base}{base!intensional}
and distinguish it from the notion of base with the embedding requirement,
which we call an \emph{extensional~base}\index{base!extensional}.
We discuss the difference between the two further in \cref{sec:int-vs-ext}.

Another variant of the notion of base that we consider is that of a
\defineCE{weak base}{base!weak}, where we weaken the specified existence
requirement for basic covering families to \emph{unspecified} existence.
We discuss this further in \cref{weak-vs-strong}.
The motivation for this may not be clear prima facie, but it turns out to be
quite important for the predicative development of the theory of spectral and
Stone locales. We discuss this in detail in \cref{chap:spec-and-stone}.

\subsection{Intensional vs.\ extensional bases}\label{sec:int-vs-ext}

We consider a variant of the notion of base that we call \emph{extensional
base}:

\begin{definition}[Extensional base]\label{defn:ext-base}
  A base $\FamEnum{i}{I}{B_i}$ is said to be
  \defineCE{extensional}{base!extensional}\index{extensional base}
  if the function $B : \ArrTy{I}{\opens{X}}$ is an \indexed{embedding}
  i.e.\ it is a subtype in the sense of \cref{defn:subtype}.
\end{definition}

The requirement that the family of basic opens be an embedding states that every
basic open has a unique representation in the base. Although this captures
exactly the standard notion of locale base, it is often more convenient to work
with bases that \emph{do have} repetitions. An example of this is the discrete
locale:

\begin{example}\label{example:discrete-locale-base-repetition}
  Let $X$ be a small type.
  The function $\beta : \ListTy{X} \to \PowSym(X)$, defined as
  \begin{equation*}
    \beta(x_0, \ldots, x_{n-1}) \quad\is\quad y \mapsto \IdTyWithType{X}{y}{x_0} \vee \cdots \vee \IdTyWithType{X}{y}{x_{n-1}},
  \end{equation*}
  gives a (small) intensional base for the frame $\PowSym(X)$. However, this is
  not an extensional base since every permutation of a list will give the same
  subset.
  One can instead work with the \emph{image} of $\beta$ to remove the
  repetitions.
  Notice, however, that the image of $\beta$ is exactly the type of
  Kuratowski finite subsets of $X$, the smallness of which requires the \VAxSR{}
  principle.
  So  taking the notion of extensional base as the standard definition would
  force us to adopt an extraneous assumption.
\end{example}

The process of taking the image to obtain an extensional base, which we
mentioned in the above example, can be generalized to arbitrary bases. We call
this the \emph{extensionalization of an intensional base}. It crucially relies
on the \indexedp{\VAxSR{}} principle.

\begin{lemma}\label{lem:base-img-small}
  For every small base $\FamEnum{i}{I}{B_i}$ on a locale $X$, the image
  of $B : \ArrTy{I}{\opens{X}}$ is a small type, assuming the \VAxSR{}
  principle.
\end{lemma}
\begin{proof}
  Let $\FamEnum{i}{I}{B_i}$ be a small base.
  By the \VAxSR{} principle (\cref{cor:set-replacement-principle}), showing
  (1) that $I$ is a small type and (2) that $\opens{X}$ is a locally small set is
  sufficient to show that $\image{B}$ is a small type. The former is immediate by
  assumption and the latter is given by \cref{prop:local-smallness-equiv}.
\end{proof}

Since replacing an index of a base with the image does not affect the joins of
the covering families given by the base, we have the following corollary:

\begin{corollary}[Base extensionalization]\label{lem:extensionalization}
  For every small intensional base $\FamEnum{i}{I}{B_i}$ on a locale $X$, the
  inclusion $\image{B} \hookrightarrow \opens{X}$, formally given by the first
  projection, is a small extensional base that contains exactly the same opens.
\end{corollary}

\subsection{Weak vs.\ strong bases}\label{weak-vs-strong}

Consider a family $\FamEnum{i}{I}{B_i}$. Neither the statement that it forms an
intensional base nor the statement that it forms an extensional base is a
proposition. This is due to the fact that there are multiple ways for a base to
give covering families with the same join.
For example, one could choose to include the bottom element or not in a
basic covering family.
Although this will not change the join of the family, the family will formally
be a different one.

In \cref{chap:spec-and-stone}, we will encounter cases of locales where an
extensional base can be shown to be unique. In the context of such locales, the
fact that forming a base is not a property will be the limiting factor
preventing us from obtaining a propositional notion of base. To address this
problem, we define the notion of \emph{weak base} where the requirement of
specified existence is relaxed to \indexedp{unspecified existence}.

\begin{definition}[Weak base]\label{defn:weak-base}
  A \defineCE{weak base}{base!weak} for a locale $X$ is a family
  $\FamEnum{i}{I}{B_i}$ such that for every $U : \opens{X}$, there is an
  unspecified directed subfamily $\FamEnum{j}{J}{B_{i_j}}$ with
  $U \IdTySym \FrmJoin_{j : J} B_{i_j}$.
\end{definition}

\begin{remark}
  It is important to note here that this notion of weak base is \emph{not}
  well behaved in general: many results that one would like to prove using a
  base
  requires the specified existence of covering families. This should therefore be
  viewed as an auxiliary technical notion that has certain uses, and is
  well behaved in the context of certain classes of locales where the structure
  of the covering of the families can be obtained from their unspecified existence.
\end{remark}

In contexts where we need to be careful about the distinction, we use the
explicit term \defineCE{strong base}{base!strong} for \cref{defn:int-base}.
Unless otherwise specified, all bases are strong. The intensional-extensional
distinction, which we specified in the context of strong bases, also applies to
weak bases.

\subsection{Directification of bases}
\label{sec:directification}

We required the basic covering families given by a base to be directed in all of
the definitions above. In the context of a frame, where we have all finite
joins, this requirement is not essential since one can always choose to work
with the closure of the base under finite joins (as in
\cref{defn:closure-under-finite-joins}), in which the basic covering families
are directed.
More specifically, here is how one directifies\index{directify} a base:
\begin{enumerate}
  \item\label{step:close-under-joins}
    Given a base $\FamEnum{i}{I}{B_i}$, we take its closure under finite joins.
  \item For each basic covering family given by the base $\FamEnum{i}{I}{B_i}$,
    we directify it also by closing it under finite joins.
    This is made possible by Step~(\ref{step:close-under-joins}).
\end{enumerate}
Step~(\ref{step:close-under-joins}) above can be skipped if the base in
consideration happens to be closed under finite joins already.

\begin{lemma}[\AgdaLink{Locales.Frame.html\#45778}]\label{lem:directification}
  For every base\/ $\FamEnum{i}{I}{B_i}$ of a locale\/ $X$, the family
  \[B^{\uparrow} \oftype \ListTy{I} \to \opens{X}\text{,}\]
  obtained by closing it under finite joins (as in \cref{defn:closure-under-finite-joins})
  forms a base with directed covering families.
\end{lemma}
\begin{proof}
  Let $U$ be an open of $X$. Denote by $\FamEnum{j}{J}{B_{i_j}}$ the basic
  covering family that $B$ gives for $U$.
  We define the directed covering family given by $B^{\uparrow}$ to be the
  subfamily $\alpha : \ListTy{J} \to \opens{X}$, defined as
  \begin{equation*}
    \alpha(j_0, \ldots, j_{n-1}) \is B_{i_{j_0}} \vee \cdots \vee B_{i_{j_{n-1}}},
  \end{equation*}
  which is obviously directed and is a subfamily of $B^{\uparrow}$.
\end{proof}

\index{closure under finite joins}
We note that the closure of a small base under finite joins is small, since
$\ListTy{I}$ inhabits the same universe as $I$ does.
Furthermore, we note that we could have defined the notion of base, in light of
Lemma~\ref{lem:directification}, without the requirement that the basic covering
families be directed.
We nevertheless prefer to include this requirement in the definition for
technical convenience.

\subsection{Examples}\label{sec:examples}

We now look at some more examples of bases, starting with the terminal locale
$\LocTerm{\UU}$.

\subsubsection{Bases for the terminal locale}

In \cref{lem:initial-frame-proto-base}, we showed how a proposition can be
decomposed into a join of decidable propositions. Having pinned down the notion
of base, we now proceed to extend \cref{lem:initial-frame-proto-base} by
showing that decidable propositions form a base for the terminal locale:%
\index{decidable proposition}

\begin{example}[%
  \AgdaLink{Locales.InitialFrame.html\#9382}%
]\label{example:term-loc-base-1}
  Recall the family $\beta : \TwoTySym \to \hprop{\UU}$
  from \cref{lem:initial-frame-proto-base},
  defined as
  \begin{align*}
    \beta(\BoolZero) &\is \FalseProp{\UU}\\
    \beta(\BoolOne)  &\is \TrueProp{\UU}
  \end{align*}
  This forms a base for the terminal locale $\LocTerm{\UU}$.
  For every proposition $P : \hprop{\UU}$,
  the family $\{ \TruePropSym \mid p : P \}$
  is a covering family for $P$ by \cref{lem:initial-frame-proto-base}.
  We apply the directification procedure from \cref{lem:directification}
  to force the basic covering families to be directed.
\end{example}

Note that the covering families given above are not directed without the use of
directification. As it is somewhat technically inconvenient to work with a
directified base, we also show how the above family $\beta : \TwoTySym \to
\hprop{\UU}$ can be shown to form a directed base directly.

\begin{example}\label{example:term-loc-base-2}
  The family $\beta : \TwoTySym \to \hprop{\UU}$ from
  \cref{example:term-loc-base-1} forms a directed base for $\LocTerm{\UU}$.
  For every proposition $P : \hprop{\UU}$, the basic covering family is given by
  \begin{equation*}
    \paren{\SigmaType{b}{\TwoTySym}{\beta(b) \le P}}
      \xrightarrow{~\Comp{\beta}{\projISym}~}
    \hprop{\UU}.
  \end{equation*}
  This family is always inhabited by empty proposition $\FalseProp{\UU}$.
  Given two $b_1, b_2 : \TwoTySym$ with $\Impl{\beta(b_1)}{P}$ and
  $\Impl{\beta(b_2)}{P}$, we have that $\Impl{\beta(b_1 \vee b_2)}{P}$, and
  hence the least upper bound $b_1 \vee b_2$ is in the family. The fact that
  it covers $P$ was given in \cref{lem:initial-frame-proto-base}.
\end{example}

\subsubsection{Base construction for the \VSierpinski{} locale}

To construct a base for the \VSierpinski{} locale, we first do some
preparation.

\begin{definition}\label{defn:upset-of-prop}
  For every proposition $P \oftype \hprop{\UU}$, its \define{upset} (or
  \define{principal filter}), denoted $\upset P$, is defined as
  \begin{equation*}
    \upset P \is Q \mapsto (\Impl{P}{Q})\text{.}
  \end{equation*}
  This defines a function $\upset P : \hprop{\UU} \to \hprop{\UU}$.
\end{definition}

\begin{lemma}\label{lem:principal-filters-are-scott-continuous}
  For every decidable proposition\/ $P \oftype \hprop{\UU}$,
  the principal filter\/ $\upset P$
  is a \indexedp{\VScottOpen{} subset}\/ $\ArrTy{\hprop{\UU}}{\hprop{\UU}}$.
\end{lemma}
\begin{proof}
  Let $P : \hprop{\UU}$ be a decidable proposition and let
  $\FamEnum{i}{I}{Q_i}$ be a directed family of propositions.
  We need to show that $\Impl{P}{\ExistsType{i}{I}{Q_i}}$ if and only if
  $\ExistsType{i}{I}{\Impl{P}{Q_i}}$.
  If $P$ is inhabited, then it is easy to see that
  $\Impl{P}{\ExistsType{i}{I}{Q_i}}$ holds, since
  $\Impl{\TruePropSym}{\ExistsType{i}{I}{Q_i}}$ iff $\ExistsType{i}{I}{Q_i}$
  iff
  $\ExistsType{i}{I}{\Impl{\TruePropSym}{Q_i}}$.
  If $P$ is \emph{not} inhabited, $\ExistsType{i}{I}{\Impl{P}{Q_i}}$ must be
  trivially true by the elimination principle for bottom, since the family
  $\FamEnum{i}{I}{Q_i}$ is inhabited. We therefore have
  \begin{equation*}
    \ExistsType{i}{I}{\Impl{P}{Q_i}} \leftrightarrow \TruePropSym \leftrightarrow \Impl{P}{\ExistsType{i}{I}{Q_i}}.
  \end{equation*}
\end{proof}

\begin{lemma}\label{lem:principal-filter-on-bottom}
  The principal filter\/ $\upset \FalseProp{\UU}$ is
  the full subset of\/ $\hprop{\UU}$.
\end{lemma}
\begin{proof}
  This is obvious since every proposition is above $\FalsePropSym$.
\end{proof}

\begin{example}\label{ex:sierp-base}
  The family $\FamEnum{b}{\TwoTySym}{\upset \beta(b)}$ forms a base for the
  \VSierpinski{} locale.
  Every \VScottOpen{} subset $S : \ArrTy{\hprop{\UU}}{\hprop{\UU}}$ can
  be expressed as a join
  \begin{equation*}
    \IdTy{S}{\FrmJoin_{b : \TwoTySym} \setof{ \upset \beta(b) \mid \beta(b) \in S } }
  \end{equation*}
  For every proposition $P \in S$, since we know (from
  \cref{example:term-loc-base-2}) that
  \[\IdTy{P}{\FrmJoin_{\stackrel{b : \TwoTySym}{\beta(b) \le P}} \beta(b)},\]
  the Scott continuity of $S$ implies
  \begin{equation*}
    \FrmJoin_{\stackrel{b : \TwoTySym}{\beta(b) \le P}} S(\beta(b))\text{,}
  \end{equation*}
  since we have $S\paren{\FrmJoin_{\stackrel{b : \TwoTySym}{\beta(b) \le P}} \beta(b)}$.
  Conversely, given a proposition $P$ such that there is some $b : \TwoTySym$
  with $\beta(b) \in S$ and $\beta(b) \le P$, it is immediate by the monotonicity of $S$
  that $P \in S$.
\end{example}

\section{Important classes of locales}\label{sec:classes}

Having pinned down the notion of base in \cref{sec:bases}, we now look at
various classes of locales that can be defined simply by requiring the existence
of a base satisfying certain conditions. This approach is inspired by the
notions of \indexedp{locally presentable} and \indexedp{locally accessible}
categories~\cite{adamek-rosicky-book}, and \VTDJMHE{}'s application of it to the
predicative development of domain theory.

\begin{definition}[{\AgdaLink{Locales.SmallBasis.html\#12277}} Spectral locale]%
\label{defn:int-spec-base}
  A locale $X$ is called \definep{spectral}\index{locale!spectral}
  if it has some small base $\FamEnum{i}{I}{B_i}$ satisfying the following three
  conditions.
  \begin{enumerate}[label={$\left({\mathsf{SP_{b}{\arabic*}}}\right)$}, leftmargin=4em]
    \item\label{item:spec-base-consists-of-compact-opens}
      Every $B_i$ in the base is compact.
    \item\label{item:spec-base-contains-top}
      It contains the top open $\One{X}$.
    \item\label{item:spec-base-coherence}
      It is closed under binary meets i.e.\ for every two $i, j \oftype I$, there
      exists some $k \oftype I$ such that $\IdTy{B_k}{B_i \meet B_j}$.
  \end{enumerate}
  A base satisfying these conditions is called a
  \emph{spectral base}\index{base!spectral}\index{spectral base}.
\end{definition}

In \cref{chap:spec-and-stone}, we will give various alternative
characterizations (using \VUniAx{}) of this notion of spectral locale. However,
we assign conceptual priority to this definition as it is an instance of a
general technique that usually results in the correct notion in the context of
predicative \VUF{}.

\begin{definition}[Zero-dimensional locale]\label{defn:zd-locale-0}
  A locale $X$ is called \emph{zero-dimensional}\index{{zero-dimensional locale}|textbf}
  if it has some base $\FamEnum{i}{I}{B_i}$ consisting of \indexedCE{clopens}{clopen}.
  We use the term \define{zero-dimensional base} to refer to such bases.
\end{definition}

\begin{definition}\label{defn:stone-base}
  A locale $X$ is called
  \definep{Stone}\index{locale!Stone|see {Stone locale}}\index{{Stone locale}}
  if it has some base $\FamEnum{i}{I}{B_i}$ satisfying the following three conditions.
  \begin{enumerate}[label={$\left({\mathsf{ST_{b}{\arabic*}}}\right)$}, leftmargin=4em]
  \item\label{item:stone-base-consists-of-compact-opens}
    Every $B_i$ in the base is compact
  \item\label{item:stone-base-contains-top}
    It contains the top open $\One{X}$.
  \item\label{item:stone-base-consists-of-clopens}
    Every $B_i$ in the base is clopen.
  \end{enumerate}
  A base satisfying the above conditions is called a \define{Stone base}.
\end{definition}

\begin{definition}[Locally compact locale]\label{defn:locally-compact}
  A locale $X$ is called \define{locally compact} if its frame is a continuous
  lattice\index{continuous lattice}, which is to say it has some base
  $\FamEnum{i}{I}{B_i}$ satisfying the condition that, for every open $U :
  \opens{X}$, the basic opens in the covering family for~$U$ are all
  \VWayBelow{}\index{way below} $U$.
\end{definition}

\begin{definition}[Regular locale]\label{defn:regular-locale}
  A locale $X$ is called
  \definep{regular}\index{regular locale}
  if it has some
  base $\FamEnum{i}{I}{B_i}$ such that for every open $U : \opens{X}$, the
  covering subfamily $\IdTy{U}{\FrmJoin_{j : J}{B_{i_j}}}$ satisfies
  $B_{i_j} \WellInside U$, for every $j : J$
\end{definition}

\begin{lemma}\label{lem:zero-dimensional-implies-regular}
  Every zero-dimensional locale is regular.
\end{lemma}
\begin{proof}
  Let $X$ be a locale with a zero-dimensional base $\FamEnum{i}{I}{B_i}$
  and consider some $U : \opens{X}$. The basic covering family
  $\IdTy{U}{\FrmJoin_{j : J} B_{i_j}}$ consists of basic opens that satisfy
  $B_{i_j} \WellInside B_{i_j}$. Since each $B_{i_j}$ is below $U$, it follows
  from \cref{lem:wi-prop-1} that each $B_{i_j}$ satisfies
  $B_{i_j} \WellInside U$.
\end{proof}

\section{Adjoint functor theorem for frames}
\label{sec:aft}

In \cref{sec:implications}, we explained that many essential constructions of
traditional locale theory, such as Heyting implications and arbitrary meets, are
not available in our predicative context. In this section, we show how this
situation can be remedied by giving a predicative proof of the \VAFT{} for
frames with small bases.
The key idea here is that our notion of base serves the r\^{o}le of the
\indexedp{solution set condition}.

\begin{definition}
  Let $(A, \le_A)$ and $(B, \le_B)$ be two preordered sets.
  Given monotone maps $f : \ArrTy{A}{B}$ and $g : \ArrTy{B}{A}$, the map $g$ is
  said to be the \define{right adjoint} of $f$
  (and $f$ the \define{left adjoint} of $g$) if it satisfies
  \begin{equation*}
    f \dashv g \quad\is\quad \text{$f(a) \le_B b \Leftrightarrow a \le_A g(b)$ for all $a : A, b : B$}\text{.}
  \end{equation*}
\end{definition}

\begin{lemma}[\AgdaLink{Locales.GaloisConnection.html\#1468}]%
\label{lem:right-adjoint-prop}
  The statement $f \dashv g$ expressing that $f$ and $g$ form an adjunction is
  a proposition. Furthermore, for every monotone map $f : \ArrTy{A}{B}$,
  its type of right adjoints,
  i.e.\ $\SigmaType{g}{B \to A}{f \dashv g}$,
  is a proposition as right adjoints are unique.
\end{lemma}

In the case of non-posetal categories, the \VAFT{} involves the so-called
solution set condition. In our predicative foundational setting, a small base
plays the r\^{o}le of the solution set, allowing us to construct the right
adjoint of a frame homomorphism. A somewhat interesting observation here is that
the unspecified existence of a \indexed{weak base}
suffices for proving the \VAFT{}.

\begin{theorem}[%
  \AgdaLink{Locales.AdjointFunctorTheoremForFrames.html\#2041} Posetal \VAFT{}%
]\label{thm:aft}
  Let $X$ and $Y$ be two locales and
  let $h : \opens{Y} \to \opens{X}$ be a monotone map of frames.
  Assume that $Y$ has an unspecified, small, and weak base.
  The map $h$ has a right adjoint if and only if it preserves the joins of small
  families.
\end{theorem}
\begin{proof}
  Let $h : \opens{Y} \to \opens{X}$ be a monotone map of frames
  and assume that $Y$ has an unspecified, small, and weak base.

  The forward direction is easy: suppose $h$ has a right adjoint $g : \opens{X}
  \to \opens{Y}$ and let $\FamEnum{i}{I}{U_i}$ be a family in $\opens{Y}$. By the
  uniqueness of joins, it is sufficient to show that $h(\bigvee_i U_i)$ is the join of
  the family $\FamEnum{i}{I}{h(U_i)}$. It is clearly an upper bound by the fact
  that $h$ is monotone. Given some other upper bound $V$ of
  $\FamEnum{i}{I}{h(U_i)}$,
  we have $h(\bigvee_i U_i) \le V$ since $h(\bigvee_i U_i) \le V \leftrightarrow \left(\bigvee_i
  U_i\right) \le g(V)$ meaning it is sufficient to show $U_i \le g(V)$ for each $U_i$.
  Since $U_i \le g(V)$ if and only if $h(U_i) \le V$, we are done as the latter can be seen to
  hold directly from the fact that $V$ is an upper bound of the family in
  consideration.

  For the converse, first note that we may appeal to the induction principle of
  propositional truncation, since the type of right adjoints of $h$ is a
  proposition by \cref{lem:right-adjoint-prop}. We therefore work with a specified
  small base $\FamEnum{i}{I}{B_i}$.
  Suppose $\IdTy{h(\bigvee_i U_i)}{\bigvee_{i : I} h(U_i)}$ for every small
  family of opens $\FamEnum{i}{I}{U_i}$. We define the right adjoint of $h$ as
  \begin{equation*}
    g(U) \quad\is\quad \bigvee \left\{ B_i \mid i \oftype I, h(B_i) \le U \right\}.
  \end{equation*}
  We need to show that $g$ is the right adjoint of $h$ i.e.\ that
  \(h(V) \le U \leftrightarrow V \le g(U)\)
  for every pair of opens $V : \opens{Y}$, $U : \opens{X}$.

  For the $(\Rightarrow)$ direction, assume $h(V) \le U$. It must be the case that
  $V = \bigvee_{j : J} B_{i_j}$ for
  some specified covering family $\FamEnum{j}{J}{i_j}$.
  This means that we just have to show $B_{i_j} \le g(U)$ for every $j : J$,
  which is the case since $h(B_{i_j}) \le h(V) \le U$.

  For the $(\Leftarrow)$ direction, assume $V \le g(U)$. This means that we have:
  \begin{align*}
    h(V) &\quad\le\quad h(g(U))\\
         &\quad=\quad h\left(\bigvee \left\{ B_i \mid i : I, h(B_i) \le U \right\}\right)\\
         &\quad=\quad \bigvee \left\{ h(B_i) \mid i : I, h(B_i) \le U \right\}\\
         &\quad\le\quad U
  \end{align*}
  so that $h(V) \le U$, as required.
\end{proof}

\subsection{Heyting implications and the open nucleus}

Using the \VAFT{}, we construct the right adjoint of a frame homomorphism on a
frame with a small (weak) base.
This will find several applications in our development.
Most importantly, it will allow us to define Heyting implications thereby
enabling us to work with the notion of \emph{open nucleus} in a predicative
foundational setting. This will be crucial to our construction of the patch
locale in \cref{chap:patch}.

\begin{definition}[\AgdaLink{Locales.AdjointFunctorTheoremForFrames.html\#5905}]\label{defn:right-adjoint-for-frame-homomorphism}
  Let $f : X \to Y$ be a continuous map of locales,
  where $Y$ has an unspecified small base.
  By Theorem~\ref{thm:aft},
  the frame homomorphism $f^* : \opens{Y} \to \opens{X}$
  has a right adjoint, denoted by
  \begin{equation*}
    f_* : \opens{X} \to \opens{Y}.
  \end{equation*}
\end{definition}

\begin{definition}[\AgdaLink{Locales.HeytingImplication.html\#2259} Heyting implication]%
\label{defn:heyting-implication}
  Let $X$ be a locale with some small base.
  For every open $U : \opens{X}$,
  the map $(\blank) \wedge U : \opens{X} \to \opens{X}$
  is a frame homomorphism by the \indexedp{frame distributivity} law.
  It thus has a \indexedp{right adjoint} as constructed in
  \cref{defn:right-adjoint-for-frame-homomorphism}.
  We denote this right adjoint by
  $U \Rightarrow (\blank) : \opens{X} \to \opens{X}$.
  The operation $(\blank) \Rightarrow (\blank)$ is known as \define{Heyting implication}.
\end{definition}

\begin{lemma}[\AgdaLink{Locales.PatchProperties.html\#10881}]
  Let\/ $X$ be a locale with a small weak base,
  and consider an open\/ $U \oftype \opens{X}$.
  The map\/ $\OpenNucleus{U} \oftype \Endomap{\opens{X}}$, defined as
  \begin{equation*}
    \OpenNucleus{U} \quad\is\quad V \mapsto U \Rightarrow V\text{,}
  \end{equation*}
  forms a nucleus.
\end{lemma}
\begin{proof}
  See \cite[\textsection II.2.4]{ptj-ss}.
\end{proof}

\begin{definition}[\AgdaLink{Locales.PatchProperties.html\#10881} Open nucleus]\label{defn:open-nucleus}
  The nucleus $\OpenNucleus{U}$ constructed in the above lemma
  is called the \define{open nucleus} induced by $U$.
\end{definition}


\chapter{Spectral and Stone locales}
\label{chap:spec-and-stone}

In the previous chapter, we presented the basics of a predicative approach to
constructive locale theory in \VUF{}.
We now start developing the theory of spectral and Stone locales in this
setting.
Our predicative approach is inspired by
\indexedp{locally presentable} categories~\cite{adamek-rosicky-book},
and their use by \VTDJMHE{}~\cite{tdj-thesis, tdj-scott-model}
in their predicative development of domain theory in \VUF{}.
The organization of this chapter is as follows:
\begin{description}[leftmargin=!,labelwidth=\widthof{Section 7.77:}]
  \item[\cref{sec:spec-intro}:] We provide a high-level overview of the notion
    of spectral locale along with some history.
  \item[\cref{sec:spec-defn}:] We discuss how the standard, impredicative
    definition of the notion of spectral locale can be translated into our
    predicative setting, and prove that this translation is equivalent to
    \cref{defn:int-spec-base}
    under the assumption of \VSRPrinciple{} and \VUniAx{}.
  \item[\cref{sec:spec-examples}:] We look at some examples, as well as
    counterexamples, of spectral locales.
  \item[\cref{sec:spec-dlat}:] We construct the frame of ideals over a
    small distributive lattice in preparation for our predicative development
    of Stone duality in \cref{sec:spec-duality}.
  \item[\cref{sec:zero-dimensional}:] We define \VZeroDimensional{} and regular
    locales, and record some of their salient properties.
  \item[\cref{sec:stone}:] We define Stone locales and start developing their
    theory.
  \item[\cref{sec:spec-duality}:]
    We prove that there is an equivalence between
    the type $\Spec$ of
    large, locally small, and small-complete spectral locales
    and the type~$\DLat{}$ of small distributive lattices.
  \item[\cref{sec:equiv-functor-part}:] We extend the type equivalence from
    \cref{sec:spec-duality} to show that there is an equivalence of categories
    between \(\DLat{}\) and \(\Spec\).
\end{description}

\section{Introduction}\label{sec:spec-intro}

In the introduction to \cref{chap:basics}, we mentioned that the idea of
studying frames as spaces can be traced back to the work of
Stone~\cite{stone-1934, stone-1936}.
Stone's work originated in the context of what are now called
\indexed{{Stone spaces}}.
In the context of classical point-set topology, these can be defined as the
spaces of \indexedCE{ultrafilters}{ultrafilter} of
Boolean algebras~\cite[\textsection 2.5]{smyth-topology}.
The remarkable insight that arose in Stone's work was that such spaces can be
studied in a \emph{completely algebraic} way, since they are fully characterized
by their Boolean algebras of clopen subsets. This is captured by
Stone's representation theorem~\cite[\textsection II.4.4]{ptj-ss}
stating that there is a dual categorical equivalence between the category of
Stone spaces and continuous maps and the category of Boolean algebras and their
homomorphisms.
This is often referred to as \indexed{Stone duality} for Boolean algebras, and
it is historically notable as the first duality of this kind to be discovered.

Stone later generalized~\cite{stone-1937} this duality from Boolean algebras to
distributive lattices (as explained in \cite{dst-spectral-spaces}). The
fundamental idea here is that, whilst Stone spaces are fully characterized by
their Boolean algebras of clopen subsets, spectral spaces are fully
characterized by their distributive lattices of compact opens. Compact open
subsets coincide with the clopen subsets in Stone spaces, so this generalization
from Stone spaces to spectral ones consists in the observation that it is really
the \emph{algebra of compact opens} that characterizes the space in
consideration. For the detailed history of the subject, we refer the reader to
the survey in \cite{dst-spectral-spaces}.

In classical point-set topology, a \indexedp{spectral space} is
defined\footnote{%
  Note that in \cite[Definition~1.1.5]{dst-spectral-spaces} the term
  \emph{quasi-compact} is used in this definition.
  We prefer to use the term \emph{compact} here.%
}~\cite[Definition~1.1.5]{dst-spectral-spaces}
as a space $X$ satisfying the following conditions:
\begin{enumerate}
  \item Space $X$ is compact and $T_0$.
  \item The collection of compact open subsets forms a base for the topology
    on $X$.
  \item Compact open subsets are closed under binary intersections.
  \item Space $X$ is \indexedp{sober}.
\end{enumerate}
In this thesis, we will work with the localic manifestation of the notion of
spectral space, which we call a \indexed{spectral locale}\/
(called \define{coherent locale} by some, including John\-stone~\cite[\textsection II.3.2]{ptj-ss}).
Classically, it can be shown that every spectral locale is spatial, but this is
equivalent to the Prime Ideal Theorem for distributive lattices in a
constructive setting~\cite[46]{ptj-the-point}. When one adopts a point-free
approach towards spectral spaces, Stone duality is provable in a completely
constructive way.

\section{Spectrality in a predicative setting}
\label{sec:spec-defn}

In \cref{sec:classes}, we gave several examples of how various classes of
locales can be defined solely by stipulating the unspecified existence of a base
subject to certain conditions.
We also explained
(1) that this approach can be considered an adaptation of locally accessible and
presentable categories to the posetal setting, and
(2) that \VTDJMHE{} drew inspiration from the literature on these to develop a
satisfactory account of domain theory~\cite{tdj-thesis, tdj-scott-model}
in constructive and predicative \VUF{}.

We recall that the definition of spectral locale (\cref{defn:int-spec-base}) we
gave in \cref{sec:classes} was the following:
\begin{itemize}
  \item a locale $X$ is said to be \defineCE{spectral}{spectral locale} if
    it has some small spectral base, where
  \item \indexed{spectral base} means a base consisting of compact opens
    that is closed under finite meets.
\end{itemize}

We also explained that we gave conceptual priority to this definition, as it is
the instance of a general technique that often leads to predicatively
well-behaved definitions. It is not a priori clear, however, that this
definition corresponds to the traditional definition of
spectral locale~\cite[Paragraph~II.3.2]{ptj-ss}.
In \cref{sec:trad-motivation} below, we motivate an alternative predicatively
well-behaved definition through the traditional definition, and then show the
equivalence of the two in \cref{sec:spec-notions}.
The proof of equivalence is not straightforward and seems to rely crucially on
both \VUniAx{} and \VTheSRPrinciple{} (which is equivalent to the existence of
small set quotients as we explained in \cref{sec:set-replacement}).

\subsection{The traditional definition from a predicative point of view}
\label{sec:trad-motivation}

The traditional definition of a \define{spectral locale} is a locale whose
compact opens form a base closed under finite
meets~\cite[\textsection II.3.2]{ptj-ss}. A careful analysis of this definition in
a predicative setting shows that it is well-behaved only within the context of
complete and small locales, among which only the trivial ones can be constructed
in a predicative setting (as we discussed in detail in \cref{sec:primer}).
Let us examine this definition in detail to elucidate why it is not satisfactory
in a predicative context.

The direct translation of the traditional definition of spectral locale would
amount to the requirement that the inclusion
\[\CompactOpens{X} \xhookrightarrow{~\projI~} \opens{X}\]
form a base closed under finite meets. When working with
$(\UU, \UU, \UU)$\nobreakdash-frames, this family is always small since the
carrier is a small set. Notice, however, that when working with large, locally
small, and small-complete frames, there is no reason for the
type~$\CompactOpens{X}$ to be small in general.
This is problematic for at least two reasons:
\begin{enumerate}
  \item Such a base would not give us the constructions that we discussed in
    \cref{sec:implications};
    it is essential that the base is small for this purpose.
  \item The natural form of Stone duality to consider for spectral locales in
    our predicative setting is that between spectral locales and \emph{small}
    distributive lattices.
    As we will explain in detail later, taking the frame of (small) ideals over
    a small distributive lattice always gives a large, locally small, and
    small-complete frame. If satisfying Stone duality in this sense is taken as a
    desideratum for a well-behaved definition of the notion of spectral locale, the
    definition permitting a large type of compact opens is not adequate.
\end{enumerate}

It turns out, as we will discuss in this chapter soon, that \emph{both} of the
above problems can be remedied simply by stipulating that the type
$\CompactOpens{X}$ be small.
This leads us to the following predicative formulation of the notion of spectral
locale:

\begin{definition}[Spectral locale]\label{defn:spectral-locale}
  A locale $X$ is called \defineCE{spectral}{spectral locale} if it satisfies
  the following conditions:
  \begin{enumerate}[label={$\left({\mathsf{SP{\arabic*}}}\right)$}, leftmargin=4em]
    \item It is compact\index{compact locale} (i.e.\ the empty meet is compact).
      \label{item:sl-compact}
    \item Compact opens are closed under binary meets.
      \label{item:sl-coherent}
    \item The type $\CompactOpens{X}$ forms a weak base.
      \label{item:weak-base}
    \item The type $\CompactOpens{X}$ is small.\index{small (type)}
      \label{item:sl-smallness}
  \end{enumerate}
\end{definition}

In addition to the motivations discussed above, we also note that we do not have
any examples of locales satisfying
Conditions~\ref{item:sl-compact}--\ref{item:weak-base},
but not Condition~\ref{item:sl-smallness}.

Finally, we record the following characterization of the order in spectral
frames:

\begin{lemma}\label{lem:spectral-yoneda}
  Let\/ $X$ be a spectral locale. The order of the frame\/ $\opens{X}$ is
  equivalent to the order defined as
  \begin{equation*}
    U \le_s V \quad\is\quad \PiTy{K}{\CompactOpens{X}}{\Impl{K \le U}{K \le V}}.
  \end{equation*}
\end{lemma}
\begin{proof}
  ($\Rightarrow$) It is immediate that $U \le V$ implies $U \le_s V$, for every
  pair of opens $U, V \oftype \opens{X}$. The nontrivial direction is the
  converse:
  ($\Leftarrow$) let $U$ and $V$ be a pair of opens with $U \le_s V$.
  By \ref{item:weak-base}, we know that there is an unspecified basic covering
  family $\FamEnum{i}{I}{K_i}$ for $U$ that consists of compact opens.
  Since we know
  $\IdTy{U}{\FrmJoin_{i : I} K_i}$,
  it suffices to show $\paren{\FrmJoin_{i : I} K_i} \le V$.
  For every $i : I$,
  we know that $K_i$ is a compact open with $K_i \le U$,
  and we thus have $K_i \le V$
  by our assumption that $U \le_s V$.
\end{proof}

\subsection{Propositionality of being spectral}
\label{sec:spec-propositionality}

We would like the type expressing the spectrality of a locale to be a
proposition,
for which we need to ensure that all four conditions in
\cref{defn:spectral-locale} are propositions.
Surprisingly, it seems to be the case that we \emph{have to} use \VUniAx{} to
force Condition~\ref{item:sl-smallness} to be a proposition, since the
propositionality of being small is logically equivalent to \VAxUA{} by
\cref{lem:being-small-is-prop}.
Moreover, to ensure that Condition \ref{item:weak-base} is a proposition, we
make use of the notion of \indexed{weak base} from \cref{defn:weak-base}.

\begin{lemma}\label{prop:spectral-is-prop}
  Assuming \VTheUniAx{}, the type expressing that\/ $X$ is spectral
  is a proposition for every locale\/ $X$.
\end{lemma}
\begin{proof}
  It is easy to see that
  Conditions \ref{item:sl-compact}--\ref{item:weak-base}
  are all propositions.
  Assuming \VUniAx{}, Condition~\ref{item:sl-smallness} is a proposition by
  \cref{lem:being-small-is-prop}.
\end{proof}

In the next section, we show that
\cref{defn:int-spec-base} and \cref{defn:spectral-locale}
are equivalent.

\subsection{Relating the two approaches}
\label{sec:spec-notions}

In \cref{sec:classes}, we defined various classes of locales as locales with
unspecified bases satisfying certain conditions.
In \cref{defn:int-spec-base} from that section, we also gave a definition of the
notion of spectral locale, using the aforementioned general technique that
applies to many other classes of locales.
Having pinned down \cref{defn:spectral-locale}, which is well motivated from the
perspective of the standard, impredicative definition, we now proceed to prove
that these two definitions are equivalent
under the assumption of \VAxUA{} and \VAxSR{}.
Our motivation for proving this equivalence is twofold:
\begin{enumerate}
  \item It captures that the choice of the base \emph{does not matter} under
    the conditions of spectrality, even in a constructive and predicative setting,
    which is a fact that is important to observe in itself.
  \item In \cref{sec:spec-duality,sec:equiv-functor-part}, we prove Stone
    duality for \cref{defn:spectral-locale}, which justifies that we have
    identified the right category of spectral locales.
    The equivalence of these two definitions is necessary for this justification
    to apply to \cref{defn:int-spec-base}, since Stone duality seems applicable
    only to \cref{defn:spectral-locale}. In other words, proving the equivalence
    of the two definitions seems to be a natural step for proving
    that \cref{defn:int-spec-base} enjoys Stone duality.
\end{enumerate}

We now start to do some preparation towards the proof of this equivalence.

\begin{lemma}\label{lem:cmp-bsc}
  For every locale\/ $X$ with a base $\FamEnum{i}{I}{B_i}$,
  every compact open of\/ $X$ falls in the base.
  In other words, for every compact open\/ $K$,
  there exists an index\/ $i \oftype I$ such that\/ $\IdTy{K}{B_i}$.
\end{lemma}
\begin{proof}
  Let $K \oftype \opens{X}$ be a compact open. As $\FamEnum{i}{I}{B_i}$ is a
  strong base, there is a specified directed family of indices
  $\FamEnum{j}{J}{i_j}$ such that
  $K = \bigvee_{j : J} B_{i_j}$. By the compactness of $K$, there is some
  $k \oftype J$ such that $K \le B_{i_k}$.
  Clearly, $B_{i_k} \le K$ is also the case since $K$ is an upper bound of the
  family $\FamEnum{j}{J}{B_{i_j}}$, which means we have $\IdTy{K}{B_{i_k}}$.
\end{proof}

\begin{corollary}[\AgdaLink{Locales.SmallBasis.html\#5651}]\label{prop:base-img-equiv}
  For every locale\/ $X$
  with a small base\/ $\FamEnum{i}{I}{B_i}$ consisting of compact opens,
  we have an equivalence of types:
  \[\Equiv{\image{B}}{\CompactOpens{X}}\text{.}\]
\end{corollary}

This equivalence is particularly useful for us since it serves the r\^{o}le of a
resizing principle for the type $\CompactOpens{X}$:

\begin{lemma}\label{lem:resizing-for-compact-opens}
  For every locale\/ $X$,
  if\/ $X$ has a small base that consists of compact opens,
  then the type\/ $\CompactOpens{X}$ is small,
  under the assumption of \indexedpCE{\VTheSRPrinciple{}}{set replacement principle}.
\end{lemma}
\begin{proof}
  Let $X$ be a locale with a small base $\FamEnum{i}{I}{B_i}$.
  It was shown in \cref{lem:base-img-small}, using \VTheSRPrinciple{},
  that having a specified small base implies the smallness of $\image{B}$.
  We have just established in \cref{prop:base-img-equiv} above that
  \[\Equiv{\CompactOpens{X}}{\image{B}}\text{,}\]
  which thus implies that $\CompactOpens{X}$ has $\image{B}$ as a small copy.
\end{proof}

\begin{lemma}\label{lem:base-to-spec}
  Assuming \VTheUniAx{},\index{univalence axiom}
  every locale with an unspecified small spectral base
  is spectral.
\end{lemma}
\begin{proof}
  First, notice that the conclusion is a proposition since being spectral is a
  proposition by Lemma~\ref{prop:spectral-is-prop} (which uses \VAxUA{}).
  This means that we may appeal to the induction principle of propositional
  truncation and assume we have a specified spectral base $\FamEnum{i}{I}{B_i}$.
  We now verify each condition from \cref{defn:spectral-locale}:
  \begin{description}
    \item[\ref{item:sl-compact}] The top element is basic and hence compact so
      Condition~\ref{item:sl-compact} holds.
    \item[\ref{item:sl-coherent}]
      Given two compact opens $K_1$ and $K_2$, they must be basic by
      \cref{lem:cmp-bsc}, meaning there exist $k_1, k_2$ such that
      $\IdTy{K_1}{B_{k_1}}$ and $\IdTy{K_2}{B_{k_2}}$.
      Because the base is closed under binary meets, we know that there are
      some $k_3$ with $\IdTy{B_{k_3}}{K_1 \meet K_2}$,
      which implies that $K_1 \meet K_2$ is compact as
      \ref{item:spec-base-consists-of-compact-opens} says that the base
      consists of compact opens.
      This concludes that Condition~\ref{item:sl-coherent} holds.
    \item[\ref{item:weak-base}]
      Consider an open $U \oftype \opens{X}$.
      We know that there is a specified small family~$\FamEnum{j}{J}{i_j}$ of
      indices such that $\IdTy{U}{\FrmJoin_{j : J} B_{i_j}}$.
      The subfamily $\FamEnum{j}{J}{B_{i_j}}$ is then clearly a small directed
      family with each $B_{i_j}$ compact, which was what we needed.
    \item \ref{item:sl-smallness}
      Follows directly from \cref{lem:resizing-for-compact-opens}.
  \end{description}
\end{proof}

\begin{lemma}[\AgdaLink{Locales.SmallBasis.html\#23581}]\label{lem:spec-gives-base}
  If a locale\/ $X$ is spectral then it has a specified, small, and extensional base.%
  \index{extensional base}
\end{lemma}
\begin{proof}
  Let $X$ be a spectral locale.
  We claim that the inclusion $\CompactOpens{X} \hookrightarrow \opens{X}$,
  which is formally given by the first projection,
  is an extensional and spectral small base.
  The fact that it is small is given by Condition~\ref{item:sl-smallness}.
  By \ref{item:sl-compact} and \ref{item:sl-coherent}, we know that this base
  contains the top open $\One{X}$ and is closed
  under binary meets.

  It remains to show that $\CompactOpens{X} \hookrightarrow \opens{X}$ forms a
  strong base.
  For the basic covering family of an open $U \oftype \opens{X}$, we pick the
  inclusion:
  \begin{equation*}
    \paren{\SigmaType{K}{\CompactOpens{X}}{\paren{K \le U}}}
    \xhookrightarrow{~\projISym~}
    \CompactOpens{X},
  \end{equation*}
  which is again small by
  \ref{item:sl-smallness} and the local smallness of the frame $\opens{X}$.
  It is clear that $U$ is an upper bound of this family so it remains to show
  that it is the least upper bound.
  Consider some $V$ that is an upper bound of this family. By
  \cref{lem:spectral-yoneda}, it suffices to show $K \le U$ implies
  $ K \le V$, for every compact open $K$.
  Every compact $K \le U$ falls in the above family by construction
  which implies $K \le V$
  as $V$ is assumed to be an upper bound of the family.
\end{proof}

\begin{lemma}[\AgdaLink{Locales.SmallBasis.html\#24477}]\label{lem:split-support}
  From every unspecified intensional spectral base on a locale\/ $X$,
  we can obtain a specified extensional spectral base.
\end{lemma}
\begin{samepage}
\begin{proof}
  Let $X$ be a locale with an unspecified intensional spectral base.
  By \cref{lem:base-to-spec}, we know that it is spectral,
  and therefore that it has a specified extensional base by \cref{lem:spec-gives-base}.
\end{proof}
\end{samepage}

In other words, the type of intensional spectral bases has
\indexed{split support} in the terminology of~\cite{keca}.
A type $X$ is said to have split support if $\TruncTy{X} \to X$.
The mere existence of a base allows us to obtain the unique, canonical base
$\CompactOpens{X} \hookrightarrow \opens{X}$, which can be thought of as the
largest possible base satisfying the conditions of spectrality.
Note here that the propositionality of being spectral,
which we discussed in \cref{sec:spec-propositionality},
is crucial since the proof uses \cref{lem:base-to-spec}.

We also record here the following fact about the closure of a spectral base
under finite joins:

\begin{lemma}\label{rmk:directification-spec}
  For every spectral base\/ $\FamEnum{i}{I}{B_i}$, its
  \indexedp{closure under finite joins} also forms a spectral base.
\end{lemma}
\begin{proof}
  We know by \cref{lem:compact-fin-join-closure} that the closure of the family
  of basic opens also consists of compact opens.
  It is easy to see that this also forms a base.
\end{proof}

We are finally ready to prove the main characterization theorem for spectral
locales, showing that various seemingly different approaches to the notion of
spectral locale are all equivalent.

\begin{theorem}\label{thm:spec-characterization}
  For every locale\/ $X$, the following are logically equivalent:
  \begin{enumerate}
    \item\label{it:II}%
      $X$ has an unspecified intensional small spectral base
      (\cref{defn:int-spec-base}).
    \item\label{it:I}%
      $X$ is spectral in the sense of \cref{defn:spectral-locale}.
    \item\label{it:IIPrime}%
      $X$ has an unspecified extensional small spectral base.
    \item\label{it:V}%
      The inclusion $\CompactOpens{X} \hookrightarrow X$ is an extensional
      spectral base, where to an open $U$ we assign the directed family of
      compact opens below it as in the construction of
      Lemma~\ref{lem:spec-gives-base}.
    \item\label{it:III}%
      $X$ has a specified intensional small spectral base.
    \item\label{it:IV}%
      $X$ has a specified extensional small spectral base.
  \end{enumerate}
\end{theorem}
\begin{proof}
  First, observe the obvious implications: we have that
  \((\ref{it:IV}) \ImplSym (\ref{it:IIPrime}) \ImplSym (\ref{it:II})\) and
  that
  ${(\ref{it:IV})} \ImplSym {(\ref{it:III})} \ImplSym (\ref{it:II})$.
  The converse of $(\ref{it:III}) \ImplSym (\ref{it:II})$ was given in
  Lemma~\ref{lem:split-support}, which concludes the equivalence
  $(\ref{it:III}) \BiImplSym (\ref{it:II})$.
  For $(\ref{it:IV}) \BiImplSym (\ref{it:V})$, observe that every
  extensional small spectral base must be equivalent to $\CompactOpens{X}$,
  since $\Equiv{\image{B}}{\CompactOpens{X}}$ by \cref{prop:base-img-equiv}.
  For $(\ref{it:I}) \BiImplSym (\ref{it:IV})$, the forward
  direction was given by Lemma~\ref{lem:spec-gives-base}. For the converse,
  recall that we know $(\ref{it:II}) \ImplSym (\ref{it:I})$ by \cref{lem:split-support},
  and that $(\ref{it:IV}) \ImplSym (\ref{it:III}) \ImplSym (\ref{it:II})$.
\end{proof}

Note that Conditions~(\ref{it:II})--(\ref{it:V}) above are all propositions,
whereas Conditions~(\ref{it:III}) and~(\ref{it:IV}) need not be propositions in
general.

\subsection{The category of spectral locales}
\label{sec:category-spec}

Having defined the notion of spectral locale, we now present the category of
spectral locales by defining the notion of \emph{spectral map}:

\begin{definition}
  A \indexedp{continuous map} $f \oftype X \to Y$ is called
  a \defineCE{spectral}{spectral map}
  if $f^*(K)$ is compact for every compact open $K$ of $Y$.
\end{definition}

We denote by $\Spec_{\UU}$
\nomenclature{\(\Spec_{\UU}\)}{category of spectral locales over base universe \VUni{\UU}}
the category of spectral locales and spectral maps over the base universe \VUni{\UU}.
If the base universe is clear from context, we drop the subscript and simply
write $\Spec$.
\nomenclature{\(\Spec\)}{category of spectral locale with base universe left implicit}

This category satisfies some desirable properties thanks to the definition of
spectral locale ensuring the existence of a small base.
For example, we have the following:

\begin{lemma}
  For every map\/  $f \oftype \ArrTy{X}{Y}$ of spectral locales,
  the right adjoint of the frame homomorphism\/
  $f^* \oftype \ArrTy{\opens{Y}}{\opens{X}}$
  exists.
\end{lemma}
\begin{proof}
  We showed in \cref{thm:spec-characterization} that spectrality implies
  having a small base.
  We can therefore construct the right adjoint of $f^*$ as in
  \cref{defn:right-adjoint-for-frame-homomorphism}.
\end{proof}

\subsubsection{Characterization in terms of perfect maps}

Using the existence of right adjoints, the characterization of spectral maps in
terms of their right adjoints can be adapted to our setting.

\begin{definition}[\AgdaLink{Locales.PerfectMaps.html\#1550}]\label{defn:spectral-map}
  A continuous map $f \oftype \ArrTy{X}{Y}$ of locales is called
  \defineCE{perfect}{perfect map}
  if the right adjoint of its frame homomorphism is \VScottContinuous{}.
\end{definition}

\begin{lemma}[\AgdaLink{Locales.PerfectMaps.html\#1776}]\label{prop:perfect-resp-way-below}
  Let $f : X \to Y$ be a perfect map where $Y$ is a locale with some small base.
  The frame homomorphism $f^*$ respects the \indexedp{way-below relation},
  that is, $U \WayBelow V$ implies $f^*(U) \WayBelow f^*(V)$, for
  every pair of opens $U, V \oftype \opens{Y}$.
\end{lemma}

A proof of this fact can be found in \cite{mhe-patch-short} and it works in our
predicative setting.

\begin{corollary}\label{cor:perfect-maps-are-spectral}
  Perfect maps are spectral as they preserve the compact opens.
\end{corollary}

\begin{lemma}
  For spectral locales\/ $X$ and $Y$,
  a continuous map\/ $f \oftype \ArrTy{X}{Y}$ is spectral
  if and only if
  it is perfect.
\end{lemma}
\begin{proof}
  The forward direction is given by
  \cref{cor:perfect-maps-are-spectral}.
  For the backward direction,
  assume $f : X \to Y$ to be a spectral map. We have to show that the right adjoint
  $f_* : \opens{X} \to \opens{Y}$ of its defining frame homomorphism is
  \VScottContinuous{}. Letting $\FamEnum{i}{I}{U_i}$ be a directed family in
  $\opens{X}$, we show $f_*(\FrmJoin_{i : I} U_i) = \FrmJoin_{i : I} f_*(U_i)$.
  The $\FrmJoin_{i : I} f_*(U_i) \le f_*(\bigvee_{i : I} U_i)$ direction is
  easy.
  For the $f_*(\bigvee_{i : I} U_i) \le \bigvee_{i : I} f_*(U_i)$ direction, we appeal
  to \cref{lem:spectral-yoneda}. Let $K$ be a compact open
  with $K \le f_*(\bigvee_{i : I} U_i)$. By the adjunction $f^* \dashv f_*$, we
  have that $f^*(K) \le \bigvee_{i : I} U_i$ and since $f^*(K)$ is compact, by the
  spectrality assumption of $f^*$, there exists some $l \oftype I$ such that $f^*(K)
  \le U_l$. Again by adjointness, $K \le f_*(U_l)$ which implies $K \le \bigvee_{i : I}
  f_*(U_i)$.
\end{proof}

\section{Examples of spectral locales}
\label{sec:spec-examples}

We now examine each of our running examples of locales to determine which
are spectral.

\subsection{The terminal locale}
\label{sec:term-loc-spec}

The first example of a spectral locale is the terminal locale:

\begin{example}\label{example:term-loc-is-spectral}
  The \indexedp{terminal locale} $\LocTerm{\UU}$
  defined in \cref{defn:terminal-locale}
  is spectral.
  We have already constructed a base for it in \cref{example:term-loc-base-1}.
  This base satisfies the definition of spectral base
  from (\cref{defn:int-spec-base}):
  \begin{description}
    \item[\ref{item:spec-base-consists-of-compact-opens}]
      It consists of compact opens since $\TrueProp{\UU}$ and $\FalseProp{\UU}$
      are both compact opens of the terminal locale (as shown in
      \cref{ex:term-loc-compact} and \cref{lem:compact-fin-join-closure}).
    \item[\ref{item:spec-base-contains-top}]
      It contains the top open $\TrueProp{\UU}$.
    \item[\ref{item:spec-base-coherence}]
      It is closed under binary meets since the conjunction of every pair of
      Booleans is a Boolean.
  \end{description}
\end{example}

\subsection{The discrete locale}
\label{sec:disc-loc-not-spec}

The discrete locale is a counterexample since it does not satisfy
Condition~\ref{item:sl-compact}.

\begin{example}\label{example:disc-loc-not-spec}
  The powerset frame $\PowSym_{\UU}(\NatTy)$ is not spectral,
  because it is not compact as discussed in \cref{ex:disc-loc-not-compact}.
\end{example}

\subsection{The \VSierpinski{} locale}

The \VSierpinski{} locale is another example of a spectral locale.
Proving this, however, is a bit more involved.
We first do some preparation.

\begin{lemma}\label{lem:principal-filters-on-decidable-props-are-compact}
  For every decidable proposition\/ $B \oftype \hprop{\UU}$, the principal filter
  map
  \[\upset B \oftype \ArrTy{\hprop{\UU}}{\hprop{\UU}}\]
  from \cref{defn:upset-of-prop} is a compact open of the \VSierpinski{} locale.
\end{lemma}
\begin{proof}
  Let $B \oftype \hprop{\UU}$ be a decidable proposition
  and
  let $\FamEnum{i}{I}{S_i}$ be a directed family of \VScottOpen{} subsets
  such that
  $\upset B \subseteq \bigcup_{i : I} S_i$.
  We proceed by case analysis on proposition~$B$:
  \begin{itemize}
    \item \emph{Case}: $\IdTy{B}{\FalsePropSym}$.
      By \cref{lem:principal-filter-on-bottom}, we know that the principal
      filter $\upset \FalsePropSym$ is the top open of the locale $\LocSierp$,
      which we know to be a compact open of the \VSierpinski{} locale
      from~\cref{example:sierpinski-is-compact}.
    \item \emph{Case}: $\IdTy{B}{\TruePropSym}$.
      In this case, we have
      $P \in \upset \TruePropSym$ if and only if $\IdTy{P}{\TruePropSym}$,
      for every proposition~$P \oftype \hprop{\UU}$.
      This implies that there is some $S_i$ such that $\TruePropSym \in S_i$.
      It must then be the case that every $P \in \upset \TruePropSym$ is
      in $S_i$, since $\IdTy{P}{\TruePropSym}$ implies $P \in S_i$.
  \end{itemize}
\end{proof}

\begin{example}\label{example:sierpinski-is-spectral}
  The
  \indexedp{\VSierpinski{} locale} $\LocSierp_{\UU}$
  from \cref{defn:sierpinski} is spectral.
  Recall from \cref{ex:sierp-base} that we constructed a base for the
  \VSierpinski{} locale as the family
  \begin{equation*}
    \FamEnum{b}{\TwoTy{}}{\upset \beta(b)},
  \end{equation*}
  where $\beta : \ArrTy{\TwoTySym}{\hprop{\UU}}$ is the function
  defined as
  $\beta(\BoolZero) \is \TrueProp{\UU}$, and
  $\beta(\BoolOne) \is \FalseProp{\UU}$. Observe that this base meets the
  the conditions from \cref{defn:int-spec-base}:
  \begin{description}
    \item[\ref{item:spec-base-consists-of-compact-opens}]
      We showed in
      \cref{lem:principal-filters-on-decidable-props-are-compact}
      that this base consists of compact opens.
    \item[\ref{item:spec-base-contains-top}]
      The top open is in the base since it is equal to $\upset \FalsePropSym$,
      as mentioned in \cref{lem:principal-filter-on-bottom}.
    \item[\ref{item:spec-base-coherence}]
      Let $b_1, b_2 \oftype \TwoTySym$.
      The meet $\beta(b_1) \meet \beta(b_2)$ falls in the base
      since $\beta$ maps the Boolean conjunction $b_1 \wedge b_2$ to it.
  \end{description}
\end{example}

\section{Distributive lattices and their spectra}
\label{sec:spec-dlat}

A distributive lattice is a poset with finite meets and finite
joins,
satisfying the distributivity law~\cite[Definition~4.4]{davey-priestley-book}.
These are also called
\emph{bounded} distributive lattices\index{bounded distributive lattice}
in the
literature~\cite[Definition~2.12]{davey-priestley-book}.
In this section, we give a predicative construction of the frame of ideals over
a small distributive lattice.
This is in preparation for \cref{sec:spec-duality,sec:equiv-functor-part} where
we develop a predicative version of the \indexedp{Stone duality} between
spectral locales and distributive lattices.

\subsection{The definition of distributive lattice}
\label{sec:dlat-defn}

Unlike our definition of the notion of frame from \cref{defn:frame}, we define
distributive lattices through their meet and join operations instead of their
orders, for reasons that we will explain in \cref{rmk:dlat-multiversal}.
The definition that we give below follows
\cite[\S{}I.1.4--\S{}I.1.5]{ptj-ss}.

\begin{definition}[\AgdaLink{Locales.DistributiveLattice.Definition.html\#569} {Distributive \VUniLattice{\UU}}\,]
\label{defn:distributive-lattice}
  A
  \defineCE{distributive\/ \VUniLattice{\UU}}{distributive \VUniLattice{\UU}}
  consists of
  \begin{itemize}
    \item a set $| L | \oftype \UU$,
    \item an element $\LatticeTop \oftype | L |$,
    \item an element $\LatticeBot \oftype | L |$,
    \item an operation $(\blank) \meet (\blank) \oftype | L | \to | L | \to | L |$
      that
      \begin{itemize}
        \item is associative,
        \item is commutative,
        \item has $\LatticeTop$ as a unit element: $\IdTy{x \meet \LatticeTop}{x}$,
          for every $x \oftype | L |$,
        \item is idempotent: $\IdTy{x \meet x}{x}$, for every $x \oftype |L|$,
        \item satisfies the \define{absorptive law\/}:
          \[\IdTy{x \wedge (x \vee y)}{x},\] for every pair of elements
          $x, y \oftype |L|$,
      \end{itemize}
    \item an operation $(\blank) \vee (\blank) \oftype | L | \to | L | \to | L |$
      that
      \begin{itemize}
        \item is associative,
        \item is commutative,
        \item has $\LatticeBot$ as a unit element:
          $\IdTy{x \vee \LatticeBot}{x}$, for every $x \oftype | L |$,
        \item is idempotent: $\IdTy{x \vee x}{x}$, for every $x \oftype | L |$,
        \item satisfies the \define{dual absorptive law\/}:
          \[\IdTy{x \vee (x \meet y)}{x},\]
          for every pair of elements $x, y \oftype |L|$.
      \end{itemize}
  \end{itemize}
\end{definition}

As usual, we engage in the abuse of notation of denoting the carrier set $|L|$
by $L$.

\begin{remark}
  We will occasionally speak of
  ``the distributive lattice structure on type~$X$''.
  By this, we refer to the algebraic structure that the carrier set $|L|$ of a
  distributive lattice $L$ is equipped with.
  We do not give a separate formal definition of this,
  but refer the reader to the discussion in \cref{sec:sip}
  to clarify how we formally implement this structure in type theory.
\end{remark}

\begin{lemma}\label{lem:orders-of-distributive-lattice}
  For every distributive\/ \VUniLattice{\UU} $L$, the relations
  \begin{align*}
    x \le_{1} y &\is \IdTy{(x \meet y)}{x},\\
    x \le_{2} y &\is \IdTy{(x \vee y)}{y}
  \end{align*}
  form equivalent partial orders with\/ \VUniHyphen{\UU}small truth values.
\end{lemma}
\begin{proof}
  It is direct from the smallness of the carrier set of $L$ that the types
  $\IdTy{x \meet y}{x}$ and $\IdTy{x \vee y}{y}$ are \VUniSmall{\UU}.
  The fact that these are equivalent is a well-known
  fact of lattice theory~\cite[Lemma~2.8]{davey-priestley-book}, but
  we nevertheless provide the proof for the sake of self-containment.

  Let $x, y : L$ be a pair of elements with $\IdTy{x \meet y}{x}$.
  \begin{align*}
    x \vee y \quad&\IdTySym\quad (x \wedge y) \vee y\\
             \quad&\IdTySym\quad y \vee (y \wedge x)\tag{\text{commutativity}}\\
             \quad&\IdTySym\quad y \tag{\text{dual absorptive law}}
  \end{align*}

  Conversely, let $x, y : L$ be a pair of elements with $\IdTy{x \vee y}{y}$.
  \begin{align*}
    x \wedge y \quad&\IdTySym\quad x \wedge (x \vee y)\\
               \quad&\IdTySym\quad x \tag{\text{absorptive law}}
  \end{align*}

  In light of the equivalence, we show just $(\le_1)$ to be a partial order.
  Reflexivity is exactly the idempotency of the operation
  $(\blank) \meet (\blank)$.
  Antisymmetry is easy to see since for every $x, y : L$, if we have $x \le_1 y$
  and $y \le_1 x$, then we have $x \IdTySym \paren{x \meet y} \IdTySym y$.
  Finally, for transitivity, let $x, y, z : L$ with
  $x \le_1 y$ and $y \le_1 z$. We have that
  \begin{align*}
    x \meet z &\quad\IdTySym\quad (x \meet y) \meet z\\
              &\quad\IdTySym\quad x \meet (y \meet z)\\
              &\quad\IdTySym\quad x \meet y\\
              &\quad\IdTySym\quad x.
  \end{align*}
\end{proof}

\begin{remark}
  Note that
  sethood is an explicit requirement in the definition of distributive lattice
  above.
  Unlike the case for frames, sethood does not follow automatically from the
  definition in this setting.
  This is because we rely on the sethood of the carrier type to conclude that
  the type families $(\le_1)$, $(\le_2)$ defined in the above lemma are
  proposition-valued.
\end{remark}

In light of the above lemma, we use the two orders interchangeably.
Given a distributive lattice $L$, we denote its order by
$(\blank) \le_L (\blank)$, or simply by $(\blank) \le (\blank)$ if the lattice
in consideration is clear from context.

\begin{remark}\label{rmk:dlat-multiversal}
  Observe here that there is a crucial difference compared to the definition of
  the notion of frame from \cref{defn:frame}.
  One could have defined a notion of distributive
  $(\UU, \VV)$\nobreakdash-lattice with
  \VUni{\UU} being the universe of the carrier set, and \VUni{\VV} being the
  universe of the order.
  However, the order can always be resized to universe \VUni{\UU} by using the
  equivalence
  $x \le y \BiImplSym x \meet y = x$
  as explained in \cref{prop:local-smallness-equiv}, which would give an
  equivalent distributive \VUniLattice{\UU} as defined above.
  This is to say that, when working with small lattices, such an ostensibly more
  general notion of $(\UU, \VV)$\nobreakdash-lattice is in fact no more general
  than the above notion of \VUniLattice{\UU}.
\end{remark}

\begin{lemma}
  For every distributive lattice\/ $L$, the operations
  $(\meet)$ and $(\vee)$ are
  meet and join operations with respect to the order\/ $(\blank) \le_L (\blank)$.
\end{lemma}
\begin{proof}
  Well-known fact of lattice theory~\cite[Theorem~2.10]{davey-priestley-book}.
\end{proof}

\begin{lemma}\label{lem:dual-distributivity}
  In every distributive lattice\/ $L$, the dual of the \indexedp{distributivity law}
  holds: \[x \vee (y \meet z) \IdTySym (x \vee y) \meet (x \vee z)\] for
  every\/ $x, y, z \oftype L$.
\end{lemma}
\begin{proof}
  Well-known fact of lattice theory~\cite[Lemma~4.3]{davey-priestley-book}.
\end{proof}

\begin{definition}[{\AgdaLink{Locales.DistributiveLattice.Homomorphism.html\#2562}}]
  A function $h \oftype \ArrTy{K}{L}$,
  from a distributive lattice $K$ into a distributive lattice $L$,
  is called a \define{distributive lattice homomorphism} if it preserves the
  finite meets and the finite joins.

  A homomorphism $s \oftype \ArrTy{K}{L}$ is said to be a
  \define{distributive lattice isomorphism}
  if it has a \indexedp{quasi-inverse} $r \oftype \ArrTy{L}{K}$
  that is also a homomorphism.
\end{definition}

By \(\DLat{\UU}\), we denote the category of distributive
\VUniLattice{\UU}s and their homomorphisms.%
\nomenclature{\(\DLat{\UU}\)}{category of distributive \VUniLattice{\UU}s}

We can simplify the characterization of distributive lattice isomorphisms by
using the style of \MHELastName{}'s \VSIP{} discussed in \cref{sec:sip}.

\begin{lemma}[{\AgdaLink{Locales.DistributiveLattice.Isomorphism.html\#8778}}]
\label{lem:dlat-isomorphism-characterization}
  A type equivalence $e \oftype \Equiv{|K|}{|L|}$, between the carrier sets
  of two distributive lattices, is said to be monotone
  if both directions are monotone maps.

  The type of monotone equivalences between distributive lattice is equivalent
  to the type of isomorphisms.
\end{lemma}
\begin{proof}
  Every equivalence preserves meets and joins as it has both a right adjoint
  and a left adjoint.
\end{proof}

\subsection{Ideals of distributive lattices}
\label{sec:dlat-ideal}

We explained in \cref{sec:subsets} that, in a predicative setting, the notion of
subset is parameterized by universes.
In other words, we work with subsets described by \VUniHyphen{\UU}propositions,
which we call \VUniHyphen{\UU}subsets.
As a result of this, one could consider a generalized notion of
\VUniHyphen{\VV}ideal over a \VUniLattice{\UU}. There may indeed be cases where
such generalized ideals are interesting to study, but for the purposes of our
work, we will not need this.
We accordingly define the following notion of ideal:

\begin{definition}[\AgdaLink{Locales.DistributiveLattice.Ideal.html\#1553} Ideal]%
\label{defn:ideal}
  An ideal of a distributive \VUniLattice{\UU} $L$
  is a subset~$I~\oftype~\Pow{\UU}{L}$
  satisfying the following conditions:
  \begin{enumerate}[label={$\left({\mathsf{I{\arabic*}}}\right)$}, leftmargin=4em]
    \item\label{item:ideal-inhabited}
      It is inhabited i.e.\ $\ExistsType{x}{L}{x \in I}$.
    \item\label{item:ideal-dc}
      It is \indexedp{\VDownwardClosed{}} i.e.\ $y \in I$ and $x \le y$ imply $x \in I$,
      for every pair of elements $x, y \oftype L$.
    \item\label{item:ideal-join-closure}
      If $x \in I$ and $y \in I$, then $(x \vee y) \in I$, for every pair of
      elements $x, y : L$.
  \end{enumerate}
\end{definition}

\begin{lemma}
  Being an ideal is a proposition.
\end{lemma}
\begin{proof}
  It is easily verified that each conjunct is a proposition.
\end{proof}

We define
\[\Ideal{}{L} \quad\is\quad \SigmaType{I}{\ArrTy{L}{\hprop{\UU}}}{\IsIdeal{I}}.\]
\nomenclature{\(\Ideal{\UU}{L}\)}{type of \VUniHyphen{\UU}ideals over lattice \(L\)}%
We will usually talk about ideals in contexts with a fixed base universe
\VUni{\UU}.
In such contexts, we will occasionally speak of
\defineCE{small ideals}{small ideal}
to put extra emphasis on the fact that our ideals are \VUni{\UU}-valued.
The reader should keep in mind, however, that we never work with large
ideals.
Finally, we also note explicitly that the type $\Ideal{}{L}$
is large, since $\hprop{\UU}$ lives in
universe $\mkern+2mu\USucc{\UU}$.

\subsubsection{Properties of ideals}

\begin{lemma}\label{lem:ideals-contain-bottom}
  Every ideal over a distributive lattice\/ $L$ contains
  the bottom element of\/~$L$.
\end{lemma}
\begin{proof}
  Immediate from \ref{item:ideal-dc}.
\end{proof}

\begin{lemma}\label{lem:ideal-closure-under-finite-joins}
  For every ideal\/ $I$, the join of a list of elements in\/ $I$ is in\/ $I$.
\end{lemma}
\begin{proof}
  Straightforward induction on lists.
\end{proof}

\subsubsection{Principal ideals}

\begin{lemma}
  For every distributive \VUniLattice{\UU} $L$ and every $x : L$, the subset
  defined as
  \begin{align*}
    &\downset x \oftype \ArrTy{L}{\hprop{\UU}}\\
    &\downset x \is y \mapsto y \le_L x
  \end{align*}%
  \nomenclature{\(\downset x\)}{principal ideal generated by $x$}%
  is an ideal.
\end{lemma}
\begin{proof}
  Let $L$ be a distributive lattice and let $x : L$.
  \begin{description}
    \item[\ref{item:ideal-inhabited}] The subset $\downset x$ is always
      inhabited by the bottom element $\LatticeBot_L$.
    \item[\ref{item:ideal-dc}]
      \VDDownwardClosure{} follows from the transitivity of the order.
    \item[\ref{item:ideal-join-closure}]
      For closure under binary joins, let $y, z \in \downset x$.
      It must be the case that $y \vee z \le x$, since $y \vee z$ is the least
      upper bound of $y$ and $z$.
  \end{description}
\end{proof}

\begin{definition}[\AgdaLink{Locales.DistributiveLattice.Ideal.html\#5008} Principal ideal]
  The ideal $\downset x$ from the above lemma is called
  the \define{principal ideal}
  generated by the element $x$.
\end{definition}

\begin{lemma}\label{lem:principial-ideal-is-an-embedding}
  The principal ideal map $\downset (\blank) \oftype \ArrTy{L}{\Ideal{}{L}}$
  is an embedding.
\end{lemma}
\begin{proof}
  Let $x, y \oftype L$ such that $\IdTy{\downset x}{\downset y}$.
  Since $x \in \downset x$ and $y \in \downset y$,
  we have that $x \le y$ and $y \le x$,
  and we thus know $\IdTy{x}{y}$ by antisymmetry.
\end{proof}

\subsubsection{Intersection of ideals}

Subset intersection specializes to a meet operation on ideals since
we have:

\begin{lemma}[\AgdaLink{Locales.DistributiveLattice.Spectrum.html\#2838}]
  For every pair of small ideals
  $I, J \oftype \ArrTy{L}{\hprop{\UU}}$,
  the intersection $I \cap J$ is a small ideal.
\end{lemma}

Thus far, we have concluded that the lattice of \VUniHyphen{\UU}ideals is a
meet-semilattice with respect to the inclusion order.
In \cref{sec:idl-frm-joins}, we will proceed to construct joins in this
meet-semilattice, thereby showing that it is a frame.

\begin{lemma}[\AgdaLink{Locales.DistributiveLattice.Spectrum-Properties.html\#7005}]\label{lem:principal-ideal-preserves-meets}
  For every pair of elements $x, y \oftype L$, we have
  $\IdTy{\Intersect{\downset x}{\downset y}}{\downset (x \meet y)}$.
\end{lemma}
\begin{proof}
  This is obvious since
  we have that,
  $z \le x \meet y$ if and only if $z \le \downset x$ and $z \le \downset y$,
  for every $z \oftype L$.
\end{proof}

\begin{corollary}\label{cor:principal-ideals-are-closed-under-intersection}
  The intersection of two principal ideals is a principal ideal.
\end{corollary}

\subsection{Joins in the lattice of ideals}
\label{sec:idl-frm-joins}

We start by doing some preparation for the construction of joins in the
\indexedp{frame of ideals}.

\begin{definition}[\AgdaLink{Locales.DistributiveLattice.Spectrum.html\#4545} Finite covering]
\label{defn:finite-covering}
  Let $L$ be a small distributive lattice.
  We define the \define{finite covering} type family
  $\CoveredBy{(\blank)}{(\blank)} \oftype \ArrTy{\ArrTy{\ListTy{L}}{\Fam{\UU}{\IdealSym(L)}}}{\UU}$
  by induction on lists as follows:
 \begin{align*}
   \epsilon   \CoveredBySym I \quad&\is\quad \UnitTySym_{\UU}\\
   x \consl t \CoveredBySym I \quad&\is\quad \paren{\SigmaType{n}{N}{x \in I_n}}
                                    \times
                                    (\Covers{I}{t})
 \end{align*}
\end{definition}

Conceptually, the type $(\Covers{I}{x_0 \consl \ldots \consl x_{m-1})}$ being
inhabited expresses that $I$ contains a finite subcover for each element $x_i$
of the list.
However, this involves nontrivial structure and is not a property so it should
be thought of as the type of all possible ways in which the family $I$ can give
a finite subcover for the list in consideration.

\begin{lemma}[\AgdaLink{Locales.DistributiveLattice.Spectrum.html\#4998}]
\label{lem:covering-spec}
  For every distributive lattice\/ $L$,
  every family of ideals\/ $\FamEnum{k}{K}{I_k}$, and
  every list\/ $x_0 \consl \ldots x_{m-1}$ of elements of\/ $L$,
  if we have\/ $\Covers{I}{(x_0 \consl \ldots x_{m-1})}$,
  then, for every\/ $j$ with $0 \le j < m$,
  there is some index\/ $k \oftype K$ such that\/~$x_j \in I_k$.
\end{lemma}
\begin{proof}
  Straightforward induction on lists.
\end{proof}

\begin{lemma}[\AgdaLink{Locales.DistributiveLattice.Spectrum.html\#5440}]\label{lem:concatenation-cover}
  For every family of ideals\/ $\FamEnum{n}{N}{I_n}$ and every pair of
  lists\/ $s, t$ of elements of\/ $L$,
  if\/ $\CoveredBy{s}{I}$\/ and\/ $\CoveredBy{t}{I}$ then\/ $\Covers{I}{s \consl t}$.
\end{lemma}
\begin{proof}
  Straightforward induction.
\end{proof}

\begin{lemma}
  For every family of ideals $\FamEnum{n}{N}{I_n}$ and
  every pair of elements $x, y \oftype L$,
  if $\Covers{I}{x}$ then $\Covers{I}{x \meet y}$.
\end{lemma}
\begin{proof}
  Immediate by the \VDownwardClosure{} of ideals
  (Condition~\ref{item:ideal-dc}).
\end{proof}

By extending this to lists, we obtain:

\begin{corollary}[\AgdaLink{Locales.DistributiveLattice.Spectrum.html\#6841}]\label{cor:covering-meets}
  For every family of ideals\/ $\FamEnum{n}{N}{I_n}$,
  every element\/ $x \oftype L$,
  and every list\/ $y_0, \ldots, y_{m-1} \oftype \ListTy{L}$,
  if\/ $\CoveredBy{(y_0, \ldots y_{m-1})}{I}$
  then\/ $\CoveredBy{(x \meet y_0), \ldots, (x \meet y_{m-1})}{I}$.
\end{corollary}

Given a list $s \is (x_0, \ldots x_{m-1})$ of elements of a lattice,
we informally abbreviate the join $(x_0 \vee \cdots \vee x_{m-1})$
to $\bigvee_{x \in s} x$.

\begin{lemma}[\AgdaLink{Locales.DistributiveLattice.Spectrum.html\#11104}]%
\label{lem:join-of-ideals}
  For every small family of ideals\/ $\FamEnum{n}{N}{I_n}$
  over a distributive lattice\/~$L$,
  the following subset of\/ $L$ is a small ideal:
  \begin{equation*}
    x \mapsto
      \Exists
        {(y_0 \consl \ldots \consl y_{m-1})}%
        {\ListTy{L}}%
        {\ProdTy{(\Covers{I}{y_0 \consl \ldots \consl y_{m-1}})}{\paren{\IdTy{x}{y_0 \vee \cdots \vee y_{m-1}}}}}
   \!\text{.}
  \end{equation*}
\end{lemma}
\begin{proof}
  Let $L$ be a distributive\/ \VUniHyphen{\UU}lattice,
  and let $\FamEnum{n}{N}{I_n}$ a family of small ideals over it.
  First observe that the type
  \(\ExistsType{s}{\ListTy{L}}{\ProdTy{(\Covers{I}{s})}{(\IdTy{x}{\bigvee_{y \in s} y})}}\)
  is a \VUniHyphen{\UU}small proposition, for every\/ $x : L$, since
  $\ListTy{L}$ is a small type and
  both conjuncts live in universe\/ \VUni{\UU}.

  We now check the conditions of being an ideal.
  \begin{description}
    \item[\ref{item:ideal-inhabited}] The above subset is always inhabited by
      the bottom element $\LatticeBot_{L}$,
      which can be expressed as the join of the empty list.
      The empty list is vacuously covered by every family of ideals.
    \item[\ref{item:ideal-dc}] Consider an element $x \oftype L$ falling in this
      subset, and let $y \oftype L$ with $y \le x$.
      We know that $\IdTy{x}{z_0 \vee \cdots \vee z_{m-1}}$, for some list
      $\CoveredBy{(z_0 \consl \ldots \consl z_{m-1})}{I}$.
      To show that $y$ also falls in the above subset,
      we pick the list $(y \meet z_0,\ \ldots,\ y \meet z_{m-1})$.
      By \cref{cor:covering-meets}, we have that
      \begin{equation*}
        \CoveredBy{(y \meet z_0)\ \ldots\ (y \meet z_{m-1})}{I}\text{.}
      \end{equation*}
      It remains to show that
      $\IdTy{y}{(y \meet z_0) \vee \cdots \vee (y \meet z_{m-1})}$.
      This follows from the fact that $y \le x$ i.e.\ $\IdTy{y \meet x}{y}$,
      since it implies:
      \begin{align*}
        y \quad&\IdTySym\quad y \meet x\\
          \quad&\IdTySym\quad y \meet (z_0 \vee \ldots \vee z_{m-1})\\
          \quad&\IdTySym\quad (y \meet z_0) \vee \cdots \vee (y \meet z_{m-1}).
      \end{align*}
    \item[\ref{item:ideal-join-closure}]
      Let $x, y \oftype L$ be a pair of elements falling in the above subset.
      We know that there are lists
      $a_0 \consl \ldots \consl a_{k-1}$
      and
      $b_0 \consl \ldots \consl b_{l-1}$
      such that
      \begin{itemize}
        \item $\IdTy{x}{a_0 \vee \cdots \vee a_{k-1}}$
              and
              $\Covers{I}{(a_0 \consl \ldots \consl a_{k-1})}$; and also
        \item $\IdTy{y}{b_0 \vee \cdots \vee b_{k-1}}$
              and
              $\Covers{I}{(b_0 \consl \ldots \consl b_{k-1})}$.
      \end{itemize}
      We obviously have
      \[x \vee y \quad\IdTySym\quad a_0 \vee \cdots \vee a_{k-1} \vee b_0 \vee \cdots \vee b_{l-1},\]
      and we know by \cref{lem:concatenation-cover}\/ that
      $\CoveredBy{(a_0 \consl \ldots a_{m-1} \consl b_{0} \consl \ldots \consl b_{n-1})}{I}.$
  \end{description}
\end{proof}

\begin{lemma}[\AgdaLink{Locales.DistributiveLattice.Spectrum.html\#11983}]%
\label{lem:meet-semilattice-of-ideals-is-small-complete}
  For every small distributive lattice\/ $L$, the meet-semilattice of small
  ideals over\/ $L$ is small complete.
\end{lemma}
\begin{proof}
  Let $\FamEnum{n}{N}{I_n}$ be a small family of ideals.
  We define the join $\FrmJoin_{n : N}I_n$
  to be the ideal constructed in \cref{lem:join-of-ideals}.
  We need to show that this is the least upper bound of the family
  $\FamEnum{n}{N}{I_n}$.

  Let $n \oftype N$.
  Too see that $I_n \subseteq \FrmJoin_{n \oftype N} I_n$,
  let $x \in I_n$.
  Clearly, $\Covers{I}{x}$ since
  $x \in I_n$ and $x$ is the join of the singleton list consisting of itself.

  Let $J$ be an ideal and assume that it is an upper bound of the family
  $\FamEnum{n}{N}{I_n}$.
  To see that $\paren{\FrmJoin_{n \oftype N}I_n} \subseteq J$,
  let $x \in \FrmJoin_{n \oftype N}I_n$.
  This means that \[\IdTy{x}{y_0 \vee \cdots \vee y_{m-1}},\]
  for some list $\CoveredBy{(y_0 \consl \ldots \consl y_{m-1})}{I}$.
  Therefore, it suffices to show by the closure of ideals under finite joins that
  each $y_k \in J$.
  Consider some $k$ with $0 \le k < m$.
  We know by \cref{lem:covering-spec} that
  there is some $n : N$ such that $y_k \in I_n$.
  It follows that $y_k \in J$
  since we have $I_n \subseteq J$
  by the assumption that $J$ is an upper bound of the family of ideals in
  consideration.
\end{proof}

\begin{lemma}[\AgdaLink{Locales.DistributiveLattice.Spectrum.html\#14800}]%
\label{lem:frame-of-ideals}
  For every small distributive lattice\/ $L$,
  the small ideals over\/ $L$ form a frame.
\end{lemma}
\begin{proof}
  We have already established
  in \cref{lem:meet-semilattice-of-ideals-is-small-complete}\/
  \/that $\Ideal{}{L}$ is
  a small-comp\-lete meet-semilattice.
  It is obvious that it is locally small since (1) the inclusion order involves
  only a quantification over the small set $L$, and (2) we are working with
  small ideals.

  It remains only to show that binary meets distributive over the joins.
  Let $I \oftype \Ideal{}{L}$ and let $\FamEnum{n}{N}{J_n}$ be a small family of
  ideals over lattice $L$.
  We show that the meet $I \meet \paren{\FrmJoin_{n \oftype N} J_n}$
  is the least upper bound of the family $\FamEnum{n}{N}{I \wedge J_n}$.
  It is easy to see that it is an upper bound. Consider some index $n : N$.
  Every $x : L$ with $x \in I \meet J_n$ obviously satisfies
  $x \in I$ and $x \in \FrmJoin_{n \oftype N} J_n$.
  To see that it is the least upper bound, let~$K$~be an ideal and assume that
  it is an upper bound of the family $\FamEnum{n}{N}{I \meet J_n}$.
  We need to show that
  $I \meet \paren{\FrmJoin_{n \oftype N} J_n} \subseteq K$.
  Let $x \in I \meet \paren{\FrmJoin_{n \oftype N} J_n}$.
  We know that $\IdTy{x}{y_0 \vee \cdots \vee y_{m-1}}$ for some
  list $\Covers{I}{(y_0 \vee \cdots \vee y_{m-1})}$.
  By the closure of ideals under finite joins, it suffices to show that
  each $y_l \in K$, for $0 \le l < m-1$.
  Notice that each $y_l$ must satisfy
    (1) $y_l \in I$ by the downward-closure of ideals since we have $y_l \le x \in I$, and
    (2) $y_l \in J_n$ for some $n : N$ (by \cref{lem:covering-spec}).
  This means that we have $y_l \in I \cap J_n$ for some $n : N$ and hence
  $y_l \in K$ since $K$ is an upper bound of the $\FamEnum{n}{N}{I \meet J_n}$.
\end{proof}

\begin{definition}[\AgdaLink{Locales.DistributiveLattice.Spectrum.html\#15386}]%
\label{defn:spectrum}
  The locale defined by the frame of ideals over
  a small distributive lattice~$L$,
  constructed in \cref{lem:frame-of-ideals},
  is called the \define{spectrum} of $L$.
  We hereafter denote this by $\Spectrum{L}$.%
  \nomenclature{\(\Spectrum{L}\)}{spectrum of small distributive lattice $L$}
\end{definition}

\begin{example}\label{example:chain-two}
  Consider the two-element chain $\mathsf{2}$, given by the \indexedp{Hasse diagram}:
  \begin{center}
    \begin{tikzpicture}
      \node[circle, draw, fill=lightgray] (a) at (0,0) {};
      \node[circle, draw, fill=lightgray] (b) at (0,1) {};

      \node[anchor=east] at (a.west) {};
      \node[anchor=east] at (b.west) {};

      \draw (a) -- (b);
    \end{tikzpicture}
  \end{center}
  Formally, this can be constructed as the type $\TwoTySym$ with meet and
  join given by Boolean conjunction and disjunction.
  The locale $\Spectrum{\mathsf{2}}$ is exactly the \indexedp{terminal locale}
  from \cref{defn:terminal-locale}.
\end{example}

\begin{example}
  Consider the three-element chain $\mathsf{3}$, given by the Hasse diagram:
  \begin{center}
    \begin{tikzpicture}
      \node[circle, draw, fill=lightgray] (a) at (0,0) {};
      \node[circle, draw, fill=lightgray] (b) at (0,1) {};
      \node[circle, draw, fill=lightgray] (c) at (0,2) {};

      \node[anchor=east] at (a.west) {};
      \node[anchor=east] at (b.west) {};
      \node[anchor=east] at (c.west) {};

      \draw (a) -- (b) -- (c);
    \end{tikzpicture}
  \end{center}
  The locale $\Spectrum{\mathsf{3}}$ is exactly the \indexedp{\VSierpinski{} locale} from
  \cref{defn:sierpinski}.
\end{example}

\begin{example}\label{example:m2}
  Consider the four-element distributive lattice
  $\mathbf{M}_2$
  given by the Hasse diagram:
  
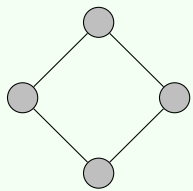
\begin{center}
    \begin{tikzpicture}
      \node[circle, draw, fill=lightgray] (a) at (0,0)  {};
      \node[circle, draw, fill=lightgray] (b) at (1,1)  {};
      \node[circle, draw, fill=lightgray] (c) at (-1,1) {};
      \node[circle, draw, fill=lightgray] (d) at (0,2)  {};

      \draw (a) -- (b);
      \draw (a) -- (c);
      \draw (b) -- (d);
      \draw (c) -- (d);
    \end{tikzpicture}
  \end{center}
  The spectrum of this lattice is the discrete locale on $\TwoTySym$.
  More precisely, when we take the copy of this lattice living in
  universe \VUni{\UU}
  and construct its frame of\/ \VUniHyphen{\UU}ideals,
  we get the frame $\Pow{\UU}{\TwoTySym}$.
\end{example}

In \cref{sec:spec-duality}, we will show that
\emph{every spectral locale arises this way},
as the spectrum of some small distributive lattice.
This is indeed what justifies the term ``spectral'' locale.
We need to first show, however, that the spectrum is itself a spectral locale.
We address this in \cref{sec:spectrum-spectrality}.

\subsection{Spectrality of the frame of ideals}
\label{sec:spectrum-spectrality}

We now verify that
the locale $\Spectrum{L}$ from the previous section
satisfies the definition of spectral locale.
We start with some preparation.

\subsubsection{Base construction for the spectrum}

\begin{lemma}%
\label{lem:top-open-as-a-principal-ideal}
  The top open of the locale\/ $\Spectrum{L}$ is the
  principal ideal\/ $\downset \LatticeTop_L$,
  for every small distributive lattice $L$.
\end{lemma}
\begin{proof}
  It is obvious that the ideal $\downset \LatticeTop_L$ contains every element
  of $L$.
  It is thus equal to the full subset of $L$ by propositional extensionality.
\end{proof}

\begin{lemma}[\AgdaLink{Locales.DistributiveLattice.Spectrum.html\#7476}]
\label{lem:directed-cover-finite-join}
  For every directed family of ideals\/ $\FamEnum{n}{N}{I_n}$
  over a small distributive lattice\/ $L$,
  and for every list\/ $x_0 \consl \ldots \consl x_{m-1}$ of elements of\/ $L$,
  if\/ $\Covers{I}{(x_0 \consl \ldots \consl x_{m-1})}$, then
  there is some index\/ $n \oftype N$
  such that\/ $(x_0 \vee \cdots \vee x_{m-1}) \in I_n$.
\end{lemma}
\begin{proof}[Proof sketch]
  Let $\FamEnum{n}{N}{I_n}$ be a directed family of ideals
  over some small distributive lattice~$L$,
  and let $x_0 \consl \ldots \consl x_{m-1}$ be a list of elements of $L$.
  We know that for each $x_k$ in the list,
  there is some index $n_k \oftype N$ with $x \in I_{n_k}$.
  By the directedness of the family $\FamEnum{n}{N}{I_n}$,
  there must be an upper bound
  of the collection of ideals we get for the $x_k$'s in the list.
  Since this upper bound contains each $x_k$,
  it must contain the join $x_0 \vee \cdots \vee x_{m-1}$
  by \cref{lem:ideal-closure-under-finite-joins}.
\end{proof}

\begin{lemma}[\AgdaLink{Locales.DistributiveLattice.Spectrum-Properties.html\#3916}]%
\label{lem:principal-ideals-are-compact}
  For every small distributive lattice\/ $L$ and every element\/ $x \oftype L$,
  the principal ideal\/ $\downset x$ is a compact open
  of the locale\/ $\Spectrum{L}$.
\end{lemma}
\begin{proof}
  Let $L$ be a small distributive lattice and let $x \oftype L$.
  To see that $\downset x$ is compact,
  let $\FamEnum{n}{N}{I_n}$ be a directed family of small ideals with
  $\downset x \subseteq \FrmJoin_{n \oftype N} I_n$.
  We need to show that there is some index $n \oftype N$ such that
  $\downset x \subseteq I_n$.
  Since $\downset x \subseteq \FrmJoin_{n \oftype N} I_n$, it must be the case that
  $x \in \FrmJoin_{n \oftype N} I_n$ as $x \in \downset x$.
  This is to say that $\IdTy{x}{y_0 \vee \cdots \vee y_{m-1}}$, for some
  list $\Covers{I}{(y_0 \consl \ldots \consl y_{m-1})}$.
  By \cref{lem:directed-cover-finite-join}, we know that there must be some
  ideal $I_n$ such that $(y_0 \vee \cdots \vee y_{m-1}) \in I_n$.
  This means that  $x \in I_n$, which was what we needed.
\end{proof}

\begin{lemma}[\AgdaLink{Locales.DistributiveLattice.Spectrum-Properties.html\#2056}]\label{lem:spec-is-compact}
  The locale\/ $\Spectrum{L}$ is compact.
\end{lemma}
\begin{proof}
  Since $\downset \One{L}$ is the top open of $\Spectrum{L}$
  (by \cref{lem:top-open-as-a-principal-ideal}),
  it suffices to show that this is compact.
  This follows directly from \cref{lem:principal-ideals-are-compact}.
\end{proof}

\begin{lemma}[\AgdaLink{Locales.DistributiveLattice.Spectrum-Properties.html\#10411}]\label{lem:spectrum-base}
  For every small distributive lattice\/ $L$, the principal ideal map
  \[\downset (\blank) : L \to \Ideal{}{L}\]
  forms a small, extensional, and strong base for the locale\/ $\Spectrum{L}$.
\end{lemma}
\begin{proof}
  We have already shown in \cref{lem:principial-ideal-is-an-embedding}
  that this map is an embedding and we know that the type $L$ is small.
  It remains to show that this map forms a strong base.

  Let $I \oftype \Ideal{}{L}$. We let the basic covering family for $I$ be
  \begin{equation*}
    \paren{\SigmaType{x}{L}{x \in I}}
      \xhookrightarrow{~\projISym~}
    L
      \xrightarrow{~\downset (\blank)~}
    \Ideal{}{L}.
  \end{equation*}
  It is easy to see that this is a small subfamily of
  $\downset (\blank) : L \to \Ideal{}{L}$.
  The fact that $I$ is an upper bound of this family is easy to see from
  the \VDownwardClosure{} of ideals (i.e.~Condition~\ref{item:ideal-dc}).
  To see that it is the \emph{least} upper bound,
  consider some ideal $J$ and suppose that
  it is an upper bound of the above family.
  For every $x \in I$,
  we know~$\downset x \subseteq J$ by the assumption that $J$ is an upper bound.
  It follows that $x \in J$ since~$x \in \downset x$.
  We thus have $I \subseteq J$, which was what we needed.
\end{proof}

\subsubsection{Spectrality of the base}

Having constructed a base for the spectrum in \cref{lem:spectrum-base}, we now
proceed to show
that this base is \indexedpCE{spectral}{spectral locale}.

\begin{lemma}[\AgdaLink{Locales.DistributiveLattice.Spectrum-Properties.html\#9763}]%
\label{lem:spec-has-spectral-base}
  The base for\/ $\Spectrum{L}$,
  constructed in Lemma~\ref{lem:spectrum-base},
  is a spectral base,
  for every small distributive lattice\/ $L$.
\end{lemma}
\begin{proof}~

  \begin{itemize}
    \item \ref{item:spec-base-consists-of-compact-opens}
      The fact that this base consists of compact opens was already established
      in \cref{lem:principal-ideals-are-compact}.
    \item \ref{item:spec-base-contains-top}
      This base contains the top open
      since the top open is equal to $\downset \LatticeTop_L$,
      as shown in~\cref{lem:top-open-as-a-principal-ideal}.
    \item \ref{item:spec-base-coherence}
      Let $x, y \oftype L$.
      By \cref{lem:principal-ideal-preserves-meets}, we know that
      $\IdTy{\Intersect{\downset x}{\downset y}}{\downset (x \meet y)}$
      meaning the meet $\Intersect{\downset x}{\downset y}$ falls in the
      base.
  \end{itemize}
\end{proof}

We have therefore established that $\Spectrum{L}$ satisfies the equivalent
conditions from \cref{thm:spec-characterization}.
Note also that we have the following:

\begin{lemma}[\AgdaLink{Locales.DistributiveLattice.Spectrum-Properties.html\#11428}]\label{lem:spec-compact-opens-are-exactly-the-principal-ideals}
  We have an equivalence
  \begin{equation*}
    \Equiv{%
      \CompactOpens{\Spectrum{L}}%
    \quad}{\quad%
      \SigmaType{I}{\Ideal{}{L}}{\ExistsType{x}{L}{(\IdTy{I}{\downset x})}}%
    }.
  \end{equation*}
  In other words, the type of compact opens of\/ $\Spectrum{L}$ is equivalent
  to its type of \indexedpCE{principal ideals}{principal ideal}.
\end{lemma}
\begin{proof}
  By \cref{prop:base-img-equiv}, the image of
  $\downset (\blank) : \ArrTy{L}{\Ideal{}{L}}$
  is equivalent to the type of compact opens of $\Spectrum{L}$,
  since it is a base consisting of compact opens
  as shown in \cref{lem:spectrum-base}.
\end{proof}

\section{\VZZeroDimensional{} and regular locales}
\label{sec:zero-dimensional}

We now consider the classes of
\indexedCE{\VZeroDimensional{}}{zero-dimensional locale}
and
\indexedCE{regular}{regular locale}
locales, in preparation for the development of the theory of
Stone locales in \cref{sec:stone}.
Some of the results we will prove about Stone locales generalize to
to regular locales, of which the \VZeroDimensional{} locales are a special case
(\cref{lem:zero-dimensional-implies-regular}).
Although we are not interested in the theory of \VZeroDimensional{} and regular
locales in itself, it will be useful in \cref{sec:stone} to have collected some
key properties of these classes of locales.

\subsection{Regular locales}
\label{sec:zd-sub-regular}

We briefly mentioned the notion of regular locale in \cref{defn:regular-locale},
where we defined it to be a locale $X$ that has an unspecified base
$\FamEnum{i}{I}{B_i}$ satisfying the condition that,
for every $U \oftype \opens{X}$,
the basic covering family for $U$ consists of
opens \VWellInside{} $U$.

We start by recalling that
\VTheWayBelowRelation{} implies \VTheWellInsideRelation{}
in \indexedpCE{regular locales}{regular locale}:

\begin{lemma}[\AgdaLink{Locales.CompactRegular.html\#29814}]
  In every regular locale $X$, if\/ $U \WayBelow V$ then\/ $U \WellInside V$,
  for every pair of o\-pens\/ $U, V \oftype \opens{X}$.
\end{lemma}
\begin{proof}
  Let $X$ be a regular locale. Since the conclusion is a proposition, we
  may appeal to the induction principle of propositional truncation and
  assume that we have a specified base $\FamEnum{i}{I}{B_i}$
  satisfying the condition of regularity.
  Let $U$ and $V$ be a pair of opens with $U \WayBelow V$, and
  let $\FamEnum{j}{J}{B_{i_j}}$ be the basic covering family for~$V$.
  Since $\IdTy{V}{\FrmJoin_{j \oftype J} B_{i_j}}$,
  there exists some $j \oftype J$
  such that $U \le B_{i_j}$, by the assumption that $U \WayBelow V$.
  We thus have $U \le B_j \WellInside V$, which implies $U \WellInside V$
  by \cref{lem:wi-prop-1}.
\end{proof}

Since an open $U$ is compact if and only if $U \WayBelow U$, and \indexedp{clopen} if and
only if $U \WellInside U$, we obtain the immediate corollary below.

\begin{corollary}[\AgdaLink{Locales.CompactRegular.html\#31092}]%
\label{cor:compact-implies-clopen-in-regular-locales}
  In every regular locale, compact opens are clopen.
\end{corollary}

When combined with~\cref{cor:clopen-implies-compact-in-compact-locale}, this
gives:

\begin{lemma}\label{lem:in-creg-locales-C-is-equivalent-to-K}
  In every compact regular locale, the compact opens coincide with the clopens.
\end{lemma}
\begin{proof}
  By \cref{cor:clopen-implies-compact-in-compact-locale}, we know that
  clopens are compact in compact locales.
  \cref{cor:compact-implies-clopen-in-regular-locales} above tells us that
  the backward implication holds in regular locales.
  Since the types expressing being clopen and being compact are both
  propositions, this logical equivalence implies that we have a type equivalence
  \(\Equiv{\CompactOpens{X}}{\Clopens{X}}\).
\end{proof}

\subsection{\VZZeroDimensional{} locales}
\label{sec:zd-sub-zd}

A \indexed{\VZeroDimensional{} space}, in the context of point-set topology, is
a topological space whose class of clopen subsets
forms a base for the
topology~\cite[Exercise~11.1]{davey-priestley-book}~\cite[Chapter~II]{hurewicz-wallman-1941}.
The notion of dimension involved here is that of \indexed{inductive dimension},
which goes back to Urysohn~\cite{urysohn-1925} and Menger~\cite{menger-1928}.
We refrain from discussing the general notion of inductive dimension for
topological spaces,
and refer the reader
to~\cite[Chapter~III]{hurewicz-wallman-1941}
for further reading on the subject.

We already mentioned the point-free analogue of this notion of
\VZeroDimensionality{}
in \cref{sec:classes},
where we defined a \VZeroDimensional{} locale
as a locale that has some small base consisting of clopens. We now revisit the
predicative manifestation of the notion of \VZeroDimensional{} locale and
examine it in detail.

In \cref{thm:spec-characterization}, we proved that several alternative
characterizations of the notion of spectral locale all coincide. The above
definition merits the question of whether an analogous characterization can be
obtained for \VZeroDimensional{} locales. Unlike the case for spectral locales,
however, it turns out that
\emph{there is no analogue} of \cref{defn:spectral-locale}
for \VZeroDimensionality{}
(as we will soon show in \cref{example:zero-dimensional-but-not-clopen-based}).
This illustrates how \cref{thm:spec-characterization} is reliant on the
compactness of the locale in consideration.

Motivated by the impredicative definition of \VZeroDimensional{} locale as a
locale in which the clopens form a base, one may consider the definition below
as a candidate for a predicative definition. We refrain from using the term
\VZeroDimensional{} for this notion, the reason for which will be explained
shortly.

\begin{definition}\label{defn:clopen-based}
  A locale $X$ is said to be \defineCE{clopen-based}{clopen-based locale} if
  \begin{enumerate}[label={$\left({\mathsf{CB{\arabic*}}}\right)$}, leftmargin=4em]
    \item the inclusion $\Clopens{X} \xhookrightarrow{~\projISym~} \opens{X}$
      forms a \indexedp{weak base}, and
    \item the type $\Clopens{X}$ is small.
  \end{enumerate}
\end{definition}

\begin{example}
  The terminal locale $\LocTerm{\UU}$ is clopen-based: observe that the base
  from \cref{example:term-loc-base-1} contains
  both of the decidable propositions $\TrueProp{\UU}$ and $\FalseProp{\UU}$.
  Furthermore, the type $\Clopens{\LocTerm{\UU}}$ is small since the type
  of decidable propositions is equivalent to the type~$\TwoTySym$.
\end{example}

It is easy to see that the property of being clopen-based is stronger than
\VZeroDimensionality{}:

\begin{lemma}\label{lem:clopen-based-implies-zero-dimensional}
  Every clopen-based locale is zero-dimensional.
\end{lemma}
\begin{proof}
  Let $X$ be a clopen-based locale.
  We show that the inclusion
  $\Clopens{X} \xhookrightarrow{~\projISym~} \opens{X}$
  forms a small, intensional, and strong base.
  This family consists of clopens by construction.
  For every open $U \oftype \opens{X}$, the covering family for it is given by
  the inclusion
  \begin{equation*}
    \paren{\SigmaType{C}{\Clopens{X}}{C \le U}}
      \xhookrightarrow{~\projISym~}
    \opens{X}\text{,}
    \tag{\dag}\label{eqn:clopens-below}
  \end{equation*}
  which is easily seen to be a small family thanks to local smallness.
  It is clear $U$ is an upper bound of this family. To see that it is the
  least upper bound, let $V$ be an upper bound of the family. We need to show
  that $U \le V$.
  By the weak base assumption, we know there is an unspecified family
  of clopens covering~$U$. As the conclusion is a proposition,
  we may appeal to the induction principle of propositional truncation and
  work with a specified
  basic covering family $\FamEnum{i}{I}{C_i}$.
  Since we
  know $\IdTy{U}{\FrmJoin_{i \oftype I} C_i}$, we just have to show that $V$ is
  above every $C_i$ in the family.
  This follows from our assumption that $V$ is an upper bound of the
  family~(\ref{eqn:clopens-below}).
\end{proof}

Seeing that the property of being clopen-based is stronger than the property of
being \VZeroDimensional{}, it is natural to ask whether we want to pick the
former or the latter as the predicative manifestation of \VZeroDimensionality{}.
We answer this through the following example%
\footnote{%
  This example was suggested to the author by Igor Arrieta
  (private communication).%
}:

\begin{example}\label{example:zero-dimensional-but-not-clopen-based}
  The locale\/ $\LocTerm{\UU}^{\neg\neg}$
  is \VZeroDimensional{}
  but not clopen-based unless \VOmegaNotNotResizing{} holds.
\end{example}
\begin{proof}
  The base from \cref{example:term-loc-base-1} also forms a base for
  $\LocTerm{\UU}^{\neg\neg}$, but the type of clopens is a large type
  since every proposition in $\hprop{\UU}^{\neg\neg}$
  is a clopen (of the locale $\LocTerm{\UU}^{\neg\neg}$).
  In other words,
  whilst the collection of clopens of $\LocTerm{\UU}^{\neg\neg}$
  does form a base,
  it is not a small collection.
  Requiring it to be small
  amounts to requiring $\hprop{\UU}^{\neg\neg}$ to be \VUniSmall{\UU},
  so it is easy to see that this locale being clopen-based gives
  \VOmegaNotNotResizing{}.
\end{proof}

It is a reasonable desideratum of the definition of \VZeroDimensionality{} that
the locale $\LocTerm{\UU}^{\neg\neg}$ satisfy it.
We thus conclude that the notion of clopen-based locale is overly restrictive in
our predicative setting and adopt \cref{defn:zd-locale-0} as the right
definition of this notion. When the locale in consideration is compact, the two
notions coincide:

\begin{lemma}\label{lem:compactness-implies-clopen-based-iff-zero-dimensional}
  A compact locale\/ $X$ is clopen-based
  if and only if
  it is \VZeroDimensional{}.
\end{lemma}
\begin{proof}
  We have already established the forward direction in
  \cref{lem:clopen-based-implies-zero-dimensional},
  which holds irrespective of compactness.
  For the backward direction,
  consider a compact and \VZeroDimensional{} locale $X$.
  By \cref{lem:in-creg-locales-C-is-equivalent-to-K}, we know that there
  is a type equivalence $\Equiv{\CompactOpens{X}}{\Clopens{X}}$
  since the \VZeroDimensionality{} of $X$ implies that it is regular.
  This means that the \VZeroDimensional{} base $\FamEnum{i}{I}{B_i}$
  of $X$ consists of compact opens,
  and we thus know that there is an equivalence
  \begin{equation*}
    \CompactOpens{X} \EquivSym \image{B} \EquivSym \Clopens{X},
  \end{equation*}
  since $\Equiv{\CompactOpens{X}}{\image{B}}$ by \cref{prop:base-img-equiv}.
  This implies that $\Clopens{X}$ is a small type so we are done.
\end{proof}

\section{Stone locales}
\label{sec:stone}

Having pinned down the right notion of
\VZeroDimensional{} locale in \cref{sec:zd-sub-regular},
we are now ready to define the class of
\emph{Stone locales}:

\begin{definition}[\AgdaLink{Locales.Stone.html\#1114}]%
\label{defn:stone-standard}
  A locale $X$ is called
  \defineCE{Stone}{Stone locale}
  if it is compact and \VZeroDimensional{}.
\end{definition}

Although this definition is perfectly sufficient for our purposes, it is
illustrative to look at alternative characterizations
as we did for spectral locales in \cref{thm:spec-characterization}.
We must also justify that the above definition is equivalent to the definition
of Stone locale from \cref{defn:stone-base}.

\begin{lemma}[\AgdaLink{Locales.CompactRegular.html\#38276}]%
\label{lem:clopen-equiv-compact-in-stone-locales}
  In every Stone locale, the compact opens coincide with the clopens.
\end{lemma}
\begin{proof}
  Every Stone locale is clearly compact and regular,
  since \VZeroDimensionality{} implies regularity
  by \cref{lem:zero-dimensional-implies-regular}.
  We thus know by \cref{lem:in-creg-locales-C-is-equivalent-to-K}
  that the compact opens coincide with the clopens.
\end{proof}

\subsection{Stone bases are spectral}
\label{sec:stone-bases-are-spectral}

We recall \cref{defn:stone-base} where we defined a Stone locale as a locale
with an unspecified Stone base.
A \indexed{Stone base} for a locale $X$ is a base $\FamEnum{i}{I}{B_i}$ such that
\begin{itemize}
  \item every basic open $B_i$ is both compact and clopen, and
  \item the top open $\FrmTop{X}$ is an element of the base.
\end{itemize}
It is easy to see that having an unspecified Stone base implies being Stone:

\begin{lemma}\label{lem:stone-base-implies-zero-dimensional-base-and-compactness}
  If a locale\/ $X$ has an unspecified Stone base then\/ $X$ is Stone as in the
  above definition i.e.\ is compact and \VZeroDimensional{}.
\end{lemma}
\begin{proof}
  \VZZeroDimensionality{} is obvious.
  For compactness, note that the top open $\FrmTop{X}$ being in the base
  implies that it is compact, since the base is stipulated to consist of compact
  opens.
\end{proof}

\begin{lemma}\label{lem:stone-base-gives-clopen-implies-compact}
  In every locale $X$ with an unspecified Stone base, every clopen is
  compact.
\end{lemma}
\begin{proof}
  Let $X$ be a locale with an unspecified Stone base.
  By \cref{lem:stone-base-implies-zero-dimensional-base-and-compactness},
  we know that $X$ is compact.
  \cref{cor:clopen-implies-compact-in-compact-locale}
  says that clopens are compact in every compact locale.
\end{proof}

\begin{lemma}\label{lem:stone-base-implies-spectral-base}
  Every Stone base is spectral.
\end{lemma}
\begin{proof}
  Let $X$ be a locale and let $\FamEnum{i}{I}{B_i}$ a specified Stone base for
  it.
  Conditions
  \ref{item:spec-base-consists-of-compact-opens}
  and
  \ref{item:spec-base-contains-top}
  from~\cref{defn:int-spec-base}
  clearly hold by construction.
  It remains to show Condition~\ref{item:spec-base-coherence}
  i.e. the fact that the base $\FamEnum{i}{I}{B_i}$ is closed under binary
  meets.
  Consider two indices $i, j \oftype I$.
  The basic opens $B_i$ and $B_j$ are both clopen, and so their meet
  $B_i \meet B_j$ must also be clopen by \cref{lem:clopen-closure}.
  Since we know
  by \cref{lem:stone-base-gives-clopen-implies-compact}
  that clopens are compact in the presence of a Stone base,
  the meet $B_i \meet B_j$ must be compact
  and must hence fall in the base,
  since compact opens are basic
  in every locale with a base (\cref{lem:cmp-bsc}).
\end{proof}

\begin{lemma}\label{lem:stone-base-characterization}
  For every locale\/ $X$ and every base\/ $\FamEnum{i}{I}{B_i}$ for\/ $X$,
  the following are equivalent:
  \begin{enumerate}
    \item\label{item:stone-base-char-I}
      the base\/ $\FamEnum{i}{I}{B_i}$ is Stone,
    \item\label{item:stone-base-char-II}
      the base\/ $\FamEnum{i}{I}{B_i}$ is spectral and also \VZeroDimensional{},
    \item\label{item:stone-base-char-III}
      the locale $X$ is compact and the base\/ $\FamEnum{i}{I}{B_i}$
      is \VZeroDimensional{}.
  \end{enumerate}
\end{lemma}
\begin{proof}
  We have already established
  $(\ref{item:stone-base-char-I}) \ImplSym (\ref{item:stone-base-char-III})$
  in \cref{lem:stone-base-implies-spectral-base,lem:stone-base-implies-zero-dimensional-base-and-compactness}.
  For the converse, first observe that the top open $\FrmTop{X}$ must fall in
  the base since compact opens are basic in every locale
  equipped with a base
  (by \cref{lem:cmp-bsc}). Furthermore,
  \cref{cor:clopen-implies-compact-in-compact-locale}
  says that clopens are compact in every compact locale, meaning each $B_i$
  must be compact in addition to being clopen.

  The implication
  $(\ref{item:stone-base-char-II}) \ImplSym (\ref{item:stone-base-char-III})$
  is direct.
  Moreover, we also have
  $(\ref{item:stone-base-char-I}) \ImplSym (\ref{item:stone-base-char-II})$,
  since it is immediate by construction that every Stone base is \VZeroDimensional{}
  and we have already shown in \cref{lem:stone-base-implies-spectral-base},
  that Stone bases are spectral.
\end{proof}

\subsection{Alternative characterization}

In our discussion of spectral locales in \cref{sec:spec-defn}, we mentioned that
the formulation of the notion of spectral locale
in \cref{defn:spectral-locale}
is the obvious translation of the standard, impredicative notion of spectral
locale into our predicative setting. Furthermore, we also stressed that the
equivalence of \cref{defn:int-spec-base} with \cref{defn:spectral-locale} is
crucial to our development.

It is natural to ask if there is a definition analogous
to \cref{defn:spectral-locale}
for Stone locales.
We have in fact already given such a definition in
\cref{lem:compactness-implies-clopen-based-iff-zero-dimensional}
where we proved that \VZeroDimensionality{} and being clopen-based coincide when
the locale at hand is compact.
The latter can be taken to be a definition similar to \cref{defn:spectral-locale}
since it amounts to the following conditions:
\begin{enumerate}[label={$\left({\mathsf{ST{\arabic*}}}\right)$}, leftmargin=4em]
  \item\label{item:stl-compact} Compactness.
  \item\label{item:stl-covering}
    The inclusion $\Clopens{X} \xhookrightarrow{~\projISym~} \opens{X}$ forming
    a weak base.
  \item\label{item:stl-smallness}
    The type $\Clopens{X}$ being small.
\end{enumerate}

\begin{lemma}\label{lem:stone-implies-stone-standard}
  Every locale that satisfies conditions
  \ref{item:stl-compact}--\ref{item:stl-smallness}
  is Stone as in \cref{defn:stone-standard}.
\end{lemma}
\begin{proof}
  This follows directly
  from \cref{lem:compactness-implies-clopen-based-iff-zero-dimensional}.
\end{proof}

\begin{lemma}[\AgdaLink{Locales.CompactRegular.html\#54758}]%
\label{lem:stone-implies-spectral}
  Every Stone locale is spectral.
\end{lemma}
\begin{proof}
  By \cref{lem:stone-implies-stone-standard},
  we know that every Stone locale is compact and has
  a specified \VZeroDimensional{} base.
  By \cref{lem:stone-base-characterization}, this specified base must be one
  that is also spectral.
\end{proof}

\begin{theorem}\label{thm:answer-stone}
  For every locale $X$, the following are logically equivalent,
  assuming \VAxUA{}\index{univalence axiom} and \VSRPrinciple{}\index{set replacement principle}.
  \begin{enumerate}
    \item\label{item:stone-VI}
     Locale\/ $X$ has an unspecified Stone base.
    \item\label{item:stone-II}
      Locale\/ $X$ is compact and \VZeroDimensional{}.
    \item\label{item:stone-IIPrimePrime}
      Locale\/ $X$ is compact and clopen-based.
    \item\label{item:stone-IIPrime}
      Locale\/ $X$ is compact and has an unspecified extensional small base of clopens.
    \item\label{item:stone-V}
      Locale\/ $X$ is compact and the
      inclusion\/ $\Clopens{X} \hookrightarrow X$ is an extensional
      base of clopens, where to an open\/ $U$ we assign the family of all
      clopens below it as in the construction of
      \cref{lem:clopen-based-implies-zero-dimensional}.
    \item\label{item:stone-III}
      Locale\/ $X$ is compact and has a specified intensional small base of clopens.
    \item\label{item:stone-IV}
      Locale\/ $X$ is compact and has a specified extensional small base of clopens.
  \end{enumerate}
\end{theorem}
\begin{proof}
  We make use of \cref{thm:spec-characterization}.
  We have already established that each of the conditions in the statement
  implies one of the equivalent conditions from \cref{thm:spec-characterization}:
  \begin{itemize}
    \item Conditions (\ref{item:stone-VI}) and (\ref{item:stone-II})
      both imply spectrality,
      which was shown in \cref{lem:stone-base-characterization}.
    \item Condition~(\ref{item:stone-IIPrimePrime})
      is equivalent to Condition~(\ref{item:stone-II}) of
      \cref{thm:spec-characterization},
      as shown in \cref{lem:compactness-implies-clopen-based-iff-zero-dimensional}.
    \item
      Conditions
      (\ref{item:stone-V}), (\ref{item:stone-III}), and (\ref{item:stone-IV})
      clearly imply, respectively, Conditions
      (\ref{item:stone-VI}), (\ref{item:stone-II}), and (\ref{item:stone-IIPrime})
      of
      \cref{thm:spec-characterization}.
  \end{itemize}
  We know by \cref{prop:base-img-equiv}
  that, for every locale $X$ with a base $\FamEnum{i}{I}{B_i}$
  consisting of compact opens,
  $\Equiv{\CompactOpens{X}}{\mathsf{image}(B)}$.
  It is also easy to see that each of the conditions in the statement
  implies \VZeroDimensionality{}, meaning
  $\Equiv{\CompactOpens{X}}{\Clopens{X}}$
  by \cref{lem:in-creg-locales-C-is-equivalent-to-K}.
  From this,
  we can observe that each condition in the statement
  is the corresponding condition (as given in the above implications)
  with the added requirement of consisting of clopens.
\end{proof}

Notice also that the first five of the above conditions are propositions, but
the last two are not in general.

\subsection{The category of Stone locales}

We can now define the category of Stone locales.
Unlike the category $\Spec$ of
spectral locales, there is no special notion of morphism for Stone locales
thanks to the following:

\begin{lemma}
  Every continuous map reflects clopens.
\end{lemma}
\begin{proof}
  Let $f \oftype \ArrTy{X}{Y}$ be a continuous map of locales and
  consider a clopen $F \oftype \opens{Y}$. We denote the complement of $F$ by $F'$.
  The application $f^*(F)$ must also be clopen, since it is complemented by
  $f^*(F')$:
  \begin{align*}
    f^*(F) \meet f^*(F')
      &\IdTySym f^*(F \meet F')
      \IdTySym f^*(\FrmBot{Y})
      \IdTySym \FrmBot{X}\text{,}\\
    f^*(F) \vee f^*(F')
    &\IdTySym f^*(F \vee F')
    \IdTySym f^*(\FrmTop{Y})
    \IdTySym \FrmTop{X}\text{.}
  \end{align*}%
\end{proof}

Since the clopens coincide with the compact opens in Stone locales, we have the
following corollary:

\begin{corollary}\label{lem:continuous-maps-are-spectral-maps}
  Every continuous map\/ $f \oftype \ArrTy{X}{Y}$ of Stone locales is a
  spectral map.
\end{corollary}

We denote by $\Stone_{\UU}$ the category of Stone locales and continuous maps
over the base universe \VUni{\UU}.%
\nomenclature{\(\Stone_{\UU}\)}{%
  the category of Stone locales over base universe \VUni{\UU}%
}
If the base universe is clear from context, we drop the subscript and simply
write $\Stone$.%
\nomenclature{\(\Stone_{}\)}{%
  the category of Stone locales with base universe left implicit%
}

\begin{lemma}
  The category\/ $\mkern+2mu\Stone_{\UU}$ is a full subcategory of\/ $\Spec_{\UU}$.
\end{lemma}
\begin{proof}
  We have already established in \cref{lem:stone-implies-spectral} that every
  Stone locale is spectral, meaning this is a subcategory.
  That it is \emph{full} is given by
  \cref{lem:continuous-maps-are-spectral-maps}.
\end{proof}

\subsection{Examples and counterexamples}

\subsubsection{The terminal locale}

\begin{example}\label{example:terminal-locale-is-stone}
  The terminal locale from \cref{defn:terminal-locale} is Stone.
  It was shown to be compact in \cref{ex:term-loc-compact},
  It was shown to be clopen-based
  in \cref{lem:clopen-based-implies-zero-dimensional},
  and it is hence \VZeroDimensional{}
  by \cref{lem:clopen-based-implies-zero-dimensional}.
\end{example}

\subsubsection{The discrete locale}

\begin{example}
  The discrete locale is not Stone in general.
  As explained in \cref{example:disc-loc-not-spec}, the \indexedp{powerset frame}
  $\PowSym_{\UU}(\NatTy)$ is not spectral so it cannot be Stone.
\end{example}

\subsubsection{The \VSierpinski{} locale}

The \indexed{\VSierpinski{} locale} is an illustrative example of a locale that
is not Stone \emph{despite} being spectral.

\begin{example}
  The \VSierpinski{} locale\/ $\LocSierp_{\UU}$ is not Stone.
\end{example}
\begin{proof}
  We explained in \cref{example:sierpinski-is-spectral} that the
  \VSierpinski{} locale has exactly one nontrivial compact open:
  the principal filter $\upset \TruePropSym$. This is
  the singleton $\{ \TruePropSym \}$ since a proposition falls in this subset
  if and only if it is inhabited.
  Suppose towards a contradiction
  that the singleton $\{ \TruePropSym \}$ has a complement
  i.e.\ there existed some $F \oftype \opens{\LocSierp_{\UU}}$
  satisfying
  \begin{align*}
    \{ \TruePropSym \} \cap F &\IdTySym \emptyset \text{,}  \tag{\dag}\label{sierp-meet-comp}\\
    \{ \TruePropSym \} \cup F &\IdTySym \opens{\LocSierp_{\UU}} \text{.} \tag{\ddag}\label{sierp-join-comp}
  \end{align*}
  We know by (\ref{sierp-join-comp}) that
  every proposition $P \oftype \hprop{\UU}$ satisfies
  either $\IdTy{P}{\TruePropSym}$ or $P \in F$.
  Combining this with (\ref{sierp-meet-comp}), we obtain that $P \in F$ if and only if
  $\neg (\IdTy{P}{\TruePropSym})$ i.e.\ $\IdTy{P}{\FalsePropSym}$.
  This is to say that $F$ is the singleton subset $\{ \FalsePropSym \}$.
  Notice, however, that the \VScottOpenness{} of the subset $F$
  implies that we also have $\TruePropSym \in F$,
  since \VScottOpen{} subsets are upward closed.
  It follows that $\IdTy{\FalsePropSym}{\TruePropSym}$, a contradiction.
\end{proof}

\begin{example}
  The \VSierpinski{} locale is spectral but \emph{not} Stone.
  Its spectrality was given in \cref{example:sierpinski-is-spectral},
  and we established in the above lemma that that it is not Stone.
\end{example}

\section{Stone duality for spectral locales}
\label{sec:spec-duality}

We showed in \cref{lem:spec-has-spectral-base} that there is a spectral locale
associated with every small distributive lattice $L$.
This is the spectrum $\Spectrum{L}$ whose defining frame is the frame of small
ideals over $L$.
In this section, we will extend this to show:
\begin{enumerate}
  \item that this is the case for every spectral locale i.e.\ \emph{every}
    spectral locale is homeomorphic the spectrum of some small distributive
    lattice, namely its small distributive lattice of compact opens,
  \item that every distributive lattice is isomorphic to the small lattice of
    compact opens of its spectrum.
\end{enumerate}
Thanks to this, we will add yet another equivalent condition to the list of
conditions from \cref{thm:spec-characterization}:
being homeomorphic to $\Spectrum{L}$ for some small distributive lattice $L$.
We will thus obtain maps
\begin{center}
  \begin{tikzpicture}
    \node (A) at (0,0) {$\Spec_{\UU}$};
    \node (B) at (4,0) {$\DLat{\UU}$,};

    \draw[->, line width=0.6pt]
    ([yshift=0.1cm]A.east) to node[midway, above] {$\DistrLattOfCompactOpensSym$} ([yshift=0.1cm]B.west);
    \draw[->, line width=0.6pt]
    ([yshift=-0.1cm]B.west) to node[midway, below] {$\Spectrum{\blank}$}([yshift=-0.1cm]A.east);
  \end{tikzpicture}
\end{center}
where $\Spec_{\UU}$ denotes the type of large, locally small, and small-complete
spectral locales, and $\DLat{\UU}$ denotes the type of small distributive
lattices (over base universe~$\UU$),
and then show that these maps form an equivalence.

Although we extend these maps to functors and show that they form a categorical
equivalence in \cref{sec:equiv-functor-part},
we note that just the type equivalence that we present in this section is quite
useful in the context of \VUF{}, since it implies that these two types are the
same in a strong sense.
For example, one can show that, for every pair $X$ and $Y$ of spectral locales,
we have an equivalence
\begin{equation*}
  \Equiv{%
    (X \cong Y)%
  }{%
    (\DistrLattOfCompactOpens{X} \cong \DistrLattOfCompactOpens{Y})%
  }\text{,}
\end{equation*}
between the type $X \cong Y$ of locale homeomorphisms
and the type $\DistrLattOfCompactOpens{X} \cong \DistrLattOfCompactOpens{Y}$ of
distributive lattice isomorphisms, and use this to show that two locales
are homeomorphic by showing that their distributive lattices of compact opens
are isomorphic.

\subsection{Transporting a lattice structure along an equivalence}

In \cref{defn:smallness}, we defined the notion of \VUniSmall{\VV}ness:
a type $X \oftype \UU$ is called \VUniSmall{\VV} if it has an equivalent copy in
universe \VUni{\VV}.
Consider a type $X$ equipped with some algebraic structure.
If we have an equivalence $e \oftype \Equiv{X}{Y}$,
we can \emph{transport} the algebraic structure through $e$ to
obtain an equivalent copy of the algebraic structure on $Y$.
If the equivalence here is between a larger universe and a smaller one, this
gives us a way to \emph{resize the algebraic structure} in consideration by
constructing an isomorphic copy in the smaller universe.

\begin{lemma}[{\AgdaLink{Locales.DistributiveLattice.Resizing.html\#10635}}]
\label{lem:transportation-of-lattice-structure}
  For every distributive\/ \VUniLattice{\UU}\/ $L$ and
  every\/ \VUniHyphen{\VV}type\/ $X$,
  if there is an equivalence\/ $\Equiv{|L|}{X}$,
  then there is a distributive lattice structure\/ $\sigma$ on\/ $X$ such that
  we have an isomorphism\/ $\Isomorphic{L}{\Pair{X}{\sigma}}$.
\end{lemma}
\begin{proof}
  Let $L$ be a distributive\/ \VUniLattice{\UU}, and let $A \oftype \VV$ be a
  type such that we have a specified equivalence $e \oftype \Equiv{|L|}{A}$.
  Denote by $s \oftype \ArrTy{|L|}{A}$ and $r \oftype \ArrTy{A}{|L|}$ the
  two directions of this equivalence.
  We define the following lattice structure on $A$:
  \begin{align*}
    \LatticeBot_A &\quad\is\quad s(\LatticeBot_L)\\
    \LatticeTop_A &\quad\is\quad s(\LatticeTop_L)\\
    a \meet_A b   &\quad\is\quad s(r(a) \meet_L r(b))\\
    a \vee_A b    &\quad\is\quad s(r(a) \vee_L r(b))
  \end{align*}
  It is easy to see that these define a distributive lattice structure on $A$.
  We denote this distributive \VUniLattice{\VV} by $L'$.
  The isomorphism between $L$ and $L'$ is given by the equivalence $e$ itself.
  It remains to show that both directions of the equivalence are homomorphisms
  of distributive lattices.
  The map $s$  preserves \(\LatticeBot_L\) and \(\LatticeTop_L\) by construction.
  To see that it preserves meets and joins, let $x, y \oftype L$.
  Since $r$ is a retraction of $s$, we have:
  \begin{align*}
    \begin{array}{rl}
      s(x \wedge_L y) &\IdTySym\,   s(r(s(x)) \wedge_L y)\\
                      &\IdTySym\,   s(r(s(x)) \wedge_L r(s(y)))\\
                      &\DefnEqSym\, s(x) \wedge_A s(y)
    \end{array}
    &
    \begin{array}{rl}
      s(x \vee_L y) &\IdTySym s(r(s(x)) \vee_L y)\\
                      &\IdTySym s(r(s(x)) \vee_L r(s(y)))\\
                      &\DefnEqSym s(x) \vee_A s(y)
    \end{array}
  \end{align*}
  Conversely, the fact that $r$ is a lattice homomorphism follows directly from
  the fact that it is a retraction of $s$.
\end{proof}

\subsection{Distributive lattice of compact opens of a spectral locale}

We now explicitly define the distributive lattice of compact opens of a spectral
locale.

\begin{lemma}[\AgdaLink{Locales.Spectrality.LatticeOfCompactOpens.html\#1683}]
\label{lem:distributive-lattice-of-compact-opens}
  For every spectral locale\/ $X$,
  the type\/ $\CompactOpens{X}$ forms a distributive lattice
  (which is a priori large).
\end{lemma}
\begin{proof}
  The frame structure on $\opens{X}$ can be restricted to operations:
  \begin{align*}
    (\blank) \meet_K (\blank)
      &\quad\oftype\quad \CompactOpens{X} \to \CompactOpens{X} \to \CompactOpens{X}\\
    (\blank) \vee_K (\blank)
      &\quad\oftype\quad \CompactOpens{X} \to \CompactOpens{X} \to \CompactOpens{X}\\
    \FrmTop{K}
      &\quad\oftype\quad \CompactOpens{X}\\
    \FrmBot{K}
      &\quad\oftype\quad \CompactOpens{X}
  \end{align*}
  This is thanks to the following:
  \begin{itemize}
    \item the bottom open $\FrmBot{X}$ is always compact by
      \cref{lem:compact-fin-join-closure},
    \item compact opens are always closed under binary joins
      (again by \cref{lem:compact-fin-join-closure}),
    \item the meet of two compact opens is compact in spectral locales
      by \ref{item:sl-coherent},
    \item the top open is compact in spectral locales by \ref{item:sl-compact}.
  \end{itemize}
\end{proof}

In \cref{defn:spectral-locale}, we required the type $\CompactOpens{X}$ of a
spectral locale $X$ to be small,
which means that, for every spectral locale $X$, we have that
$\CompactOpens{X}$ is equivalent to a small type.
This means that we can \emph{resize} the distributive lattice of compact opens
from the above lemma.

\begin{lemma}[%
  \AgdaLink{Locales.Spectrality.LatticeOfCompactOpens-Duality.html\#3481}%
]\label{lem:resizing-of-K}
  For every spectral locale\/ $X$,
  the large distributive lattice\/ $\CompactOpens{X}$ from the above lemma
  has an isomorphic small copy.
\end{lemma}
\begin{proof}
  We appeal to \cref{lem:transportation-of-lattice-structure} with
  the equivalence stipulated in Condition~\ref{item:sl-smallness}.
\end{proof}

\begin{remark}
  In the formalization, we carefully distinguish between the large and small
  copies of the distributive lattice of compact opens.
  In our presentation here, however, we abstract over this technical detail,
  as it makes the presentation rather cumbersome.
  We will hereafter make implicit use of the fact that the distributive lattice
  $\CompactOpens{X}$ has a \VUniSmall{\UU} copy.
\end{remark}

The upshot here is that the conditions
\ref{item:sl-compact}--\ref{item:sl-smallness}
ensure, when combined, that the type $\CompactOpens{X}$ forms a small
distributive lattice.
We denote by \[\DistrLattOfCompactOpensSym \oftype \ArrTy{\Spec_{\UU}}{\DLat{\UU}}\]
the map sending a spectral locale to its small distributive lattice of compact
opens, as constructed in the above lemma.

Technically, the above map is obtained as the composition:
\begin{center}
  \begin{tikzpicture}
    \node (A) at (0,0) {$\Spec_{\UU}$};
    \node (B) at (3,0) {$\DLat{\USucc{\UU}}$};
    \node (C) at (6,0) {$\DLat{\UU}$,};

    \draw[->, line width=0.6pt]
    (A.east) to node[midway, above] {} (B.west);
    \draw[->, line width=0.6pt]
    (B.east) to node[midway, above] {} (C.west);
  \end{tikzpicture}
\end{center}
where the first map \(\ArrTy{\Spec_{\UU}}{\DLat{\USucc{\UU}}}\) takes a spectral
locale to its a priori \emph{large} distributive lattice of compact opens, and
the second one takes this distributive lattice to its small copy.

\begin{remark}\label{rmk:large-and-small-K}
We will sometimes have to distinguish between the two maps above.
When such technical precision is required, we will use the notations
\begin{itemize}
  \item \(\DistrLattOfCompactOpensLargeSym\) for the map \(\ArrTy{\Spec_{\UU}}{\DLat{\USucc{\UU}}}\), and
  \item \(\DistrLattOfCompactOpensSmallSym\) for the composition
    \(\ArrTy{\Spec_{\UU}}{\DLat{\UU}}\).
\end{itemize}
The reader is advised to bear this distinction in mind, which will play a
particularly important r\^{o}le in \cref{sec:equiv-functor-part}.
\end{remark}

\subsection{Object part of Stone duality}

So far, we have constructed functions:
\begin{center}
  \begin{tikzpicture}
    \node (A) at (0,0) {$\Spec_{\UU}$};
    \node (B) at (4,0) {$\DLat{\UU}$,};

    \draw[->, line width=0.6pt]
    ([yshift=0.1cm]A.east) to node[midway, above] {$\DistrLattOfCompactOpensSym$} ([yshift=0.1cm]B.west);
    \draw[->, line width=0.6pt]
    ([yshift=-0.1cm]B.west) to node[midway, below] {$\Spectrum{\blank}$}([yshift=-0.1cm]A.east);
  \end{tikzpicture}
\end{center}
between the type of spectral locales and the type of distributive lattices.
We now show that these form an equivalence of types.

The use of the term ``equivalence'' requires some caution here as we are
constructing only the object part at this point.
To be more specific, we will prove two results:
\begin{enumerate}
  \item For every \VUniLattice{\UU} $L$, there is a distributive lattice
    isomorphism $L \cong \CompactOpens{\Spectrum{L}}$.
  \item For every spectral locale $X$, there is a locale homeomorphism
    $X \cong \Spectrum{\CompactOpens{X}}$.
\end{enumerate}
Under the assumption of \VAxUA{}, the above isomorphisms can be turned into
equalities using the \VSIP{}, establishing that $\CompactOpens{\blank}$
is a retraction of $\Spectrum{\blank}$ and vice versa.
This allows us to extend the above to a \indexed{type equivalence}:
\begin{equation*}
  \Equiv{\Spec_{\UU}}{\DLat{\UU}}\text{.}
\end{equation*}

\begin{lemma}[\AgdaLink{Locales.Spectrality.LatticeOfCompactOpens-Duality.html\#31023}]%
\label{lem:K-spec}
  Every small distributive lattice\/ $L$ is isomorphic
  to the lattice\/ $\CompactOpens{\Spectrum{L}}$.
\end{lemma}
\begin{proof}
  Let $L$ be a small distributive lattice.
  Recall that we have a map
  \[\downset (\blank) \oftype \ArrTy{L}{\CompactOpens{\Spectrum{L}}}\]
  since principal ideals generated by the elements of $L$
  are compact in $\Spectrum{L}$ (by \cref{lem:principal-ideals-are-compact}).
  We claim that this map forms a distributive lattice isomorphism.
  Recall that, in \cref{lem:spec-compact-opens-are-exactly-the-principal-ideals}
  we showed that for every compact open $K \oftype \Spectrum{L}$,
  we have that there exists an element $x \oftype L$ satisfying $\IdTy{K}{\downset x}$.
  Because the principal ideal map is an \indexedp{embedding} by
  \cref{lem:principial-ideal-is-an-embedding},
  we know that its fibres are propositions.
  This concludes that it is an equivalence as its fibres are contractible.
  By \cref{lem:dlat-isomorphism-characterization}, it suffices to show that
  both sides of this equivalence are monotone to establish an isomorphism.
  We already showed in \cref{lem:principal-ideal-preserves-meets},
  that the principal ideal map preserves meets,
  which implies that it is monotone.
  The other direction is easy to see.
\end{proof}

\begin{lemma}[\AgdaLink{Locales.Spectrality.LatticeOfCompactOpens-Duality.html\#20350}]%
\label{lem:spec-K}
  Every spectral locale\/ $X$ is homeomorphic to\/ $\Spectrum{\CompactOpens{L}}$.
\end{lemma}
\begin{proof}
  Let $X$ be a spectral locale over base universe \VUni{\UU}.
  We know that there is a map~\(\phi \oftype \ArrTy{\opens{X}}{\Ideal{\UU}{\CompactOpens{X}}}\),
  defined
  (\AgdaLink{Locales.Spectrality.LatticeOfCompactOpens-Duality.html\#6500}) by
  \begin{equation*}
    \phi(U) \quad\is\quad \setof{ C \oftype \CompactOpens{X} \mid C \le U }\text{,}
  \end{equation*}
  which is a small collection thanks to the smallness of $\CompactOpens{X}$.
  Note that this is the principal ideal restricted to the small
  collection \(\CompactOpens{X}\).
  We claim that this forms a frame isomorphism $\opens{X} \cong {\opens{\Spectrum{L}}}$.
  The inverse of this is simply the map sending an ideal
  $I \subseteq \CompactOpens{X}$
  to the join of the family
  \(\paren{\SigmaType{K}{\CompactOpens{X}}{K \in I}} \hookrightarrow \CompactOpens{X}\).
  We denote this map by $\vartheta \oftype \Ideal{\UU}{\CompactOpens{X}} \to \opens{X}$.
  It suffices to show that $\vartheta$ and $\phi$ are inverses.

  Let $I \subseteq \CompactOpens{X}$ be a \VUniHyphen{\UU}ideal over the
  distributive \VUniLattice{\UU} of compact opens of $X$.
  We first show that $I \IdTySym \phi(\vartheta(I))$.
  We use antisymmetry.
  The nontrivial direction is $\phi(\vartheta(I)) \subseteq I$.
  Let $K_1$ be a compact open with $K_1 \le \FrmJoin I$.
  Since ideals are directed, we know that there exists some compact $K_2 \in I$
  with $K_1 \le K_2$.
  By the \VDownwardClosure{} of $I$, we have that $K_1 \in I$.
  For the other direction, consider an open $U \oftype \opens{X}$.
  The fact that $U  \IdTySym \vartheta(\phi(U))$ follows
  directly from Condition~(\ref{it:V}) of \cref{thm:spec-characterization}.
\end{proof}

It is easily verified that  frame isomorphisms give a
\indexed{standard notion of structure} for frames
as discussed in \cref{sec:sip}.
We thus obtain the following from the \indexedp{\VSIP{}}:

\begin{lemma}[\AgdaLink{Locales.SIP.FrameSIP.html\#8073}]
\label{lem:locale-sip}
  For every pair of locales $X$ and $Y$, there is an equivalence
  \begin{equation*}
    \Equiv{(X \cong Y)}{(\IdTy{X}{Y})}\text{,}
  \end{equation*}
  where $X \cong Y$ denotes the type of homeomorphisms betwen $X$ and $Y$.
\end{lemma}

Similarly for distributive lattice, we can show:

\begin{lemma}[\AgdaLink{Locales.SIP.DistributiveLatticeSIP.html\#9635}]
\label{lem:dlat-sip}
  For every pair of distributive lattices $K$ and $L$, there is an equivalence
  \begin{equation*}
    \Equiv{(K \cong L)}{(\IdTy{K}{L})}\text{,}
  \end{equation*}
  where $K \cong L$ denotes the type of isomorphisms betwen $K$ and $L$.
\end{lemma}

From these, we get a type equivalence:

\begin{theorem}[\AgdaLink{Locales.StoneDuality.ForSpectralLocales.html\#4829}]
\label{thm:stone-duality}
  The type\/ $\Spec_{\UU}$ is equivalent to the type\/ $\DLat{\UU}$.
\end{theorem}
\begin{proof}
  We turn the isomorphism given by \cref{lem:K-spec}
  and the homeomorphism given by \cref{lem:spec-K}
  into equalities using \cref{lem:locale-sip} and \cref{lem:dlat-sip}.
\end{proof}

In the context of \VUF{}, the above is a rather strong statement.
It implies that these two mathematical structures satisfy exactly the same
properties.

\section{Categorical part of Stone duality}
\label{sec:equiv-functor-part}

So far, we have shown that the maps
\begin{center}
  \begin{tikzpicture}
    \node (A) at (0,0) {$\Spec_{\UU}$};
    \node (B) at (4,0) {$\DLat{\UU}$,};

    \draw[->, line width=0.6pt]
    ([yshift=0.1cm]A.east) to node[midway, above] {$\DistrLattOfCompactOpensSym$} ([yshift=0.1cm]B.west);
    \draw[->, line width=0.6pt]
    ([yshift=-0.1cm]B.west) to node[midway, below] {$\SpectrumSym$}([yshift=-0.1cm]A.east);
  \end{tikzpicture}
\end{center}
form an equivalence of \emph{types}.
In this section, we show that this can be extended to an equivalence of
categories.
Before we proceed to the proof, however, we first discuss the definition of the
notion of categorical equivalence in \VHoTTUF{}.

\subsection{Categorical equivalences in \VHoTTUF{}}

Recall that the standard definition~\cite[Definition~7.25]{awodey-cat-theory}
of an equivalence between categories \(\mathbb{C}\) and \(\mathbb{D}\) is as a
pair of functors
\begin{equation*}
  \begin{tikzpicture}
    \node (A) at (0,0) {$\mathbb{C}$};
    \node (B) at (2,0) {$\mathbb{D}$};

    \draw[->, line width=0.6pt]
    ([yshift=0.1cm]A.east) to node[midway, above] {$E$} ([yshift=0.1cm]B.west);
    \draw[->, line width=0.6pt]
    ([yshift=-0.1cm]B.west) to node[midway, below] {$F$}([yshift=-0.1cm]A.east);
  \end{tikzpicture}
\end{equation*}
equipped with natural isomorphisms
\(\alpha \oftype \ArrTy{\mathbf{1}_{\mathbb{C}}}{F \circ E}\)
and
\(\beta \oftype \ArrTy{\mathbf{1}_{\mathbb{D}}}{E \circ F}\).
It is a well-known fact of
category theory~\cite[Proposition~7.26]{awodey-cat-theory}
that
a functor \(E \oftype \ArrTy{\mathbb{C}}{\mathbb{D}}\) being part of an
equivalence in the above sense
is equivalent to it being fully faithful and essentially surjective.

In a constructive foundational setting, some caution must be exercised regarding
whether to formulate the notion of an essentially surjective functor using
specified or unspecified existence.
In the \VHoTTBook{}, the formulation using specified existence
is called a \emph{split} essentially surjective functor:%
\index{essentially surjective}\index{split essentially surjective}

\begin{definition}[%
  {Split essentially surjective functor~\cite[Definition~9.4.4]{hott-book}}%
]
  \ \ \ A~functor \(F \oftype \ArrTy{\mathbb{C}}{\mathbb{D}}\)
  is called \define{split essentially surjective} if,
  for every object \(D \oftype \mathbb{D}\),
  there is a \emph{specified} object \(C \oftype \mathbb{C}\)
  such that
  \(\Isomorphic{F(C)}{D}\).
\end{definition}

For the sake of self-containment, we also provide the definitions of full and
faithful functors.

\begin{definition}[{\cite[Definition~9.4.4]{hott-book}}]
  Let \(F \oftype \ArrTy{\mathbb{C}}{\mathbb{D}}\) be a functor.
  If the mapping
  \[F \oftype \ArrTy{\HomSet{C_1}{C_2}}{\HomSet{F(C_1)}{F(C_2)}}\]
  is an \indexedp{embedding} for every pair of objects \(C_1\) and \(C_2\),
  functor \(F\) is called \define{faithful}.
  Similarly, \(F\) is called \define{full} if this mapping is
  \indexedp{surjective}, for every pair of objects~\(C_1\) and~\(C_2\).
  If \(F\) is both full and faithful, it is called \define{fully faithful}.
\end{definition}

\begin{samepage}
\begin{lemma}[{\cite[Lemma~9.4.5]{hott-book}}]
\label{lem:equiv-of-equivs}
  For every functor \(E \oftype \ArrTy{\mathbb{C}}{\mathbb{D}}\), the
  following are equivalent:
  \begin{enumerate}
    \item \(E\) is part of an equivalence of categories.
    \item \(E\) is fully faithful and split essentially surjective.
  \end{enumerate}
\end{lemma}
\end{samepage}

For our purposes, it will be convenient to make use of the above
characterization of categorical equivalence when showing that
the categories \(\Spec_{\UU}\) and \(\DLat{\UU}\) are equivalent.
More specifically, we will establish the categorical equivalence of interest
by showing that the map
\(\DistrLattOfCompactOpensSym \oftype \ArrTy{\Spec_{\UU}}{\DLat{\UU}}\)
extends to a fully faithful and split essentially surjective functor.

\subsection{%
  The functorial action of
  \texorpdfstring{\(\DistrLattOfCompactOpensSmallSym\)}{K⁻}%
}

Recall that a \indexedp{spectral map} \(f \oftype \ArrTy{X}{Y}\) is a continuous
function whose defining frame homomorphism
maps the \indexedpCE{compact opens}{compact open} of \(Y\) to the compact opens
of \(X\).
More formally, the spectrality of a continuous map \(f \oftype \ArrTy{X}{Y}\)
is witnessed by some function
\[\varphi \oftype \PiTy{V}{\opens{Y}}{(V\ \mathsf{is\ compact} \ImplSym f^*(V)\ \mathsf{is\ compact})}\text{.}\]
We define the map
\(\DistrLattOfCompactOpensSym_f \is \Pair{K}{\kappa} \mapsto \Pair{f^*(K)}{\varphi_K(\kappa)}\text{,}\)
where \(\Pair{K}{\kappa}\) is a pair consisting of an open \(K \oftype \opens{Y}\)
equipped with some proof \(\kappa\) that it is compact.
The function \(\DistrLattOfCompactOpensSym_f\) maps such a pair to the pair
\(\Pair{f^*(K)}{\varphi_K(\kappa)}\), where \(\varphi_K(\kappa)\) is the witness that
the result~\(f^*(K)\) is compact.
This obviously gives a lattice homomorphism
\begin{equation*}
  \DistrLattOfCompactOpensSym_f \oftype \ArrTy{\DistrLattOfCompactOpens{Y}}{\DistrLattOfCompactOpens{X}}
  \text{.}
\end{equation*}
We thus have a mapping from the \indexedp{Hom-set} of \(\Spec_{\UU}\) to the
Hom-set of \(\DLat{\USucc{\UU}}\):
\begin{equation*}
  \DistrLattOfCompactOpensSym_{(\blank)}
  \oftype
  \ArrTy{\HomSet{X}{Y}}{\HomSet{\DistrLattOfCompactOpens{Y}}{\DistrLattOfCompactOpens{X}}}\text{.}
\end{equation*}
Although this is morally the definition of the \indexedp{functorial action} of
\(\DistrLattOfCompactOpensSym \oftype \ArrTy{\Spec_{\UU}}{\DLat{\UU}}\)
on the morphisms of \(\Spec_{\UU}\),
we also need to account for the fact that we want to obtain a morphism in the
category \(\DLat{\UU}\) of \emph{small} of distributive lattices.
Recalling the distinction between
\(\DistrLattOfCompactOpensLargeSym\) and \(\DistrLattOfCompactOpensSmallSym\),
which we discussed \cref{rmk:large-and-small-K},
the issue here is that the above definition is for
the map~\(\DistrLattOfCompactOpensLargeSym \oftype \ArrTy{\Spec_{\UU}}{\DLat{\USucc{\UU}}}\),
and not for
the map~\(\DistrLattOfCompactOpensSmallSym \oftype \ArrTy{\Spec_{\UU}}{\DLat{\UU}}\).

Given spectral locales \(X\) and \(Y\), we know by the definition of spectrality
that we have equivalences
\(e_X \oftype \Equiv{\DistrLattOfCompactOpensLarge{X}}{\DistrLattOfCompactOpensSmall{X}}\)
and
\(e_Y \oftype \Equiv{\DistrLattOfCompactOpensLarge{Y}}{\DistrLattOfCompactOpensSmall{Y}}\),
which are homomorphisms of distributive lattices as we showed in \cref{lem:resizing-of-K}.
We will use these isomorphisms to transport the functorial action
from
\(\HomSet{\DLoCOLargeSym(Y)}{\DLoCOLargeSym(X)}\) to \(\HomSet{\DLoCOSmallSym(Y)}{\DLoCOSmallSym(X)}\).

In the remainder of this section, we will use the notations
\begin{center}
  \begin{tikzpicture}
    \node (A) at (0,0) {$\DistrLattOfCompactOpensLarge{X}$};
    \node (B) at (3,0) {$\DistrLattOfCompactOpensSmall{X}$};

    \draw[->, line width=0.6pt]
    ([yshift=0.1cm]A.east) to node[midway, above] {$s_X$} ([yshift=0.1cm]B.west);
    \draw[->, line width=0.6pt]
    ([yshift=-0.1cm]B.west) to node[midway, below] {$r_X$}([yshift=-0.1cm]A.east);
  \end{tikzpicture}
  \quad
  \begin{tikzpicture}
    \node (A) at (0,0) {$\DistrLattOfCompactOpensLarge{Y}$};
    \node (B) at (3,0) {$\DistrLattOfCompactOpensSmall{Y}$,};

    \draw[->, line width=0.6pt]
    ([yshift=0.1cm]A.east) to node[midway, above] {$s_Y$} ([yshift=0.1cm]B.west);
    \draw[->, line width=0.6pt]
    ([yshift=-0.1cm]B.west) to node[midway, below] {$r_Y$}([yshift=-0.1cm]A.east);
  \end{tikzpicture}
\end{center}
to refer to the two directions of these equivalences.
The diagram in \cref{fig:diagram-of-resizing-maps} summarizes this state of
affairs.

\begin{figure}[h!]
  \caption{%
    Spectral locales \(X\) and \(Y\) and the resizing maps for their
    distributive lattices of compact opens.%
  }
  \label{fig:diagram-of-resizing-maps}
  \begin{center}
  \begin{tikzcd}
    \opens{Y} \arrow[r, "f^*"]
  & \opens{X}\\
    \DistrLattOfCompactOpensLarge{Y} \arrow[u, hook', "\projISym"]
                                     \arrow[r, "\DLoCOLargeSym_f"]
                                     \arrow[d, "s_Y", bend left=30]
  & \DistrLattOfCompactOpensLarge{X} \arrow[u, hook, "\projISym", swap]
                                \arrow[d, "s_X", bend left=30]\\
    \DistrLattOfCompactOpensSmall{Y} \arrow[u, "r_Y", bend left=30]
                                     \arrow[r, "\DLoCOSmallSym_{f}"]
  & \DistrLattOfCompactOpensSmall{X} \arrow[u, "r_X", bend left=30]
  \end{tikzcd}
  \end{center}
\end{figure}

\begin{samepage}
\begin{definition}[%
  Functorial action of \(\DLoCOSmallSym\)%
]\label{defn:functorial-action-of-K}
  Let \(f \oftype \ArrTy{X}{Y}\) be a \indexedp{spectral map}, whose spectrality is
  formally witnessed by a proof\index{functorial action}
  \[\varphi \oftype \PiTy{V}{\opens{Y}}{(V\ \mathsf{is\ compact} \ImplSym f^*(V)\ \mathsf{is\ compact})}\text{.}\]
  The \indexedp{functorial action}
  \(\DLoCOSmallSym_f \oftype \ArrTy{\DistrLattOfCompactOpensSmall{Y}}{\DistrLattOfCompactOpensSmall{X}}\)
  is defined as
  \(\DLoCOSmallSym_f \is \Comp{s_X}{\Comp{\DLoCOLargeSym_f}{r_Y}}\text{,}\)
  which amounts to the map
  \(K \mapsto s_X\Pair{f^*(\projI(r_Y(K)))}{\varphi(\projII(r_Y(K)))}\).
\end{definition}
\end{samepage}

Although we gave the fully formal definition above, we will generally not be as
precise. For example, we will often omit applications of the projections for the
sake of readability.

\begin{lemma}
  For every spectral map\/ \(f \oftype \ArrTy{X}{Y}\) between spectral
  locales\/ \(X\) and\/ \(Y\), the map\/
  \(%
    \DistrLattOfCompactOpensSmallSym_f
    \oftype
    \ArrTy{\DistrLattOfCompactOpensSmall{Y}}{\DistrLattOfCompactOpensSmall{X}}%
  \)
  is a homomorphism of distributive lattices.
\end{lemma}
\begin{proof}
  In \cref{lem:resizing-of-K}, we established that the equivalences
  \(\Equiv{\DistrLattOfCompactOpens{Y}}{\DistrLattOfCompactOpensSmall{Y}}\)
  and \(\Equiv{\DistrLattOfCompactOpens{X}}{\DistrLattOfCompactOpensSmall{X}}\)
  are homomorphisms of distributive lattices.
  The desired result follows from this combined with the fact that
  \(f^*\) is a frame homomorphism.
\end{proof}

We have thus completed the definition of the functor
\(\DLoCOSmallSym \oftype \ArrTy{\Spec_{\UU}}{\DLat{\UU}}\).

\subsection{%
  The functor
  \texorpdfstring{\(\DistrLattOfCompactOpensSmallSym\)}{K⁻}
  gives a categorical equivalence
}
\label{sec:cat-equiv-proof}

We are now ready to establish the desired categorical equivalence by showing
that the functor \(\DLoCOSmallSym\) is
(1) fully faithful
and
(2) split essentially surjective.

\begin{lemma}
  The
  functor\/ \(\DistrLattOfCompactOpensSmallSym \oftype \ArrTy{\Spec_{\UU}}{\DLat{\UU}}\)
  is \indexedp{split essentially surjective}.
\end{lemma}
\begin{proof}
  We need to show that, for every small distributive lattice \(L\),
  there is a specified spectral locale \(X\) such that
  \(\Isomorphic{\DistrLattOfCompactOpensSmall{X}}{L}\).
  This was already established in \cref{lem:K-spec}, where we showed that
  the distributive lattice of compact opens of \(\Spectrum{L}\), which is small,
  is isomorphic to \(L\).
\end{proof}

Before we proceed to showing that the functor in consideration is fully
faithful, we first record the following lemma in preparation.

\begin{lemma}[\AgdaLink{Locales.CompactRegular.html\#60471}]%
\label{lem:characterisation-of-continuity}
  A monotone endomap\/ $h \oftype \Endomap{\opens{X}}$
  on a spectral locale\/ $X$ is
  \VScottContinuous{} whenever it satisfies the following condition:
  for every\/ $U \oftype \opens{X}$
  and compact\/ $K \oftype \opens{X}$ with\/ $K \le h(U)$,
  there is some compact\/ $K' \le U$ such that\/ $K \le h(K')$.
\end{lemma}
\begin{proof}
  It suffices to show the relation
  $h\paren{\FrmJoin_{i \oftype I} U_i } \leq \FrmJoin_{i \oftype I} h(U_i)$ holds.
  Since $X$ is spectral,
  let $\FamEnum{j}{J}{K_j}$ be a small family of compact opens such that
  $\FrmJoin_{j \oftype J} K_j \IdTySym h\paren{\FrmJoin_{i \oftype I} U_i}$.
  For any $j \oftype J$, we have $K_j \leq h\paren{\FrmJoin_{i \oftype I} U_i}$,
  so by assumption there is a compact open $K\leq  \FrmJoin_{i \oftype I} U_i$
  such that $K_j \leq h(K)$. By compactness of $K$ there is an $i \oftype I$
  such that $K \leq U_i$,
  and so $K_j\leq h(K)\leq h(U_i)\leq \FrmJoin_{i:I} h(U_i)$
  and so we can take $K' \is K_j$.
\end{proof}

Since the map \((\blank) \vee (\blank)\) is obviously \indexedp{\VScottContinuous{}} on both
arguments, we have the following useful corollary:

\begin{corollary}\label{cor:bin-join-continuity-characterization}
  Let\/ \(X\) be spectral locale and consider two opens\/ \(U, V \oftype \opens{X}\).
  For every compact open\/ \(K\) with\/ \(K \le U \vee V\), there exist compact
  opens\/ \(K_1\) and \(K_2\)
  satisfying\/ \(K \le K_1 \vee K_2\), and also\/ \(K_1 \le U\) and\/ \(K_2 \le V\).
\end{corollary}

\begin{lemma}
  Let\/ \(X\) and\/ \(Y\) be two spectral locales and
  let\/ \(h \oftype \ArrTy{\DistrLattOfCompactOpensSmall{Y}}{\DistrLattOfCompactOpensSmall{X}}\)
  be a homomorphism of distributive lattices.
  The map
  \begin{equation*}
    \SpectrumSym^*_h(V) \is \FrmJoin
     \setof{ r_X(h(K)) \mid K \oftype \DLoCOSmallSym(Y),\, r_Y(K) \le V }
  \end{equation*}
  is a frame homomorphism\/ \(\ArrTy{\opens{Y}}{\opens{X}}\).
\end{lemma}
\begin{proof}
  The fact that the map \(\SpectrumSym^*_h\) preserves finite meets is easy to see.
  We focus on (1) the preservation of joins and (2) spectrality.

  (1) Let \(\FamEnum{i}{I}{V_i}\) be a directed family of opens in \(Y\).
  It is a well-known fact of lattice theory that a function preserves arbitrary
  joins if and only if it preserves finite joins as well as directed ones.
  If we assume that the family is directed, we can easily see that we have
  \[
     \FrmJoin
       \setof{
         r_X(h(K)) \mid K \oftype \DLoCOSmallSym(Y),\, r_Y(K) \le \left(\FrmJoin_{i \oftypes I} V_i\right)
       }
     \IdTySym
     \FrmJoin_{i \oftypes I} \SpectrumSym^*_h\left(V_i\right)
  \]
  by the \indexedp{compactness} of \(r_Y(K)\).
  For the case of binary joins, consider two opens \(V_1\) and \(V_2\).
  We need to show that
  \(%
    \IdTy{
      \SpectrumSym^*_h(V_1 \vee V_2)
    }{
      \SpectrumSym^*_h(V_1) \vee \SpectrumSym^*_h(V_2)
    }\text{.}%
  \)
  The \[\SpectrumSym^*_h(V_1) \vee \SpectrumSym^*_h(V_2) \le \SpectrumSym^*_h(V_1 \vee V_2)\]
  direction is easy to see, so we focus on the other direction.
  We have to show that
  \(%
    \FrmJoin \{ r_X(h(K)) \mid K \oftype \DLoCOSmallSym(Y),\, r_Y(K) \le V_1 \vee V_2 \}%
  \)
  is below the join
  {\footnotesize
  \[%
    \paren{\FrmJoin
      \setof{ r_X(h(K)) \mid K \oftype \DLoCOSmallSym(Y),\, r_Y(K) \le V_1 }%
    }
    \vee
    \paren{%
      \FrmJoin
        \setof{ r_X(h(K)) \mid K \oftype \DLoCOSmallSym(Y),\, r_Y(K) \le V_2 }%
    }\text{.}
  \]
  }
  First, observe that this is the same as the join
  \[%
    \paren{%
      \FrmJoin
        \setof{
          r_X(h(K_1 \vee K_1)) \mid K_1, K_2 \oftype \DLoCOSmallSym(Y),\,
          r_Y(K_1) \le V_1,\, r_Y(K_2) \le V_2
        }%
    }
  \]
  by distributivity.
  Let \(K \oftype \DLoCOSmallSym(Y)\) such that \(r_Y(K) \le V_1 \vee V_2\).
  We know by \cref{cor:bin-join-continuity-characterization} that
  there exist compact \(K_1, K_2\)
  such that \(r_Y(K_1) \le V_1\) and \(r_Y(K_2) \le V_2\) as well as
  \(r_Y(K) \le K_1 \vee K_2\).
  This implies that
  \(r_X(h(K)) \le r_X(h(K_1 \vee K_2))\) and also \(r_Y(K_1) \le V_1\) and
  \(r_Y(K_2) \le V_2\), which was exactly what we needed.
\end{proof}

\begin{samepage}
\begin{lemma}\label{lem:spec-on-compact-opens}
  Let\/ \(X\) and\/ \(Y\) be two spectral locales and
  let\/ \(h \oftype \ArrTy{\DLoCOSmallSym(Y)}{\DLoCOSmallSym(Y)}\)
  be a distributive lattice homomorphism between them.
  We then have that\/ \[\SpectrumSym^*_h(K) \IdTySym r_X(h(s_Y(K)))\text{,}\]
  for every\/ \(K \oftype \DLoCOSmallSym(Y)\).
\end{lemma}
\end{samepage}
\begin{proof}
  Let \(h \oftype \ArrTy{\DLoCOSmallSym(Y)}{\DLoCOSmallSym(Y)}\) be
  a distributive lattice homomorphism between the distributive lattices of
  compact opens of spectral locales \(X\) and \(Y\),
  and consider some~\(K \oftype \DLoCOSmallSym(Y)\).
  We first show \(\SpectrumSym^*_h(K) \le r_X(h(s_Y(K)))\).
  Let \(K' \oftype \DLoCOSmallSym(Y)\) such that~\(r_Y(K') \le K\).
  Since we have \(r_Y \dashv s_Y\), we know that \(K' \le s_Y(K)\), from which
  it follows that \(r_X(h(K')) \le r_X(h(s_Y(K)))\) since \(r_X \circ h\) is monotone.
  For the \[r_X(h(s_Y(K))) \le \SpectrumSym^*_h(K)\] direction,
  it suffices to observe that we have\/ \(r_Y(s_Y(K)) \le K\),
  since\/ \(r_Y(s_Y(K)) \IdTySym K\).
\end{proof}

\begin{corollary}
  For every distributive lattice
  homomorphism\/ \(h \oftype \ArrTy{\DLoCOSmallSym(Y)}{\DLoCOSmallSym(Y)}\),
  where\/ \(X\) and\/ \(Y\) are spectral locales,
  the frame homomorphism\/ \(\SpectrumSym^*_h\) preserves compact opens,
  since\/ \(r_X(h(s_Y(K)))\)
  is a compact open for every compact open\/ \(K \oftype \opens{Y}\).
  This means that\/ \(\SpectrumSym_h\) is a spectral map.
\end{corollary}

\begin{lemma}
  The
  map\/ \(\SpectrumSym_{(\blank)} \oftype \ArrTy{\HomSet{\DLoCOSmallSym(Y)}{\DLoCOSmallSym(X)}}{\HomSet{X}{Y}}\)
  is the quasi-inverse of the map
  \(%
    \DLoCOSmallSym_{(\blank)} \oftype \ArrTy{\HomSet{X}{Y}}{\HomSet{\DLoCOSmallSym(Y)}{\DLoCOSmallSym(X)}}%
  \)
\end{lemma}
\begin{proof}
  We start by showing that
  \(\SpectrumSym_{(\blank)}\) is a retraction of \(\DLoCOSmallSym_{(\blank)}\).
  Let \(f \oftype \ArrTy{X}{Y}\) be a spectral map and let
  \(V \oftype \opens{Y}\).
  We need to show that \(\SpectrumSym_{\DLoCOSmallSym_f}(V) \IdTySym f^*(V)\).
  First, observe that
  \begin{align*}
    \SpectrumSym_{\DLoCOSmallSym_f}(V)
  &\DefnEqSym
    \FrmJoin \setof{
      r_X(\DLoCOSmallSym_f(K)) \mid K \oftype \DLoCOSmallSym(Y),\, r_Y(K) \le V
    }\\
  &\DefnEqSym
    \FrmJoin \setof{
      r_X(s_X(f^*(r_Y(K)))) \mid K \oftype \DLoCOSmallSym(Y),\, r_Y(K) \le V
    }\\
  &\IdTySym
    \FrmJoin \setof{
      f^*(r_Y(K)) \mid K \oftype \DLoCOSmallSym(Y),\, r_Y(K) \le V
    }
  \end{align*}
  We know that \(V\) can be decomposed into a join
  \(\FrmJoin \setof{ K \mid K \oftype \opens{Y},\, K \le V }\)
  , and it is easily verified that
  \begin{equation*}
    f^*\left(\FrmJoin \setof{ K \mid K \oftype \opens{Y},\, K \le V }\right)
    \IdTySym
    \FrmJoin
      \setof{
        f^*(r_Y(K)) \mid K \oftype \DLoCOSmallSym(Y),\, r_Y(K) \le V
      }\text{.}
  \end{equation*}
  We now show that \(\SpectrumSym_{(\blank)}\) is a section of \(\DLoCOSmallSym_{(\blank)}\).
  Let \(h \oftype \ArrTy{\DLoCOSmallSym(Y)}{\DLoCOSmallSym(X)}\) be a
  homomorphism of distributive lattices.
  We need to show that \(\DLoCOSmallSym_{\SpectrumSym_{h}}(K) \IdTySym h(K)\),
  for every \(K \oftype \DLoCOSmallSym(Y)\).
  First, observe the definitional equality
  \begin{align*}
    \DLoCOSmallSym_{\SpectrumSym_{h}}(K)
  &\quad\DefnEqSym\quad s_X(\SpectrumSym^*_h(r_Y(K)))\\
  &\quad\DefnEqSym\quad
     s_X\left(\FrmJoin \setof{ r_X(h(K')) \mid K' \oftype \DLoCOSmallSym(Y),\, r_Y(K') \le r_Y(K) }\right)
    \tag{\dag}\label{eqn:k-cancels-spec-1}
  \end{align*}
  We first show that the join in \((\dag)\) is below \(h(K)\), for which it
  suffices to show
  \begin{equation*}
    \left(\FrmJoin \setof{ r_X(h(K')) \mid K' \oftype \DLoCOSmallSym(Y),\, r_Y(K') \le r_Y(K) }\right)
    \le
    r_X(h(K))\text{,}
  \end{equation*}
  since \(r_X\) is the right adjoint of \(s_X\).
  It suffices to show that
  \(r_X(h(K')) \le r_X(h(K))\),
  for every \(K' \oftype \DLoCOSmallSym(Y)\) with \(r_Y(K') \le r_Y(K)\).
  Consider such a \(K' \oftype \DLoCOSmallSym(Y)\)
  satisfying~\(r_Y(K') \le r_Y(K)\).
  Since \(r_Y\) is an order embedding, we have that \(K \le K\)
  and it follows that \(r_X(h(K')) \le r_X(h(K))\) as \(r_X \circ h\) is clearly
  a monotone map.
  We now address the other direction of the equality in consideration, namely,
  \begin{equation*}
    h(K)
      \le
    s_X\left(\FrmJoin \setof{ r_X(h(K')) \mid K' \oftype \DLoCOSmallSym(Y),\, r_Y(K') \le r_Y(K) }\right)\text{.}
  \end{equation*}
  As \(s_X\) is the right adjoint of \(r_X\), it suffices to show
  \begin{equation*}
    r_X(h(K))
      \le
    \FrmJoin \setof{ r_X(h(K')) \mid K' \oftype \DLoCOSmallSym(Y),\, r_Y(K') \le r_Y(K) }\text{,}
  \end{equation*}
  which is immediate.
\end{proof}

\begin{corollary}
  The functor\/ \(\DLoCOSmallSym \oftype \ArrTy{\Spec_{\UU}}{\DLat{\UU}}\)
  is \indexedp{fully faithful}.
\end{corollary}


\chapter{The patch locale}
\label{chap:patch}

We defined the categories $\Spec$ and $\Stone$
in \cref{chap:spec-and-stone},
and explained that the latter is a full subcategory of the former.
In this chapter, we delve deeper into the relationship between these two
categories and examine how a spectral locale can be
universally transformed
into a Stone one in a predicative setting.

In an impredicative foundational setting,
it is well known that the category~$\Stone$
is a \indexed{coreflective subcategory} of
the category\/ $\Spec$~\cite{mhe-patch-short,mhe-patch-full}.
This coreflection amounts to the fact that every spectral locale can be
universally transformed into a Stone one.
The coreflector involved here is what is known as the patch locale
of a spectral locale:
the localic manifestation of the \indexedp{patch topology} on a \indexedp{spectral space}
(also called the \indexed{constructible topology} in the
literature~\cite[\S 1.3.11]{dst-spectral-spaces}).
This can be thought of as the universal locale in which all compact opens of a
spectral locale are transformed into clopens.
Formally, this is captured by the universal property below.

\begin{definition}[\AgdaLink{Locales.UniversalPropertyOfPatch.html\#48926} Universal property of the patch locale]\label{defn:ump-of-patch}
  Let $A$ be a spectral locale.
  A Stone locale $A_p$
  is said to be a \define{patch locale} of $A$
  if there is a specified spectral map $\PatchCounit \oftype \ArrTy{A_p}{A}$,
  such that,
  for every continuous map $f \oftype \ArrTy{X}{A}$,
  from a \indexedp{Stone locale}~$X$ into $A$,
  there exists a continuous map $\bar{f} \oftype \ArrTy{X}{A_p}$,
  unique with the property that $\IdTy{f}{\PatchCounit \CompSym \bar{f}}$.
  This is summarized by the diagram:
  \begin{center}
    \begin{tikzcd}
      & A_p \arrow[d, "\PatchCounit"]\\
      X \arrow[r, "f", swap] \arrow[ur, dashed, "\bar{f}"] & A
    \end{tikzcd}
  \end{center}
\end{definition}

It is not a priori clear that the patch locale of every spectral locale can be
constructed in a constructive and predicative foundational setting,
since all proofs of this coreflection in the literature seem to use some form of
impredicativity.
The focus of this chapter is on showing that every spectral locale can be
coreflected into a Stone one as given in the above universal property.
To achieve this, we follow an idea due to
\MHELastName{}~\cite{mhe-patch-short, mhe-patch-full}, that the patch locale can
be defined by the frame of \VScottContinuous{} nuclei on a spectral locale.
This description of the patch locale turns out to be amenable to constructive
and predicative reasoning.
Although our construction is mostly the same, \MHELastName{}'s proof of the
universal property~\cite{mhe-patch-short} of the patch locale relies on the
existence of the frame of \emph{all} nuclei, which can easily be constructed in
an impredicative foundational setting.
In our predicative setting, where the frame of all nuclei does not seem to be
available, the proof of the universal property is new, and turns out to be more
complicated.

The organization of this chapter is as follows:
\begin{description}[leftmargin=!,labelwidth=\widthof{Section 7.77:}]
  \item[\cref{sec:nuclei-meet-semilattice}:]
    We present the poset of all nuclei, and show that it forms a
    meet-semilattice.
    Furthermore, we discuss the problem that this poset does not seem to be
    locally small without some form of propositional resizing.
  \item[\cref{sec:frame-of-nuclei-impredicatively}:]
    We explain that the meet-semilattice of all nuclei
    from \cref{sec:nuclei-meet-semilattice}
    can be shown to be \emph{complete} in an impredicative setting, and we
    survey the impredicative approaches to constructing joins in this lattice.
    Furthermore, we discuss the problem that the construction of these joins
    does not seem to be predicatively possible.
  \item[\cref{sec:frame-of-scott-continuous-nuclei}:]
    We discuss a sub-meet-semilattice of the meet-semilattice of nuclei,
    namely the sublattice consisting of \emph{\VScottContinuous{}} nuclei.
    Furthermore, we show that \MHELastName{}'s construction of
    joins~\cite{mhe-properly-injective} in this frame can easily be translated into
    our predicative setting.
    We thus conclude that this sublattice is small-complete
    \emph{even in} a predicative setting, and that it forms a frame.
  \item[\cref{sec:patch-locale}:]
    We prove that the locale defined by
    the frame of \VScottContinuous{} nuclei
    from~\cref{sec:frame-of-scott-continuous-nuclei}
    satisfies the universal property of the patch locale from
    \cref{defn:ump-of-patch}, hence establishing that $\Stone$ is a coreflective
    subcategory of $\Spec$ in a predicative setting.
  \item[\cref{sec:examples-of-patch}:]
    We present several examples of the patch locale and discuss its
    lattice-theoretic interpretation through Stone duality.
\end{description}

\section{Meet-semilattice of nuclei on a locale}
\label{sec:nuclei-meet-semilattice}

\subsection{Pointwise ordering of nuclei}

Let $X$ be a locale over base universe \VUni{\UU} and consider
the type $\Nucleus{X}$,%
\nomenclature{\(\Nucleus{X}\)}{the type of nuclei on locale \(X\)}
which lives in universe~\VUni{\USucc{\UU}}.
This type forms a poset when ordered pointwise:

\begin{lemma}
  For every locale\/ $X$ (over base universe\/ \VUni{\UU}),
  the pointwise ordering,
  \begin{equation*}
    j \le k \quad\is\quad \PiTy{U}{\opens{X}}{j(U) \le k(U)},
  \end{equation*}
  is a partial order with large truth values i.e.\ truth values living in
  universe\/ \VUni{\UU}.
\end{lemma}

\begin{definition}\label{defn:poset-of-nuclei}
  The \define{poset of nuclei}\/ on a locale $X$ is
  the type $\Nucleus{X}$
  partially ordered by the above order.
\end{definition}

\begin{remark}\label{rmk:poset-of-nuclei-is-not-locally-small}
  Although the above order seems to have necessarily large
  truth values,
  we do not have a resizing taboo establishing this at the time of writing.
  We conjecture that the local smallness of this poset gives a form of resizing.
\end{remark}

\subsection{Meets of nuclei}

\begin{lemma}\label{defn:nuclei-semilattice}
  For every locale\/ $X$,
  the poset of nuclei on\/ $X$
  has finite meets.
\end{lemma}
\begin{proof}
  Let $X$ be a locale.
  The top nucleus is defined as the constant map with value~$\One{X}$ and the
  meet of two nuclei as $j \meet k \is U \mapsto j(U) \meet k(U)$.
  It is easy to see that $j \meet k$ is the greatest lower bound of
  the nuclei $j$ and $k$, so it remains to show that
  $j \meet k$ satisfies the conditions of being a nucleus.

  Inflationarity can be seen to be satisfied from the inflationarity
  of $j$ and $k$, combined with the fact that $j(U) \meet k(U)$ is the
  greatest lower bound of $j(U)$ and~$k(U)$.
  To see that meet preservation holds, let $U, V \oftype \opens{X}$.
  \begin{align*}
    (j \meet k)(U \meet V)
    &\quad\DefnEqSym\quad   j (U \meet V) \meet k (U \meet V)\\
    &\quad\IdTySym\quad     j(U) \meet j(V) \meet k(U) \meet k(V)
                            && [\text{meet-preservation}]\\
    &\quad\IdTySym\quad     (j(U) \meet k(U)) \meet (j(V) \meet k(V))
                            && [\text{commutativity}]\\
    &\quad\DefnEqSym\quad   (j \meet k)(U) \meet (j \meet k)(V)
  \end{align*}
  For idempotency, let $U \oftype \opens{X}$.
  \begin{align*}
    \quad&\hphantom{\IdTySym}\quad\;\; (j \meet k)((j \meet k)(U))\\
    \quad&\DefnEqSym\quad  j (j(U) \meet k(U)) \meet k(j(U) \meet k(U))\\
    \quad&\IdTySym\quad    j(j(U)) \meet j(k(U)) \meet k(j(U)) \meet k(k(U))
                           &&[\text{meet-preservation}]\\
    \quad&\le\quad         j(j(U)) \meet k(k(U))\\
    \quad&\IdTySym\quad    j(U) \meet k(U)
                           && [\text{idempotency}]\\
    \quad&\DefnEqSym\quad  (j \meet k)(U)
  \end{align*}
\end{proof}

\section{Nuclei form a frame, impredicatively}
\label{sec:frame-of-nuclei-impredicatively}

In the traditional development of locale theory in impredicative foundations,
the \indexed{frame of nuclei} is a standard construction.
\VPTJLastName{}~\cite[\textsection II.2.5]{ptj-ss}
shows that the poset of nuclei forms a frame by
constructing arbitrary meets and Heyting implications in it.
He remarks in passing that
``joins in [the frame of nuclei] are harder to describe explicitly''%
~\cite[51]{ptj-ss}.
Let us first examine this difficulty involved in the explicit description of
these joins, before looking at the details of \VPTJLastName{}'s construction.

Following the case of meets, one might attempt to define the join of a family
of nuclei
$\FamEnum{i}{I}{k_i}$ as the pointwise join, i.e.
\begin{equation*}
  \paren{\FrmJoin_{i \oftype I} k_i}(x) \quad\is\quad \FrmJoin_{i \oftype I} k_i(x).
\end{equation*}
However, this fails to be idempotent in general, and is not inflationary when
the family in consideration is empty.
There are several approaches to constructing these joins in the literature, but
all of these involve some kind of impredicativity. We first discuss the
aforementioned approach by \VPTJLastName{}.

Suppose we were working with complete and small locales, the theory of which can
be developed impredicatively as explained in \cref{sec:implications}.
Let $L$ be a complete and small frame.
If $L$ is meet-complete, then $L$ is also join-complete since we could easily
construct the joins as
\begin{equation*}
  \paren{\FrmJoin_{i \oftype I} x_i}
    \quad\is\quad
      \bigwedge
        \setof{ y \oftype L \mid y\ \text{is an upper bound of}\ \FamEnum{i}{I}{x_i} },
\end{equation*}
which would be a small collection when the carrier set in consideration is small.
Therefore, showing meet-completeness is sufficient to establish
join-completeness in the context of complete, small lattices.
Observe here that we must also have local smallness for this to be a small
collection.
As discussed in~\cref{prop:local-smallness-equiv}, local smallness
follows automatically from the smallness of the carrier set in the context of
complete, small lattices.
In a predicative foundational setting, such a construction of joins from meets
does not seem to be possible, since the above construction seems fundamentally
reliant on the meet over the entire carrier set.
Even if this collection were somehow small, we would still have the problem
that the order does not have small truth values in general
(see \cref{rmk:poset-of-nuclei-is-not-locally-small}).

Taking advantage of the definability of joins this way, \VPTJLastName{}
constructs meets to establish join-completeness.
He defines~\cite[\textsection II.2.5]{ptj-ss} the meets pointwise:
\begin{equation*}
  \paren{\BigMeet{i}{I}{k_i}} \quad\is\quad x \mapsto \BigMeet{i}{I}{k_i(x)}\text{.}
\end{equation*}
Moreover, to show that this join-complete lattice is a frame, \VPTJLastName{}
proceeds to construct Heyting implications. The existence of Heyting
implications implies that the frame distributivity law holds, i.e.
\begin{equation*}
  x \meet \paren{\FrmJoin_{i : I} y_i} \quad\IdTySym\quad \FrmJoin_{i : I} x \meet y_i,
\end{equation*}
for every $x \oftype L$ and every small family $\FamEnum{i}{I}{y_i}$.
This is because
$x \meet (\blank)$ is the left adjoint of $x \ImplSym (\blank)$, and left
adjoints always preserve colimits.
Therefore, the existence of the right adjoint of $x \meet (\blank)$ implies that
it preserves the colimits of $L$.
There is a second use of impredicativity at this point of the argument, since
\VPTJLastName{}'s
construction of Heyting implications~\cite[52]{ptj-ss}
also uses a meet over a large collection. He defines Heyting implications as
\begin{equation*}
  (j \Rightarrow k)(x)
    \quad\is\quad
      \BigMeetSym \setof{ (j(y) \Rightarrow k(y)) \mid y \oftype L, x \le y }\!\text{,}
\end{equation*}
using the Heyting implications of the underlying frame $L$.
In a predicative context, there is no reason for this collection to be small in
general.

There are several other constructions of joins of nuclei in the literature.
It seems to be the case, however, that all of these use some form of
impredicativity (or classical principles):
\begin{itemize}
  \item
    Banaschewski~\cite{banaschewski-tychonoff}
    and
    Wilson~\cite{wilson-thesis}
    both use constructions involving large meets, similar to \VPTJLastName{}'s
    construction discussed above.
  \item Simmons~\cite{simmons-cantor-bendixson} uses
    ordinals and \indexedp{transfinite induction},
    and his proof seems to be reliant on classical reasoning
    (though we have not carefully analyzed it).
  \item \MHELastName{}~\cite{mhe-joins} uses \VPataraiaFPT{},
    which also does not seem to be predicatively available.
    \VTDJMHE{} show \cite[Theorem~52]{tdj-mhe-aspects} that the main lemma used
    in the proof of \indexedp{\VPataraiaFPT{}} \emph{implies} a form of propositional
    resizing.
\end{itemize}
In light of the fact that all of the above constructions involve some form of
impredicativity, we conjecture that the meet-semilattice of nuclei cannot be
shown to be small-complete in our setting, unless a form of
\indexedp{propositional resizing} holds.

In the next section,
we will start working towards the predicative construction of the patch locale,
the defining frame of which is the aforementioned frame of
\VScottContinuous{} nuclei.

\section{The frame of \VScottContinuous{} nuclei}
\label{sec:frame-of-scott-continuous-nuclei}

We now restrict attention to
those nuclei that are \VScottContinuous{} and show that,
whilst joins in the lattice of \emph{all} nuclei seem to involve an inherent
impredicativity,
joins in this sublattice can be constructed in a completely predicative manner.
Furthermore, this sublattice consisting of \VScottContinuous{} nuclei also does
not suffer from the problem of not being locally small, which we discussed in
\cref{rmk:poset-of-nuclei-is-not-locally-small}.

As mentioned before, our motivation for constructing the frame of
\VScottContinuous{} nuclei is to define the patch locale of a spectral locale,
the universal property of which was given in the introduction to this chapter
(\cref{defn:ump-of-patch}).
The fact that the patch locale can be constructed this way
is due to \MHELastName{}~\cite{mhe-patch-short, mhe-patch-full}.

\subsection{Meet-semilattice of Scott continuous nuclei}
\label{sec:meet-semilattice-of-scott-continuous-nuclei}

We now refine the meet-semilattice
from \cref{sec:nuclei-meet-semilattice}
to those nuclei that are \VScottContinuous{}.

\begin{definition}
  A nucleus $j \oftype \Endomap{\opens{X}}$
  is called \define{\VScottContinuous{}}\index{nucleus!{\VScottContinuous{}}}
  if it preserves joins of small directed families.
\end{definition}

\begin{example}
  For every open $U \oftype \opens{X}$ of a locale $X$,
  the \indexedp{closed nucleus}~$\ClosedNucleus{U}$ (from \cref{example:closed-nucleus})
  is \VScottContinuous{} since the map $U \vee (\blank)$ preserves all joins,
  and in particular the directed ones.
\end{example}

\begin{example}[\AgdaLink{Locales.PatchProperties.html\#9959}]%
\label{example:open-nucleus-is-scott-continuous}
  For every compact open $K \oftype \CompactOpens{X}$ of a spectral locale $X$,
  the open nucleus $\OpenNucleus{K}$
  is \VScottContinuous{}.
\end{example}
\begin{proof}
  Let $K \oftype \CompactOpens{X}$ be a compact open of some spectral locale $X$.
  We need to show that the map $K \Rightarrow (\blank)$ is \VScottContinuous{}.
  As we know that this is a nucleus, we know that it is monotone since nuclei
  are always monotone (\cref{lem:prenuclei-are-monotone}).
  We show that this map is \VScottContinuous{} by showing that it satisfies the
  sufficient condition from \cref{lem:characterisation-of-continuity}.
  Let $V$ and $K_1$ be two opens of $X$
  and suppose that $K_1$ satisfies $K_1 \le K \Rightarrow V$.
  We need to show that there exists some compact $K_2 \le K$
  such that $K_1 \le K \Rightarrow K_2$.
  Pick $K_2 \is K \meet K_1$.
  We know that this is compact since compact opens are closed under binary meets
  in spectral locales.
  It remains to check $K_2 \le V$ and $K_1 \le K \Rightarrow K_2$.
  The former holds directly by the assumption that $K_1 \le K \Rightarrow V$,
  since this implies $K_1 \meet K \le V$.
  The latter is immediate.
\end{proof}

\subsubsection{Poset of \VScottContinuous{} nuclei}

We now show that the order of the poset of \VScottContinuous{} nuclei on a
spectral locale is small, in contrast with the case of all nuclei.

\begin{lemma}\label{lem:nuclei-basic-order}
  For every pair of nuclei\/ $j, k \oftype \Endomap{\opens{X}}$ on some
  locale\/ $X$, if
  \begin{enumerate}
    \item the locale\/ $X$ is spectral, and
    \item the nuclei\/ $j$ and\/ $k$ are \VScottContinuous{},
  \end{enumerate}
  then the
  pointwise ordering\/ $\PiTy{U}{\opens{X}}{j(U) \le k(U)}$
  is\/ \VUniSmall{\UU}.
\end{lemma}
\begin{proof}
  Let $X$ be a spectral locale and
  let $j, k \oftype \Endomap{\opens{X}}$
  be a pair of \VScottContinuous{} nuclei on $X$.
  We define the following small version of the pointwise order:
  \begin{equation*}
    j \le_s k \quad\is\quad \PiTy{K}{\CompactOpens{X}}{j(K) \le k(K)}\text{,}
  \end{equation*}
  which involves only a small quantification thanks to the smallness of the
  type of compact opens.
  We claim that $j \le k$ if and only if $j \le_s k$.
  The forward direction is trivially true,
  so it remains to show that $j \le_s k$ implies $j \le k$.
  Suppose $j \le_s k$.
  To see that $j \le k$, let $U \oftype \opens{X}$.
  As we have a base consisting of compact opens,
  we know that there is a directed basic covering family
  $\FamEnum{n}{N}{B_n}$ for $U$.
  Therefore, we just have to show
  \begin{equation*}
    j\paren{\FrmJoin_{n \,{\oftype}\, N} B_n}
    \quad\le\quad
    k\paren{\FrmJoin_{n \,{\oftype}\, N} B_n}
    \!\text{.}
  \end{equation*}
  By the \VScottContinuity{} of the nuclei $j$ and $k$, it suffices to show
  \begin{equation*}
    \FrmJoin_{n \,{\oftype}\, N} j(B_n)
    \quad\le\quad
    \FrmJoin_{n \,{\oftype}\, N} k(B_n)\text{.}
  \end{equation*}
  It follows directly from our assumption of $j \le_s k$
  that $j(B_n) \le k(B_n)$ for each~$n\oftype N$, so we are done.
\end{proof}

In summary, we have now established:

\begin{corollary}\label{cor:poset-of-scott-continuous-nuclei-is-locally-small}
  For every spectral locale\/ $X$,
  the poset of \VScottContinuous{} nuclei on\/ $X$
  is large and locally small.
\end{corollary}

\subsubsection{Meets of \VScottContinuous{} nuclei}

\begin{lemma}\label{prop:sc-nuclei-semilattice}
  The poset of \VScottContinuous{} nuclei on every locale
  forms a meet-semilattice.
\end{lemma}
\begin{proof}
  Let $X$ be a locale.
  The construction is the same as the one from
  Lemma~\ref{defn:nuclei-semilattice}.
  The top element is the constant map with value $\One{X}$, which is trivially
  \VScottContinuous{},
  so it remains to show that the meet of two \VScottContinuous{} nuclei
  is \VScottContinuous{}.
  Let $j$ and $k$ be a pair of \VScottContinuous{} nuclei,
  and consider a small directed family~$\FamEnum{n}{N}{U_n}$.
  We then have:
  \begin{align*}
       (j \meet k) \paren{\FrmJoin_{i \oftype I} U_i}
  &\quad\DefnEqSym\quad j \paren{\FrmJoin_{n \oftype N} U_n} \meet k \paren{\FrmJoin_{n \oftype N} U_n}
     && \\
  &\quad\IdTySym\quad \paren{\FrmJoin_{n \oftype N} j(U_n)} \meet \paren{\FrmJoin_{n \oftype N} k(U_n)}
     && \quad[\text{Scott continuity of $j$ and $k$}]\\
  &\quad\IdTySym\quad \FrmJoin_{(m, n) \oftype N \times N} j(U_m) \meet k(U_n)
     && \quad[\text{distributivity}]\\
  &\quad\IdTySym\quad \FrmJoin_{n \oftype N} j(U_n) \meet k(U_n)
     && \quad[\text{\dag}] \\
  &\quad\DefnEqSym\quad \FrmJoin_{n \oftype N} (j \meet k)(U_n)
     && \quad[\text{meet preservation}]
  \end{align*}
  For the step (\dag) above, we use antisymmetry.
  The $(\ge)$ direction is immediate.
  For the $(\le)$ direction, we need to show that
  \[\FrmJoin_{(m, n) \oftype N \times N} j(U_m) \meet k(U_n)
    \quad\le\quad \FrmJoin_{n \oftype N} j(U_n) \meet k(U_n)\text{,}\]
  for which it suffices to show that
  $\FrmJoin_{n \oftype N} j(U_n) \meet k(U_n)$
  is an upper bound of the family
  \[\{ j(U_m) \meet k(U_n) \}_{(m, n) \oftype N \times N}\text{.}\]
  Let $m, n \oftype N$ be a pair of indices.
  As $\FamEnum{n}{N}{U_n}$ is a directed family, there must exist
  some $o \oftype N$ such that
  $U_o$ is an upper bound of $U_m$ and $U_n$.
  Using the monotonicity of the nuclei $j$ and $k$, we get
  $j(U_m) \meet k(U_n)
  \le j(U_o) \meet k(U_o)
  \le \FrmJoin_{n \oftype N} j(U_n) \meet k(U_n)$,
  which was what we needed.
\end{proof}

\subsection{Joins}
\label{sec:joins}

Unlike the frame of \emph{all} nuclei, the meet-semilattice of
\VScottContinuous{} nuclei does admit small joins in a predicative setting.
We now look at the construction of these joins, which is the nontrivial
component of the construction of the patch frame.

Let $\FamEnum{i}{I}{k_i}$ be a small family of \VScottContinuous{} nuclei.
The problem that the pointwise join
\[U \mapsto \FrmJoin_{i \,{\oftype}\, I} k_i(U)\]
is not idempotent, which we discussed in
\cref{sec:frame-of-nuclei-impredicatively}, is also present in the context of
\VScottContinuous{} nuclei.
When the family in consideration is \emph{directed}, however, the pointwise join
is idempotent, and it thus gives the join in the poset of \VScottContinuous{}
nuclei:

\begin{lemma}[\AgdaLink{Locales.PatchProperties.html\#3911}]
\label{lem:directed-computed-pointwise}
  For every directed family\/ $\FamEnum{i}{I}{k_i}$
  of \VScottContinuous{} nuclei,
  the join is computed pointwise, that is,
  \[\paren{\FrmJoin_{i \oftype I} k_i}(U) \quad\IdTySym\quad \FrmJoin_{i \oftype I} k_i(U)\text{,}\]
  for every open\/ $U$.
\end{lemma}
\begin{proof}
  The argument given in the paragraph preceding
  \cite[Lemma~3.1.8]{mhe-properly-injective} works in our setting.
\end{proof}

The situation regarding arbitrary joins is more complicated.

\subsubsection{Directification of a family of nuclei}

A construction of the join, due to \MHELastName{}~\cite{mhe-properly-injective},
is based on the idea that, if we start with a family $\FamEnum{i}{I}{k_i}$ of
nuclei, we can directify it by considering the family with
index type $\ListTy{I}$,
defined by
\begin{equation*}
  (i_0i_1 \cdots i_{n-1})
    \mapsto
      k_{i_{n-1}} \CompSym \cdots \CompSym k_{i_1} \circ k_{i_0}\text{.}
\end{equation*}
This family is easily seen to be directed.
We refer to this as the \indexed{family of finite compositions}, and define the
following function $(\blank)^*$ implementing it:

\begin{definition}[%
  \AgdaLink{Locales.PatchLocale.html\#12073} Family of finite compositions%
]
\label{defn:family-of-finite-compositions}
  Given a small family $\FamEnum{i}{I}{k_i}$ of nuclei on some locale $X$,
  its \define{family of finite compositions} is the family
  $k^*$,
  defined as
  \begin{equation*}
    k^*(i_0\cdots i_{n-1})
    \quad\is\quad
      k_{i_{n-1}} \CompSym \cdots \CompSym k_{i_0}\text{.}
  \end{equation*}
\end{definition}

A finite composition $k^*_s$ is \emph{not} a nucleus in general.
It is, however, always a \indexedp{prenucleus}:

\begin{lemma}[\AgdaLink{Locales.PatchLocale.html\#12394}]%
\label{lem:star-prenucleus}
  For every family\/ $\FamEnum{i}{I}{k_i}$ of nuclei on a locale,
  the finite composition\/~$k^*_{s}$ is a \indexedp{prenucleus},
  for every\/~$s \oftype \ListTy{I}$.
\end{lemma}
\begin{proof}
  Let $\FamEnum{i}{I}{k_i}$ be a family of nuclei on some locale $X$.
  We proceed by list induction.

  \emph{Base case:} the empty list $\emptyl$.
  We are done since $k_{\emptyl} \DefnEqSym \FnIdSym$, and it is easy to see
  that the identity function $\FnIdSym$ is a prenucleus.

  \emph{Induction step:} consider a list $i \consl s$,
  for some $i \oftype I$, $s \oftype \ListTy{I}$, and
  assume inductively that the finite composition $k_{s}$ is a prenucleus.
  We need to show that $\Comp{k_{s}}{k_i}$ is a prenucleus.
  For meet preservation, let $U, V \oftype \opens{X}$.
  \begin{align*}
    (\Comp{k_{s}}{k_i})(U \meet V)
    &\quad\DefnEqSym\quad k_{s}(k_i(U \meet V))                                \\
    &\quad\IdTySym\quad k_{s}(k_i(U) \meet k_i(V))
    && [\text{$k_i$ preserves meets}]                                          \\
    &\quad\IdTySym\quad k_{s}(k_i(U)) \meet k_{s}(k_i(V))
    && [\text{inductive hypothesis}]                                           \\
    &\quad\DefnEqSym\quad (k_{s} \CompSym k_i)(U) \meet (k_{s} \CompSym k_i)(V)
  \end{align*}
  For inflationarity, consider some $U \oftype \opens{X}$.
  We know that both $k_i$ and $k_{s}$ are prenuclei, and thus
  that they are inflationary.
  It follows that we have
  $U \le k_i(U) \le k_{s}(k_i(U))$.
\end{proof}

\begin{lemma}\label{prop:star-ub}
  For every
  nucleus\/ $j$
  and
  every small family\/ $\FamEnum{i}{I}{k_i}$ of nuclei
  on a locale,
  if\/ $j$ is an upper bound of the family\/ $\FamEnum{i}{I}{k_i}$,
  then it is also an upper bound of
  the family\/~$\FamEnum{s}{\ListTy{I}}{k^*_{s}}$
  of its finite compositions.
\end{lemma}
\begin{proof}
  Let $j$ and $\FamEnum{i}{I}{k_i}$ be, respectively,
  a nucleus and a family of nuclei on a locale,
  and suppose that $j$ is an upper bound of $\FamEnum{i}{I}{k_i}$.
  We show by list induction that $j$ is an upper bound of the family of finite
  compositions of $\FamEnum{i}{I}{k_i}$.

  \emph{Base case:} the empty list $\emptyl$.
  We have that
  $k^*_{\emptyl}(U) \DefnEqSym \FnIdSym(U) \DefnEqSym U \le j(U)$.

  \emph{Induction step:} consider a list $i \consl s$,
    for some $i \oftype I$, $s \oftype \ListTy{I}$.
    Assume inductively that $k_s \le j$.
    We need to show that~$k^*_{is} \le j$.
    We know that~$k^*_{s}$ is monotone
    since it is a prenucleus by \cref{lem:star-prenucleus} and prenuclei are
    monotone by \cref{lem:prenuclei-are-monotone}.
    The result then follows since we have:
    \begin{align*}
      k^*_{s}(k_i(U))
        \quad&\le\quad      k^*_{s}(j(U))  && [\text{monotonicity of $k^*_{s}$}]\\
        \quad&\le\quad      j(j(U))       && [\text{inductive hypothesis}]\\
        \quad&\IdTySym\quad j(U)          && [\text{idempotency of $j$}]
    \end{align*}
\end{proof}

\begin{lemma}[\AgdaLink{Locales.PatchLocale.html\#15650}]%
\label{prop:star-sc}
  For every family\/ $\FamEnum{i}{I}{k_i}$
  of \VScottContinuous{} nuclei on a locale,
  the prenucleus\/ $k^*_s$ is \VScottContinuous{}
  for every\/ $s \oftype \ListTy{I}$.
\end{lemma}
\begin{proof}
  Finite compositions of \VScottContinuous{} functions
  are \VScottContinuous{}.
\end{proof}

\begin{lemma}\label{lem:star-dir}
  For every family\/ $\FamEnum{i}{I}{k_i}$ of nuclei on a locale,
  the family of finite compositions
  is directed.
\end{lemma}
\begin{proof}
  The family $\FamEnum{s}{\ListTy{I}}{k^*_{s}}$ of finite compositions
  is always inhabited by
  the identity nucleus, given by $k^*_{\emptyl}$.
  For every pair of finite compositions $k^*_{s}$ and $k^*_t$,
  the upper bound is given by the nucleus $k^*_{s \append t}$,
  which is equal to $\Comp{k^*_{t}}{k^*_{s}}$.
  The fact that this is an upper bound of $k^*_{s}$ and $k^*_{t}$ follows from
  the monotonicity of $k^*_{t}$ and the inflationarity~of~$k^*_{s}$.
\end{proof}

\begin{lemma}[\AgdaLink{Locales.PatchLocale.html\#25387}]\label{lem:delta-gamma}
  Let\/ $j$ and\/ $\FamEnum{i}{I}{k_{i}}$ be, respectively,
  a nucleus and a family of nuclei
  on a locale\/~$X$.
  Consider the family of finite compositions over
  the family\/ $\FamEnum{i}{I}{j \meet k_i}$.
  Each finite composition\/ $(j \meet k)^*_{s}$ is a lower bound of\/
  $k_{s}$ and\/ $j$, for every\/ $s \oftype \ListTy{I}$.
\end{lemma}

\subsubsection{Construction of the joins}

Having collected some key properties of the family of finite compositions,
we are now ready to construct the join operation in the
poset of \VScottContinuous{} nuclei.
The construction here is due to \cite{mhe-properly-injective}.

\begin{theorem}[%
  \AgdaLink{Locales.PatchLocale.html\#17527} Join of \VScottContinuous{} nuclei%
]\label{defn:sc-join}
  For every small family\/ $\FamEnum{i}{I}{k_i}$ of \VScottContinuous{} nuclei
  on a spectral locale\/ $X$,
  the join can be calculated as
  \begin{equation*}
    \paren{\FrmJoin_{i \,{\oftype}\, I} k_i}(U) \quad\is\quad
      \FrmJoin_{s \,{\oftype}\, \ListTy{I}} k^*_{s}(U)\text{.}
  \end{equation*}
\end{theorem}
\begin{proof}
  Let $\FamEnum{i}{I}{k_i}$ be a small family of nuclei on a locale $X$.
  It must be checked that this is
  (1) indeed the join, and
  (2) is a \VScottContinuous{} nucleus
  i.e.\ it is inflationary, binary-meet-preserving, idempotent, and \VScottContinuous{}.

  We start by checking (2).
  The inflationarity property is direct.
  For meet preservation, consider some $U, V \oftype \opens{X}$. We have:
  \begin{align*}
    \paren{\FrmJoin_{i \oftype I} k_i}(U \meet V)
  &\quad\DefnEqSym\quad
    \FrmJoin_{s \oftype \ListTy{I}} k^*_{s}(U \meet V)\\
  &\quad\IdTySym\quad
    \FrmJoin_{s \oftype \ListTy{I}} k^*_{s}(U) \meet k^*_{s}(V)
    && [\text{Lemma~\ref{prop:star-sc}}]\\
  &\quad\IdTySym\quad
    \FrmJoin_{s,\ t \oftype \ListTy{I}} k^*_{s}(U) \meet k^*_{t}(V)
    && [\dag]\\
  &\quad\IdTySym\quad
    \paren{\FrmJoin_{s \oftype \ListTy{I}} k^*_{s}(U)}
    \meet
    \paren{\FrmJoin_{t \oftype \ListTy{I}} k^*_{t}(V)}
    && [\text{distributivity}]\\
  &\quad\DefnEqSym\quad
     \paren{\FrmJoin_{i \oftype I} k_i}(U) \meet \paren{\FrmJoin_{i \oftype I} k_i}(V)
  \end{align*}
  The step ($\dag$) above uses antisymmetry.
  The $(\le)$ direction is direct, whereas
  for the $(\ge)$ direction, we show that
  $\FrmJoin_{s \oftype \ListTy{I}} k^*_{s}(U) \meet k^*_{s}(V)$
  is an upper bound of the family
  $\FamEnum{s,\mkern+2mu t}{\ListTy{I}}{k^*_{s}(U) \meet k^*_{t}(V)}.$
  Consider arbitrary $s, t \oftype \ListTy{I}$.
  By the directedness of the family of finite compositions we know that
  there exists some
  $u \oftype \ListTy{I}$
  such that the prenucleus $k^*_{u}$ is an upper bound of the prenuclei
  $k^*_{s}$ and $k^*_{t}$.
  We then have:
  \begin{equation*}
    k^*_{s}(U) \meet k^*_{t}(V)
    \quad\le\quad
    k^*_{u}(U) \meet k^*_{u}(V)
    \quad\le\quad
    \FrmJoin_{s \oftype \ListTy{I}} k^*_{s}(U) \meet k^*_{s}(V).
  \end{equation*}
  For idempotency, let $U \oftype \opens{X}$.
  \begin{align*}
    \paren{\FrmJoin_{i} k_i}\paren{\paren{\FrmJoin_{j} k_j}(U)}
  &\quad\DefnEqSym\quad
     \FrmJoin_{s \oftype \ListTy{I}}
       k_{s}\paren{\FrmJoin_{t \oftype \ListTy{I}} k_{t}(U)}\\
  &\quad\IdTySym\quad
     \FrmJoin_{s \oftype \ListTy{I}} \FrmJoin_{t \oftype \ListTy{I}}
       k^*_{s}(k^*_{t}(U))
    &&\quad [\text{Lemma~\ref{prop:star-sc}}]\\
  &\quad\le\quad
    \FrmJoin_{s,\mkern+2mu t \oftype \ListTy{I}} k^*_{s}(k^*_{t}(U))\\
  &\quad\le\quad
    \FrmJoin_{s \oftype \ListTy{I}} k^*_{s}(U)
    &&\quad [\ddag]\\
  &\quad\DefnEqSym\quad
    \paren{\FrmJoin_{i \oftype I} k_i}(U)
  \end{align*}
  For the step (\ddag) above, it suffices to show that
  $\FrmJoin_{s \oftype \ListTy{I}} k^*_{s}(U)$ is an upper bound
  of the family
  $\FamEnum{s,\mkern+2mu t}{\ListTy{I}}{k^*_{s}(k^*_{t}(U))}.$
  The prenucleus $k^*_{t \append s}$ is an upper bound of $k^*_{s}$ and $k^*_{t}$
  (as in Lemma~\ref{lem:star-dir}).
  We have that
  \(k_{s}(k_{t}(U)) \DefnEqSym k_{t \append s}(U) \le \FrmJoin_{s \oftype
    \ListTy{I}}k_{s}(U)\).

  For \VScottContinuity{}, let $\FamEnum{n}{N}{U_n}$ be a directed family
  of opens of $X$. We have:
  \begin{align*}
    \paren{\FrmJoin_{i \oftype I} k_i}\paren{\FrmJoin_{n \oftype N} U_n}
  &\quad\DefnEqSym\quad
     \FrmJoin_{s \oftype \ListTy{I}} k^*_{s}\paren{\FrmJoin_{n \oftype N} U_n}\\
  &\quad\IdTySym\quad
     \FrmJoin_{s \oftype \ListTy{I}} \FrmJoin_{n \oftype N} k^*_{s}(U_n)
     && [\text{Lemma~\ref{prop:star-sc}}]    \\
  &\quad\IdTySym\quad
    \FrmJoin_{n \oftype N} \bigvee_{s \oftype \ListTy{I}} k^*_{s}(U_n)
    && [\text{joins commute}] \\
  &\quad\DefnEqSym\quad
    \FrmJoin_{n \oftype N} \paren{\FrmJoin_{i \oftype I} k_i}(U_n).
  \end{align*}

  We now check that the above definition of the join gives the least upper bound
  of the family $\FamEnum{i}{I}{k_i}$.
  The fact that $\FrmJoin_i k_i$
  is an upper bound of the family $\FamEnum{i}{I}{k_i}$
  is easy to verify.
  To see that it is \emph{the least} upper bound,
  consider a \VScottContinuous{} nucleus~$j$,
  and assume that it is an upper bound of $\FamEnum{i}{I}{k_i}$.
  Let $U \oftype \opens{X}$.
  We need to show that $(\FrmJoin_i k_i)(U) \le j(U)$.
  Since $j$ is an upper bound of the family $\FamEnum{i}{I}{k_i}$, it follows
  from Lemma~\ref{prop:star-ub} that is also an upper bound of the family
  $\FamEnum{s}{\ListTy{I}}{k^*_s}$ of finite compositions.
  This is to say that $k^*_{s}(U) \le j(U)$ for every
  $s \oftype \ListTy{I}$.
  It follows that
  \(%
    \paren{\FrmJoin_{i \oftype I} k_i}(U)
    \DefnEqSym
    \FrmJoin_{s \oftype \ListTy{I}}{k^*_s(U)} \le j(U)\text{.}%
  \)
\end{proof}

We have thus established that the meet-semilattice of nuclei over a spectral
locale is small-complete.
To conclude that it is a frame, it remains just to show that the distributivity
law holds.

\subsubsection{The distributivity law}

We appeal to Lemma~\ref{lem:delta-gamma} to prove the distributivity law.

\begin{lemma}[%
  \AgdaLink{Locales.PatchLocale.html\#29815} Distributivity%
]\label{prop:distributivity-of-sc-nuclei-joins}
  For every nucleus\/ $j$ and every small family\/ $\FamEnum{n}{N}{k_n}$ of
  \VScottContinuous{} nuclei, we have the equality:
  \begin{equation*}
    j \meet \paren{\FrmJoin_{n \oftype N} k_n} \quad\IdTySym\quad \FrmJoin_{n \oftype N} j \meet k_n.
  \end{equation*}
\end{lemma}
\begin{proof}
  Let $j$ and $\FamEnum{n}{N}{k_n}$ be,
  respectively,
  a nucleus and a small family of \VScottContinuous{} nuclei. To see that
  $j \meet \FrmJoin_{n \oftype N}{k_n}$ is extensionally equal to the join
  $\FrmJoin_{n \oftype N} j \meet k_n$, let $U \oftype \opens{X}$.
  \begin{align*}
    \paren{j \meet \paren{\FrmJoin_{i \oftype I} k_i}}(U)
  \quad&\DefnEqSym\quad
    j(U) \meet \paren{\FrmJoin_{i \oftype I} k_i}(U)\\
  \quad&\DefnEqSym\quad
    j(U) \meet \paren{\FrmJoin_{s \oftype \ListTy{I}} k^*_s(U)}\\
  \quad&\IdTySym\quad
    \FrmJoin_{s \oftype \ListTy{I}} j(U) \meet k^*_s(U) && [\text{distributivity}]\\
  \quad&\DefnEqSym\quad
    \FrmJoin_{s \oftype \ListTy{I}} (j \meet k^*_s)(U)\\
  \quad&\DefnEqSym\quad
    \paren{\FrmJoin_{i \oftype I} j \meet k_i}(U)     && [\text{\dag}]
  \end{align*}
  For the step labelled (\dag), we need to show that join of the family
  $s \mapsto j(U) \meet k^*_s(U)$
  is equal to the join of the family of finite compositions of the family
  \[s \mapsto (j \meet k_{(\blank)})^*_s(U)\text{.}\]
  One direction of this follows from \cref{lem:delta-gamma} whereas the other
  can be shown by straightforward induction.
\end{proof}

We have thus established:
\begin{lemma}[\AgdaLink{Locales.PatchLocale.html\#36348}]
  For every spectral locale\/ $X$,
  the poset of \VScottContinuous{} nuclei on\/ $X$ forms a frame.
\end{lemma}
\begin{proof}
  The construction of finite meets was
  given in \cref{prop:sc-nuclei-semilattice},
  and local smallness in
  \cref{cor:poset-of-scott-continuous-nuclei-is-locally-small}.
  Joins were constructed in \cref{defn:sc-join}.
  That finite meets distribute over small joins was established in
  \cref{prop:distributivity-of-sc-nuclei-joins} above.
\end{proof}

\begin{remark}
  The construction of this frame could in fact be generalized from spectral
  locales to \emph{all} locales, except the resulting
  frame of \VScottContinuous{} nuclei would not always be locally small.
  To ensure local smallness, a \emph{specified} small base seems to be
  necessary.
  Since we will be restricting attention to the case of spectral locales anyway,
  we refrain from presenting these results at the highest level of generality for
  the sake of brevity.
  The point where the assumption of spectrality becomes absolutely necessary is
  when proving that the frame of \VScottContinuous{} nuclei satisfies the
  universal property of the patch frame, which we address in the next section.
\end{remark}

\section{The patch locale}
\label{sec:patch-locale}

In the previous section, we presented the construction of the frame of
\VScottContinuous{} nuclei over a spectral locale $X$.
We now proceed to show that the locale defined by this frame
satisfies the universal property of the patch locale on $X$.

\begin{definition}[Patch locale of a spectral locale]\label{defn:patch}
  Let $X$ be a spectral locale.
  We denote by $\Patch{X}$ the locale defined by the frame of
  \VScottContinuous{} nuclei on $X$.%
  \nomenclature{\(\Patch{X}\)}{patch locale on a spectral locale \(X\)}
\end{definition}

Although our proof of the universal property will have to to wait until
\cref{sec:ump-of-patch}, we already adopt the notation of $\Patch{X}$ for the
locale defined by the frame of \VScottContinuous{} nuclei.
Moreover, we also remind the reader at this point that the patch locale
is large, locally small, and \VSmallComplete{},
like every locale that we work with in this thesis.
This is to say that, for every spectral locale $X$ in the
category~$\Loc_{\UU}$,
the locale $\Patch{X}$ is also an object falling in $\Loc_{\UU}$.

\subsection{Base construction for the patch locale}
\label{sec:patch-base}

We now construct a small base for the patch locale
using the notions of closed and open nuclei, which we first briefly recall.
Let $U$ be an open of a locale.
\begin{enumerate}
  \item The \indexedp{closed nucleus} induced by $U$ is the map $V \mapsto U \vee V$
    (\cref{example:closed-nucleus});
  \item The \indexedp{open nucleus} induced by $U$ is the map
    $V \mapsto U \Rightarrow V$
    (\cref{defn:open-nucleus}).
\end{enumerate}

\begin{lemma}\label{lem:top-nucleus-in-terms-of-closed-and-open-nuclei}
  The top nucleus on a locale\/ $X$,
  which we defined in \cref{defn:nuclei-semilattice} as
  the map\/ $U \mapsto \FrmTop{X}$,
  is equal to each of the following:
  \begin{itemize}
    \item the closed nucleus\/ $\ClosedNucleus{\FrmTop{X}}$,
    \item the open nucleus\/ $\OpenNucleus{\FrmBot{X}}$.
  \end{itemize}
\end{lemma}
\begin{proof}
  Follows directly from the identities $\FrmTop{X} \vee U \IdTySym \FrmTop{X}$
  and $(\FrmBot{X} \Rightarrow U) \IdTySym \FrmTop{X}$, which hold for every open
  $U \oftype \opens{X}$.
\end{proof}

Using the closed and open nuclei, we construct a base for $\Patch{X}$ by
considering the family
\begin{equation*}
  \ProdTy{\CompactOpens{X}}{\CompactOpens{X}} \xrightarrow{~\gamma~} \opens{\Patch{X}},
\end{equation*}
defined by
\begin{equation*}
  \gamma(K_1, K_2) \is \ClosedNucleus{K_1} \meet \OpenNucleus{K_2}.
\end{equation*}
It is easy to see that this family is small.
We now do some preparation towards showing that this forms a base.

\begin{lemma}[\AgdaLink{Locales.PatchProperties.html\#39946}]%
\label{lem:johnstones-lemma}
  For every \VScottContinuous{} nucleus\/ $j \oftype \Endomap{\opens{X}}$
  on a locale\/ $X$, we have that
  \begin{equation*}
    j
    \quad\IdTySym\quad
    \FrmJoin_{K \oftype \CompactOpens{X}} \ClosedNucleus{j(K)} \meet \OpenNucleus{K}
    \quad\IdTySym\quad
    \FrmJoin_{\stackrel{K_1,K_2 \oftype \CompactOpens{X}}{K_1 \le j(K_2)}} \ClosedNucleus{K_1} \meet \OpenNucleus{K_2}.
  \end{equation*}
\end{lemma}
\begin{proof}
  The second equality in the statement is clear, so let us show the first one.
  We use the fact, proved in \cite[Lemma II.2.7]{ptj-ss}, that for every nucleus
  $j$ on any locale,
  \begin{equation*}
    \IdTy{j}{\FrmJoin_{U \oftype \opens{X}} \ClosedNucleus{j(U)} \meet \OpenNucleus{U}}\text{.}
  \end{equation*}
  Suppose additionally that $X$ is spectral and the nucleus $j$
  is \VScottContinuous{}.
  The inequality
  $\FrmJoin_{K \oftype \CompactOpens{X}} \ClosedNucleus{j(K)} \meet \OpenNucleus{K}
  \leq \FrmJoin_{U: \opens{X}} \ClosedNucleus{j(U)} \wedge \OpenNucleus{U} \IdTySym j$
  is trivial, so let us show the reverse one.
  Let $K \oftype \CompactOpens{X}$ and notice that
  \[%
    \paren{\ClosedNucleus{j(K)} \meet \OpenNucleus{K}}(K)
    \IdTySym
    (j(K) \vee K) \meet
    (K \Rightarrow K)
    \IdTySym
    j(K)\text{.}
  \]
  Therefore
  \[%
    j(K)
    \IdTySym
    \paren{\ClosedNucleus{j(K)} \meet \OpenNucleus{K}}(K)
    \leq
    \paren{\FrmJoin_{K' \oftype \CompactOpens{X}} \ClosedNucleus{j(K')} \meet \OpenNucleus{K'}}(K)\text{,}%
  \]
  so the required relation follows from Lemma~\ref{lem:nuclei-basic-order}.
\end{proof}

\begin{lemma}[\AgdaLink{Locales.PatchProperties.html\#40534}]%
\label{lem:base-for-patch}
  For every spectral locale\/ $X$, the above family
  \begin{equation*}
    \ProdTy{\CompactOpens{X}}{\CompactOpens{X}} \xrightarrow{~\gamma~} \opens{\Patch{X}},
  \end{equation*}
  forms a small, strong base.
\end{lemma}
\begin{proof}
  Let $j$ be a \VScottContinuous{} nucleus over a locale $X$.
  For the basic covering family for $j$, we pick the family
  \begin{equation*}
    \paren{\SigmaType{K_1, K_2}{\CompactOpens{X}}{\paren{K_1 \le j(K_2)}}}
    \xrightarrow{~\Comp{\gamma}{\projISym}~}
    \opens{\Patch{X}}\text{,}
  \end{equation*}
  which is easily seen to be small.
  The fact that
  \(%
    j
    \IdTySym
    \FrmJoin \setof{\ClosedNucleus{K_1} \meet \OpenNucleus{K_2} \mid K_1 \le j(K_2)}%
  \)
  was already established in \cref{lem:johnstones-lemma}.
  These basic covering families do not seem to be directed, so we work with
  their directifications as explained in \cref{sec:directification}.
\end{proof}

\subsection{Patch is Stone}\label{sec:patch-stone}

We prove that the patch locale is Stone by showing
(1) that it is compact, and
(2) that the base constructed in \cref{lem:base-for-patch}
is a \VZeroDimensional{} one.\index{zero-dimensional locale}

\begin{lemma}[\AgdaLink{Locales.PatchProperties.html\#23745}]
  The open nucleus\/ $\OpenNucleus{U}$ is the Boolean complement of the closed
  nucleus\/~$\ClosedNucleus{U}$, for every open\/ $U$ of a spectral locale\/ $X$.
\end{lemma}
\begin{proof}
  The argument from \cite[6]{mhe-patch-short} works in our setting, once it
  is ensured that Heyting implications always exist as discussed
  in \cref{defn:heyting-implication}.
  We provide the full-proof here for the sake of self-containment.

  Let $X$ be locale and consider an open $U \oftype \opens{X}$.
  We claim that $\OpenNucleus{U}$ is both the meet-complement and the
  join-complement of $\ClosedNucleus{U}$.
  To show that it is the meet-complement, we need to show that
  $(\ClosedNucleus{U} \meet \OpenNucleus{U})(V) \IdTySym \FrmBot{}(V) \DefnEqSym V$,
  for every~$V\oftype\opens{X}$.
  Observe that
  \begin{align*}
      (\ClosedNucleus{U} \meet \OpenNucleus{U})(V)
    \quad&\DefnEqSym\quad
      \ClosedNucleus{U}(V) \meet \OpenNucleus{U}(V)\\
    \quad&\IdTySym\quad
      (U \vee V) \meet (\Impl{U}{V})\\
    \quad&\IdTySym\quad
      (U \meet (\Impl{U}{V})) \vee (V \meet (\Impl{U}{V}))\\
    \quad&\le\quad
      V \vee V\\
    \quad&\IdTySym\quad
      V\text{.}
  \end{align*}
 To show that it is the join-complement, we need to show
 \[(\OpenNucleus{U} \vee \ClosedNucleus{U})(V) \IdTySym \FrmTop{}(V) \DefnEqSym \FrmTop{X}\text{,}\]
 for every $V \oftype \opens{X}$.
 Let $V \oftype \opens{X}$.
 It follows from the construction of the join from \cref{defn:sc-join} that
 the pointwise binary join satisfies
 \begin{equation*}
   \OpenNucleus{U}(V) \vee \ClosedNucleus{U}(V)
   \le
   (\OpenNucleus{U} \vee \ClosedNucleus{U})(V)\text{.}
 \end{equation*}
 Therefore, it suffices to show
 $\FrmTop{X} \le \OpenNucleus{U}(V) \vee \ClosedNucleus{U}(V)$,
 which we derive below.
 \begin{align*}
   \FrmTop{X}
     \quad&\le\quad        \Impl{U}{(U \vee V)}\\
     \quad&\IdTySym\quad \OpenNucleus{U}(U) \meet \OpenNucleus{U}(V)\\
     \quad&\IdTySym\quad \FrmTop{X} \meet \OpenNucleus{U}(V)\\
     \quad&\le\quad \OpenNucleus{U}(V)\\
     \quad&\le\quad \OpenNucleus{U}(V) \vee \ClosedNucleus{U}(V)\text{.}
 \end{align*}
\end{proof}

\begin{corollary}\label{cor:closed-and-open-nuclei-are-clopens}
  For every compact open\/ $K \oftype \opens{X}$ of a spectral locale\/ $X$,
  the closed nucleus $\ClosedNucleus{K}$ and the open nucleus $\OpenNucleus{K}$
  are \indexedCE{clopens}{clopen} as they complement each other.
\end{corollary}

In Lemma~\ref{lem:eps-perfect}, we prove that the map whose inverse image sends
an open $U$ to the closed nucleus $\ClosedNucleus{U}$ is perfect. Before
Lemma~\ref{lem:eps-perfect}, we record an auxiliary result:

\begin{lemma}\label{lem:eps-ra-bot}
  Let $X$ be a spectral locale.
  The right adjoint $\PatchCounit_* \oftype \opens{\Patch{X}} \to \opens{X}$ to the
  closed nucleus formation operation $\ClosedNucleus{\blank}$ is
  given by $\PatchCounit_*(j) \IdTySym j(\FrmBotSym)$,
  for every \VScottContinuous{} nucleus\/ $j$ on~$X$.
\end{lemma}

\begin{lemma}[\AgdaLink{Locales.PatchProperties.html\#21412}]%
\label{lem:eps-perfect}
  The closed nucleus formation operation is a perfect frame homomorphism
  \[\ClosedNucleus{\blank} \oftype \opens{X} \to \opens{\Patch{X}}\text{.}\]
\end{lemma}
\begin{proof}
  It is easy to see that this is a frame homomorphism,
  so we show that it is perfect.
  We have to show that the right adjoint $\PatchCounit_*$ of $\ClosedNucleus{\blank}$ is
  \VScottContinuous{}.
  Let~$\FamEnum{i}{I}{k_i}$ be a directed family of \VScottContinuous{} nuclei.
  Thanks to Lemma~\ref{lem:eps-ra-bot}, it suffices to show
  \[\left(\bigvee_{i \oftype I} k_i\right)(\FrmBotSym) = \FrmJoin_{i \oftype I} \PatchCounit_*(k_i).\] By
  Lemma~\ref{lem:directed-computed-pointwise}, we have
  that $\paren{\bigvee_{i \oftype I} k_i}(\FrmBotSym) = \FrmJoin_{i \oftype I}
  k_i(\FrmBotSym)$.
\end{proof}

\begin{lemma}[%
  \AgdaLink{Locales.PatchProperties.html\#41985}%
]\label{prop:patch-is-compact}
  The locale\/ $\Patch{X}$ is compact,
  for every spectral locale\/ $X$.
\end{lemma}
\begin{proof}
  Recall that the top open
  $\FrmTopSym$ of $\Patch{X}$ is the top nucleus defined as
  \[\FrmTopSym \is U \mapsto \FrmTop{X}\text{.}\]
  Since $\PatchCounit^*$ is a frame
  homomorphism, we know that $\FrmTopSym \IdTySym \PatchCounit^*(\FrmTop{X})$,
  meaning it suffices to show
  $\PatchCounit^*(\FrmTop{X}) \WayBelow \PatchCounit^*(\FrmTop{X})$.
  By \cref{prop:perfect-resp-way-below}, it suffices to show
  $\FrmTop{X} \WayBelow \FrmTop{X}$
  which is immediate by the assumption that $X$ is spectral.
\end{proof}

\begin{lemma}[\AgdaLink{Locales.PatchProperties.html\#45117}]
  For every spectral locale\/ $X$, the locale\/ $\Patch{X}$ is Stone.
\end{lemma}
\begin{proof}
  Let $X$ be a spectral locale.
  We showed in \cref{prop:patch-is-compact} that $\Patch{X}$ is compact.
  It suffices to show that it is \VZeroDimensional{}, for which
  we show that the base constructed in \cref{lem:base-for-patch} consists
  of clopens.
  For every pair $K_1, K_2$ of compact opens of $X$,
  we know by \cref{cor:closed-and-open-nuclei-are-clopens}
  that the closed nucleus $\ClosedNucleus{K_1}$ and the open nucleus
  $\OpenNucleus{K_2}$ are both clopens.
  It follows from \cref{lem:clopen-closure} that the meet
  \(\ClosedNucleus{K_1} \meet \OpenNucleus{K_2}\)
  is clopen, so we are done.
\end{proof}

\subsection{The universal property of the patch locale}%
\label{sec:ump-of-patch}

We are finally ready to show that the locale $\Patch{X}$ satisfies the universal
property of the patch locale from \cref{defn:ump-of-patch}.

\begin{lemma}
\label{lem:adjoint-to-induced-map}
  For every spectral map\/ $f \oftype \ArrTy{X}{A}$
  from a Stone locale\ $X$ into a spectral locale\/ $A$,
  define a map\/ $\bar{f}^* \oftype \ArrTy{\opens{\Patch{A}}}{\opens{X}}$ by
  \begin{equation*}
    \bar{f}^*(j) \quad\is\quad \FrmJoin_{K \oftype \CompactOpens{A}} f^*(j(K)) \meet \neg f^*(K).
  \end{equation*}
  Then, the map\/ $\bar{f}_* \oftype \ArrTy{\opens{X}}{\opens{\Patch{A}}}$
  defined by
  \begin{equation*}
    \bar{f}_*(V) \quad\is\quad \Comp{f_*}{\Comp{\ClosedNucleus{V}}{f^*}}
  \end{equation*}
  is the right adjoint of\/ $\bar{f}^*$.
\end{lemma}
\begin{proof}
  First, note that $\Comp{f_*}{\Comp{\ClosedNucleus{V}}{f^*}}$ is indeed a
  \VScottContinuous{} nucleus,
  and both $\bar{f}^*$ and $\bar{f}_*$ are clearly monotone.
  Let us first prove the forward implication.
  Assume that for $j \oftype \opens{\Patch{A}}$ and $V \oftype \opens{X}$
  the relation $\bar f^*(j) \leq V$ holds.
  By \cref{lem:nuclei-basic-order}, in order to show that
  $j \leq \bar{f}_*(V)$,
  it suffices to show that for every compact open $K \oftype \CompactOpens{A}$,
  the inequality $j(K) \leq \bar{f}_*(V)(K)$ holds.
  Hence, let $K \oftype \CompactOpens{A}$, and note that
  \begin{equation*}
    f^*(j(K)) \meet \neg f^*(K) \leq \bar f^*(j) \leq V
  \end{equation*}
  Notice that $f^*(K)$ is clopen, by the fact that $f^*$ is a spectral map and
  Lemma~\ref{lem:compactness-implies-clopen-based-iff-zero-dimensional}.
  It is therefore complemented in the lattice $\opens{X}$, and so
  we have
  \[f^*(j(K)) \leq V \vee f^*(K) \IdTySym \ClosedNucleus{V}(f^*(K))\text{,}\]
  which by adjunction yields
  \[j(K)\leq f_*(\ClosedNucleus{V}(f^*(K))) \IdTySym \bar f_{*}(K)\text{,}\]
  as required.

  Let us now show the reverse implication.
  Let $ j \oftype \opens{\Patch{A}}$ and $V \oftype \opens{X}$
  and assume that $j \leq \bar f_*(V)$.
  Once again, by the definition of the ordering on $\opens{\Patch{A}}$, for all
  $K \oftype \CompactOpens{A}$
  we have $j(K) \leq f_*(V \vee f^*(K))$, which by adjunction
  equivalently yields $f^*(j(K))\leq V \vee f^*(K)$.
  Since $f^*(K)$ is clopen, and hence complemented in the lattice $\opens{X}$
  it follows that $f^*(j(K)) \meet \neg f^*(K) \leq V$.
  Hence, $\bar f^*(j) \leq V$.
\end{proof}

\begin{theorem}[\AgdaLink{Locales.UniversalPropertyOfPatch.html\#48926}]
\label{thm:main}
  For every spectral map\/ $f \oftype \ArrTy{X}{A}$
  from a Stone locale into a spectral locale,
  there exists a unique continuous map\/ $\bar{f} : X \to \Patch{A}$
  satisfying\/ $\PatchCounit \circ \bar{f} \IdTySym f$,
  as illustrated in the following diagram in\/ $\Spec$:
  \begin{center}
    \begin{tikzcd}
      & \Patch{A} \arrow[d, "\PatchCounit"]\\
      X \arrow[r, swap, "f"] \arrow[ur, dashed, "\bar{f}"] & A
    \end{tikzcd}
  \end{center}
\end{theorem}
\begin{proof}
 Assume that a locale map $\bar f : X \to \Patch{A}$ satisfies the condition in
 the theorem. Then, for every $j : \opens{\Patch{A}}$, we have
 \begin{align*}
   \bar{f}^*(j)
 &\quad=\quad
   \bar{ f }^* \left( \FrmJoin_{K : \CompactOpens{A}}\ClosedNucleus{j(K)} \meet \OpenNucleus{K}\right)\\
 &\hspace{6cm} [\text{Lemma~\ref{lem:johnstones-lemma}}]\\
 &\quad=\quad
   \FrmJoin_{K : \CompactOpens{A}} \bar f^*\left( \ClosedNucleus{j(K)} \meet \OpenNucleus{K}\right)\\
   &\hspace{6cm} [\text{$\bar{f}^*$ preserves small joins}]\\
 &\quad=\quad  \FrmJoin_{K : \CompactOpens{A}} \bar f^*\left( \ClosedNucleus{j(K)}\right) \wedge \neg \bar f^*\left(\ClosedNucleus{K}\right)\\
  &\hspace{6cm} [\text{$\bar f^*$ preserves meets and complements}]\\
 &\quad=\quad  \FrmJoin_{K : \CompactOpens{A}}  f^*\left( j(K)\right) \meet \neg  f^*\left( K\right)\\
   &\hspace{6cm} [\text{commutativity of the diagram}]
 \end{align*}
 and hence $\bar f$ is uniquely determined.
 If we now define a monotone map
 $\overline{f}^* \oftype \opens{\Patch{A}} \to \opens{X}$ by
 \[%
    \bar{f}^*(j) \quad\is\quad \bigvee_{K : \CompactOpens{A}} f^*(j(K)) \wedge \neg f^*(K),%
 \]
  it is easy to show it preserves the top element (namely the top nucleus with
constant value $\FrmTop{A}$) because $\FrmBot{A}$ is compact. It also preserves binary (pointwise) meets as
 \begin{align*}
   \hphantom{\IdTySym}\quad
     &\bar{f}^*(j_1) \meet \bar{f}^*(j_2)\\
   \IdTySym\quad
     &\FrmJoin_{K_1,K_2 \oftype \CompactOpens{A}}f^*\paren{j_1(K_1)} \meet \neg  f^*\paren{K_1} \meet f^*\paren{j_2(K_2)} \meet \neg  f^*\paren{K_2}\\
   &[\text{distributivity}]\\
   \IdTySym\quad
     &\FrmJoin_{K_1,K_2 : \CompactOpens{A}}
       f^*\paren{j_1(K_1) \meet j_2(K_2)} \meet \neg  f^*\paren{K_1} \meet\neg  f^*\paren{K_2}\\
   &[\text{$f^*$ preserves binary meets}]\\
   \leq\quad &\FrmJoin_{K_1,K_2 : \CompactOpens{A}}
           f^*\left( j_1(K_1\vee K_2) \meet j_2(K_1\vee K_2)\right) \meet \neg  f^*\left( K_1\right)  \meet \neg  f^*\paren{K_2}\\
     &[\text{monotonicity}]\\
  \IdTySym\quad
    &\FrmJoin_{K_1,K_2 \oftype \CompactOpens{A}}  f^*\left( (j_1\meet j_2)(K_1\vee K_2)\right) \meet \neg  f^*\left( K_1\right)  \meet \neg  f^*\paren{K_2}\\
  \IdTySym\quad &\FrmJoin_{K_1,K_2 \oftype \CompactOpens{A}}  f^*\left( (j_1 \meet j_2)(K_1\vee K_2)\right) \meet \neg  f^*\left( K_1 \vee K_2\right)\\
    &[\text{De Morgan law}]\\
  \IdTySym\quad &\bar f^*(j_1 \meet j_2)\\
     & [\text{$\CompactOpens{X}$ closed under $(\blank) \vee (\blank)$}]
  \end{align*}

  Moreover, Lemma~\ref{lem:adjoint-to-induced-map} and Theorem~\ref{thm:aft}
  ensure that $\bar{f}^*$ preserves small joins and so it is a frame homomorphism.

  Let us finally show that $\bar{f}$ makes the diagram commute.
  Since compact opens form a small base of $A$, it suffices to show that
  $\bar{f}^*(\ClosedNucleus{K}) \IdTySym f^*(K)$ for every
  $K \oftype \CompactOpens{A}$.
  Let $K \oftype \CompactOpens{A}$ and note that
  \begin{align*}
    \bar f^*(\ClosedNucleus{K})
  &\quad\IdTySym\quad
     \FrmJoin_{K' \oftype \CompactOpens{A}}
       f^*\left(K \vee K'\right) \meet \neg  f^*\left( K'\right)\\
  &\quad\IdTySym\quad
     \FrmJoin_{K' \oftype \CompactOpens{A}}
       f^*\left( K\right) \meet \neg  f^*\left( K'\right)  && [\text{$f^*$ preserves binary joins}]\\
  &\quad\IdTySym\quad
     f^*(K) \meet \FrmJoin_{K' \oftype \CompactOpens{A}}  \neg  f^*\paren{K'} && [\text{distributivity}]\\
  &\quad\IdTySym\quad
     f^*(K) \meet \FrmTop{X} && [\FrmBot{X}\ \text{is compact}]\\
 \end{align*}
 as required.
\end{proof}

\section{Examples of patch}
\label{sec:examples-of-patch}

To make things a bit more concrete, we now look at some examples of the patch
transformation.

\subsubsection{Patch of the terminal locale}

\begin{example}
  The patch locale $\Patch{\LocTerm{\UU}}$ of the \indexedp{terminal locale}
  is homeomorphic to the terminal locale itself,
  since $\LocTerm{\UU}$ is \indexedp{Stone}.
\end{example}

\subsubsection{Patch of \VSierpinski{}}

\begin{example}\label{example:patch-of-sierpinski}
  The locale\/ $\Patch{\LocSierp_{\UU}}$
  is homeomorphic the \indexedp{discrete locale}
  on type\/~$\TwoTy{\UU}$.
\end{example}
\begin{proof}
  It was given in \cref{example:m2} that the frame $\Pow{\UU}{\TwoTySym}$ is the
  frame of ideals over the lattice $\mathbf{M}_2$.
  It is therefore sufficient by duality (\cref{thm:stone-duality})
  to show that the lattice
  $\CompactOpens{\Patch{\LocSierp_{\UU}}}$ is isomorphic to
  $\mathbf{M}_2$.
  From the base construction for $\Patch{\LocSierp_{\UU}}$,
  and the fact that that the compact opens coincide with the basic opens
  in spectral locales (\cref{prop:base-img-equiv}),
  it can be seen that $\CompactOpens{\Patch{\LocSierp_{\UU}}}$ is
  the following distributive lattice:
  
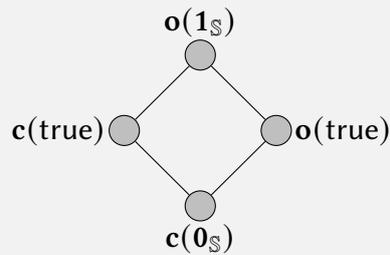
\begin{center}
    \begin{tikzpicture}
      \node[circle, draw, fill=lightgray] (a) at (0,0)  {};
      \node[circle, draw, fill=lightgray] (b) at (1,1)  {};
      \node[circle, draw, fill=lightgray] (c) at (-1,1) {};
      \node[circle, draw, fill=lightgray] (d) at (0,2)  {};

      \node[below=0.1cm] at (a) {$\ClosedNucleus{\FrmBotSym_{\LocSierp}}$};
      \node[left=0.1cm]  at (c) {$\ClosedNucleus{\mathsf{true}}$};
      \node[right=0.1cm] at (b) {$\OpenNucleus{\mathsf{true}}$};
      \node[above=0.1cm] at (d) {$\OpenNucleus{\FrmTopSym_{\LocSierp}}$};

      \draw (a) -- (b);
      \draw (a) -- (c);
      \draw (b) -- (d);
      \draw (c) -- (d);
    \end{tikzpicture}
  \end{center}
  where $\mathsf{true} \oftype \opens{\LocSierp}$ is the nontrivial open
  of \VSierpinski{} constructed in \cref{prop:sierpinski-universal-property}.
  This is clearly the same lattice as $\mathbf{M}_2$ from \cref{example:m2}.
\end{proof}


\chapter{Topology of Scott domains}
\label{chap:scott}

In this chapter, we investigate the point-free topology of domains~\cite{clad},
with a particular focus on the class of domains known as
\indexedCE{Scott domains}{Scott domain}.
Our motivation for this is twofold:
\begin{enumerate}
  \item\label{item:slsd}
    Scott domains provide a rich source of examples of spectral locales
    through their \indexedCE{Scott topologies}{Scott topology}, which are
    always spectral.
  \item The patch topology on the Scott topology of a Scott domain is a
    particularly illustrative example of the patch construction, giving a
    \emph{computational intuition} on the patch topology through its
    relationship to the so-called
    \indexed{Lawson topology} in domain theory~\cite[Definition~III-1.5]{clad}.
\end{enumerate}
In addition to these, whether the topology of domains can be carried out
predicatively is also a natural question in light of the recent developments of
\VTDJMHE{}~\cite{tdj-scott-model,tdj-thesis}, showing that many nontrivial
constructions of domain theory can be developed in constructive and predicative
\VUF{}.

The organization of this chapter is as follows:
\begin{description}[leftmargin=!,labelwidth=\widthof{Section 7.77:}]
  \item[\cref{sec:domain-theory-primer}:]
    We provide a brief summary of \VTDJMHE{}'s constructive and predicative
    development of domain theory in \VUF{}.
  \item[\cref{sec:scott-locale}:] We define the \indexedp{Scott topology} of a
    \VDCPO{} and show that it is a large, locally small, and small-complete
    frame when the \VDCPO{} in consideration is algebraic.
  \item[\cref{sec:sharp-elements}:] We summarize \TDJLastName{}'s notion of
    \indexed{sharp element} in the context of Scott domains, and we then give a
    topological characterization of it in terms of the Scott topology of the Scott
    domain in consideration.
  \item[\cref{sec:spectral-points}:] We present and discuss the notion of
    \indexed{spectral point}
    in the context of the Scott topology.
  \item[\cref{sec:sharp-elements-and-spectral-points}:]
    We show that the spectral points of the Scott locale of a Scott domain~$D$
    are in bijective correspondence with the sharp elements of $D$.
  \item[\cref{sec:points-of-patch}:] Extending the result from
    \cref{sec:sharp-elements-and-spectral-points},
    we show that the sharp elements of a Scott domain $D$ are in bijective
    correspondence with the points of the patch locale of the Scott locale of
    $D$.
    This is because the points of the patch of a spectral locale $X$ are exactly
    the spectral points of $X$.
\end{description}

\section{Primer on domain theory in \VUFAbbr{}}
\label{sec:domain-theory-primer}

As the work we present in this chapter builds on
\VTDJMHE{}'s~\cite{tdj-scott-model,tdj-mhe-cad}
development of constructive and predicative domain theory in \VUF{},
we provide a brief summary of their work.

\subsection{Definition of \VDCPO{}}

In \cref{defn:frame}, we gave a definition of the notion of frame with all
universes generalized.
After this, we explained \VTDJMHE{}'s result (\cref{thm:tdj-taboo}) that
complete and small frames cannot be constructed without a form of propositional
resizing,
and we accordingly adopted the
convention of working \emph{only} with large, locally small, and small-complete
frames over a fixed base universe.
Since exactly the same situation applies to \VDCPO{}s,
we omit their definition with generalized universes altogether, and directly
provide the definition for the well-behaved case of large, locally small, and
directed small-complete \VDCPO{}s.

\begin{definition}[\emph{cf.}\ {\cite[Definition~3.2.7]{tdj-thesis}}]%
\label{defn:dcpo}
  A \VDCPO{} (over base universe \VUni{\UU}) consists of
  \begin{itemize}
    \item a type $|D| \oftype \USucc{\UU}$,
    \item a partial order
      $(\blank) \sqsubseteq (\blank) \oftype |D| \to |D| \to \hprop{\UU}$,
    \item an operation $\DcpoJoin (\blank) \oftype \DirFam{\UU}{D} \to D$,
      giving the least upper bound $\DcpoJoin_{i \oftype I} x_i$ of every directed
      small family $\FamEnum{i}{I}{x_i}$ of elements of $D$.
  \end{itemize}
\end{definition}

As in the case of frames, the sethood of the carrier type
follows automatically from the partial order thanks to
\cref{lem:antisymmetry-gives-sethood}, and it hence does not have to be
explicitly stipulated.

We adopt the convention of using the distinct order symbol $(\sqsubseteq)$ to
differentiate the underlying poset of a \VDCPO{} than that of a frame.
We sometimes use the term \indexed{information order} to refer to the
order of a \VDCPO{}.
Furthermore, we adopt the usual abuse of notation of denoting the carrier set
$|D|$ by $D$.

\subsection{Pointed \VDCPO{}s}

\begin{definition}[Pointed \VDCPO{}~{\cite[Definition~3.2.11]{tdj-thesis}}]
  A \VDCPO{} is called
  \defineCE{pointed}{pointed \VDCPO{}}
  if it has a bottom element.
\end{definition}

We denote the bottom element of a pointed \VDCPO{} $D$ by $\DcpoBot{D}$, often
dropping the subscript and writing simply $\DcpoBotSym$.
\nomenclature{\(\DcpoBot{D}\)}{bottom element of \VDCPO{} $D$}

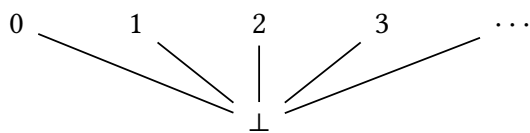
\begin{figure}[h!]
  \centering
  \begin{tikzcd}
    0 \ar[drr,dash] & 1 \ar[dr,dash] & 2 \ar[d,dash] & 3 \ar[dl,dash] & \cdots \ar[dll,dash] \\
    & & \DcpoBotSym
  \end{tikzcd}
  \caption{The \textbf{classical} picture of the flat domain of natural numbers.}
  \label{fig:flat-domain-of-nat}
\end{figure}

The archetypical example of a pointed \VDCPO{} is the
\indexed{flat domain} of natural numbers,
the classical picture of which is given in the Hasse diagram in
\cref{fig:flat-domain-of-nat}.
There is a bottom element $\DcpoBotSym$, and all other
elements besides $\DcpoBotSym$ are incomparable.
It is helpful to think of the elements of this \VDCPO{} as possibly undefined
computations.
The bottom element represents a computation failing to produce a value
e.g.\ due to nontermination.
An element above $\DcpoBotSym$ is said to be \emph{defined}, and it is thought
of as a computation successfully terminating with
a value in $\NatTy$.

More generally, the flat \VDCPO{} obtained by adjoining a bottom element to a
set~$X$ is called the \indexed{lifting of the set\/ $X$}.
The construction of the lifting of a set $X$ as $\SumTy{X}{\UnitTySym}$,
pictured in \cref{fig:flat-domain-of-nat},
cannot be shown to be directed-complete
unless the \VLPO{} holds~\cite[Proposition~3.4.1]{tdj-thesis}.
\VMHECMK{}~\cite{mhe-cmk-partial} gave a construction of the lifting \VDCPO{}
that lends itself to a constructive (and predicative) development:

\begin{example}[{\emph{cf.}\ \cite[Proposition~3.4.14]{tdj-thesis}}]
  The lifting of a set $X \oftype \UU$ is the type
  \begin{equation*}
    \Lift{\UU}{X} \quad\is\quad \SigmaType{P}{\hprop{\UU}}{(\ArrTy{P}{X})}\text{,}
  \end{equation*}
  equipped with the partial order
  \(%
    \Pair{P}{\varphi} \sqsubseteq \Pair{Q}{\psi}
      \is
        \SigmaType{f}{\Impl{P}{Q}}{(\ExtEq{{\varphi}}{\Comp{\psi}{f}})}\text{.}%
  \)
  This forms a pointed \VDCPO{}.
  The bottom element is the false proposition~$\FalseProp{\UU}$ with the
  unique function $\FalsePropSym \to X$.

  For each inhabitant $x \oftype X$, its lifting into $\Lift{\UU}{X}$ is
  given by the tuple
  \((\TruePropSym, \TTUnit \mapsto x)\).
  This defines a map, denoted $\eta_X \oftype \ArrTy{X}{\Lift{\UU}{X}}$.
\end{example}

The flat domain of natural numbers can then be simply defined as:

\begin{example}[Flat domain of $\NatTy$]
  The \emph{flat domain of natural numbers}\index{flat domain} is
  the \VDCPO{}~$\Lift{\UZero}{\NatTy}$.
\end{example}

\begin{example}\label{example:sierpinski-dcpo}
  The flat domain $\Lift{\UU}{\UnitTySym}$
  is isomorphic to $\hprop{\UU}$
  considered as a \VDCPO{}.
  It is easy to see that the carrier type is just
  $\hprop{\UU}$ since $\Equiv{(\ArrTy{P}{\UnitTySym})}{\UnitTySym}$, for
  every proposition~$P \oftype \hprop{\UU}$.
  Moreover, since every pair of functions
  $\varphi, \vartheta \oftype \ArrTy{P}{\UnitTy}$
  are equal to each other,
  the order $(P, \varphi) \sqsubseteq (Q, \psi)$ amounts to $\Impl{P}{Q}$.
\end{example}

\subsection{Finite elements of domains}
\label{sec:finite-elements}

\VTTheWayBelowRelation{} from \cref{defn:way-below} can be transferred directly
to the context of \VDCPO{}s.
In fact, this relation was first formulated in the context of domains and was
subsequently adapted to point-free topology
(see \cite[76]{clad} for an encyclopedic history).

\begin{definition}[{\cite[Definition~4.2.1]{tdj-thesis}}]
  An element $x$ of a \VDCPO{} is said to be \defineCE{\VWayBelow{}}{way below}
  an element $y$ if
  for every small, directed family $\FamEnum{i}{I}{z_i}$ of elements
  with~$y \sqsubseteq \DcpoJoin_{i \oftype I} z_i$, there is some $i \oftype I$
  such that $x \sqsubseteq z_i$.
\end{definition}

Similar to the notion of \indexedp{compact open} in a frame, which is defined as
an open that is \VWayBelow{} itself, we have a notion of compact element in a
\VDCPO{}.
We will follow the domain-theoretic convention of using the term
\indexed{finite element} in the context of domains.
The formulation of the notion of finite element is exactly the same as
Condition~(\ref{itm:compact-dirfam})
from \cref{prop:compactness}:

\begin{definition}[{\cite[Definition~4.2.3]{tdj-thesis}}]
  An element $x$ of a \VDCPO{} is called
  \indexedCE{finite}{finite element} if
  it is \VWayBelow{} itself.
\end{definition}

We write
\(\FiniteElements{D}\)
\nomenclature{\(\FiniteElements{D}\)}{type of finite elements of \VDCPO{} \(D\)}
for the type of finite elements of a \VDCPO{}
\(D\), overloading the notation we use for locales.
This slight abuse of notation is justified by the fact that an open \(U\) of
a locale \(X\) is \indexedp{compact} if and only if it is a finite element of
the underlying \VDCPO{} of the frame~\(\opens{X}\).

It is important to observe here that the characterization of compactness in
frames from \cref{prop:compactness} does not apply to \VDCPO{}s.
Some of the equivalent formulations from \cref{prop:compactness} rely on the
existence of all small joins even to be formulated, and even in the presence of
all joins, some of the equivalences rely crucially on the frame
\indexedp{distributivity law}.

\begin{lemma}[{\cite[Example~4.2.4]{tdj-thesis}}]\label{lem:bottom-is-finite}
  For every pointed \VDCPO{}\/ $D$, the bottom element\/ $\DcpoBot{D}$ is
  finite.
\end{lemma}
\begin{proof}
  The argument from \cref{lem:compact-fin-join-closure} works here as well.
\end{proof}

\begin{example}[\emph{cf.}\ {\cite[Example~4.2.6]{tdj-thesis}}]
  Every element $\eta(x)$ is finite in the lifting \VDCPO{} $\Lift{\UU}{X}$
  over some small set $X$.
\end{example}

The class of
\indexedCE{continuous \VDCPO{}s}{continuous \VDCPO{}}~\cite[Theorem~I-4.17]{clad}
is of special interest in domain theory.
The so\nobreakdash-called \indexedp{continuity condition} on a \VDCPO{} $D$
captures the idea that every element $x \oftype D$ is determined by the elements
that are finite relative to $x$.
The class of \indexedCE{algebraic \VDCPO{}s}{algebraic \VDCPO{}}
are a subclass of continuous \VDCPO{}s where the continuity condition is
stricter.
The \indexedp{algebraicity condition} states that every computation in a
\VDCPO{}~$D$ is expressible as the join of the (globally) finite computations
below it.

For the purposes of our work, we will only be interested in algebraic domains,
and particularly in the further special case of
\indexedpCE{Scott domains}{Scott domain}.
We thus refrain from presenting continuous \VDCPO{}s here and restrict our
attention to algebraic ones.
The reader interested in the details of the predicative treatment of continuous
\VDCPO{}s is referred
to \cite[\textsection 4.4]{tdj-thesis}.

\subsection{Algebraic \VDCPO{}s}

The definition of algebraic \VDCPO{} that we provide below differs from
those in~\cite{tdj-thesis,tdj-scott-model,tdj-mhe-cad}.
We work with an equivalent reformulation of \VTDJMHE{}'s definition, our
motivation for which is twofold:
\begin{enumerate}
  \item to remain close to our approach to spectral locales,
  \item for technical convenience whilst constructing the Scott locale
    of a Scott domain in~\cref{sec:scott-locale}.
\end{enumerate}

%
In more recent work~\cite[\textsection 5.3]{tdj-mhe-cad},
\VTDJMHE{} discuss the fact that
the split support result we gave in \cref{lem:split-support} applies also to the
case of algebraic \VDCPO{}s
(formalized in \VTypeTopology{} by~\textcite{type-topology-compact-basis}).
In fact, \VTDJMHE{} prove a lemma~\cite[Lemma~5.23]{tdj-mhe-cad} showing that a
formulation analogous to our
Condition~(\ref{it:V}) from \cref{thm:spec-characterization}
is possible through this split support result, but they do not give an explicit
definition based on this.
We turn \VTDJMHE{}'s Lemma~5.23 from \textit{op.~cit.}\ into an explicit
definition here:

\begin{definition}[Algebraic \VDCPO{}]\label{defn:algebraic-dcpo}
  A \VDCPO{} $D$ (over base universe \VUni{\UU})
  is called \define{algebraic \VDCPO{}} if it satisfies the
  following two conditions:
  \begin{enumerate}[label={$\left({\mathsf{AD{\arabic*}}}\right)$}, leftmargin=4em]
    \item\label{item:ad-covering}
      For every $x \oftype D$, the family
      \(\paren{\SigmaType{c}{\CompactOpens{D}}{c \sqsubseteq x}} \xhookrightarrow{~\projISym~} D\)
      is directed and gives~$x$ as its join.
    \item\label{item:ad-smallness}
      The type $\CompactElements{D}$ is \VUniSmall{\UU}.
  \end{enumerate}
\end{definition}

Condition~\ref{item:ad-covering} above says that the type $\CompactElements{D}$
forms a basis (with certain kinds of basic covering families), which we could
make precise by formulating an auxiliary notion of basis for \VDCPO{}s.
However, as our aim here is merely to summarize predicative domain theory to the
extent that we need it, we refrain from addressing such questions in detail.
\cref{defn:algebraic-dcpo} above
is sufficient for the purposes of our work.

\begin{remark}
  The above definition amounts to
  Definition~4.8.1 from \TDJLastName{}'s thesis~\cite{tdj-thesis},
  combined with the assertion that the basic family in consideration is the
  inclusion
  \[\CompactElements{D} \xhookrightarrow{~\projISym~} D\text{.}\]
  Note, however, that \VTDJLastName{}~\cite{tdj-thesis} uses a more general
  definition of \VDCPO{} which permits the underlying poset not to be locally
  small.
  Accordingly, the definition from \emph{op.~cit.}\ also stipulates local
  smallness for basic elements
  (Condition~(iii) of \cite[Definition~4.8.1]{tdj-thesis}).
  We do not have such a condition here as that is built into \cref{defn:dcpo}.
\end{remark}

\begin{example}[\emph{cf.}\ {\cite[Example~4.9.4]{tdj-thesis}}]%
\label{example:lifting-is-algebraic}
  The lifting \VDCPO{} \(\Lift{\UU}{X}\) is algebraic,
  for every set \(X \oftype \UU\).
\end{example}

Before we give the proof of this example, we prove some auxiliary lemmas.

\begin{lemma}[{\textit{cf.}\ \cite[Example~4.9.2]{tdj-thesis}}]\label{lem:lifting-basis}
  For every small set\/ $X$, every element of~$\Lift{\UU}{X}$ is decomposable
  into a directed join
  of elements in the family
  \(\beta \oftype \ArrTy{\SumTy{\UnitTySym}{X}}{\Lift{\UU}{X}}\text{,}\)
  defined as
  \begin{align*}
    \beta(\SumInlSym(\TTUnit)) &\is \DcpoBotSym\text{,}\\
    \beta(\SumInrSym(x))       &\is \eta(x)\text{.}
  \end{align*}
\end{lemma}

\begin{lemma}\label{lem:lifting-base-equivalence}
  The map\/
  $\beta \oftype \ArrTy{\SumTy{\UnitTySym}{X}}{\CompactOpens{\Lift{\UU}{X}}}$
  from \cref{lem:lifting-basis} is an equivalence, meaning we have\/
  \(\Equiv{\SumTy{\UnitTySym}{X}}{\CompactElements{\Lift{\UU}{X}}}\text{.}\)
\end{lemma}

\begin{corollary}\label{cor:finite-elements-form-a-small-type}
  The type\/ $\CompactElements{\Lift{\UU}{X}}$ of finite elements of the lifting
  is small,
  for every small set\/ $X$.
\end{corollary}

\begin{proof}[Proof of \cref{example:lifting-is-algebraic}]
  Let $X$ be a small set.
  \begin{description}
    \item[\ref{item:ad-covering}]
      Follows directly from \cref{lem:lifting-basis}.
    \item[\ref{item:ad-smallness}]
      Given in \cref{cor:finite-elements-form-a-small-type}.
  \end{description}
\end{proof}

\subsection{Scott domains}
\label{sec:scott-domains}

In the classical development of domain theory, a \emph{Scott domain} is defined
as an algebraic pointed \VDCPO{} that is \VBoundedComplete{}.
We now develop the notion of Scott domain in the context of \VTDJMHE{}'s
constructive and predicative development.
To ensure that our definition of Scott domain is amenable to constructive
reasoning, we include an additional \emph{decidability condition} that holds in
all examples of interest.

For every pair of
elements $x, y \oftype D$ of a domain $D$,
we define the proposition
\begin{equation*}
  \Bounded{x}{y} \quad\is\quad \ExistsType{u}{D}{(\ProdTy{x \sqsubseteq u}{y \sqsubseteq u})}\text{,}
  \nomenclature{\(\Bounded{x}{y}\)}{\(x\) and \(y\) are bounded above}
\end{equation*}
expressing that elements $x$ and $y$ are
\define{bounded above} (or \define{\VUpperBounded{}}).

\begin{definition}[%
  \AgdaLink{DomainTheory.BasesAndContinuity.ScottDomain.html\#2457}
  Scott domain%
]\label{defn:scott-domain}
  A \define{Scott domain} is a pointed algebraic \VDCPO{}~$D$ that satisfies
  the following two conditions:
  \begin{enumerate}[label={$\left({\mathsf{SD{\arabic*}}}\right)$}, leftmargin=4em]
    \item\label{item:bc} \emph{\VBoundedCompleteness{}}: every small family
      that has an upper bound, also has a least upper bound,
    \item\label{item:dec-ub}
      \emph{decidability of \VUpperBoundedness{} for compact elements}: the
      type $\Bounded{c}{d}$ is decidable, for every pair of compact elements
      $c, d \oftype \CompactElements{D}$.
  \end{enumerate}
\end{definition}

\begin{lemma}[\AgdaLink{Locales.ScottLocale.Properties.html\#4791}]%
\label{lem:not-bounded}
  For every pair of elements\/ $x$ and\/ $y$ of a Scott domain\/ $D$,
  if there does not exist an upper bound for $x$ and $y$,
  then we have \(\upset x \meet \upset y \IdTySym \emptyset\).
\end{lemma}
\begin{proof}
  Let $x, y \oftype D$ be a pair of elements of some Scott domain $D$
  and assume that they are not bounded above.
  Consider some $z \oftype D $ with $z \in \upset x \cap \upset y$.
  This is to say $x \sqsubseteq z$ and $y \sqsubseteq z$, meaning $z$ is an
  upper bound of $x$ and $y$.
  This contradicts the assumption that $x$ and $y$ are not bounded above.
\end{proof}

We will see that, in quite a few examples, these conditions satisfy the
decidability condition by virtue of being trivially true.

\begin{lemma}\label{lem:small-complete-implies-scott-domain}
  Every small-complete poset automatically satisfies the
  conditions~\ref{item:bc} and \ref{item:dec-ub}.
\end{lemma}
\begin{proof}
  Let $L$ be a small-complete lattice.
  Condition~\ref{item:bc} holds trivially in the presence of least upper bounds
  for all small families.
  Furthermore, small-completeness implies that every pair of elements has a
  least upper bound, meaning for every pair of elements $c$ and $d$
  the proposition $\Bounded{c}{d}$ is trivially true and hence decidable.
\end{proof}

\subsubsection{Examples of Scott domains}

\begin{example}
  The lifting $\Lift{\UU}{\UnitTySym}$ is a Scott domain
  since we know by \cref{example:sierpinski-dcpo} that $\Lift{\UU}{\UnitTySym}$
  is the initial frame $\hprop{\UU}$ qua \VDCPO{}, and is thus small-complete.
  \cref{lem:small-complete-implies-scott-domain} then implies that it is
  a Scott domain.
\end{example}

\begin{example}
  The domain $\PowSym(\NatTy)$ is a Scott domain, as we know that it
  is small-complete by \cref{ex:powerset-frame}.
\end{example}

\section{Scott locale of an algebraic \VDCPO{}}
\label{sec:scott-locale}

Having summarized the preliminaries of constructive and predicative domain
theory in \VUF{}, we now begin to develop the topology of domains.
We start by defining the \define{Scott topology} over a \VDCPO{}.

\begin{definition}[%
  \AgdaLink{DomainTheory.Topology.ScottTopology.html\#2199} \VScottOpen{}%
]\label{defn:scott-open}
  Let $D$ be \VDCPO{} (over base universe \VUni{\UU})
  and consider a
  subset~$S \oftype \ArrTy{D}{\hprop{\UU}}$.
  The subset $S$ is called \define{\VScottOpen{}} if
  \begin{enumerate}[label={$\left({\mathsf{SO{\arabic*}}}\right)$}, leftmargin=4em]
    \item\label{item:so-uc}
      it is \indexedp{\VUpwardClosedNH{}}, and
    \item\label{item:so-ia}
      it is \indexedp{inaccessible} by small, directed families, that is,
      \begin{equation*}
        \DcpoJoin_{i \,{\oftype}\, I} x_i \in
          S \ImplSym \ExistsType{i}{I}{x_i \in S}\text{,}
      \end{equation*}
      for every small and directed family $\FamEnum{i}{I}{x_i}$ of elements
      of $D$.
  \end{enumerate}
\end{definition}

\begin{remark}
  It is possible to generalize the above definition as to define
  the \VScottOpen{} \VUniHyphen{\VV}subsets of a \VUniHyphen{\UU}\VDCPO{}.
  However, this generalization is not necessary for our purposes, and nor is its
  mathematical meaning clear.
  The reader interested in the generalization of universes is referred to the
  \VAgda{} formalization where some of the results we present here are carried
  out at a higher level of universe generality.
\end{remark}

As explained by
\VMBSLastName{}~\cite[\textsection 2.2]{smyth-topology},
it is often useful to think of the \VScottOpen{} subsets of a \VDCPO{} $D$ as
the \emph{finitely verifiable} properties
of the elements of $D$, which can be thought of as representing computations.
Under this interpretation,
\begin{itemize}
  \item Condition~\ref{item:so-uc} says that,
    once a computation $x$ has been verified to satisfy a finitely verifiable
    property $U$,
    then every further refinement of the computation~$x$ will also satisfy
    the property $U$.
  \item Condition~\ref{item:so-ia} says that,
    if the limit of a sequence of increasingly more refined computations
    satisfies a finitely verifiable property~$U$, then
    there must be some computation in the sequence that \emph{already} satisfies
    the property~$U$.
    In other words, it should always be possible to judge that the limit
    computation satisfies the property on the basis of its finite approximations.
\end{itemize}

\subsection{Frame structure}

We now show that the poset of \VScottOpen{}s has finite meets and small joins.

\begin{lemma}\label{lem:scott-locale-finite-meets}
  \VScottOpen{} subsets are closed under finite meets.
  In other words, the full subset\/ $x \mapsto \TruePropSym$
  is \VScottOpen{}
  and the intersection of two \VScottOpen{} subsets
  is \VScottOpen{}.
\end{lemma}
\begin{proof}
  Trivial and well-known fact of domain theory.
  The proof is unchanged in our setting.
\end{proof}

\begin{lemma}\label{lem:scott-frame-small-complete}
  The union of a small family of \VScottOpen{} subsets is \VScottOpen{}.
\end{lemma}
\begin{proof}
  Let $\FamEnum{i}{I}{U_i}$ be a small family of \VScottOpen{} subsets
  over some \VDCPO{} $D$.
  \begin{description}
    \item[\ref{item:so-uc}:] Let $x \in {\bigcup_{i \oftype I} U_i}$ and consider
      some $y \sqsupseteq x$.
      By the definition of subset union, there is some $U_i$ with $x \in U_i$.
      It follows from the \VUpwardClosure{} of~$U_i$ that $y \in U_i$,
      and we thus have $y \in \bigcup_{i \oftype I} U_i$.
    \item[\ref{item:so-ia}:]
      Let $\FamEnum{j}{J}{x_j}$ be a directed family of elements of $D$ such that
      \[\paren{\DcpoJoin_{j \oftype J} x_j} \in {\bigcup_{i \oftype I} U_i}\text{.}\]
      We know that there is some $U_i$ with
      \(\paren{\DcpoJoin_{j \oftype J} x_j} \in U_i\),
      and it
      thus follows from the inaccessibility of $U_i$ by directed joins
      that there is some \(x_j \in U_i\).
      This implies that \(x_j \in \bigcup_{i \oftype I} U_i\) by definition,
      which was what we needed.
  \end{description}
\end{proof}

We have thus established that the poset of \VScottOpen{} subsets over every
\VDCPO{} has finite meets and is small-complete.

\subsection{Local smallness}

In \cref{ex:powerset-frame}, we defined the discrete locale over a set $X$
and discussed the issue that it is not locally small unless the set $X$ in
consideration is small.
We formally showed this with the resizing taboo we gave in
\cref{cor:local-smallness-resizing-taboo},
establishing that the local smallness of the powerset frame does not hold in
general without a form of propositional resizing.

A similar situation is present in
the poset of \VScottOpen{} subsets over a \VDCPO{}.
Although this poset
is small-complete,
it does not seem to be locally small in general.
It \emph{is}, however, locally small
when the \VDCPO{} in consideration is algebraic.

\begin{lemma}[%
  \AgdaLink{Locales.ScottLocale.ScottLocalesOfAlgebraicDcpos.html\#4316}%
]\label{lem:scott-frame-local-smallness}
  For every \VDCPO{}\/ $D$ (over some base universe\/ \VUni{\UU}),
  if\/ $D$ is algebraic
  then the inclusion order
  has\/ \VUniHyphen{\UU}small truth values.
\end{lemma}
\begin{proof}
  Let $D$ be an algebraic \VDCPO{} and
  consider two \VScottOpen{}s \(U, V \oftype \ArrTy{D}{\hprop{\UU}}\).
  We define the following small version of the inclusion order:
  \begin{equation*}
    U \subseteq_s V \quad\is\quad \PiTy{c}{\CompactOpens{D}}{c \in U \ImplSym c \in V}\text{.}
  \end{equation*}
  We need to show that $U \subseteq_s V$ implies $U \subseteq V$.
  Suppose $U \subseteq_s V$ and let $x \in U$.
  Since the \VDCPO{} $D$ is algebraic, we know that
  \(\IdTy{x}{\DcpoJoin \setof{ c \oftype \CompactElements{D} \mid c \sqsubseteq x }}\).
  The fact that
  \(\paren{\DcpoJoin \setof{ c \oftype \CompactElements{D} \mid c \sqsubseteq x }} \in U\)
  implies, thanks to~\ref{item:so-ia}, that
  there is some \(c \in U\) with~\(c \sqsubseteq x\).
  We thus have \(c \in V\) by our assumption of \(S \subseteq_s V\),
  which implies $x \in V$
  by the \VUpwardClosure{} of \VScottOpen{} subsets.
\end{proof}

\begin{lemma}[%
  \AgdaLink{Locales.ScottLocale.ScottLocalesOfAlgebraicDcpos.html\#6567}%
]\label{lem:scott-opens-form-frame}
  For every algebraic \VDCPO{}\/ $D$,
  the poset of \VScottOpen{} subsets of\/ $D$ forms a frame.
\end{lemma}
\begin{proof}
  The existence of finite meets and small joins
  was established in
  \cref{lem:scott-locale-finite-meets,lem:scott-frame-small-complete}.
  We have just shown local smallness in
  \cref{lem:scott-frame-local-smallness} above.
  The frame distributivity law holds automatically since this is a subframe of
  the powerset frame.
\end{proof}

\begin{definition}[%
  \AgdaLink{Locales.ScottLocale.ScottLocalesOfAlgebraicDcpos.html\#6992}%
]
  The frame constructed in the above lemma is called
  the \define{frame of \VScottOpen{} subsets} of an algebraic \VDCPO{} $D$
  and is denoted $\ScottFrame{D}$.
  \nomenclature{\(\ScottFrame{D}\)}{%
    the frame of \VScottOpen{} subsets of algebraic \VDCPO{} $D$%
  }
  We refer to the locale that this defines as the \define{Scott locale} of an
  algebraic \VDCPO{} $D$
  and denote it $\ScottLocale{D}$.
  \nomenclature{\(\ScottLocale{D}\)}{Scott locale of an algebraic \VDCPO{} $D$}
\end{definition}

\subsection{Elementary properties of the Scott locale}

In this section, we collect some key properties of the Scott locale.

\begin{lemma}\label{lem:upset-of-bottom-is-the-top-scott-open}
  For every pointed \VDCPO{}\/ $D$, the principal filter\/ $\upset \DcpoBot{D}$
  is the top element of the frame\/ $\ScottFrame{D}$.
\end{lemma}
\begin{proof}
  It is easy to see that every element of $D$ falls in $\upset \DcpoBot{D}$
  by virtue of $\DcpoBot{D}$ being the bottom element.
\end{proof}

\begin{lemma}[%
  \AgdaLink{Locales.ScottLocale.Properties.html\#3029}%
]\label{lem:contains-bot-implies-contains-top}
  For every \VScottOpen{} subset\/ $U \oftype \ArrTy{D}{\hprop{\UU}}$
  of a pointed \VDCPO{}\/ $D$,
  the subset\/ $U$ contains the bottom element\/~$\DcpoBot{D}$
  if and only if
  it is the top element of the frame\/ $\ScottFrame{D}$.
\end{lemma}
\begin{proof}
  For every pointed \VDCPO{} $D$
  and
  \VScottOpen{} subset $U$,
  it is easy to see that,
  if $U$ contains $\DcpoBot{D}$
  then $U$ contains every element of $x$ by \VUpwardClosure{}.
\end{proof}

\begin{lemma}[%
  \AgdaLink{DomainTheory.Topology.ScottTopologyProperties.html\#2483}%
]\label{lem:upset-of-finite-element-iff-scott-continuous}
  For every element\/ $x \oftype D$ of an algebraic \VDCPO{}\/ $D$,\/
  the element\/ $x$ is
  finite\index{finite|see {finite element}}
  if and only if
  the principal filter\/ $\upset x$ is \VScottOpen{}.\index{principal filter}
\end{lemma}
\begin{proof}
  Let $x \oftype D$ be an element of a \VDCPO{} $D$.
  It is easy to see that $\upset x$ being
  inaccessible by directed joins is exactly the finiteness of $x$,
  since it says,
  for every directed family $\FamEnum{i}{I}{y_i}$
  with $\DcpoJoin_{i \oftype I} y_i \in \upset x$,
  there is some $i \oftype I$ with $y_i \in \upset x$.
  The ($\Leftarrow$) direction immediately follows from this.
  For the ($\Rightarrow$) direction, observe that $\upset x$ always satisfies
  Condition~\ref{item:so-uc} by construction.
\end{proof}

\begin{lemma}[%
  \AgdaLink{DomainTheory.Topology.ScottTopologyProperties.html\#4881}%
]\label{lem:scott-open-decomposition}
  Every \VScottOpen{} subset\/ $U \oftype \ArrTy{D}{\hprop{\UU}}$,
  of an algebraic \VDCPO{} D,
  is expressible as the union of the principal filters
  generated by the finite elements that it contains.
\end{lemma}
\begin{proof}
  Let $D$ be an algebraic \VDCPO{}, and consider a \VScottOpen{} subset
  \(U \oftype \ArrTy{D}{\hprop{\UU}}\).
  We claim that
  \begin{equation*}
    \IdTySpaces{%
      U%
    }{%
      \bigcup \setof{ \upset c \mid c \in U, c\ \text{finite}}%
    }\!\text{.}
  \end{equation*}
  It is easy to see that the ($\supseteq$) direction holds
  since the \VUpwardClosure{} of \VScottOpen{} subsets implies that
  $\upset x \subseteq U$, for every $x \in U$.

  For the ($\subseteq$) direction, let $x \in U$.
  We need to show that there is a finite $c \in U$ with $x \in \upset c$.
  As the \VDCPO{} in consideration is algebraic, we know that
  \[%
    \IdTy{%
      x%
    }{%
      \DcpoJoin \setof{c \oftype \CompactElements{D} \mid c \sqsubseteq x}%
    }
    \!\text{.}%
  \]
  The inaccessibility of $U$ by directed joins then implies that
  there is some finite $c \in U$
  such that~$c \sqsubseteq x$, which is to say $x \in \upset c$.
\end{proof}

\begin{lemma}[%
  \AgdaLink{Locales.ScottLocale.Properties.html\#3576}%
]\label{lem:principal-filter-of-finite-element-is-compact}
  For every finite element\/ $c$ of a \VDCPO{},
  the principal filter\/ $\upset c$ is a compact open of
  the frame\/ $\ScottFrame{D}$.
\end{lemma}
\begin{proof}
  Let $D$ be a \VDCPO{} and consider a finite element $c \oftype D$.
  To see that the principal filter $\upset c$ is compact,
  let $\FamEnum{i}{I}{U_i}$ be a directed family of \VScottOpen{} subsets such
  that~$\upset c \subseteq \bigcup_{i \oftype I} U_i$.
  As $c \in \upset c$, we have that $c \in \bigcup_{i \oftype I} U_i$
  i.e.\ $c \in U_i$ for some $i \oftype I$.
  It follows from the \VUpwardClosure{} of $U_i$ that $\upset c \subseteq U_i$,
  which was what we needed.
\end{proof}

When the \VDCPO{} in consideration is pointed,
the principal filter $\upset \DcpoBotSym$ gives the top \VScottOpen{} as we
explained in \cref{lem:upset-of-bottom-is-the-top-scott-open}. Combining this
with the above lemma, we obtain:

\begin{lemma}[\AgdaLink{Locales.ScottLocale.Properties.html\#4323}]
  For every pointed \VDCPO{}\/ $D$, the locale\/ $\ScottLocale{D}$ is compact.
\end{lemma}
\begin{proof}
  By \cref{lem:upset-of-bottom-is-the-top-scott-open},
  the top open of $\ScottLocale{D}$ is $\upset \DcpoBot{D}$,
  which is compact by \cref{lem:principal-filter-of-finite-element-is-compact}
  since $\DcpoBot{D}$ is always a finite element (\cref{lem:bottom-is-finite}).
\end{proof}

\subsection{Base construction for the Scott locale over a Scott domain}

We showed in \cref{lem:scott-open-decomposition} that every \VScottOpen{} subset
can be decomposed as the union of the principal filters it contains.
We now extend this result and show that the principal filters on the compact
elements in fact form a basis for the Scott locale.

\cref{lem:scott-open-decomposition} suggests that the base of
$\ScottLocale{D}$ should be given by the family
\[\upset(\blank) \oftype \ArrTy{\CompactElements{D}}{\ScottFrame{D}}\]
of principal filters on finite elements of $D$.
Our primary motivation for this base construction is to establish that
\(\ScottLocale{D}\)
is a spectral locale for every Scott domain~\(D\),
and it turns out to be more convenient to work with all finite unions of such
principal filters for this purpose.

\begin{lemma}[\AgdaLink{Locales.ScottLocale.ScottLocalesOfScottDomains.html\#20038}]%
\label{lem:scott-locale-base}
  For every Scott domain\/ $D$,
  the family\/
  \(\gamma \oftype \ArrTy{\ListTy{\CompactElements{D}}}{D}\),
  defined by
  \begin{equation*}
    \gamma(c_0 \consl \ldots \consl c_{n-1})
      \quad\is\quad
        \upset c_0 \cup \cdots \cup \upset c_{n-1}\text{,}
  \end{equation*}
  forms a strong, directed base for the locale\/ $\ScottLocale{D}$.
\end{lemma}
\begin{proof}
  Let $D$ be a Scott domain
  and
  consider a \VScottOpen{} subset $U \oftype \ArrTy{D}{\hprop{\UU}}$.
  As the basic covering family for $U$, we pick the basic covering
  index family
  \begin{equation*}
    \setof{%
      (c_0 \consl \ldots \consl c_{n-1}) \oftype \ListTy{\CompactElements{D}}
      \mid
      \PiTy{i}{\NatTy}{0 \le i < n \ImplSym c_i \in U}%
    }\text{,}
  \end{equation*}
  of all lists of compact elements contained in $U$.
  Formally, this gives the basic covering:
  \begin{equation*}
    \paren{%
    \SigmaType{%
    s%
    }{%
      \ListTy{\CompactElements{D}}%
    }{%
      \PiTy{c}{\CompactElements{D}}{c \in s \ImplSym \upset c \in U}%
    }%
    }
    \xhookrightarrow{~\Comp{\gamma}{\projISym}~}
    \opens{\ScottLocale{D}}
  \end{equation*}
  It easily follows from \cref{lem:scott-open-decomposition} that
  this family gives $U$ as its join.

  It is also easy to see that this family is directed.
  It is always inhabited by the empty list $\emptyl$ which
  gives $\gamma(\emptyl) \IdTySym \emptyset$. Furthermore, given a pair
  of lists
  $s \is c_0 \consl \ldots \consl c_{m-1}$ and $t \is d_0 \consl \ldots \consl d_{n-1}$
  of finite elements of $D$,
  the least upper bound of the basic \VScottOpen{}s $\gamma(s)$ and $\gamma(t)$
  is in the family since it is give by $\gamma(s \append l)$.
\end{proof}

Observe that the base constructed above is small, since
(1) the type $\CompactElements{D}$ is a small type, and
(2) the type of lists over a small type is small.

By appealing to \cref{lem:cmp-bsc} on the above base, we obtain:

\begin{corollary}\label{cor:scott-open-decomposition}
  Every compact \VScottOpen{}\/ $K$ is equal to a union\/
  $\upset c_0 \cup \cdots \upset c_{n-1}$
  for some list\/ $c_0 \consl \ldots c_{n-1}$ of finite elements of\/ $D$,
  for every Scott domain $D$.
\end{corollary}

\subsection{Spectrality of the Scott locale over a Scott domain}
\label{sec:spec-scott-locale-over-scott-domain}

As we have already constructed a base for the Scott locale, we can easily
establish its spectrality by showing that this base is spectral as in
\cref{defn:int-spec-base}.

\begin{lemma}[%
  \AgdaLink{Locales.ScottLocale.ScottLocalesOfScottDomains.html\#22026}%
]\label{lem:scott-locale-base-consists-of-compact-opens}
  The base constructed in \cref{lem:scott-locale-base} consists of compact
  opens.
  In other words,
  for every list\/ $c_0 \consl\ldots\consl c_{n-1}$ of finite elements of
  an algebraic \VDCPO{}\/ \(D\),
  the \VScottOpen{} subset given by
  the union $\upset c_0 \cup \cdots \cup \upset c_{n-1}$
  is compact.
\end{lemma}
\begin{proof}
  We know by \cref{lem:compact-fin-join-closure} that compact opens are
  closed under finite joins.
\end{proof}

We now would like to show that the base
\(\gamma \oftype \ArrTy{\ListTy{\CompactElements{D}}}{\ScottFrame{D}}\)
is closed under binary intersections, for which it is crucial that the \VDCPO{}
in consideration is a Scott domain.

\begin{lemma}\label{lem:principal-filter-reverses-joins-and-antitone}
  For every Scott domain $D$, the principal filter map
  \[\upset(\blank) \oftype \ArrTy{\CompactElements{D}}{\opens{\ScottLocale{D}}}\]
  is (1) antitone, and (2) reverses joins, to the extent that they exist.
\end{lemma}

\begin{lemma}[\AgdaLink{Locales.ScottLocale.ScottLocalesOfScottDomains.html\#18262}]%
\label{lem:scott-base-coherence}
  For every Scott domain\/ $D$, the base for $\ScottLocale{D}$, constructed
  in \cref{lem:scott-locale-base}, is closed under binary intersections.
\end{lemma}
\begin{proof}[Proof sketch]
  Let $D$ be a Scott domain,
  and consider two lists $c_0 \consl \ldots \consl c_{m-1}$
  and $d_0 \consl \ldots \consl d_{m-1}$ of finite elements.
  We need to show that
  $\gamma(c_0 \consl \ldots \consl c_{m-1}) \cap \gamma(d_0 \consl \ldots \consl d_{n-1})$
  falls in the image of $\gamma$.
  It can be shown by straightforward induction that
  \begin{equation*}
    \gamma(c_0 \consl \ldots \consl c_{m-1}) \cap \gamma(d_0 \consl \ldots \consl d_{n-1})
    \IdTySym
    \gamma(c_0 \consl \ldots \consl c_{m-1} \consl d_0 \consl \ldots \consl d_{n-1})\text{.}
  \end{equation*}
\end{proof}

\begin{theorem}[\AgdaLink{Locales.ScottLocale.ScottLocalesOfScottDomains.html\#24945}]%
\label{thm:scott-is-spectral}
  For every Scott domain\/ $D$, its Scott locale\/ $\ScottLocale{D}$ is
  spectral.
\end{theorem}
\begin{proof}
  Let $D$ be a Scott domain.
  \begin{description}[leftmargin=!,labelwidth=\widthof{(SPB..1)}]
    \item[\ref{item:spec-base-consists-of-compact-opens}]
      Given in \cref{lem:scott-locale-base-consists-of-compact-opens}.
    \item[\ref{item:spec-base-contains-top}]
      We know that the top \VScottOpen{} is $\upset \DcpoBot{D}$.
      It is thus easy to see that the base contains it, since $\DcpoBot{D}$ is
      always compact and thus
      $\gamma(\DcpoBotSym) \IdTySym \upset \DcpoBotSym \IdTySym \FrmTop{\ScottFrame{D}}$.
    \item[\ref{item:spec-base-coherence}]
      Given in \cref{lem:scott-base-coherence}.
  \end{description}
\end{proof}

\subsection{The universal property of \VSierpinski{}}
\label{sec:universal-property-of-sierpinski}

In \cref{sec:sierpinski}, we defined the \indexedp{\VSierpinski{} locale} and postponed the
proof of its universal property.
Now that we have the necessary topological framework for
\indexedpCE{algebraic \VDCPO{}s}{algebraic \VDCPO{}}
in place,
we shall now revisit this question and prove
the universal property of the \indexedp{\VSierpinski{} locale}.
This also serves as an illustrative example of the technique of presenting a
locale through a \VDCPO{}.

Recall that in \cref{defn:sierpinski} we defined the
\indexedp{\VSierpinski{} locale} \(\LocSierp_{\UU}\)
as the locale given by the frame of \VScottContinuous{} endomaps on the
\indexedp{initial frame} \(\hprop{\UU}\).
An alternative description of this frame is as the Scott locale of the lifting
\VDCPO{} \(\Lift{\UU}{\UnitTySym}\).

\begin{lemma}\label{lem:sierpinski-alternative-construction}
  The frame\/ \(\ScottFrame{\Lift{\UU}{\UnitTySym}}\) is isomorphic
  to the frame of \VScottContinuous{} maps on
  the initial frame\/ \(\hprop{\UU}\).
\end{lemma}
\begin{proof}
  The type \(\Lift{\UU}{\UnitTySym}\) is defined as
  \(\SigmaType{P}{\hprop{\UU}}{\ArrTy{P}{\UnitTySym}}\),
  which is clearly equivalent to~\(\hprop{\UU}\).
  We now show that a function \(\Endomap{\hprop{\UU}}\) is \VScottContinuous{}
  if and only if it is \VScottOpen{} (in the sense of \cref{defn:scott-open}).
  Let \(S \oftype \ArrTy{\hprop{\UU}}{\hprop{\UU}}\)
  and let \(\FamEnum{i}{I}{P_i}\) be a directed family of propositions.

  (\(\Rightarrow\)) Suppose \(S\) is \VScottContinuous{}.
  It is a well-known fact of domain theory that \VScottContinuous{} maps are
  monotone, which is exactly the \VUpwardClosure{} of \(S\),
  viewed as a subset of~\(\hprop{\UU}\).
  To see that \(S\) is inaccessible by directed joins,
  suppose \(S\left( \FrmJoin_{i \oftypes I} P_i \right)\) holds.
  This implies, by the \VScottContinuity{} of \(S\),
  that \(\FrmJoin_{i \oftypes I} S(P_i)\) holds, which says exactly that
  some \(P_k\) holds.

  (\(\Leftarrow\)) Conversely, suppose \(S\) is a \VScottOpen{} of
  \(\hprop{\UU}\) qua \VDCPO{}.
  We need to show that
  \[S\left(\FrmJoin_{i \oftypes I}{P_i}\right) \IdTySym \FrmJoin_{i \oftypes I} S(P_i)\text{.}\]
  The \VUpwardClosure{} of \(S\) is exactly the fact that it is monotone,
  which immediately gives
  \(\FrmJoin_{i \oftypes I} S(P_i) \le S\left(\FrmJoin_{i \oftypes I}{P_i}\right)\).
  For the other direction, suppose we know
  \(S\left(\FrmJoin_{i \,\oftype\, I}{P_i}\right)\text{.}\)
  By the inaccessibility of \(S\) by directed joins, we have that there exists
  some \(k \oftype I\) such that \(S(P_k)\),
  which is to say that we have \(\FrmJoin_{i \oftypes I} S(P_i)\).

  Finally, it is verified easily that the orders of the two frames are the same,
  from which we conclude that the two are isomorphic frames.
\end{proof}

Recall that we denote by \(\SierpTruth\) the singleton open of the
\VSierpinski{} locale, which is simply the identity map
\(\SierpTruth(P) \is P\),
asserting a given proposition to be true.

\begin{lemma}
  The map\/ \(\SierpTruth\) is the principal filter\/ \(\upset \TruePropSym\).
\end{lemma}
\begin{proof}
  It is an easy observation that a proposition \(P\) holds if and only if
  it is below the true proposition \(\TruePropSym\).
\end{proof}

Since principal filters on finite elements of \VDCPO{}s are compact opens of the
Scott locale
(as shown in \cref{lem:principal-filter-of-finite-element-is-compact}),
it follows that:

\begin{corollary}\label{cor:truth-is-compact}
  The singleton open\/ \(\SierpTruth\) is compact.
\end{corollary}

\begin{lemma}\label{lem:sierp-base-trichotomy}
  An open of \VSierpinski{} is compact if and only if it is either
  the top open\/~\(\FrmTop{\LocSierp}\),
  the singleton open\/ \(\SierpTruth\), or
  the bottom open\/ \(\FrmBot{\LocSierp}\).
\end{lemma}
\begin{proof}
  We know that all three of these are compact:
  the bottom open \(\FrmBot{\LocSierp}\) is compact in every locale,
  the fact that \(\FrmTop{\LocSierp}\) is compact was established in
  in \cref{example:sierpinski-is-compact}, and
  we showed in \cref{cor:truth-is-compact} that \(\SierpTruth\) is compact.
  We now show that \emph{every} compact open of the \VSierpinski{} locale is
  one of the above.
  Firstly, we already know by \cref{prop:base-img-equiv} that the compact opens
  are exactly the opens falling in the base.
  It thus suffices to show that every basic open is one of the above three
  opens.
  Recall the base construction for the \indexedp{Scott locale}
  from \cref{lem:scott-locale-base}:
  the base of \(\ScottLocale{\Lift{\UU}{\UnitTySym}}\) consists of finite unions
  of principal filters on the basic elements of the \VDCPO{} in consideration.
  We know that \(\Lift{\UU}{\UnitTySym}\) has exactly two finite elements:
  \(\TrueProp{\UU}\) and \(\FalseProp{\UU}\).
  We now proceed by induction to show that each finite union of principal filters
  on these basic elements is one of the above three opens.
  \begin{itemize}
    \item Base case: the empty list. The empty union is \(\FrmBot{\LocSierp}\).
    \item Inductive step, case (1): \(\upset \TruePropSym \cup \upset c_{1} \cup \cdots \cup \upset c_{n-1}\).
      The principal filter \(\upset \TruePropSym\) is exactly the open
      \(\SierpTruth\), as established in \cref{cor:truth-is-compact}.
      By the inductive hypothesis, we know that \(\upset c_{1} \cup \cdots \cup \upset c_{n-1}\)
      is one of the three opens of interest.
      If it is \(\FrmBot{\LocSierp}\), the union is
      \(\SierpTruth \cup \FrmBot{\LocSierp} \IdTySym \SierpTruth\).
      If it is \(\FrmTop{\LocSierp}\), the union is simply
      \(\SierpTruth \cup \FrmTop{\LocSierp} \IdTySym \FrmTop{\LocSierp}\).
      If it is \(\SierpTruth\), the union is
      \(\SierpTruth \cup \SierpTruth \IdTySym \SierpTruth\).
    \item Inductive step, case (2):
      \(\upset \FalsePropSym \cup \upset c_1 \cup \cdots \cup \upset c_{n-1}\).
      We know by \cref{lem:upset-of-bottom-is-the-top-scott-open}
      that~\(\upset \FalsePropSym \IdTySym \FrmTop{\LocSierp}\),
      so the union is simply \(\FrmTop{\LocSierp}\).
  \end{itemize}
\end{proof}

We are now ready to give the proof of \cref{prop:sierpinski-universal-property},
in which we stated that our construction of the \VSierpinski{} locale has the
universal property of \VSierpinski{} (from~\cref{defn:sierp-universal-property}).

\begin{proof}[Proof of \cref{prop:sierpinski-universal-property}]
  Let \(X\) be a locale and let \(U \oftype \opens{X}\).
  We need to show that there is
  a continuous map \(f \oftype \ArrTy{X}{\LocSierp}\)
  unique with the property that \(\IdTy{U}{f^*(\SierpTruth)}\).

  We define this map as follows: first, we define a family
  \begin{equation*}
    \alpha
     \oftype
      \PiTy{S}{\opens{\LocSierp}}{\ArrTy{\SumTy{S(\TruePropSym)}{S(\FalsePropSym)}}{\opens{X}}}\text{,}\\
  \end{equation*}
  which gives \(U\) if \(\TruePropSym \in S\) and \(\FrmTop{X}\) if \(\FalsePropSym \in S\).
  Formally, this is encoded as the function:
  \begin{align*}
    \alpha_S(\SumInlSym(p)) &\is U\\
    \alpha_S(\SumInrSym(q)) &\is \FrmTop{X}\text{.}
  \end{align*}
  We then define the desired map \(f \oftype \ArrTy{X}{\LocSierp}\) as
  \(f(S) \is \FrmJoin \setof{ \alpha_S(p) \mid p \oftypes \SumTy{S(\TruePropSym)}{S(\FalsePropSym)} }\).

  The fact that \(f\) satisfies the desired property of
  \(\IdTy{U}{f^*(\SierpTruth)}\) is obvious:
  \(\SierpTruth(\TruePropSym)\) always holds, so we have
  \(\alpha_{\SierpTruth}(\SumInlSym(\TTUnit)) \IdTySym U\).

  Moreover, it is easy to see that \(f^*\) preserves the top element of
  \(\opens{\LocSierp}\) as well as the joins. We focus on the preservation
  of binary meets:
  let \(S, T \oftype \Endomap{\hprop{\UU}}\) be two \VScottContinuous{} maps.
  We need to show that \(f^*(S \cap T) \IdTySym f^*(S) \meet f^*(T)\).
  The \(f^*(S \cap T) \le f^*(S) \meet f^*(T)\) direction is trivial by monotonicity
  so we focus on the \(f^*(S) \meet f^*(T) \le f^*(S \cap T)\) direction.
  First, observe that we have the following equality thanks to distributivity:
  \begin{align*}
    &\qquad\; f^*(S) \meet f^*(T)\\
    &\DefnEqSym\quad
    \left(\FrmJoin \setof{ \alpha_S(p) \mid p \oftypes \SumTy{S(\TruePropSym)}{S(\FalsePropSym)}}\right)
    \meet
    \left(\FrmJoin \setof{ \alpha_S(p) \mid p \oftypes \SumTy{T(\TruePropSym)}{T(\FalsePropSym)}}\right)\\
    &\IdTySym\quad
      \FrmJoin
        \setof{
          \alpha_{S}(p) \meet \alpha_{T}(q) \mid p \oftypes \SumTy{S(\TruePropSym)}{S(\FalsePropSym)},\,
                                   q \oftypes \SumTy{T(\TruePropSym)}{T(\FalsePropSym)}
        } \tag{\dag}\label{eqn:sierp-1}
  \end{align*}
  It suffices to show that this is below \(f(S \cap T)\), which expands to
  \begin{equation}
    \FrmJoin \setof{
      \alpha_{(S \cap T)}(p) \mid p \oftypes \SumTy{(S \cap T)(\TruePropSym)}{(S \cap T)(\FalsePropSym)}
    } \tag{\ddag}\label{eqn:sierp-2}
  \end{equation}
  We show that each element of the family in (\ref{eqn:sierp-1}) is below some
  element of the family in~(\ref{eqn:sierp-2}).
  Let
  \(p \oftype \SumTy{S(\TruePropSym)}{S(\FalsePropSym)}\)
  and
  \(q \oftype \SumTy{T(\TruePropSym)}{T(\FalsePropSym)}\).
  We proceed by case analysis on the proofs \(p\) and \(q\).
  \begin{itemize}
    \item Case: \(p \IdTySym \SumInlSym(p')\) and \(\IdTy{q}{\SumInlSym(q')}\).
      This means that have \(\TruePropSym \in S\) and \(\TruePropSym \in T\),
      which implies
      \(\alpha_S(\SumInlSym(p')) \meet \alpha_S(\SumInlSym(q'))
        \IdTySym
        U \meet U
        \IdTySym
        U\text{.}
      \)
      In this context, it is easy to show that \(U\) is below
      (\ref{eqn:sierp-2}) since \(r \is \Pair{p'}{q'}\) constitutes proof that
      \(\TruePropSym \in S \cap T\), meaning we have
      \(U \le \alpha_{(S \cap T)}(\SumInlSym(r)) \DefnEqSym U\).
    \item Case: \(p \IdTySym \SumInlSym(p')\) and \(\IdTy{q}{\SumInrSym(q')}\).
      This means that we have \(\TruePropSym \in S\) and \(\FalsePropSym \in T\),
      which implies
      \(
        \alpha_S(\SumInlSym(p')) \meet \alpha_T(\SumInrSym(q'))
        \IdTySym
        U \meet \FrmTop{X}
        \IdTySym
        U\text{.}
      \)
      By \cref{lem:contains-bot-implies-contains-top}, we know that a
      \VScottOpen{} subset containing \(\FalsePropSym\) must also contain
      \(\TruePropSym\).
      Pairing this with \(p' \oftype S(\TruePropSym)\),
      we obtain a proof \(r \oftype \TruePropSym \in S \cap T\), and thus
      we can easily conclude
      \(U \le \alpha_{(S \cap T)}(\SumInlSym(r)) \DefnEqSym U\).
    \item Case: \(p \IdTySym \SumInrSym(p')\) and \(\IdTy{q}{\SumInlSym(q')}\).
      The same argument as in the previous case applies here as well.
    \item Case: \(p \IdTySym \SumInrSym(p')\) and \(\IdTy{q}{\SumInrSym(q')}\).
      This means that we have
      \(\FalsePropSym \in S\) and \(\FalsePropSym \in T\),
      which implies
      \(
        \alpha_S(\SumInrSym(p')) \meet \alpha_T(\SumInrSym(q'))
        \IdTySym
        \FrmTop{X} \meet \FrmTop{X}
        \IdTySym
        \FrmTop{X}\text{.}
      \)
      Observe that \(\Pair{p'}{q'}\) constitutes a proof that
      \(\FalsePropSym \in S \cap T\).
      The desired result follows from this since we have
      \(\FrmTop{X} \le \alpha_{(S \cap T)}(\SumInrSym(p', q')) \DefnEqSym \FrmTop{X}\).
  \end{itemize}

  We now proceed to show that the continuous map \(f\) is unique.
  Let \(g \oftype \ArrTy{X}{\LocSierp}\) be some other continuous map satisfying
  the property that \(\IdTy{g^*(\SierpTruth)}{U}\).
  We show that \(\IdTy{f^*(S)}{g^*(S)}\), for every~\(S \oftype \opens{\LocSierp}\).
  The \(f^*(S) \le g^*(S)\) direction is easy to see, so we focus on
  the \(g^*(S) \le f^*(S)\) direction.
  We established in \cref{lem:sierpinski-alternative-construction}
  that the \VSierpinski{} locale is \(\ScottLocale{\Lift{\UU}{\UnitTySym}}\),
  which we know to be always spectral when the \VDCPO{} in consideration is
  a Scott domain.
  Thanks to this, we know that \(S\) can be decomposed as
  \(\IdTy{S}{\FrmJoin_{i \oftypes I} B_i}\),
  where \(\FamEnum{i}{I}{B_i}\) is a family of compact \VScottOpen{}s.
  Because the map~\(g^*\) is a frame homomorphism and thus monotone,
  it remains to show that~\(g(B_i) \le f^*(S)\), for each~\(i \oftype I\).
  We appeal to \cref{lem:sierp-base-trichotomy}, which implies that each
  basic open \(B_i\) is either
  the top open \(\FrmTop{\LocSierp}\),
  the singleton open \(\SierpTruth\), or the bottom open~\(\FrmBot{\LocSierp}\).
  \begin{itemize}
    \item Case: \(\IdTy{B_i}{\FrmTop{\LocSierp}}\).
      This is to say that \(S \IdTySym \FrmTop{\LocSierp}\).
      The desired result follows, since we have
      \( g^*(B_i) \IdTySym \FrmTop{\LocSierp} \IdTySym f^*(\FrmTop{X}) \IdTySym f^*(S) \).
    \item Case: \(\IdTy{B_i}{\SierpTruth}\).
      We know \(g^*(\SierpTruth) \IdTySym U\), and
      we already showed that \(U \IdTySym f^*(\SierpTruth)\).
      This gives
      \(g^*(B_i) \IdTySym g^*(\SierpTruth) \IdTySym f^*(\SierpTruth) \IdTySym f^*(B_i) \le f^*(S)\).
    \item Case: \(\IdTy{B_i}{\FrmBot{\LocSierp}}\).
      Observe that we have \(g^*(\FrmBot{\LocSierp}) \IdTySym \FrmBot{X} \le f^*(S)\).
  \end{itemize}
\end{proof}

\section{Sharp elements}
\label{sec:sharp-elements}

\TDJLastName{}~\cite{tdj-sharp-elements} formulated a notion of
\indexed{sharp element} in the context of his constructive development of domain
theory.
A sharp element~\cite[Definition~5.8]{tdj-sharp-elements} of
an algebraic \VDCPO{} $D$
is an element $x \oftype D$ whose principal ideal $\downset x$ admits decidable
membership for finite elements.

Although this notion can be formulated for \VDCPO{}s in general, its description
in the context of algebraic \VDCPO{}s is much more concrete and straightforward.
As algebraic \VDCPO{}s are our primary focus here,
we will study sharp elements directly in the context of algebraic \VDCPO{}s.

\subsection{Sharp elements of algebraic \VDCPO{}s}

\begin{definition}[%
  \AgdaLink{Locales.LawsonLocale.SharpElementsCoincideWithSpectralPoints.html\#5474}
  {\emph{cf.}\ \cite[Proposition~5.11]{tdj-sharp-elements}}%
]
  An element $x \oftype D$ of an algebraic \VDCPO{} $D$ is called \define{sharp}
  if $c \sqsubseteq x$ is a decidable proposition for every \indexedp{finite element} $c \oftype D$.
  In other words, an element is sharp if it admits decidable membership in
  \indexedpCE{principal filters}{principal filter} generated by finite elements.
\end{definition}

\begin{lemma}
  For every element $x \oftype D$ of an algebraic \VDCPO{} $D$,
  the proposition expressing that $x$ is sharp
  is small.
\end{lemma}
\begin{proof}
  Directly follows from local smallness and the stipulation that
  $\CompactElements{D}$ is a small type
\end{proof}

\subsubsection{Examples of sharp elements}

\begin{example}
  The sharp elements of $\PowSym(\NatTy)$ qua \VDCPO{} are exactly the decidable
  subsets of $\NatTy$.
\end{example}

\begin{example}
  The sharp elements of the \VSierpinski{} \VDCPO{} $\Lift{\UU}{\UnitTySym}$
  are exactly
  the trivial propositions $\FalsePropSym$ and $\TruePropSym$.
\end{example}
\begin{proof}
  Let $P$ be a sharp element of $\hprop{\UU}$ qua \VDCPO{}.
  Since $\TrueProp{\UU}$ is a finite element of $\hprop{\UU}$, we have that
  $\TrueProp{\UU} \ImplSym P$ is decidable.
  If $\TrueProp{\UU} \ImplSym P$, then $\IdTy{P}{\TruePropSym}$.
  Otherwise, we have
  $\neg (\IdTy{P}{\TruePropSym})$, which means $\IdTy{P}{\FalsePropSym}$.
\end{proof}

\begin{example}
  The sharp elements of the flat domain $\Lift{\UU}{\NatTy}$ are those
  elements $(P, \varphi)$ such that $\eta(n) \sqsubseteq (P, \varphi)$ is
  decidable, for every $n \oftype \NatTy$.
\end{example}
\begin{samepage}
\begin{proof}
  We know by \cref{lem:lifting-base-equivalence} that the finite elements
  of $\Lift{}{\NatTy}$ are $\DcpoBotSym$ along with those of the form
  $\eta(n)$ for $n \oftype \NatTy$.
  This is to say sharp elements of $\Lift{}{\NatTy}$ are those elements
  $x$ such that $\bot \sqsubseteq x$ and $\eta(n) \sqsubseteq x$ are decidable.
  The former is trivially true meaning $x$ is sharp if and only if
  $\eta(n) \sqsubseteq x$ is decidable, for every $n \oftype \NatTy$.
\end{proof}
\end{samepage}

\subsection{Topological characterization of sharp elements}

We now give a new characterization of the sharp elements of
an algebraic \VDCPO{} $D$
in terms of the Scott opens of $D$.
We first observe that every element $x \oftype D$ that admits decidable
membership in compact \VScottOpen{}s is sharp:

\begin{lemma}\label{lem:decidable-membership-implies-sharpness}
  For every element\/ $x \oftype D$ of an algebraic \VDCPO{}\/ $D$,
  we have that\/ $x$ is sharp
  whenever the proposition\/ $x \in K$ is decidable for every
  compact \VScottOpen{}\/ $K \oftype \ArrTy{D}{\hprop{\UU}}$.
\end{lemma}
\begin{proof}
  Let $x \oftype D$ be an element that admits decidable membership in
  compact \VScottOpen{}s.
  Let $c$ be a finite element of $D$.
  Since we know by \cref{lem:principal-filter-of-finite-element-is-compact}
  that $\upset c$ is a compact \VScottOpen{}, the proposition
  $x \in \upset c$
  is decidable
  which is to say that $x$ is a sharp element.
\end{proof}

In fact, it turns out that the converse is also true:

\begin{lemma}[%
  \AgdaLink{Locales.LawsonLocale.SharpElementsCoincideWithSpectralPoints.html\#9416}%
]\label{lem:sharp-element-topological-characterization}
  An element $x \oftype D$ of an algebraic \VDCPO{} is sharp
  if and only if
  it admits decidable membership in compact \VScottOpen{}s.
\end{lemma}
\begin{proof}
  We established the ($\Leftarrow$) direction
  in \cref{lem:decidable-membership-implies-sharpness}.
  To see that the ($\Rightarrow$) direction also holds,
  let $x \oftype D$ be a sharp element and let
  \(K\oftype\ArrTy{D}{\hprop{\UU}}\)
  be a compact \VScottOpen{}
  In \cref{lem:scott-locale-base}, we showed that the family
  \[%
    (c_0 \consl \ldots \consl \ldots c_{n-1})
      \mapsto
        \upset c_0 \cup \cdots \cup \upset c_{n-1}%
  \]
  forms a base for the locale $\ScottLocale{D}$.
  We thus know by \cref{cor:scott-open-decomposition} that there must be a list
  $c_0 \consl \ldots \consl c_{n-1}$ such that
  $\IdTy{K}{\upset c_0 \cup \cdots \cup \upset c_{n-1}}$.
  Since the proposition $x \in \upset c_i$ is decidable for each
  $i$ with $0 \le i < n$, it follows that $x \in K$ is decidable,
  since decidable propositions are closed under finite joins.
\end{proof}

\section{Spectral points of the Scott locale}
\label{sec:spectral-points}

The notion of a space in point-set topology is defined as a set $X$ of points
equipped with a topology i.e.\ a subframe of the powerset frame $\PowSym(X)$.
In other words, the opens are defined in terms of the points.
We previously discussed that, in point-free topology, this conceptual state of
affairs is reversed: we start with the opens, and define points in their terms
when the need arises.
Up until this section, we have not had any motivation to work with the
\indexedCE{points of a locale}{point}.
We now start discussing points, as they arise naturally in our study of the
Lawson topology of a Scott domain in a constructive setting.

Before we delve into the technical details,
we provide a conceptual summary of the notion of point in the context of
point-free topology.
Consider a point $x \in X$ of a topological space $X$.
We can collect together all open neighbourhoods of $x$,
which can be thought of as the class of all
\indexedCE{observable properties}{observable property}
of $x$.
We denote this by $\NeighbourhoodFilter{x}$:
\begin{equation*}
  \NeighbourhoodFilter{x}
    \quad\is\quad
      \setof{ U \in \opens{X} \mid x \in U }\!\text{.}
\end{equation*}
When we view $\NeighbourhoodFilter{x}$ as a lattice under the inclusion order,
we can think of $\NeighbourhoodFilter{x}$ as the class of observations about $x$
that get increasingly better, eventually converging around $x$.
This satisfies three properties:
\begin{enumerate}
  \item \emph{\VUUpwardClosure{}}: for every $U \in \mathcal{F}_x$ and every
    $V \supseteq U$, we have that $V \in \mathcal{F}_x$.
  \item \emph{Closure under binary intersections}: for every pair of opens $U, V
    \in \mathcal{F}_x$, the intersection $U \cap V$ is also in $\mathcal{F}_x$.
  \item \emph{Inaccessibility by unions}: if we have
    $\paren{\bigcup_{i \oftype I} U_i} \in \mathcal{F}_x$ for a
    family $\FamEnum{i}{I}{U_i}$ of opens,
    then there is some $i \oftype I$ such that $U_i \in \mathcal{F}_x$.
\end{enumerate}
These three properties define what is called a \indexed{completely prime
filter} of the frame $\opens{X}$ of open subsets of $X$.
If there is a bijection between the set $X$ of points and the class of
completely prime filters of the frame $\opens{X}$, then the space $X$ is said to
be \indexedCE{sober}{sober!see {sober space}}.
In the point-free setting where one works with an abstract frame of opens,
one takes a completely prime filter as the definition of a point of a locale.
For further reading on points in point-free topology, we refer the reader to
\cite[\textsection II.3.1]{picado-and-pultr}.

\subsection{The notion of point in locale theory}

In the literature on locale theory, there are three common definitions of the
notion of point of a
locale $X$~\cite[\textsection II.1.3]{ptj-ss}:
\begin{enumerate}
  \item\label{item:point-cpf}
    as a completely prime filter of the frame $\opens{X}$, on which we
    elaborated above,
  \item\label{item:point-map}
    as a continuous map $\ArrTy{\LocTerm{\UU}}{X}$, and
  \item\label{item:point-pe}
    as a prime element of the frame $\opens{X}$.
\end{enumerate}
Whilst Conditions~(\ref{item:point-map}) and (\ref{item:point-cpf})
are equivalent constructively,
the equivalence of Condition~(\ref{item:point-pe}) to the other two is
fundamentally classical.
We take (\ref{item:point-map}) as our primary formulation of the notion of
point:
\begin{definition}
  A \define{point} of a locale $X$ (over base universe $\UU$)
  is a continuous map \[\ArrTy{\LocTerm{\UU}}{X}\text{,}\]
  which is the same thing as a frame homomorphism $\opens{X} \to \hprop{\UU}$.
\end{definition}

We denote the type of points of a locale by $\Point{X}$.%
\nomenclature{\(\Point{X}\)}{type of points of locale \(X\)}

\subsubsection{Points are exactly the completely prime filters}

\begin{definition}[Filter]
  A subset $F \subseteq L$ of a frame $L$ is called a \define{filter} if it is
  \begin{itemize}
    \item \VUpwardClosedNH{}, and
    \item closed under finite meets.
  \end{itemize}
\end{definition}

\begin{definition}[Completely prime filter]
  A filter $F \subseteq L$ in a frame $L$ is said to be
  \define{completely prime} or (\define{small-completely prime} for extra clarity)
  if
  for every small family~$\FamEnum{i}{I}{x_i}$
  with $(\bigvee_{i \oftype I} x_i) \in F$,
  there is some $i \oftype I$ such that $x_i \in F$.
\end{definition}

\begin{lemma}\label{lem:bijection-points-and-cpfs}
  Points of a locale\/ $X$ amount to the
  completely prime filters of\/ $\opens{X}$.
\end{lemma}
\begin{proof}
  Let $X$ be a locale over base universe \VUni{\UU}.
  For every point $p \oftype \ArrTy{\LocTerm{\UU}}{X}$, it is easy to see that
  its defining frame homomorphism $p^* \oftype \ArrTy{\opens{X}}{\hprop{\UU}}$ is
  a completely prime filter, as it preserves finite meets and small joins.
  Conversely, every completely prime filter
  $F \oftype \ArrTy{\opens{X}}{\hprop{\UU}}$
  is a frame homomorphism. Since $F$ contains the top element, we have
  $F(\FrmTop{X}) \IdTySym \TruePropSym$.
  Furthermore, the \VUpwardClosure{} of $F$ says exactly that it is monotone
  when viewed as a frame homomorphism.
  That $F$ is closed under binary meets means
  $F(U) \wedge F(U) \le F(U \wedge V)$,
  which implies by monotonicity
  that $F(U \meet V) \IdTySym F(U) \meet F(V)$.
  Finally, the complete primality condition says exactly that
  \begin{equation*}
    F\paren{\FrmJoin_{i \oftype I} U_i} \le \FrmJoin_{i \oftype I} F(U_i)\text{,}
  \end{equation*}
  from which it follows by monotonicity that
  \begin{equation*}
    F\paren{\FrmJoin_{i \oftype I} U_i} \IdTySym \FrmJoin_{i \oftype I} F(U_i)\text{.}
  \end{equation*}
\end{proof}

We will often use the letter $p$ for variables ranging over the set
$\ArrTy{\LocTermSym}{X}$
of points of a locale $X$.
In light of the equivalence above, however, we will also use the variable letter
$F$, when we are taking the view of a point simply as a subset
$F \oftype \ArrTy{\opens{X}}{\HPropSym}$.

\subsection{Spectral points}

In \cref{defn:spectral-map}, we defined a spectral map as a map that reflects
compact opens. We now look at points that are spectral when considered as
continuous maps.

\begin{definition}
  A point\index{point!spectral}
  $\ArrTy{\LocTermSym}{X}$ is called \defineCE{spectral}{spectral point}
  if it is a spectral map of locales.
\end{definition}

We denote by $\SpecPt{X}$%
\nomenclature{\(\SpecPt{X}\)}{type of spectral points of locale \(X\)}
the type of spectral points of a locale $X$.

\begin{lemma}\label{lem:spectral-points-are-decidable-for-finite-elements}
  A point\/ $p \oftype \ArrTy{\LocTerm{\UU}}{X}$
  is spectral if and only if
  the frame homomorphism\/ $p^* \oftype \ArrTy{\opens{X}}{\hprop{\UU}}$
  is a subset admitting decidable membership of compact opens of\/ $X$.
\end{lemma}
\begin{proof}
  Let $p \oftype \ArrTy{\LocTermSym}{X}$ be a point.
  Recall that the map $p$ being spectral amounts to the statement
  that~$p^*(K)$ is a compact open of the initial frame $\hprop{\UU}$
  whenever $K$ is a compact open of $X$.
  We know by \cref{example:terminal-locale-is-stone} that the terminal locale
  $\LocTerm{\UU}$ is Stone, and we thus know
  that its compact opens coincide with its clopens
  since \cref{lem:clopen-equiv-compact-in-stone-locales} says that the compact
  opens coincide with the clopens in Stone locales.
  It is therefore easy to see that $p^*$ preserves compact opens if and only if
  it maps the compact opens of $X$ to the clopens of $\LocTerm{\UU}$,
  which are exactly the decidable propositions.
\end{proof}

\section{Sharp elements and spectral points}
\label{sec:sharp-elements-and-spectral-points}

We now turn to the relationship between the two notions we have discussed so
far: sharp elements and spectral points. It turns out that the sharp elements
of a Scott domain $D$ coincide with the spectral points
$\ArrTy{\LocTermSym}{\ScottLocale{D}}$.
Before we get to the proof of this, however, we will need to do some preparation.

\subsection{Points of the Scott locale of Scott domain}
\label{sec:points-of-the-scott-locale}

We will need the following auxiliary lemma.

\begin{lemma}[\AgdaLink{Locales.LawsonLocale.CompactElementsOfPoint.html\#12213}]%
\label{lem:finite-elements-in-point-directed}
  Let\/ $D$ be a Scott domain.
  For every point\/ $F$ of\/ $\ScottLocale{D}$,
  the family
  \begin{equation*}
    \setof{ c \oftype \CompactElements{D} \mid \upset c \in F }
  \end{equation*}
  is small and directed.
\end{lemma}
\begin{proof}
  Let $D$ be a Scott domain and consider a point $F$ of $\ScottLocale{D}$.
  It is easy to see that the family
  \(\setof{ c \oftype \CompactElements{D} \mid \upset c \in F }\)
  is small, as the quantification over
  $\CompactElements{D}$ and the proposition $\upset c \in F$ are both small.

  We now show that this family is directed.
  The bottom element $\DcpoBot{D}$ of the domain is
  a finite element (\cref{lem:bottom-is-finite})
  and always inhabits the above family since we always have that
  $\IdTy{\upset\DcpoBot{D}}{\FrmTop{\ScottFrame{D}}}$,
  and we also know
  $\FrmTop{\ScottFrame{D}} \in F$ since $F$ is a filter.
  To see that this family is directed,
  let $b, c \oftype \CompactOpens{D}$ satisfying
  $\upset b \in F$ and $\upset c \in F$.

  We need to show that there is some compact $d$ that is above both $b$ and $c$
  and satisfies $\upset d \in F$.
  We appeal to our assumption that $\Bounded{b}{c}$ is a decidable proposition
  (Condition~\ref{item:dec-ub} from \cref{defn:scott-domain}) and proceed by
  case analysis.

  (\textbf{Case 1}) There exists an upper bound $u$ of $b$ and $c$.
  By \VBoundedCompleteness{},
  the least upper bound of $b$ and $c$ must then exist, for which
  we write $b \vee c$. This is finite as finite elements are always
  closed under binary joins. It remains to show that
  $\upset (b \vee c) \in F$. The principal filter operation $\upset(\blank)$
  reverses
  joins so we have $\upset (b \vee c) = \upset b \wedge \upset c$.
  It then remains to show $(\upset b \wedge \upset c) \in F$, which holds
  if and only if
  $F(\upset b) \wedge F(\upset c)$, by the fact that $F$ preserves meets.
  Notice that this means we are done because we already know that both conjuncts
  $F(\upset b)$ and $F(\upset c)$ hold.

  (\textbf{Case 2}) There does not exist an upper bound of $b$ and $c$.
  We then know by Lemma~\ref{lem:not-bounded} that
  $\upset b \wedge \upset c \IdTySym \FrmBot{\ScottFrame{D}}$.
  Because we know $\upset b \in F$ and $\upset c \in F$, it must be the case that
  $F(\upset b \wedge \upset c)$ holds since
  $F(\upset b \wedge \upset c) = F(\upset b) \wedge F(\upset c)$ by
  meet preservation.
  This means that $F(\FrmBot{\ScottFrame{D}})$ holds, which is a contradiction
  since we know that $F(\FrmBot{\ScottFrame{D}}) \IdTySym \FalsePropSym$
  as $F$ is a frame homomorphism.
\end{proof}

The idea above is that, by collecting together all the finite elements
consistent with the property expressed by the filter $F$, we obtain a directed
family that gives the specified computation in the limit.

\begin{lemma}\label{lem:equiv-of-domain-elements-and-scott-points}
  We have an equivalence\/ \(\Equiv{D}{\Point{\ScottLocale{D}}}\),
  for every Scott domain\/ $D$.
\end{lemma}
\begin{proof}
  Let $D$ be a Scott domain.
  We define maps
  \begin{center}
    \begin{tikzpicture}
      \node (A) at (0,0) {$D$};
      \node (B) at (4,0) {$\Point{\ScottLocale{D}}$};

      \draw[->, line width=0.6pt]
      ([yshift=0.1cm]A.east) to node[midway, above] {\(\PointOfSym_{(\blank)}\)} ([yshift=0.1cm]B.west);
      \draw[->, line width=0.6pt]
      ([yshift=-0.1cm]B.west) to node[midway, below] {\(\nu\)} ([yshift=-0.1cm]A.east);
    \end{tikzpicture}
  \end{center}
  as follows:
  \begin{align*}
    \PointOfSym_{x}(U) \quad&\is\quad x \in U\\
    \nu(F) \quad&\is\quad \DcpoJoin \setof{ c \oftype  \CompactElements{D} \mid \upset c \in F }\!\text{.}
  \end{align*}
  We start by checking that these functions are well-defined.
  The fact that the directed join above exists follows from the directedness of the
  family $\setof{ c \oftype  \CompactElements{D} \mid \upset c \in F }$, which was
  established in \cref{lem:finite-elements-in-point-directed}.
  The fact that $\PointOfSym_{x} \oftype \opens{\ScottLocale{D}} \to \HPropSym$
  defines a frame homomorphism for every $x \oftype D$ is easy to see.
  Let $x \oftype D$.
  It is immediate that it preserves the top
  element since $\PointOfSym_{x}(\FrmTopSym) = x \in \FrmTopSym = \top$.
  Given two \VScottOpen{} subsets~$U$ and $V$ of $D$,
  we have that $\PointOfSym_{x}(U \cap V) \DefnEqSym x \in U \cap V$
  if and only if
  $\PointOfSym_{x}(U)$ and $\PointOfSym_{x}(V)$ hold,
  which concludes that this map also preserves meets.
  It is similarly easy to see that it preserves joins.
  Let $\FamEnum{i}{I}{U_i}$ be a small family of \VScottOpen{} subsets.
  We have that
  $\PointOfSym_{x}(\FrmJoin_{i \oftype I} U_i)\DefnEqSym x \in \FrmJoin_{i \oftype I}U_i$
  if and only if $\exists i \oftype I.\ x \in U_i$
  if and only if $\exists i \oftype I.\ \PointOfSym_{x}(U_i)$,
  which is the same thing as $\FrmJoin_{i : I} \PointOfSym_{x}(U_i)$.

  We now show that $\nu$ is a retraction of $\PointOf{\blank}$ and vice versa.
  Let $x \oftype D$. We need to show that $\Comp{\nu}{\PointOfSym_{x}} \IdTySym x$.
  This follows directly from the fact that $D$ is an algebraic domain:
  \begin{align*}
    \nu(\PointOfSym_{x})
    \quad&\IdTySym\quad   \DcpoJoin \setof{ c \oftype \CompactOpens{D} \mid \upset c \in \PointOfSym_{x}}\\
    ~ \quad&\IdTySym\quad \DcpoJoin \setof{ c \oftype \CompactOpens{D} \mid x \in \upset c }\\
    ~ \quad&\IdTySym\quad \DcpoJoin \setof{ c \oftype \CompactOpens{D} \mid c \sqsubseteq x }\\
    ~ \quad&\IdTySym\quad x.
  \end{align*}

  To see that $\PointOf{\blank}$ is a retract of $\nu$,
  let $F$ be point of $\ScottLocale{D}$.
  We need to show that $\PointOfSym_{\nu(F)}(U) \IdTySym F(U)$.
  By \cref{lem:scott-open-decomposition}, it suffices to show
  \begin{equation*}
    \PointOfSym_{\nu(F)}(U) \IdTySym F\paren{\FrmJoin \setof{ \upset c \mid c \in U, c\ \text{finite} }}\text{,}
  \end{equation*}
  and because $F$ is a frame homomorphism, it remains to show
  \begin{equation*}
    \paren{ \FrmJoin \setof{ c \oftype \CompactElements{D} \mid \upset c \in F } } \in U
    \quad\BiImplSym\quad
    \FrmJoin \setof{ F(\upset c) \mid c \in U, c\ \text{finite} }\!\text{.}
  \end{equation*}
  For the ($\Leftarrow$) direction, suppose there is some finite $c \in U$ with
  $\upset c \in F$.
  Since
  \begin{equation*}
    c \sqsubseteq \paren{ \FrmJoin \left\{ c \oftype \CompactElements{D} \mid \upset c \in F \right\} }\text{,}
  \end{equation*}
  we know that
  $\left(\FrmJoin \left\{ c \oftype \CompactElements{D} \mid \upset c \in F \right\}\right) \in U$
  by the \VUpwardClosure{} of $U$.
  For the ($\Rightarrow$) direction, suppose we know
  $\left( \FrmJoin \paren{ c \oftype \CompactElements{D} \mid \upset c \in F} \right) \in U$.
  Because $U$ is inaccessible by directed joins, there must be
  some finite $c \in U$ with $\upset c \in F$,
  which means the proposition
  $\FrmJoin \left\{ F(\upset c) \mid c \in U, c\ \text{finite} \right\}$
  holds.
\end{proof}

\subsection{Spectral points of a Scott locale}

We now refine the above equivalence to an equivalence between sharp elements and
spectral points.

\begin{lemma}\label{lem:below-sharp}
  Let $D$ be a Scott domain.
  For every finite $c \oftype D$ and every spectral point
  $F \oftype \opens{\ScottLocale{D}} \to \HPropSym$,
  we have that
  \[\upset c \in F \quad\BiImplSym\quad c \sqsubseteq \nu(F)\text{.}\]
\end{lemma}
\begin{proof}
  Let $c$ be a finite element of $D$, and
  $F \oftype \opens{\ScottLocale{D}} \to \HPropSym$,
  a spectral point of $\ScottLocale{D}$.
  The ($\Leftarrow$) direction is immediate: if
  $\upset c \in F$,
  then
  \[c \sqsubseteq \nu(F) \equiv \DcpoJoin \{ b \oftype \CompactElements{D} \mid \upset b \in F \}\text{.}\]
  For the $(\Rightarrow)$ direction, suppose that
  $c \sqsubseteq \FrmJoin \{ b \oftype \CompactElements{D} \mid \upset b \in F \}$
  holds, which is to say
  \begin{equation*}
    \left(\DcpoJoin \{ b : \CompactElements{D} \mid \upset b \in F \}\right)
    \in
    \upset c
    \text{.}
  \end{equation*}
  Because $\upset c$ is a Scott open and is hence inaccessible by directed
  joins,
  we know that there is some compact $b \oftype D$
  with $\upset b \in F$ and $b \in \upset c$.
  The point $F$ is monotone (as it is a frame homomorphism),
  which means $F(\upset b) \ImplSym F(\upset c)$, so it suffices
  to show~$\upset b \le \upset c$.
  Since we already know $c \sqsubseteq b$,
  it follows that $\upset b \le \upset c$ holds since
  the principal filter operation $\upset (\blank)$ is antitone
  (\cref{lem:principal-filter-reverses-joins-and-antitone}).
\end{proof}

\begin{theorem}[%
  \AgdaLink{Locales.LawsonLocale.SharpElementsCoincideWithSpectralPoints.html\#20767}%
]\label{thm:sharp-equiv-spectral-points}
  For every Scott domain\/ $D$,
  we have a type equivalence
  \[\Equiv{\SpecPt{\ScottLocale{D}}}{\Sharp{D}}\text{.}\]
\end{theorem}
\begin{proof}
  The equivalence is given by the bijection from the construction of
  \cref{lem:equiv-of-domain-elements-and-scott-points}.
  It remains to show:
  \begin{enumerate}
    \item that $\PointOfSym_{x}$ is spectral point for every
      sharp element $x \oftype D$, and
    \item that $\nu(F)$ is a sharp element, for every spectral point
      $F$ of $\ScottLocale{D}$.
  \end{enumerate}

  For (1), let $x \oftype D$ be a sharp element.
  We need to show that $\PointOfSym_x$ is a spectral point.
  By \cref{lem:spectral-points-are-decidable-for-finite-elements},
  it suffices to show that the frame homomorphism $\PointOfSym_{x}^*$ is a
  subset admitting decidable membership for compact opens.
  Let $K$ be a compact \VScottOpen{} of $D$.
  We know that $\PointOfSym_{x}(K) \DefnEqSym x \in K$ is decidable by our
  topological characterization of sharp elements
  (\cref{lem:sharp-element-topological-characterization}), stating that
  sharp elements are exactly the elements admitting decidable membership in
  compact \VScottOpen{}s.

  For (2), let $F$ be a spectral point of $\ScottLocale{D}$.
  To see that $\nu(F)$ is a sharp element, let~$c \oftype \CompactElements{D}$.
  We know by \cref{lem:below-sharp} that $c \sqsubseteq \nu(F)$ if and only if
  $\upset c \in F$,
  which we know to be a decidable proposition by
  \cref{lem:spectral-points-are-decidable-for-finite-elements}.
\end{proof}

\section{Sharp elements and the points of patch}
\label{sec:points-of-patch}

In this section, we extend the equivalence from
\cref{thm:sharp-equiv-spectral-points}
as to prove that the set~$\Sharp{D}$ of sharp elements of a Scott domain $D$
is in bijection with the set
\[\ArrTy{\LocTermSym}{\Patch{\ScottLocale{D}}}\]
of points of its \indexedp{patch locale}.

\subsection{Spectral points coincide with the points of patch}

\begin{lemma}\label{lem:sharp-elements-and-spectral-points-of-patch-coincide}
  For every spectral locale\/ $X$, we have a type equivalence
  \begin{equation*}
    \Equiv{\SpecPt{X}}{\SpecPt{\Patch{X}}}\text{.}
  \end{equation*}
\end{lemma}
\begin{proof}
  Let $X$ be a spectral locale.
  For every spectral point $p \oftype \ArrTy{\LocTermSym}{X}$,
  we know by the universal property of the patch locale (\cref{thm:main})
  that there exists a point
  $\bar{p} \oftype \ArrTy{\LocTermSym}{\PatchSym(X)}$
  of patch,
  unique with the property that $\IdTy{p}{\PatchCounit \CompSym \bar{p}}$.
  Conversely, for every point $p \oftype \ArrTy{\LocTermSym}{\PatchSym(X)}$
  of the patch locale, the map $\Comp{\PatchCounit}{p}$ is a spectral point
  of~$X$,
  since we know that $\PatchCounit$ is spectral by \cref{lem:eps-perfect}.
  We thus have maps
  \begin{center}
  \begin{tikzpicture}
    \node (A) at (0,0) {$\SpecPt{X}$};
    \node (B) at (4,0) {$\SpecPt{\Patch{X}}$};

    \draw[->, line width=0.6pt]
    ([yshift=0.1cm]A.east) to node[midway, above] {\(\alpha\)} ([yshift=0.1cm]B.west);
    \draw[->, line width=0.6pt]
    ([yshift=-0.1cm]B.west) to node[midway, below] {\(\beta\)} ([yshift=-0.1cm]A.east);
  \end{tikzpicture}
  \end{center}
  defined as
  $\alpha \is p \mapsto \bar{p}$
  and
  $\beta \is p \mapsto \Comp{\PatchCounit}{p}$.

  The universal property of $\Patch{X}$ directly implies
  that these maps form a bijection.
  For every spectral point~$p \oftype \ArrTy{\LocTermSym}{X}$,
  the existence part of the universal property gives exactly that
  $p \IdTySym \Comp{\PatchCounit}{\bar{p}} \DefnEqSym \beta(\alpha(p))$.
  Conversely,
  let $p \oftype \ArrTy{\LocTermSym}{\Patch{X}}$ be a spectral point
  of the patch of $X$.
  We need to show $p \IdTySym \overline{\Comp{\PatchCounit}{p}}$,
  which follows immediately from the uniqueness of
  $\overline{\Comp{\PatchCounit}{p}}$.
\end{proof}

Recall \cref{lem:continuous-maps-are-spectral-maps} stating that every
continuous map of Stone locales is spectral.
When combined with the above,
this gives the corollary below,
since the terminal locale~$\LocTermSym$ is Stone
by \cref{example:terminal-locale-is-stone}.

\begin{corollary}\label{cor:spectral-points-and-patch-points-coincide}
  For every spectral locale\/ $X$, we have a type equivalence
  \begin{equation*}
    \Equiv{\SpecPt{X}}{\Point{\Patch{X}}}\text{.}
  \end{equation*}
\end{corollary}

\subsection{Summary of characterizations of sharp elements}

We finally combine all of the above equivalences to derive the main result of
interest: the sharp elements of a Scott domain $D$ are in bijection with the
points of $\Patch{\ScottLocale{D}}$.

\begin{theorem}[%
  \AgdaLink{Locales.LawsonLocale.PointsOfPatch.html\#11351}%
]\label{thm:sharp-equiv-main}
  For every Scott domain\/ $D$, we have equivalences
  \begin{equation*}
    \Sharp{D}
    \EquivSym
    \SpecPt{\ScottLocale{D}}
    \EquivSym
    \SpecPt{\Patch{\ScottLocale{D}}}
    \EquivSym
    \Point{\Patch{\ScottLocale{D}}}
    \text{.}
  \end{equation*}
\end{theorem}
\begin{proof}
  We established the first equivalence in
  \cref{thm:sharp-equiv-spectral-points},
  and the other two in
  \cref{lem:sharp-elements-and-spectral-points-of-patch-coincide}
  and \cref{cor:spectral-points-and-patch-points-coincide} above.
\end{proof}


\chapter{Conclusion}
\label{chap:conclusion}

We conclude by providing a brief summary of our contributions as well as a
discussion of avenues for further research.
Furthermore, we discuss related work in \cref{sec:related-work}, which, for us,
mostly consists of the subject known as \indexed{formal topology}.
Accordingly, we provide a brief introduction to formal topology and compare it
with the approach that we have presented in this thesis.

\section{Summary}

We have presented a new approach to the constructive and predicative study of
point-free topology in \VUF{}.
We argued that,
if one works under suitable size assumptions,
a considerable amount of locale theory can be carried out without resorting to
the use of impredicative principles.
To support this case, we have developed several nontrivial constructions of
locale theory.
To be more specific, we have presented the following developments:
\begin{enumerate}
  \item We developed the basic properties of salient classes of locales,
    including spectral, Stone, zero-dimensional, and regular locales.
    We have focused particularly on the first two,
    and demonstrated that their theory is well behaved in
    the constructive and predicative setting of \VUF{}.
  \item We showed that various constructions of locale theory,
    which are traditionally constructed by impredicative means,
    can be carried out in a predicative setting under the assumption
    of a small base.
    The primary example of this has been our construction of the right adjoint
    of a frame homomorphism,
    which has served a crucial r\^{o}le in our development.
    The two main applications of this have been Heyting implications
    and the notion of \indexedp{perfect map}.
    The former allowed
    us to work with the notion of \indexedp{open nucleus}
    whereas we used the latter in several proofs.
  \item We gave a refined, predicative version of the Stone duality for spectral
    locales, in terms of the category of small distributive lattices and the category
    of large, locally small, and small-complete spectral locales.
  \item We exhibited the category $\Stone$ of Stone locales as a coreflective
    subcategory of the category $\Spec$ of spectral locales and spectral maps,
    by constructing the patch locale of a spectral locale
    and proving its universal property.
    We explained how the proof of the universal property
    becomes more complicated in a predicative setting,
    where one does not have access to the frame of all nuclei
    --- a construction that seems to be inherently impredicative.
  \item Finally, we investigated the point-free topology of Scott domains,
    and used our predicative framework to establish a new result in
    constructive domain theory:
    the set $\Sharp{D}$ of sharp elements of a Scott domain $D$
    is in bijection with the set of points of $\Patch{\ScottLocale{D}}$
    i.e.\ the patch locale of the Scott locale of $D$.
\end{enumerate}

In the work that we have presented, we worked only with locales and introduced
the assumption of a base only when it became necessary.
This is in contrast with the standard approach to the predicative treatment
of point-free topology in type theory,
where one restricts attention to small presentations of locales in terms of
their basic opens, called \emph{formal topologies}.
In \cref{sec:comparison-with-formal-topology}, we discuss and compare this with
our approach.

\section{Related work}
\label{sec:related-work}

In this section, we provide a brief survey of related work on the constructive
development of point-free topology.
For us, the most relevant line of research in this direction is the subject
known
as \indexed{formal topology}~\cite{sambin-1987,negri-2002,cssv-inductively-generated}.

\subsection{Formal topology}
\label{sec:comparison-with-formal-topology}

As explained by Negri~\cite{negri-2002}: ``\indexedp{formal topology} was
introduced in a series of lectures by \VPMLFullName{} in 1985 as an answer to the question of
whether it is possible to generalize the concept of Scott domain so as to
comprise all thinkable topological spaces''.
This idea was then developed further through the collaboration of Giovanni
Sambin with \VPMLLastName{}; see the introduction of \cite{sambin-1987} for a
detailed history.

Sambin explicitly remarks~\cite[1]{sambin-domains} that the development of
topology in type theory was one of the original motivations for the subject:
\begin{quote}
  \emph{%
    What is formal topology? A good approximation to the correct answer is:
    formal topology is topology as developed in (Martin-L\"{o}f's) type
    theory.%
  }
\end{quote}
For the sake of self-containment, we provide a brief introduction to the key
ideas of formal topology.

\subsubsection{Brief summary of formal topology}

It is an idea familiar from Stone duality that certain classes of spaces are
captured completely by their bases consisting of certain classes of opens
e.g.\ compact opens or clopens.
For spectral spaces, for example, this allows one to obtain a description of the
space in consideration in terms of the finitary algebra of compact opens, which
is always a small collection in practice.
This technique is highly convenient for a predicative development; when one
works with the small algebra of basic opens generating the spectral space in
consideration, the theory becomes much more amenable to predicative reasoning.
For example, although describing the product of two locales predicatively is an
open problem, the product readily admits an easy description for spectral
locales as one can simply construct the coproduct of the distributive lattices
of compact opens of the spectral locales in consideration.

Formal topology arises as a generalization of this technique: one axiomatizes a
notion of a small algebra of basic opens, using which one presents a locale.
Although there are various definitions in the literature, this is usually a
meet-semilattice $B$ equipped with a cover relation
\begin{equation*}
  (\blank) \lhd (\blank) \subseteq B \times \PowSym(B)\text{,}
\end{equation*}
subject to certain axioms
(see \cite[Definition~2.1]{negri-2002} for a complete definition).
Given a basic open $b \in B$, the subsets $S \subseteq B$ such that $b \lhd S$
are intended to be the collections of basic opens with $b \le \bigvee S$ in the
presented frame.
This way, the cover relation governs the manner in which the basic opens are to
be combined as to generate a topology.

More formally, the frame \emph{presented by} a formal topology is the so-called
frame of \indexedCE{saturated subsets}{saturated subset}~\cite[23]{negri-2002}
of the formal topology in consideration. One takes the class $\downset B$ of
\VDownwardClosed{} subsets of $B$, and considers the subclass of sets that are
fixed points with respect to the
nu\-cle\-us~$\mathcal{C} \oftype \downset B \to \downset B$,\index{nucleus}
defined by
\begin{equation*}
  \mathcal{C}(U) \is \setof{ b \in B \mid b \lhd U }\text{.}
\end{equation*}
From a predicative point of view, the topology here is a large one since the
subsets of~$B$ in consideration form a large collection.
Since this large collection is generated by a \emph{small collection of basic
opens}, however, this allows one to circumvent a great deal of size problems.
In summary, formal topology is the branch of point-free topology where one works
with small presentations of frames, given by a generalized notion of an algebra
of basic opens.

\subsubsection{Inductively generated formal topologies}

Although formal topologies enable a predicative development of point-free
topology to a considerable degree of sophistication,
they do not solve all size problems.
Historically, one of the most salient open problems in formal topology has been
the predicative construction of the product of two formal topologies.
To address this issue (among other motivations),
\citeauthor{cssv-inductively-generated}~\cite{cssv-inductively-generated}
have developed the notion of an
\indexed{inductively generated formal topology}.
In the literature, the term ``formal topology'' is used to refer to both
inductively generated formal topologies as well as formal topologies that are
not necessarily inductively generated.

The key idea of inductively generated formal topologies is to require the cover
relation
\((\blank) \lhd (\blank)\)
to be an inductive type family.
Although there are various technical mechanisms to achieve this, this
stipulation is usually made using a variant of
\indexedCE{indexed \(\WTySym\)-types}{indexed \(\WTySym\)-type}
(which we briefly mentioned in \cref{sec:w-types}).

When working with inductively generated formal topologies, for example, it
becomes easy to describe the product of two formal topologies. However, this
comes at the cost of restricting the class of spaces that can be presented by
one's notion of formal topology.
In~\cite[\textsection 4.7]{cssv-inductively-generated}, the authors give an
example of a formal topology that is \emph{not} inductively generated.

At the time of writing, the problem of defining the product of two arbitrary
formal topologies remains open according to Coquand (personal communication).
This construction might be possible to carry out in our foundational setting
through the use of suitably powerful \indexedpCE{higher inductive types}{higher
inductive type},
and should be investigated further.

\subsubsection{Comparison of our approach with formal topology}

In this thesis, we took an alternative approach to the predicative development
of point-free topology that is slightly different from the strand of research on
formal topology.
Instead of delineating a category of formal topologies to embody a predicative
manifestation of the notion locale, we worked directly with locales under
certain size requirements ensuring that it is predicatively well behaved.
To be more specific, we worked with large, locally small, and small-complete
frames under the assumption of a small base (subject of \cref{sec:bases}),
drawing inspiration from the work of
\VTDJMHE{}~\cite{tdj-scott-model,tdj-mhe-cad,tdj-thesis}.

Instead of requiring all locales to be equipped with a small base, however, we
assumed the existence of a suitable small base \emph{only when} necessary.
In other words, we introduced small bases into our theory to the extent that we
needed them, rather than replace our notion of space with some presentation
given by an algebra of basic opens.
It is easy to see that every frame presented by a small formal topology will
meet these size requirements
(see \cite[\textsection 4.4]{tosun-msc}).

This approach has several benefits and drawbacks compared to inductively
generated formal topologies:
\begin{itemize}
  \item Our theory is not as predicatively well behaved as that of inductively
    generated formal topologies. Most saliently, our notion of locale with a small
    base suffers from the same problem that it is unclear if the product of two
    locales can be predicatively defined.
  \item As we work with the general notion of a locale, we can investigate the
    general constructions that \emph{do work well} without any restrictions.
  \item In our development of the theory of spectral locales, we gave a
    definition of spectrality (\cref{defn:spectral-locale}),
    using the \VTheUniAx{},
    where a notion of small base arises naturally.
    We thus obtained a definition that is equivalent to working with a
    formal topology
    (it is easy to see that the cover from \cref{defn:finite-covering}
    is an inductively generated cover in the sense of formal topology),
    but this was well motivated starting from the impredicative definition.
  \item Finally, we also chose to start a predicative investigation of
    point-free topology from first principles due to the novelties provided by
    \VUTT{}.
    It is conceivable that some of the problems motivating formal topology
    can be circumvented in other ways in \VUF{}, such as through the
    use of suitable higher inductive types giving certain resizing principles.
    The work of Coquand and the author~\cite{tosun-msc,coq-tosun-book-chapter}
    showed that, in \VUF{}, one needs certain higher inductive types to work
    with inductively generated formal topologies in the first place.
    So the ramifications of higher inductive types for the type-theoretic,
    predicative development of locale theory merit further study, and prematurely
    restricting one's attention to formal topologies might narrow the scope of such
    an investigation.
\end{itemize}

\subsubsection{Review of relevant work in formal topology}

\paragraph{Type-theoretic formalization of Cederquist}

One of the earliest formalizations of formal topology was developed by
Cederquist in his PhD thesis~\cite{cederquist-thesis}.
Cederquist's work is focused particularly on the applications of point-free
topology to constructive
analysis~\cite{cederquist-heine-borel,ccn-hahn-banach}.
In his licentiate thesis~\cite{cederquist-licentiate}
(which is part of the PhD thesis~\cite{cederquist-thesis}),
Cederquist studies the points of the Scott topology of a Scott domain in a
formal-topological framework, as part of his formalized library of formal topology.
This has parallels with our investigation from
\cref{sec:points-of-the-scott-locale}, and the exact relationship to the
formal-topological framework should be studied further.
Cederquist's formalization is developed in the \textsc{ALF} proof assistant~\cite{alf}.

\paragraph{Formal topology in \VUFAbbr{}}

In the context of \VUF{}, the first formalization of formal topology
was undertaken by the author's MSc work~\cite{tosun-msc,coq-tosun-book-chapter}.
A completely predicative account of inductively generated formal topologies was
developed,
and was fully formalized by the author~\cite{github-formal-topology-in-uf}.

\paragraph{Coquand and Zhang's description of patch}

\citeauthor{coq-zhang}~\cite{coq-zhang} give a predicative description of the
patch topology of a stably compact locale in terms of the entailment relation
presenting the locale.
Their approach to presentations is closely related to formal topology, although
they do not explicitly work with formal topologies.

In our work, we exhibit the patch construction only as the coreflector of Stone
locales into the category of spectral locales (and spectral maps). We leave it
as further work to extend this to the coreflection of compact regular locales
into the category of stably compact locales.
\citeauthor{coq-zhang}, on the other hand, describe the patch locale for the
more general case of stably compact locales, and use this to derive the patch of
a spectral locale as a special case.

Their work is an alternative to \MHELastName{}'s predicative description of the
patch locale that we used in \cref{sec:patch-locale}.
Adapting this to our setting would be a promising direction of future research.
We believe that it will be possible to use the approach presented in this thesis
instead of the formal-topological approach of~\cite{coq-zhang}.

\subsection{Other related works}

\subsubsection{Constructive but impredicative developments of locale theory}

Johnstone's development of point-free topology in \cite{ptj-ss} is classical.
Although he carefully marks his uses of the \indexedp{\VAC{}} with an asterisk,
he freely uses the \indexedp{law of excluded middle} in his development.
Unlike \cite{ptj-ss}, however, Johnstone develops some locale theory in
\cite[\S{}C]{elephant-vol-1} in a completely constructive way.
Other noteworthy developments of locale theory that are constructive in the
sense of the \indexedp{internal language} of a topos are
Manuell's PhD thesis~\cite{manuell-thesis} and his lecture
notes~\cite{manuell-pt-and-cm-2023}, and the PhD thesis of
Wrigley~\cite[\S{}A.II]{wrigley-thesis}.

\subsubsection{Other constructive and predicative developments}

In \cref{sec:comparison-with-formal-topology}, we explained that formal topology
is the standard approach to the development of constructive and predicative
locale theory in the context of type theory.
In the context of constructive set theory, however, there is a line of work on
constructive and predicative%
\footnote{%
  We remind the reader once again that set theorists use the term ``constructive''
  to refer to what we call ``constructive \emph{and} predicative'' in this thesis.%
}
topology, originating in the work of Aczel~\cite{aczel-2006};
we do not attempt to provide a survey here.
In this context, one works with a notion of \indexed{set-generated}
locale~\cite[Definition~11]{aczel-2006}, which is very similar to what we called
large and small-complete locales.
The precise relationship of set-generated locales to the predicative notions of
locales that we worked with deserves further investigation.

\section{Future work}

\subsection{Extension to stably compact locales}

In our work, we exhibited the patch construction only as the coreflector of the
category of Stone locales into the category of spectral locales (and spectral
maps).
As shown by \MHELastName{}~\cite{mhe-patch-short,mhe-patch-full},
the same patch construction also exhibits the category $\mathbf{CReg}$
of \indexedpCE{compact regular locales}{compact regular locale} (and continuous maps)
into the category $\mathbf{SC}$
of \indexedpCE{stably compact locales}{stably compact locale} (and perfect maps).
A promising direction for further research is
to develop this coreflection
in the framework that we have described in this thesis.
In fact, we already gave a candidate for the definition of
the notion of \indexedp{regular locale} in \cref{defn:regular-locale}.

We believe that this extension of the coreflection will be predicatively
possible, although it might present further difficulties compared to the case
for spectral locales.
For a stably compact locale $X$,
the basic covering families for an open $U$ consist of opens that
are not globally compact, but only compact
\emph{relative to} $U$. This means that there is no single type,
such as the type of all compact opens,
giving the base.
So the situation in this direction will be similar to \TDJLastName{}'s investigation
of continuous lattices~\cite[\textsection 4.4]{tdj-thesis}

\subsection{The patch locale and the Lawson topology}
\label{sec:lawson-topology}

Our investigation of the topology of domains in \cref{chap:scott}
focused mostly on the
\indexedp{Scott topology} of a Scott domain.
In the literature on domain theory, there is another crucial topology
considered on domains:
the \indexed{Lawson topology}~\cite[Definition~III-1.1]{clad}.

In a classical setting, the relationship between the Lawson topology and the
patch of the Scott topology is straightforward: the space
$\LawsonLocale{D}$ is homeomorphic to $\Patch{\ScottLocale{D}}$.
In a constructive setting, however, the situation here is more complicated.
Constructively, it cannot be the case in general that
\begin{equation}
  \LawsonLocale{D} \cong \Patch{\ScottLocale{D}} \tag{\textreferencemark}
  \label{eqn:classically-patch-gives-lawson}
\end{equation}
This is because $\LawsonLocale{D}$ is defined as a topological space,
given by the set of points~$D$
equipped with a topology given as a certain class of subsets of $D$.
The above homeomorphism would thus imply that
$\Patch{\ScottLocale{D}}$ is always a \indexed{spatial locale},
since it asserts that it is homeomorphic to a point-set topological space.

Consider the powerset \VDCPO{} $\PowSym(\NatTy)$, which is a Scott domain.
By taking the patch of its Scott locale, we get the Cantor locale
$\mathbf{2}^{\mathbb{N}}$:
\begin{equation*}
  \mathbf{2}^{\mathbb{N}} \IdTySym \Patch{\ScottLocale{\PowSym(\NatTy)}}\text{.}
\end{equation*}
So the homeomorphism in (\ref{eqn:classically-patch-gives-lawson})
immediately gives the spatiality of the Cantor space
as a special case.
Unfortunately, however, there is a well known result in constructive mathematics
due to
\textcite{mpf-rjg-formal-spaces}
stating that the \emph{spatiality of the Cantor space}\index{spatial}\index{Cantor space} is equivalent
to \indexedp{the Fan Theorem}, which is known to be independent of constructive mathematics.
We thus know that (\ref{eqn:classically-patch-gives-lawson}) cannot hold in
general.

We conjecture that an analogue to the above homeomorphism
can obtained constructively (and predicatively)
if the Lawson topology is defined on the set $\Sharp{D}$ of sharp elements of
$D$, instead of all its elements.
The result we proved in \cref{thm:sharp-equiv-main}
is the first step in this direction,
since it shows that the points of $\Patch{\ScottLocale{D}}$
are exactly the sharp elements.
We believe it will be possible to show that this equivalence of points
is a homeomorphism with respect to a suitable definition
of the Lawson topology on the set \(\Sharp{D}\).
However, the construction of the Lawson topology is rather involved and we have
not formulated a suitable definition at the time of writing.
We thus leave this as further work.


\backmatter%
\printbibliography[heading=bibintoc]%
\printnomenclature%
\printindex

\end{document}